# COSMIC-LAB:
# CHEMICAL AND KINEMATICAL PROPERTIES
# OF BLUE STRAGGLER STARS
# IN GALACTIC GLOBULAR CLUSTERS

Dottorando
**Loredana Lovisi**

Relatori
**Chiar.mo Prof. F. R. Ferraro**
**Dott. A. Mucciarelli**
**Dott.ssa B. Lanzoni**

Coordinatore
**Chiar.mo Prof. L. Moscardini**

*Volere è Potere*

# Abstract

Blue straggler stars (BSSs) are brighter and bluer (hotter) than the main-sequence (MS) turnoff and they are known to be more massive than MS stars. Two main scenarios for their formation have been proposed: collision-induced stellar mergers (COL-BSSs), or mass-transfer in binary systems (MT-BSSs). Depleted surface abundances of C and O are expected for MT-BSSs, whereas no chemical anomalies are predicted for COL-BSSs. Both MT- and COL-BSSs should rotate fast, but braking mechanisms may intervene with efficiencies and time-scales not well known yet, thus preventing a clear prediction of the expected rotational velocities.

Within this context, an extensive survey is ongoing by using the multi-object spectrograph FLAMES@VLT, with the aim to obtain abundance patterns and rotational velocities for representative samples of BSSs in several Galactic GCs. A sub-population of CO-depleted BSSs has been identified in 47 Tuc, with only one fast rotating star detected (Ferraro et al. 2006).

For this PhD Thesis work I analyzed FLAMES spectra of more than 130 BSSs in four GCs: M4, NGC 6397, M30 and $\omega$ Centauri. This is the largest sample of BSSs spectroscopically investigated so far. Hints of CO depletion have been observed in only 4-5 cases (in M30 and $\omega$ Centauri), suggesting either that the majority of BSSs have a collisional origin, or that the CO-depletion is a transient phenomenon. Unfortunately, no conclusions in terms of formation mechanism could be drawn in a large number of cases, because of the effects of radiative levitation. Remarkably, however, this is the first time that evidence of radiative levitation is found in BSSs hotter than 8200 K. Finally, we also discovered the largest fractions of fast rotating BSSs ever observed in any GCs: 40% in M4 and 30% in $\omega$ Centauri.

While not solving the problem of BSS formation, these results provide invaluable information about the BSS physical properties, which is crucial to build realistic models of their evolution.

# Contents

# Chapter 1

# Introduction

This Ph.D. Thesis has been carried out in the framework of the Cosmic-Lab project. This 5-year project (PI: F. R. Ferraro) has been funded by the European Research Council and has the aim to undestand the complex interplay between dynamics and stellar evolution by using Galactic globular clusters (GCs) as cosmic laboratories and blue straggler stars (BSSs), millisecond pulsars and intermediate-mass black holes as probe particles. In fact, the ultra-dense cores of GCs are the ideal place for generating such exotic objects and most of them are thought to be the by-products of the evolution of binary systems, originated and/or hardened by stellar interactions. In a more general framework, GCs are crucial for a number of issues: firstly, they are the oldest stellar clusters in our Galaxy and they can be used as tracers of the structure and history of the Milky Way; secondly, they are the simplest stellar population where to test the predictions of theoretical models of stellar evolution; finally, their high density (especially in the central regions) favour stellar interactions between both single stars and (more likely) binary systems. All these reasons make GCs very important laboratories both from a theoretical and an observational point of view. In particular, studying the nature of exotica like BSSs can be a powerful diagnostic of the dynamical evolution of the host cluster and can bring crucial information on the progenitor stellar population and binary systems (seeBailyn 1995, and reference therein). This topic has received strong impulse in the recent years. This Ph.D. Thesis is focused, in particular, on the chemical and kinematical properties of the BSS population in different GCs.

## 1.1  Blue Straggler stars: "state of the art"

BSSs are commonly defined as stars brighter and bluer (hotter) than the main sequence (MS) turnoff (TO) located along an extension of the MS in the color-magnitude diagram (CMD, see Figure 1.1). Thus, they mimic a rejuvenated population and their existence has been a puzzle for many years. First discovered in 1953 in the GC M3 (Sandage 1953), they are currently observed in all GCs properly investigated. Indeed, the BSS sequence is a typical feature of the CMD of any GCs, even if the BSS existence cannot be explained only through the normal stellar evolution theory. According to their position in the CMD and also from direct measurements (Shara et al. 1997), they appear more massive than normal MS stars, thus indicating that a process able to increase the initial mass of a single star must be at work. Their formation mechanisms are not yet completely understood, but at the moment there are two main leading scenarios: BSSs could be generated by mass-transfer activity in a binary system (MT-BSSs; McCrea 1964; Zinn & Searle 1976), possibly up to the complete coalescence of the two companions, or they may form by collision-induced stellar mergers (COL-BSSs; Hills & Day 1976). Hence, BSSs certainly represent the link between standard stellar evolution and cluster dynamics. Moreover, they are able to give information about the dynamical history of the cluster, the role of the dynamics on stellar evolution, the amount of binary sistems and the role of binaries in the cluster evolution.

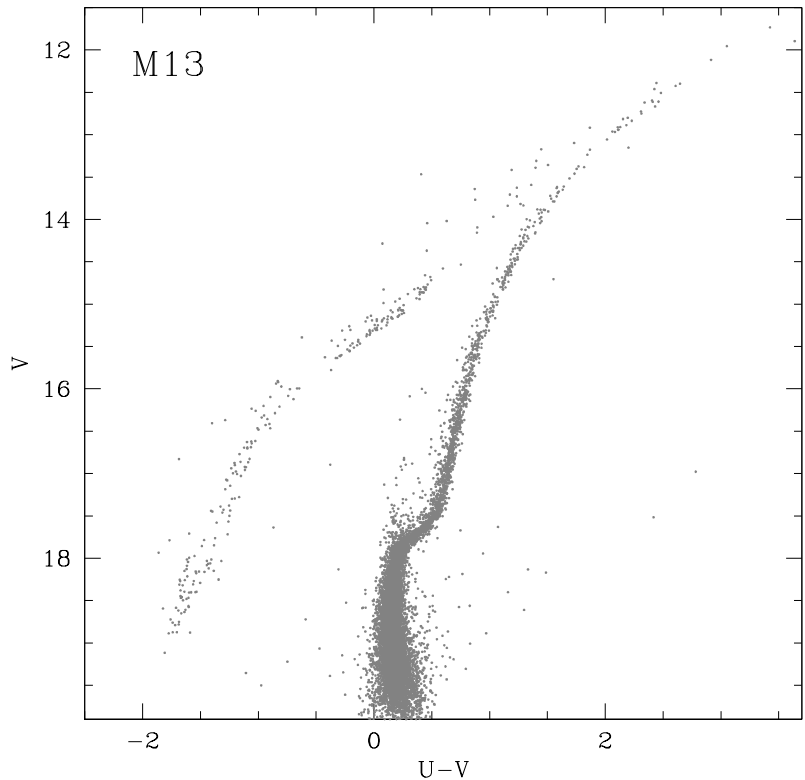

Figure 1.1: CMD of M13 in the optical plane: the BSS sequence lies in the brighter and bluer region above the TO.

## 1.2 The UV approach

The observational and interpretative scenario of BSSs has significantly changed in the last years. For almost 40 years since their discovery, BSSs have been detected only in the outer regions of GCs or in relatively loose clusters, thus generating the idea that low-density environments were their natural habitats. However, this was just an observational bias and, starting from the early '90s, high resolution studies allowed to properly image and discover BSSs also in the highly-crowded central regions of dense GCs. In particular, the advent of the Hubble Space Telescope (HST) represented a real improvement in BSS studies, thanks to its unprecedented spatial resolution and imaging/spectroscopic capabilities in the ultraviolet (UV; see Paresce et al. 1991; Ferraro & Paresce 1993; Guhathakurta et al. 1994). The study of the BSS population in the UV CMDs represents a true turning point: in fact, the optical CMD of old stellar populations is dominated by the cool stellar component, so that the observation and the construction of complete samples of hot stars (as BSSs, other by-products of binary system evolution, extreme blue horizontal branch stars, etc.) is difficult in this plane. Moreover, BSSs can be easily mimicked by photometric blends of sub-giant branch (SGB) and red giant branch (RGB) stars in the optical CMD. Thus, the systematic study of BSSs in the optical bands still remains problematic even if using HST, especially in the central regions of high density clusters. Instead, at UV wavelenghts RGB stars are faint, while BSSs are among the brightest objects. In particular, BSSs define a narrow, nearly vertical sequence spanning $\sim 3$ mag in the UV plane (see Figure 1.2), thus being much more easily recognizable. In the mean time, BSS-like blends are much less severe at these wavelengths because of the relative faintness of SGB and RGB stars. Indeed, the ($m_{255}$, $m_{255}$-U) plane is ideal for selecting BSSs even in the cores of the densest GCs, and its systematic use allowed to put the BSS study into a more quantitative basis than ever before.

## 1.3 BSS specific frequency and primordial binary fraction

In the meanwhile, also the spatial resolution and the quality of optical photometry improved and the first catalogs of BSSs have been published by Fusi Pecci et al. (1992) (see also Ferraro et al. 1995). The most recent collection of BSSs in the optical bands counts nearly 3000 candidates in 56 GCs (Piotto et al. 2004). These works have significantly contributed to form the nowadays commonly accepted idea that BSSs are a normal stellar population in GCs, since they are present in all the properly observed clusters. However, according to Fusi Pecci et al. (1992), BSSs in different environments could have different origins. In particular, BSSs in loose GCs might be produced by the coalescence of primordial binaries, while in high density GCs, BSSs might arise mostly from stellar interactions, particularly those which involve binaries. While the suggested mechanisms of BSS formation could be separately at work in clusters with different densities (Ferraro et al. 1995, 1999a), there is evidence that they could also act simultaneously within the same cluster (as in the case of M3; see Ferraro et al. 1993, 1997). Moreover, Piotto et al. (2004) studied the BSS population in 56 GCs finding that the total number of BSSs varies by only a factor 10 from cluster to cluster and it is largely independent of both the total cluster mass (at odds with what expected in the case of BSS formed through the MT channel only) and the stellar collision rate (at odds with what expected in case of only stellar collisions).

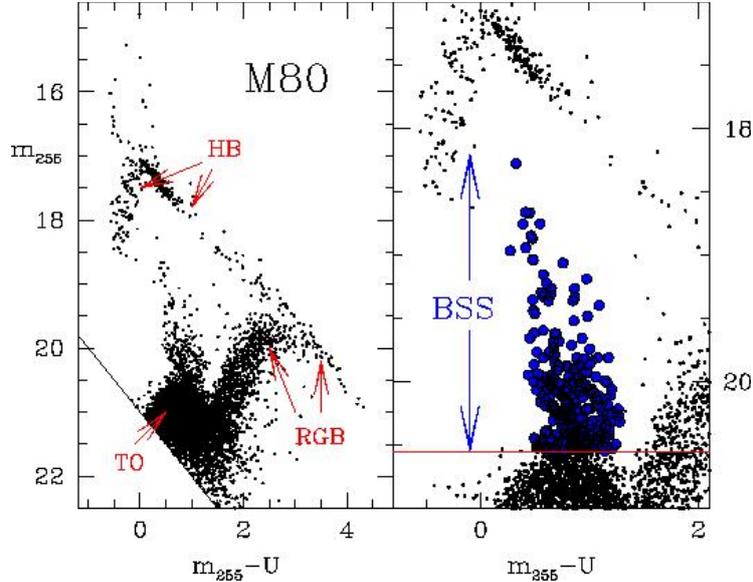

Figure 1.2: CMD of M80 in the UV plane, adapted from Ferraro et al. (1999a). On the left panel all the most prominent sequences are labelled. On the right panel a zoom on the BSS region, that appears almost vertical spanning ∼ 3 mag, is presented.

A number of other interesting results have been obtained from direct cluster-to-cluster comparisons. For this purpose, the BSS specific frequency can be an useful tool. This quantity has been defined as the number of BSSs counted in a given region of the cluster, normalized to the number of cluster stars (generally the horizontal branch stars, HBs) in the same region, adopted as reference. The BSS specific frequency has been found to largely vary from cluster to cluster. For the six GCs considered in Ferraro et al. (2003), the BSS frequency varies from 0.07 to 0.92, and does not seem to be correlated with central density, total mass, velocity dispersion or any other obvious cluster property. The most puzzling case is provided by the two "twin" clusters M3 and M13: they are very similar in terms of mass and central density, but they harbor a quite different BSS population. In particular, the specific frequency in M13 is the lowest ever measured in a GC (0.07) and it turns out to be 4 times lower than that measured in M3 (0.28). The paucity of BSSs in M13 suggests either that the primordial population of binaries in M13 was poor, or that most of them were destoyed. Alternatively, the mechanism producing BSSs in the central region of M3 is more efficient than that in M13, because the two systems are in different dynamical evolutionary phases.

On the other hand, the most surprising result has been found for NGC 288 and M80. For both these clusters, that have very different central density values (the highest for M80 and the lowest for NGC 288 among all Galactic GCs), the largest BSS specific frequency has been found. This suggests that the two formation channels can have comparable efficiency in producing BSSs in their respectively typical environment.

Given the role of binary systems in both the MT and the COL scenarios, one of the most important ingredient necessary to properly understand the BSS formation process certainly is the fraction of primordial binaries in each cluster. By analysing the color distribution of MS stars, Sollima et al. (2007) derived the fraction of binary systems in a sample of 13 low-density GCs. The estimated global fractions of binary systems range from 10 to 50 per cent depending on the cluster. Interestingly, this fraction has been found to nicely correlate with the BSS frequency (Sollima et al. 2008, see Figure 1.3). This is the cleanest evidence ever obtained that the unperturbed evolution of primordial binaries is the dominant BSS formation process in low density environments. Moreover, these results have been confirmed also in other GCs by Milone et al. (2012). According to Knigge et al. (2009) a clear correlation exists between the number of BSSs found in cluster cores and the core stellar mass,whereas no correlation has been found with the collisional parameter, suggesting that most of the BSSs in GCs of any density are generated by MT. Nevertheless, according to Leigh et al. (2012), the relation between the BSS number and the binary fraction in the core is weaker than expected. However, the progenitor binaries may themselves have strongly suffered from collisions.

## 1.4   The BSS radial distribution

Starting from 1993, systematic studies of the BSS radial distribution have been performed. The first pilot analysis was carried on M3 with the attempt of sampling the BSS population over the entire extension of the cluster. By combining UV HST observations of the central region (Ferraro et al. 1997) and wide field ground-based observations in the optical bands (Buonanno et al. 1994; Ferraro et al. 1993), the BSS radial distribution of M3 was studied all over its radial extent (r ∼ 6'). The resulting distribution (Ferraro et al., 1997) was completely unexpected: BSSs appeared to be more centrally concentrated than RGB stars in the central regions, and less concentrated in the cluster outskirts.

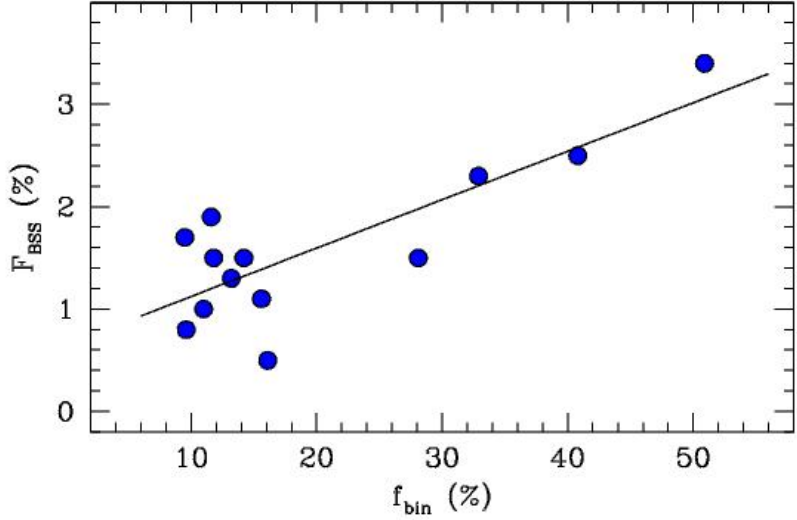

Figure 1.3: The BSS specific frequency well correlates with the binary fraction found by Sollima et al. (2007) in 13 loose GCs.

For further investigating such a surprising result, the surveyed area was divided in a number of concentric annuli and the number of BSS and RGB stars normalized to the sampled luminosity in each annulus, according to the following relations

$$R_{BSS} = \frac{\frac{N_{BSS}}{N_{BSS,tot}}}{\frac{L_S}{L_{S,tot}}} \tag{1.1}$$

and

$$R_{RGB} = \frac{\frac{N_{RGB}}{N_{RGB,tot}}}{\frac{L_S}{L_{S,tot}}} \tag{1.2}$$

respectively. The result is shown in Figure 1.4. The BSS radial distribution is bimodal: it reaches a maximum at the centre of the cluster, shows a clear-cut dip in the intermediate region (at 100" < r < 200"), and rises again in the outer region (out to r ∼ 10'). While the bimodality detected in M3

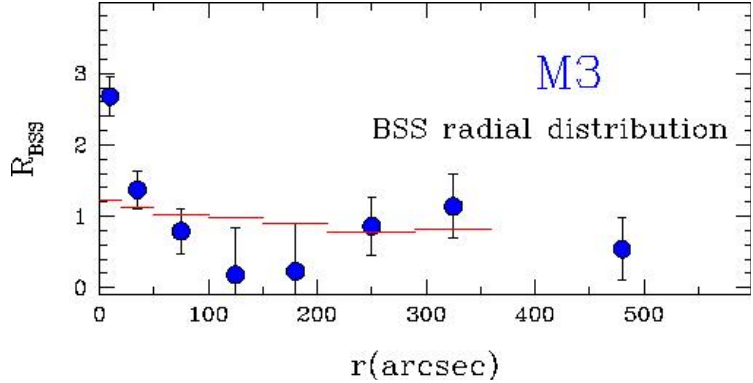

Figure 1.4: The BSS radial distribution of M3 derived by Ferraro et al. (1997). The horizontal segments, which stay constant around 1, show the relative frequency of RGB stars used as reference population.

was considered for years to be peculiar, subsequent results demonstrated that this is not the case. In fact, the same observational strategy has been applied to a number of other clusters and bimodal distributions have been detected in several cases (see Table 1.1 for references). Originally Ferraro

| Cluster ID | References |
|---|---|
| 47Tuc | Ferraro et al. (2004) |
| AM1 | Beccari et al. (2012) |
| Eridanus | Beccari et al. (2012) |
| M5 | Warren et al. (2006); Lanzoni et al. (2007a) |
| M53 | Beccari et al. (2008) |
| M55 | Zaggia et al. (1997); Lanzoni et al. (2007b) |
| M75 | Contreras Ramos et al. (2012) |
| NGC 1904 | Lanzoni et al. (2007c) |
| NGC 2419 | Dalessandro et al. (2008a) |
| NGC 6229 | Sanna et al. (2012) |
| NGC 6388 | Dalessandro et al. (2008b) |
| NGC 6752 | Sabbi et al. (2004) |
| $\omega$ Cen | Ferraro et al. (2006a) |
| Palomar 3 | Beccari et al. (2012) |
| Palomar 4 | Beccari et al. (2012) |
| Palomar 14 | Beccari et al. (2011) |

Table 1.1: Cluster ID and references for the studied BSS radial distributions.

et al. (1997) argued that the bimodal distribution of BSSs in M3 could be the signature of the two formation mechanisms acting simultaneously in the same cluster: the external BSSs would be generated from MT activity in primordial binaries, while the central BSSs would be generated by stellar collisions leading to mergers. Sigurdsson et al. (1994) suggested another explanation:

according to these authors, all the BSSs were formed in the core by direct collisions and then ejected to the outer regions by the recoil of the interactions. The BSSs kicked out to a few core radii would rapidly drift back to the centre of the cluster due to mass segregation, thus leading to the central BSS concentration and a paucity of BSSs in the intermediate regions. More energetic recoils would kick the BSSs to larger distances and, since these stars require much more time to drift back toward the core, they may account for the overabundace of BSSs in the cluster outskirts. Nevertheless, models of BSS dynamical evolution in a number of specific clusters (Mapelli et al., 2004, 2006; Lanzoni et al., 2007a,c) demonstrate that the observed BSS bimodal distribution cannot be explained within a pure collisional scenario in which all the BSSs are generated in the core through stellar interactions. In fact, all the BSSs kicked outward because of collision recoils, sink back to the core in only 1-2 Gyr. An accurate reproduction of the external upturn can be obtained only by requiring that a sizable ($\sim 20\%$ - $40\%$) fraction of BSSs is generated in the peripheral regions, where primordial binaries can evolve in isolation and experience mass transfer processes without suffering significant interactions with other cluster stars.

While the bimodal BSS radial distribution is a common feature of most of the analysed clusters, some exceptions exist: NGC 1904 and M75 do not present any external upturn, whereas $\omega$ Cen, NGC 2419 and Palomar 14 show a completely flat BSS radial distribution (see Figure 1.5). In particular, by using a proper combination of high-resolution HST data and wide-field ground-based observations sampling the entire radial extension of $\omega$ Centauri, Ferraro et al. (2006a) detected in this cluster the largest population of BSSs ever observed in any stellar system: more than 300 candidates have been identified. At odds, with all the GCs previously surveyed, the BSS population in $\omega$ Centauri has been found not to be centrally segregated with respect to the other cluster stars. This is the cleanest evidence ever found that $\omega$ Cen is not fully relaxed, even in the central regions, and it suggests that the observed BSSs are the progeny of primordial binaries, whose radial distribution is not yet significantly altered by the dynamical evolution of the cluster. A similar result has been found by Dalessandro et al. (2008a) in NGC 2419. As in the case of $\omega$ Cen, this evidence indicates that NGC 2419 is not yet relaxed even in the central regions. This observational fact is in agreement with the estimated half-mass relaxation time, which is of the order of the cluster age. Finally, by using the same technique, Beccari et al. (2011) studied the BSS distribution in the loose GC Palomar 14. The BSS radial distribution turns out to be indistinguishable from that of the other stars and demonstrates that Palomar 14 is dynamically young, still far from having established energy equipartition even in its innermost regions. This is also confirmed by the dynamical simulations of Mapelli et al. (2004, 2006) and Lanzoni et al. (2007a,c), showing that the radial position of the dip of the BSS radial distribution sensibly depends on the dynamical friction efficiency within the cluster, while flat BSS distributions indicate little or no dynamical evolution of the system. Moreover, recent results (see Ferraro et al. 2012) show that GCs can be grouped into a three distinct families on the basis of the radial distribution of BSSs. This grouping corresponds well to an effective ranking of the dynamical stage reached by stellar systems, thereby permitting a direct measure of the cluster dynamical age purely from observed properties.

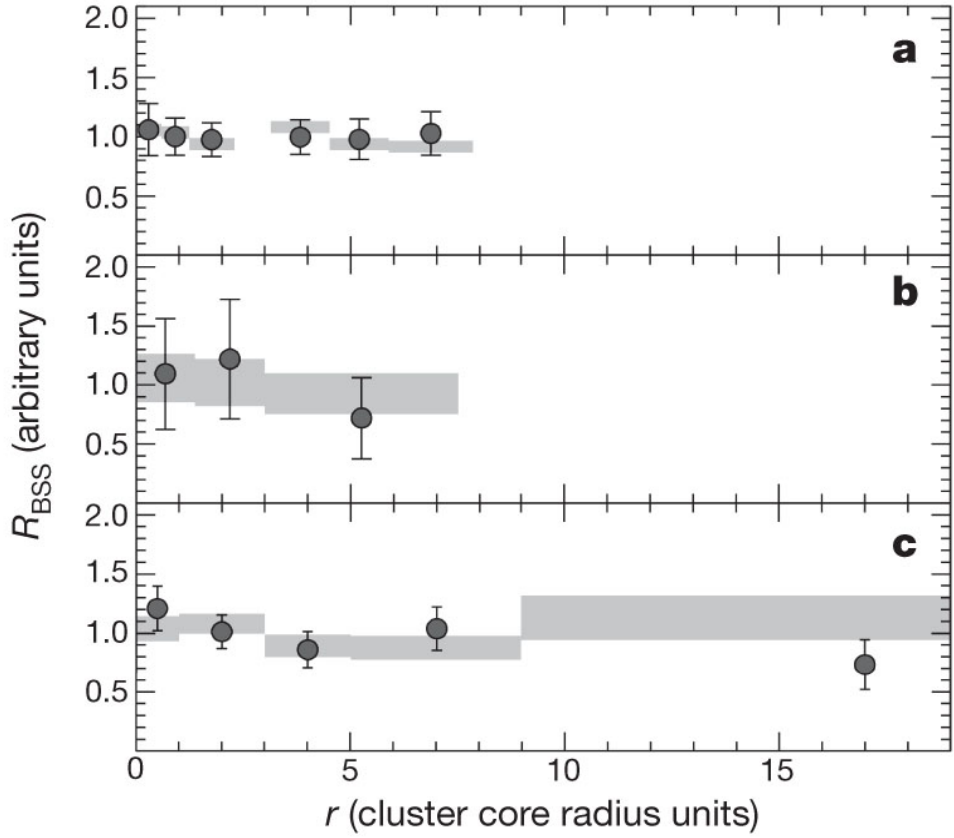

Figure 1.5: Flat BSS radial distribution of $\omega$ Centauri (top panel), Palomar 14 (central panel) and NGC 2419 (bottom panel) demonstrating that these GCs are dinamically unevolved (Ferraro et al. 2012, see also Ferraro et al. 2006a; Dalessandro et al. 2008a; Beccari et al. 2012).

## 1.5   The BSS double sequences

Thanks to the very high photometric accuracy of the available data, an exceptional discovery has been recently performed in M30: by using 44 high-resolution WFPC2 HST images of this cluster, Ferraro et al. (2009a) revealed the presence of two well distinct sequences of BSSs (see Figure 1.6). The two sequences are similarly populated (24 stars in the blue and 21 stars in the red sequence) and almost parallel. The detected BSSs are more concentrated towards the cluster centre than the parent cluster stars (as expected) and the red BSSs appear more concentrated than the blue ones. In particular, no red BSSs are observed at an angular distance r > 30" from the cluster centre. Moreover, the blue-BSS sequence is well fit by evolutionary models of COL-BSSs formed 1-2 Gyr ago (Sills et al. 2009), while the other one is too red to be properly reproduced by COL isochrones of any age. Instead, the red-BSS sequence well corresponds to the position of the low-luminosity boundary delineated by a population of binaries that are experiencing an active phase of MT (Tian et al. 2006). Because of normal stellar evolution, in a few Gyrs the COL products will populate the region between the two sequences and MT-BSSs will move toward redder and brighter regions. The fact that we currently see two well-separated sequences therefore indicates that the same, short-lived event (instead of a continuous formation process) recently generated the two BSS populations. On the other hand, the shape of the star density profile indicates that M30 already experienced the core collapse (CC) event, which is known to significantly increase the gravitational interaction rate among the stars. Based on this evidence, the following scenario has been suggested: 1-2 Gyr ago M30 underwent the core collapse; this has boosted both the rate of direct stellar collisions, and the hardening MT processes in binary systems; as a result, two different BSS sequences are now observed, the blue one formed by COL-BSSs and the red one by MT-BSSs. Finally the BSS variability has been tested, finding five candidate variables (labelled with triangles in Figure 1.6): on the basis of the light-curve characteristics, the three brightest variables have been classified as W Uma contact binaries, whereas the two faintest candidates have quite scattered light curves that prevent a reliable classification. In the Galactic field W Uma objects are bynary systems losing orbital momentum because of magnetic braking. These shrinking binary systems, initally detached, evolve to the semi-detached and contact stages (when mass-transfer starts) and finally merge into a single star (Vilhu 1982). In dense cluster environments stellar interactions can drive binaries toward merger: these systems could resonably be expected to display W Uma characteristics, although the evolutionary time scales could be very different. Among the variable stars found in M30, 3 out 5 are on the red sequence and among the W Uma 2 out 3 are red BSSs.

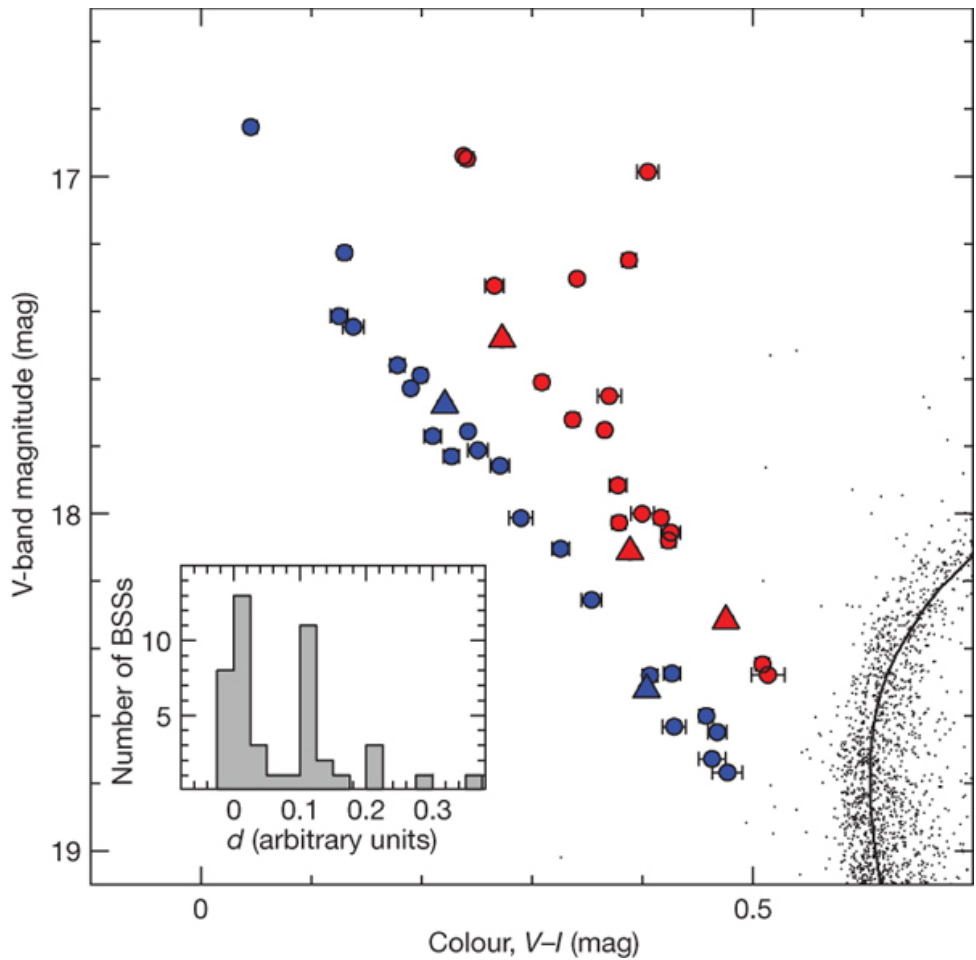

Figure 1.6: Double BSS sequence in M30, from Ferraro et al. (2009a): the red sequence is thought to be formed by MT-BSSs, whereas the blue one by COL-BSSs. The BSS formation could have been boosted by a short lived event (like core collapse) occurred 1-2 Gyrs ago. Triangles label variable stars.

An extensive survey of candidate CC GCs started in the last few months with the WFPC3@HST. The very inital analysis revealed the presence of a double BSS sequence in NGC 362 (Dalessandro et al. 2013, in preparation). Also in this case, the two sequences are similarly populated and the red BSSs are more centrally concentrated than the blue BSSs. Even if the dynamical status of NGC 362 is still uncertain, the observed density profile is compatible with that of a collapsed core (Vesperini & Trenti, 2010). The analysis of the variability allowed the identification of 2 SX Phe stars (pulsating variable stars with short periods) and the new discovery of 4 W Uma stars. 5 out 6 variables are on the red sequence and 3 of them are W Uma. Finally, other preliminary studies reveal hints for a double BSS sequence also in NGC 6397 (Contreras Ramos et al. 2013, in preparation) and M15 (Beccari et al. 2013, in preparation).

## 1.6 Evolved BSSs

BSSs are known to be more massive than normal MS stars, with masses up to twice the TO one. These objects are expected to evolve off the MS, as other normal stars, and thus "contaminating" the cluster evolutionary sequences in the CMD. The easiest sequence for searching evolved-BSSs (E-BSSs) should be the HB (Renzini & Fusi Pecci 1988), where stars are placed according to their mass. As a normal star, a BSS in this sequence should also be experiencing the core helium burning phase but in the reddest and brightest part of the HB, according to its mass. Following this track Ferraro et al. (1997) identified 19 candidates E-BSSs in the HB of M3 and observed that the radial distribution was the same as for the unevolved BSSs. Similar results have been obtained with the large BSS population of M80 (Ferraro et al., 1999a).

Even if the HB should be the best sequence where to identify the E-BSSs, other promising results have been obtained on the asymptotic giant branch (AGB). Firstly, Wylie et al. (2006) detected peculiar s- and r-process elements enhancement on the surface of seven AGB stars in the external region (r > 10') of 47 Tuc and they proposed that at least part of these AGB stars might be formed through mass transfer along the evolutionary path of a 1.4 + 0.5 $M_\odot$ binary system. In the CMD of the same cluster, Beccari et al. (2006) measured also a significant excess of stars in the AGB region (see Figure 1.7) and found that they are more centrally segregated than the RGB and HB populations. Within 1' from the cluster centre they observe $\sim$ 53 AGB stars, while only $\sim$ 38 of such objects are predicted on the basis of the HB star number counts and the post-MS evolutionary time-scales (Renzini & Fusi Pecci 1988): this makes an excess of $\sim$ 40%. Because of the typical low stellar mass along the AGB ($\sim$ 0.6 $M_\odot$), this excess and the higher central concentration are hardly understandable in terms of a mass segregation effect. Instead they could be explained by a population of more massive objects that, given the large sample of BSSs in 47 Tuc, most probably are the BSS descendants. Indeed, the comparison with theoretical tracks (Pietrinferni et al. 2006) and collisional models (Sills et al. 2009) shows that the AGB population of 47 Tuc can be significantly contaminated by BSS descendants ($\sim$ 1.2 - 1.5 $M_\odot$) that are currently experiencing the first ascending RGB.

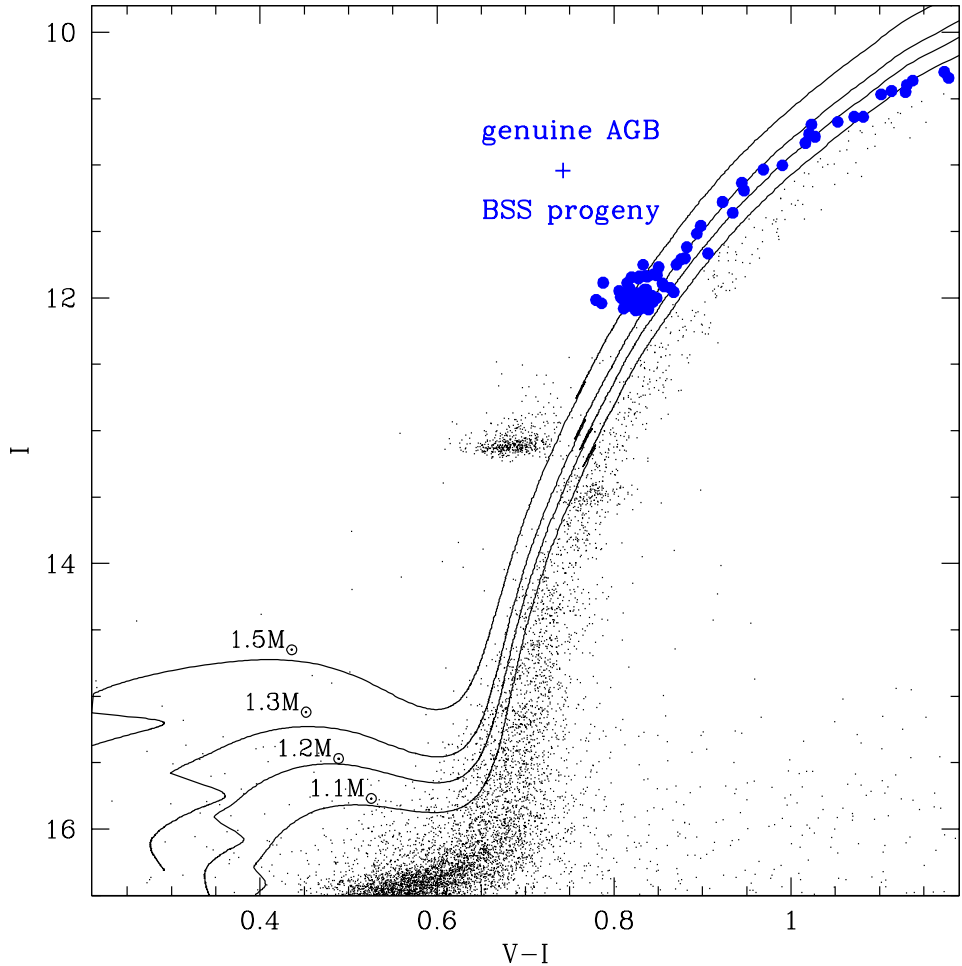

Figure 1.7: CMD of 47 Tuc in the region of the AGB, adapted from Beccari et al. (2006). The excess of the AGB population could be a signature of the presence of E-BSSs experiencing the RGB phase.

## 1.7  Signatures of the formation mechanisms: the case of 47 Tuc

BSSs are a clear example of how dynamics can affect the (otherwise) normal stellar evolution. In turn, studying these objects can bring fundamental information about the relaxation and dynamical processes occurring in dense stellar systems. Indeed, distinguishing COL-BSSs from MT-BSSs is crucial to use these stars as dynamical probes. However, in spite of the increasing number of observational results collected in the last decades, this task seemed to be far from being achieved till recently. In fact, the bimodal radial distribution of BSSs discovered in many GCs is suggestive of a tight link with the dynamical friction efficiency within the cluster, but it does not allow to clearly separate COL- from MT-BSSs. Also the connection with the binary fraction does not seem a simple task, mainly because the progenitor binaries may themselves have strongly suffered from COL.

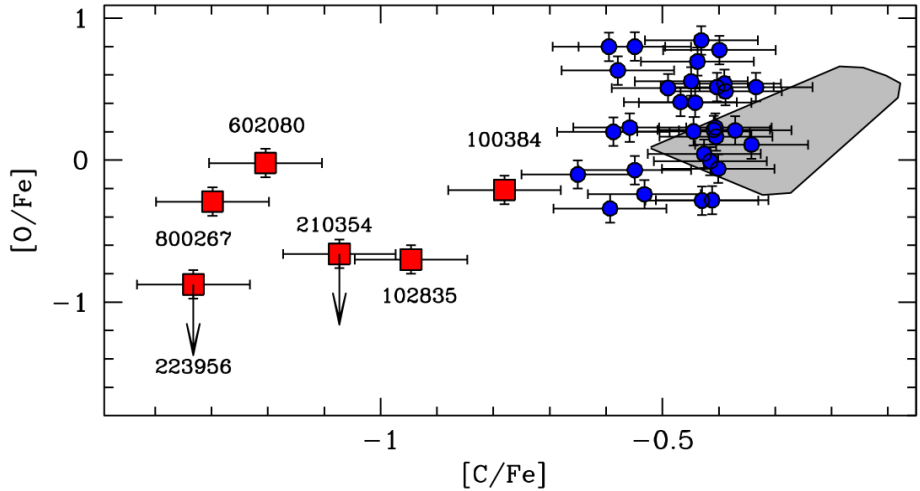

Figure 1.8: [O/Fe] versus [C/Fe] for BSSs in 47 Tuc from F06. The grey region marks the abundances for TO stars obtained by Carretta et al. (2005). Blue dots are BSSs with C and O abundances consistent with the TO, whereas red squares are the CO depleted-BSSs.

In this framework, the spectroscopic analysis seems to be the most promising way to discriminate between the two formation channels. From the spectroscopic analysis of the BSSs, both chemical abundances and rotational velocities can be derived. Concerning rotational velocities, the theoretical scenario is quite complex. MT-BSSs are expected to have high rotational velocities because of the angular momentum transfer (Sarna & De Greve, 1996). Unfortunately accurate simulations are lacking, mostly because of the difficulty in following the evolution of a hydrodynamic system (the mass transfer between binary components) for the lenght of time required for the system to merge. This time-scale is unknown but certainly large, of the order of half a billion years (Rahunen, 1981). According to some authors (Benz & Hills, 1987), also COL-BSSs should rotate fast. Nevertheless, braking mechanisms (like magnetic braking and disk locking) may intervene (both for MT- and COL-BSSs) with efficiencies and time-scales not well known yet (see Leonard & Livio 1995 and Sills et al. 2005). For these reasons any observation of the BSS rotational velocities cannot be univocally interpreted in terms of formation mechanisms. From a chemical point of view, simulations predict different chemical patterns for MT- and COL-BSSs: in fact, hydrodynamical simulations by Lombardi et al. (1995) have shown that a very little mixing is expected to occur between the inner cores and the outer envelopes of the colliding stars, whereas signatures of mixing with incomplete CN-burning products are expected at the surface of BSSs formed via the MT channel, since the gas at the BSS surface is expected to come from deep regions of the donor star, where the CNO burning was occurring (Sarna & De Greve 1996). Thus, it should be possible to distinguish MT-BSSs from COL-BSSs by looking at their C and O abundances and searching for some depletion in the MT-BSSs atmosphere.

In this context the advent of 8-10 meters class telescopes equipped with multiplexing capability spectrographs has given a new impulse to the study of the BSS properties. By using the multi-object spectrograph FLAMES at ESO Very Large Telescope (VLT), an extensive survey has been performed in order to obtain abundance patterns and rotational velocities for representative numbers of BSSs in a sample of GCs. The first result of this research has lead to an exciting discovery (Ferraro et al. 2006b, hereafter F06): by measuring the surface abundance patterns of 43 BSSs in 47 Tuc, a sub-population of BSSs with a significant depletion of carbon (C) and oxygen (O), with respect to the dominant population has been discovered (see Figure 1.8). This evidence have been interpreted as the presence of CNO burning products on the BSS surface, coming from the core of a deeply peeled parent star, as expected in the case of the MT formation channel. Thus, this could be the first detection of a chemical signature pointing to the MT formation process for BSSs in a GC.

Concerning rotational velocities, the observations in 47 Tuc have shown that most of the BSSs are slow rotators (v $\sin(i) < 10$ km s$^{-1}$), with velocities in agreement with those measured in unperturbed TO stars (Lucatello & Gratton 2003). The high-rotation tail of the distribution consists of 10 objects with v $\sin(i) > 10$ km s$^{-1}$, with only one having a really large rotational velocity (v $\sin(i) \sim 80$ km s$^{-1}$). No correlation has been found between CO depletion and rapid rotation, but the CO-depleted BSSs and the few BSSs with v $\sin(i) > 10$ km s$^{-1}$ appear to be "less evolved" than the others: they all lie within a narrow strip at the faint-end of the BSS luminosity distribution in the CMD. Moreover, 3 W Uma binary systems have been found in the BSS sample and two of them are also CO-depleted. The third one is the fastest rotating BSS in the sample and unfortunately no abundance was derived because of the high rotational velocity.

A possible interpretative scenario has been suggested by F06 and it is shown in Figure 1.9. In the early stage of mass transfer in W Uma systems (Stage-1), the transferred mass could come from the unprocessed material and the resulting star would have normal C-O abundances. As the transfer continues reaching into the region of CNO processing, first C and then both C and O would appear to be depleted (Stage-2). Thus it is possible to find depleted C, normal O BSSs/W Uma stars.

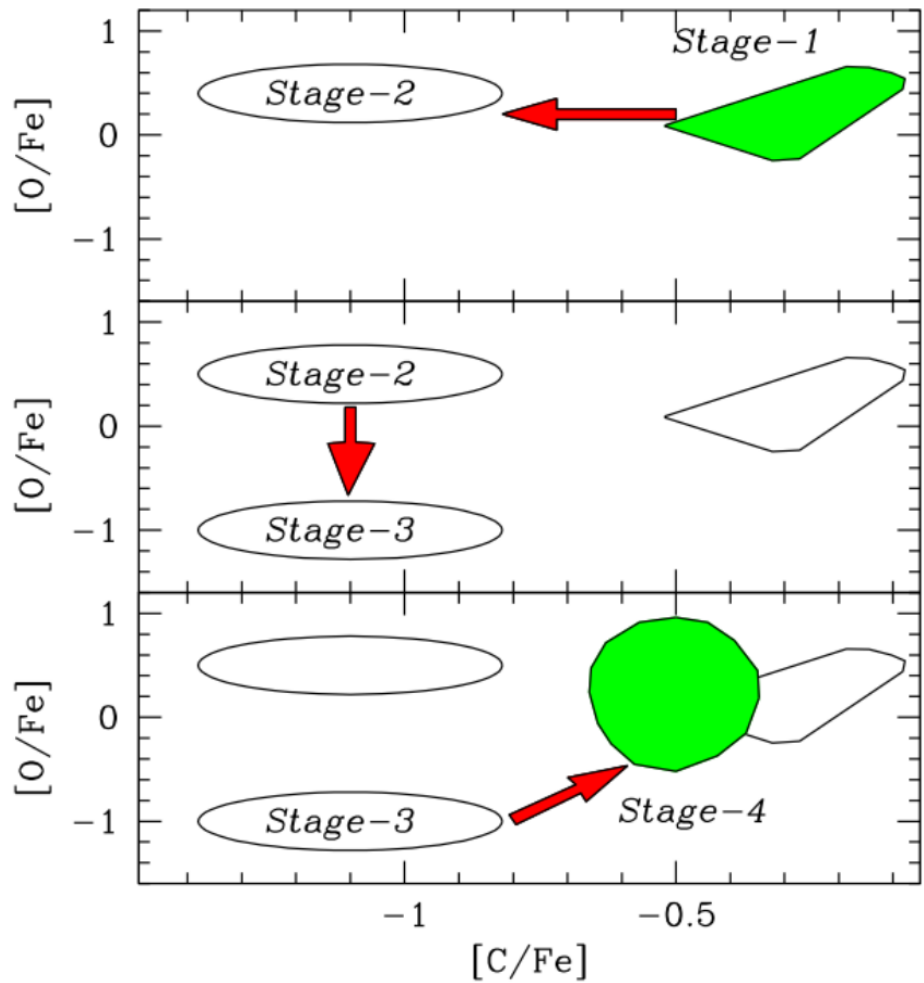

Figure 1.9: A possible scenario, proposed by F06, for the evolution of C and O abundances in MT-BSSs.

After the merger, the star would appear as a CO-depleted non variable BSS (Stage-3). Finally (Stage-4) the rotation slows down and a mixing process could convert C and O abundances back to normal. In the 47 Tuc sample, there could be 2 or 3 stars in Stage-2 and 4 in Stage-3. Classical MT binaries would also result in BSSs with CO depletion, perhaps with a low mass He White Dwarf companion. Since the donor star is evolving off the MS, the transferred mass might be more heavily processed. The resulting BSS would be very similar to a Stage-3 W Uma BSS.

## 1.8  Lithium: an other possible diagnostic?

Lithium (Li) is a very fragile element, which is destroyed when the temperature is larger than $\sim 2.5 \times 10^6$ K through the 7 Li(p,$\alpha$)4 He reaction. Hence, whenever Li is produced in stellar interiors during hydrostatic burnings, it is immediately disintegrated. The discovery of a constant Li abundance in unevolved, Population II stars, the so-called Spite plateau (Spite & Spite, 1982), has been interpreted as the signature of the primordial abundance of Li, produced during the Big Bang nucleosynthesis. When the lithium abundance is expressed as A(Li) = log[n(Li)/n(H)]+12, the Spite plateau turns out to be between 2.1 and 2.4, depending on the adopted T scale (Bonifacio & Molaro, 1997; Charbonnel & Primas, 2005; Asplund et al., 2006; Bonifacio et al., 2007; Aoki et al., 2009).

From the theoretical point of view, the BSS surface Li abundance could provide clues on their formation mechanism. In fact, smooth particle hydrodynamical simulations of stellar collisions (Lombardi et al., 1995; Lombardi, Rasio & Shapiro, 1996; Sills et al., 1997) predict that COL-BSSs are not fully mixed. Additionaly, Sills et al. (1997) suggest that the COL-BSSs are not able to develop surface convective layers that eventually could be able to mix some re-processed material from the inner stellar regions and to bring the surface Li to the inner part of the star, where it can be easily destoyed. Consequently, COL-BSSs are predicted not to burn significant amount of Li and we can expect that they exhibit the same Li abundance of the stars they formed from (thus, the value of the Spite Plateau) or slightly lower if some Li has been burned. On the other hand, MT-BSSs are expected to be virtually Li-free. In fact, during the evolution of the binary system, the material accreted on the BSS surface can come from the inner regions of the donor star where Li has been already destroyed, so that the resulting BSS should be completely Li-free. Recently, Glebbeck et al. (2010) compared the Li abundance expected for MT- and COL-BSSs: in case of collision processes (reasonably assuming that the BSS progenitors are MS stars with a Li content equal to the value of the Spite Plateau) the final product have a surface Li abundance lower by ∼ 1 dex than the original value. On the other hand, BSSs formed through mass transfer show a very large depletion ($\Delta(Li)$ ∼5 dex) with respect to the MS Li content. Up to the present, the lithium abundance for BSSs has been obtained just for some object in the field (Preston & Sneden, 2000; Carney, Latham & Dodson, 2005) and in the intermediate-age open cluster M67 (Shetrone & Sandquist, 2000). All these studies provide upper limits suggesting Li depletion in the photosphere of these BSSs. Unfortunately, no result exists in literature on the Li abundances of BSSs in GCs.

Interesting enough, Ryan et al. (2001) suggest a fashinating link between the BSSs and the ultra-Li-depleted (ULD) Halo stars, a class of dwarf field stars that exhibit surface Li abundances lower than the Spite Plateau. To date 9 ULD stars have been identified (see Ryan et al. 2001 and references therein) and only A(Li) upper limits are provided. The authors proposed that the ULD stars are BSSs presently hidden on the MS and they call them *blue-stragglers-to-be*. Hence, searching for Li depletion (unespected in unevolved Halo stars) could be an efficient method to detect BSSs (or *blue-stragglers-to-be*) among the MS stars.

## 1.9 FLAMES

Within the framework described in the previous sections, this Thesis is devoted to the detailed investigation of the chemical and kinematical properties of BSSs in GCs. All the objects studied in this work have been observed by using FLAMES at the ESO VLT. FLAMES is a multi-object, intermediate and high resolution spectrograph mounted at the Nasmyth A platform of UT2. FLAMES can access targets over a large field of view (25′ diameter). It consists of three main components:

- Fibre Positioner (OzPoz) hosting two plates

- A medium-high resolution optical spectrograph, GIRAFFE, with three types of feeding fibre systems : MEDUSA, IFU, ARGUS.

- A link to the UVES spectrograph (Red Arm) via 8 single fibres of 1″ entrance aperture.

### 1.9.1 The OzPoz fibre positioner

The OzPoz fibre positioner is able to host up to four plates, but only two are used in the FLAMES configuration: while one plate is observing the other positions the fibres for the subsequent observations, therefore limiting the dead time between one observation and the next to less than 15 minutes, including the telescope preset and the acquisition of the next field. Each of these two plates feeds GIRAFFE and the red arm of the UVES spectrographs. Plate One is hosting 132 GIRAFFE MEDUSA buttons, 30 GIRAFFE IFU buttons (15 objects plus 15 sky), 8 UVES buttons. With Plate One it is possible to use UVES and GIRAFFE simultaneously. Plate Two is hosting the same buttons as above, a central GIRAFFE IFU "Argus" facility and 15 Argus-sky buttons. The minimum object separation is $10.5''$. This minimum distance between two targets is entirely limited by the size of the magnetic buttons. OzPoz is able to position the fibres with an accuracy of better than $0.1''$ .

### 1.9.2 The GIRAFFE spectrograph

GIRAFFE is a medium-high resolution (R=7500-30000) spectrograph for the entire visible range 370-900 nm. GIRAFFE is aimed at carrying out intermediate and high resolution spectroscopy of galactic and extragalactic objects. The name comes from the first design, where the spectrograph was standing vertically on a platform. GIRAFFE is equipped with two gratings, high and low resolution. Filters are used to select the required spectral range. Each object can be only observed in a single echelle order at once. The light comes from one of the 6 available slits and passes through the order sorting filter. It is then reflected into a double pass collimator and goes to the grating. After an intermediate spectrum is formed, the light is finally re-imaged on the CCD. The fibre system feeding GIRAFFE consists of the following components:

- The MEDUSA fibers, which allow up to 132 separate objects (including sky fibres) to be observed in one go. Two separate sets of MEDUSA fibers exists, one per positioner plate. Each fiber has an aperture of $1.2''$ on the sky.

- The IFU: each IFU (deployable Integral Field Unit) consists of a rectangular array of 20 microlenses of $0.52''$ each, giving an aperture of $2 \times 3''$. For each plate there are 15 IFU units dedicated to objects and another 15 dedicated to sky measurements. In the latter, only the central fibre is present.

- ARGUS: the large integral field unit ARGUS is mounted at the centre of one plate of the fibre positioner and consists of a rectangular array of $22 \times 14$ microlenses. Two magnification scales are available: "1:1" with a sampling of 0.52 arcsec/microlens and a total aperture of $11.5 \times 7.3$ arcsec, and "1:1.67" with 0.3 arcsec/microlens and a total aperture of $6.6 \times 4.2$ arcsec. In addition, 15 ARGUS sky fibres can be positioned in the $25'$ field.

### 1.9.3 The UVES spectrograph

UVES is the high resolution spectrograph. It was designed to work in long slit mode but it has been possible to add a fibre mode (6 to 8 fibres, depending on setup and/or mode) fed by the FLAMES positioner to its Red Arm only. Only the three standard UVES Red setups are offered, with central wavelength of 520, 580 and 860 nm respectively. All 16 fibres coming from the two positioner plates are mounted on two parallel slits. With an aperture on the sky of $1''$, the fibres project onto 5 UVES pixels giving a resolving power of 47000. For faint objects and depending on the spectral region, one or more fibres can be devoted to recording the sky contribution. In addition, for the 580 nm setup only, a separate calibration fibre is available to acquire simultaneous ThAr calibration spectra. This allows very accurate radial velocity determinations. In this configuration, 7 fibres remain available for targets on sky.

# Chapter 2

# BSSs in M4

– Based on the results of Lovisi et al. 2010 ApJ 719, L121

M4 (NGC 6121) is the closest Galactic GC (2.1 kpc, Harris 1996, 2010 edition). For this reason it has been deeply studied both photometrically and spectroscopically. In spite of this, its dynamical state still remains not well understood. In fact, M4 shows a surface brightness profile compatible with a flat core, consistent with a system that has not yet suffered core collapse. Nevertheless, recent results cast some doubts on this conclusion. Heggie & Giersz (2008) performed Monte Carlo simulations for the dynamical evolution of M4, by including the treatment of two-body relaxation, three- and four-body interactions and by involving primordial binaries and those formed dynamically, the Galactic tidal field and the internal evolution of both single and binary stars. The authors obtain a set of initial parameters for the cluster which, after 12 Gyr of evolution, gives a model with a satisfactory match to the M4 surface-brightness profile, velocity-dispersion profile and luminosity function. The most surprising result of this study is that, according to this model, M4 is actually a post-core collapse cluster where the core radius is sustained by binary burning.

In this framework, the knowledge of the binary frequency is one of the key parameters in dynamical models of GCs. Through their formation and destruction, binaries play a fundamental role in the dynamical evolution of a GC, especially during the core collapse phase. In fact, they can be seen as an efficient heating source that can halt the core collapse (Hut et al., 1992a,b). Additionally, encounters between binary systems and other single or multiple stars can disrupt the wider binaries and, on the other hand, can lead to the formation of BSSs and also millisecond pulsars and X-ray binaries.

The most complete work on the research of binary systems in the GC M4, has been performed by Sommariva et al. (2009), who observed 2469 stars located within the red giant branch tip and one magnitude below the TO all over the cluster extension. Thanks to at least two observing epochs with a temporal separation of about three years, they searched for variations in radial velocities among these stars and found 57 binary candidates. The corresponding total binary fraction is f = 2.3 ± 0.3%, that is significantly smaller than the previous results presented by Cote & Fischer (1996). Nevertheless, according to the authors, the two results are consistent within the errors. The binary fraction inside the cluster core radius is f = 4.1 ± 2.1%, whereas outside it is f = 2.2 ± 0.5%. It is worth noticing that the sample found by Sommariva et al. (2009) consists in only spectroscopic binary candidates and does not include the total population of binary systems: based on Gaussian statistics it is likely that their sample is incomplete and includes some false detections. Based on the assumption that an additional epoch could be helpful for confirming the candidate binariety and for estimating the completeness of the sample, the authors calculate how many binary candidates they would be able to find among stars with three epochs. This simple test shows that the binary fraction from two epochs suffers from an incompleteness at least of 40%. Accounting for this incompleteness, and considering the number of targets with two and three epochs observed inside and outside the core, the total binary fraction in the whole sample increases up to f = 3.0 ± 0.3%, whereas inside and outside the cluster core becomes f = 5.1± 2.3%, and f = 3.0 ± 0.4%, respectively. Unfortunately, no studies have been performed to estimate whether the binary fraction derived by Sommariva et al. (2009) is sufficient to explain the results obtained by Heggie & Giersz (2008).

By considering the nearness to the Sun, the special dynamical state and the presence in literature of extensive studies on this cluster, we selected M4 as one of the best targets where performing a detailed analysis of the BSS population. Part of the following results (radial and rotational velocities, Fe, C and O abundances for all the GIRAFFE targets, see Section 2.1) are discussed in Lovisi et al. (2010, hereafter L10) whereas some others (Li abundances both for the GIRAFFE and UVES targets) are still in preparation and they will be published in Mucciarelli et al. (2013).

## 2.1 Observations

The observations were performed at the ESO-VLT by using the multi-object high-resolution spectrograph FLAMES (Pasquini et al. 2002) in the UVES+GIRAFFE combined mode. The sample includes 20 BSSs, 53 SGBs and 38 RGBs observed with GIRAFFE. 3 BSSs were also observed with UVES: 2 of them (namely #2000075 and #2000085) are in common with the GIRAFFE dataset, whereas for the other 1 (#52720) just the UVES spectrum is available. The observations with UVES were performed in service mode between April and May 2004, whereas GIRAFFE observations were performed in visitor mode during three nights in June 2008. The spectroscopic target selection has been performed on a photometric catalog obtained by combining ACS@HST data for the central region (i.e., for the innermost 100″) and WFI@ESO observations for the outer region. This strategy allows to cover the entire extension of the cluster and to mantain high accuracy and resolution in the most central and crowded regions. The gravity centre of M4 ($\alpha$(J2000) = $16^{\rm h}\,23^{\rm m}\,35.03^{\rm s}$, $\delta$(J2000) = $-26^{\rm o}\,31'\,33.89''$) was determined from the absolute positions of individual stars in the ACS dataset, following the approach described in Lanzoni et al. (2007a). This is $\sim 7''$ south-west from the centre quoted by Harris (1996, 2010 edition). We have also taken into account the stellar proper motions for the wide-field sample (Anderson et al. 2006), and discarded all the stars having a source with a comparable or a brighter luminosity at a distance smaller than $3''$. The selected BSS sample represents $\sim 70\%$ of the entire population within $800''$ ($\approx 11$ core radii) from the cluster centre, with $\sim 50\%$ of them being located at radial distances $r < 100''$. The RGB and SGB stars have been selected from the WFI sample at distances $100'' < r < 800''$. Figure 2.1 shows the CMD for the ACS and WFI dataset, with the position of all the analysed targets, whereas in Figure 2.2 the BSS location with respect to the cluster centre is shown. Three different setups were used for the GIRAFFE observations: HR15 (with spectral resolution R=17000), HR18 (R=18300) and HR22 (R=11700), suitable to sample the H$\alpha$ line, the OI triplet at $\lambda \simeq 7774$ Å, and the CI line at $\lambda \simeq 9111$ Å, respectively. The 580 Red Arm (R=45000) has been used for the UVES observations. The spectral coverage of the UVES arm and HR15 setup allows to sample also the region of the Li doublet at 6708 Å. Exposure times amount to one hour for the HR15 setup, two hours each for the HR18 and HR22, and 5 hours for the UVES setups. Spectra have been pre-reduced by using the standard ESO pipelines, including bias subtraction, flat-fielding, wavelength calibration and spectrum extraction. The accuracy of the wavelength calibration has been checked by measuring the wavelength position of a number of emission telluric lines (Osterbrock et al. 1996). Then, we subtracted the mean sky spectrum from each stellar spectrum. By combining the exposures, we finally obtained GIRAFFE median spectra with signal-to-noise ratios S/N$\simeq 50-100$ for the selected BSS and SGB stars, and S/N$\simeq 100-300$ for the RGBs; for the UVES median spectra we obtained a S/N$\simeq 30-130$.

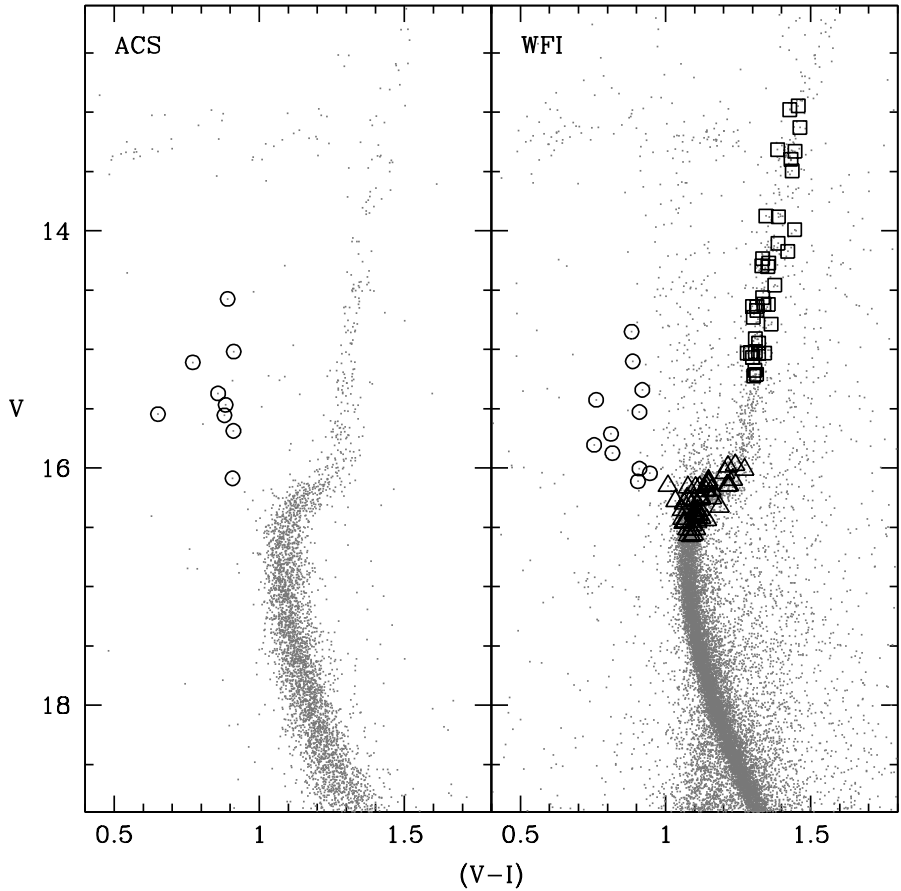

Figure 2.1: CMD for the ACS and WFI datasets. All the analysed targets are shown: triangles and squares label SGB and RGB stars respectively, whereas circles are the BSSs.

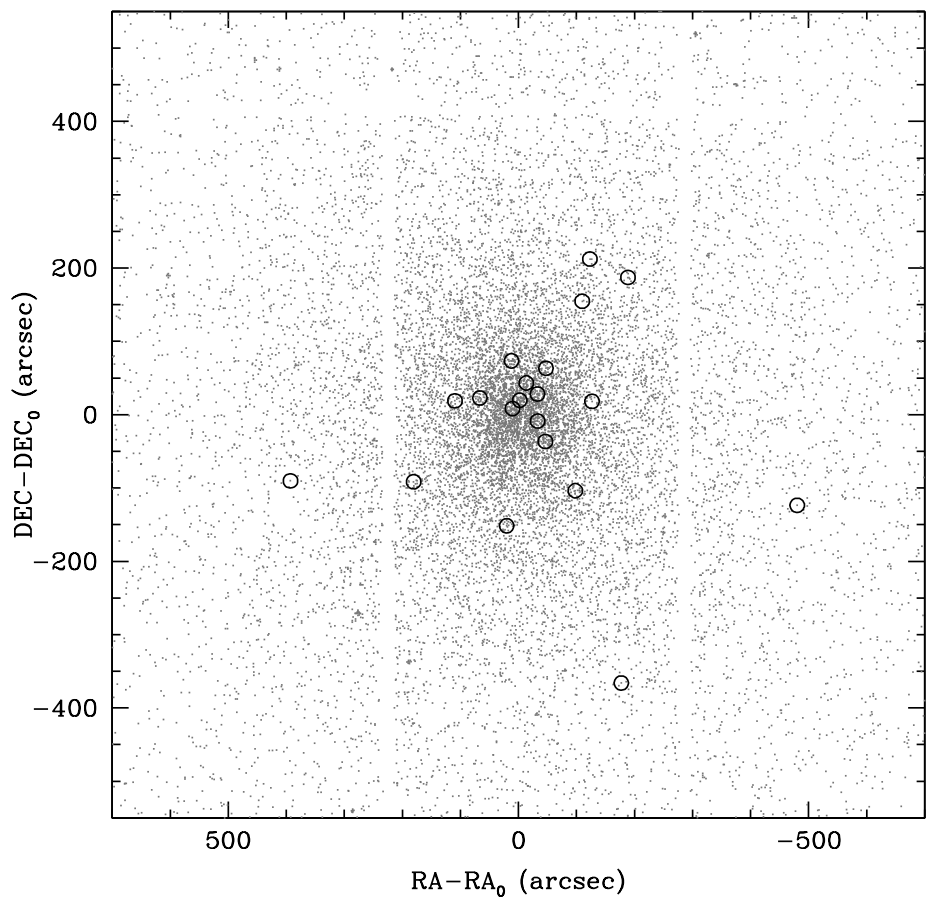

Figure 2.2: The position with respect to the cluster centre is shown for all the observed BSSs (empty circles). The sample covers the entire cluster extension both in the inner and the outer regions.

## 2.2 Analysis and results

The procedure adopted to derive the radial and rotational velocities, and the [O/Fe], [C/Fe] and [Fe/H] abundance ratios for the observed sample is summarized below. Table 2.1 lists the values obtained for the BSSs, together with the adopted temperatures (T) and gravities ($\log g$).

### 2.2.1 Radial velocities

Radial velocities (RVs) were measured by using the IRAF task *fxcor* that performs the Fourier cross-correlation between the target spectra and a template of known radial shift, following the prescriptions by Tonry & Davis (1979). As a template for the samples of BSSs, SGBs and RGBs we used the corresponding spectra with the highest S/N ratio, for which we computed the RV by measuring the wavelength position of a few tens of metallic lines. RV values obtained from the three different GIRAFFE setups and the UVES arm are consistent each other within the uncertainties, which are of the order of $0.5\,\mathrm{km\,s^{-1}}$ for the SGB stars and most of the BSSs (see Table 2.1), and $0.15\,\mathrm{km\,s^{-1}}$ for the RGBs. For the fast rotating stars (see Section 2.2.3) the RV has been estimated from the centroid of the H$\alpha$ line, which is almost unaffected by rotation.

Figure 2.3 shows the derived distribution of RVs for the giants (RGBs+SGBs) and the BSSs. The mean RV of the total sample is $\langle RV \rangle = 71.28 \pm 0.50\,\mathrm{km\,s^{-1}}$, with a dispersion $\sigma = 5.26\,\mathrm{km\,s^{-1}}$. The distribution for the giant stars is peaked at nearly the same value, $\langle RV \rangle = 71.25 \pm 0.43$ ($\sigma = 4.08$) $\mathrm{km\,s^{-1}}$, which we adopt as the systemic velocity of M4. This is in good agreement with previous determinations (Peterson et al. 1995a; Harris 1996, 2010 edition; Marino et al. 2008; Sommariva et al. 2009; Lane et al. 2010). The average of the BSS RV distribution is also similar, $\langle RV \rangle = 71.40 \pm 2.01 \, \mathrm{km\,s^{-1}}$, but the dispersion is larger ($\sigma = 9 \, \mathrm{km\,s^{-1}}$) due to the discordant values measured for five of these stars (see Table 2.1).

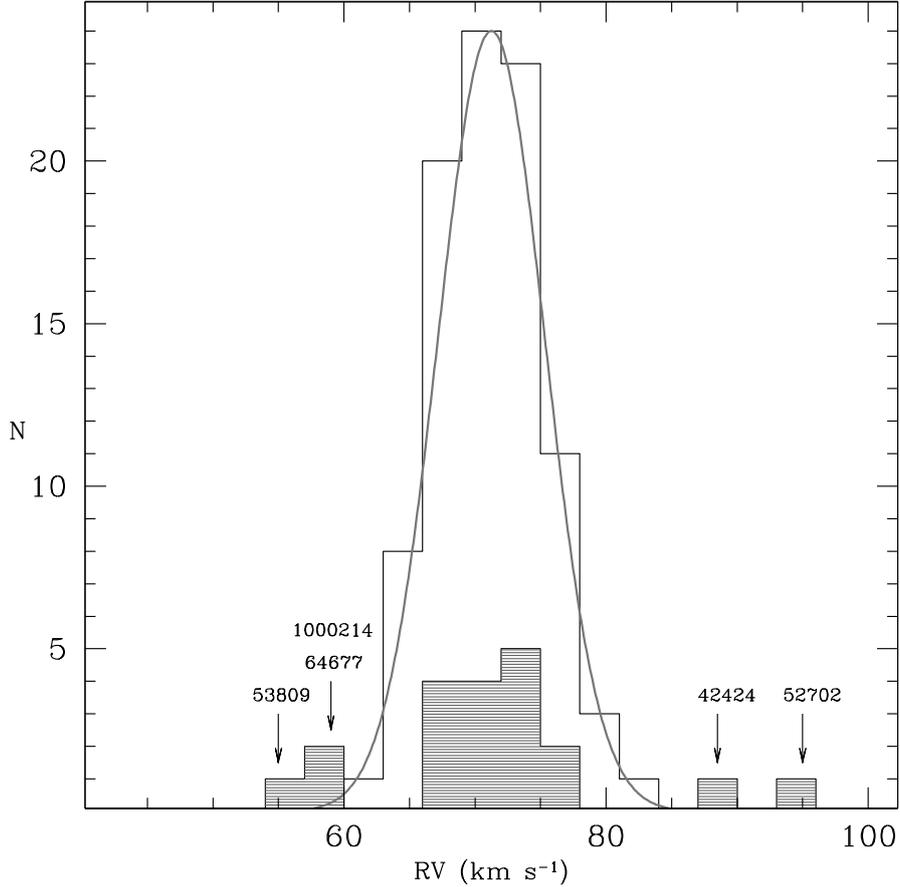

Figure 2.3: Radial velocity distribution for the observed stars in M4. The empty histogram represents the distribution obtained for SGB and RGB stars, the shaded histogram is for BSSs. The solid curve is the gaussian best-fit to the giant radial velocity distribution. BSSs displaying discordant radial velocities with respect to the cluster distribution are labelled with their identification number (see Table 2.1).

## 2.2.2 Atmospheric parameters

The effective temperature for each BSS and SGB star has been derived by fitting the wings of the H$\alpha$ line, which are insensitive to other parameters like metal content and gravity. As shown in Figure 3 of Fuhrmann et al. (1993), H$\alpha$, H$\beta$, H$\gamma$ and H$\delta$, are very sensitive to the stellar atmospheric parameters. In particular, all the Balmer lines are strongly T sensitive with the profile variation increasing from Hα to Hδ. Nevertheless, not all the Balmer lines can be safely used as T indicators. In fact, the depression of the line wing for increasing T is strongest for the higher series members, such as Hγ and Hδ, but the metal line blending is increasing from Hα to Hδ, thus making the analysis of an unperturbed Balmer line profile considerably more difficult for Hγ and Hδ. Additionally, the global flux distribution of FG-type stars plus the CCD sensitivity are shifted toward the red, thus allowing a higher S/N for Hα than for the other lines. Hence the use of Hα and Hβ should be favoured, while the higher series members may be more difficult to analyse. The variation with gravity is rather small as compared with temperature for all the Balmer lines and, in particular, Hα is largely insensitive to gravity. The response to varying the metal abundance is more complicated: at T ≃ 6000 K, the Hα line is rather insensitive to metal abundance, but the higher series members become quite shallow as soon as the metal abundance gets lower than [M/H] ≃ −1. Finally, for hotter stars the residual sensitivity of the Hα wings to [M/H] vanishes. Also in this case, as for gravity, the metallicity effect on the Hα profile is irrelevant. Unfortunately, the high sensitivity of the Hα line to T variations is restricted to a limited range of temperature, between ∼ 5500 K and ∼8000 K. In fact, it is well known that below 5500 K the Hα wings are not sensitive to T variations, whereas above 8000 K, the Stark effect starts dominating the Hα profile and the T dependence decreases. Moreover, there are indications that for T higher than ∼ 8500 K the behaviour of the Hα profile is completely inverted, so that the Hα wings decrease with higher temperatures. This is perfectly visible in Figure 2.4: the integral of the Hα wings flux with respect to the continuum level for a sample of synthetic spectra with different T is plotted as a function of the temperature. For lower temperatures, the variation of the Hα wings with T is lower, whereas for high temperatures (≳8000 K) the trend is completely inverted.

Keeping in mind all these reasons, we computed spectroscopic temperatures by using the Hα profile only for BSSs and SGBs. For all these stars, we firstly accurately normalized the spectrum in a region close to the Hα. This procedure is very important and it should be exectuted very carefully: in fact, a bad spectrum normalization could alter the Hα wing profiles thus providing a wrong T value. After the normalization, we performed a $\chi^2$ minimization between the observed Hα profile and a grid of synthetic spectra computed with different temperatures, by using SYNTHE code (Kurucz 1993). Typical uncertainties for the temperatures are of the order of ∼ 50 K for the BSSs, and ∼ 100 − 150 K for the SGBs. T values for the RGBs have been computed photometrically by projecting the target position in the CMD on a grid of evolutionary tracks extracted from the *BaSTI Library* (Pietrinferni et al. 2006) and the results are shown in

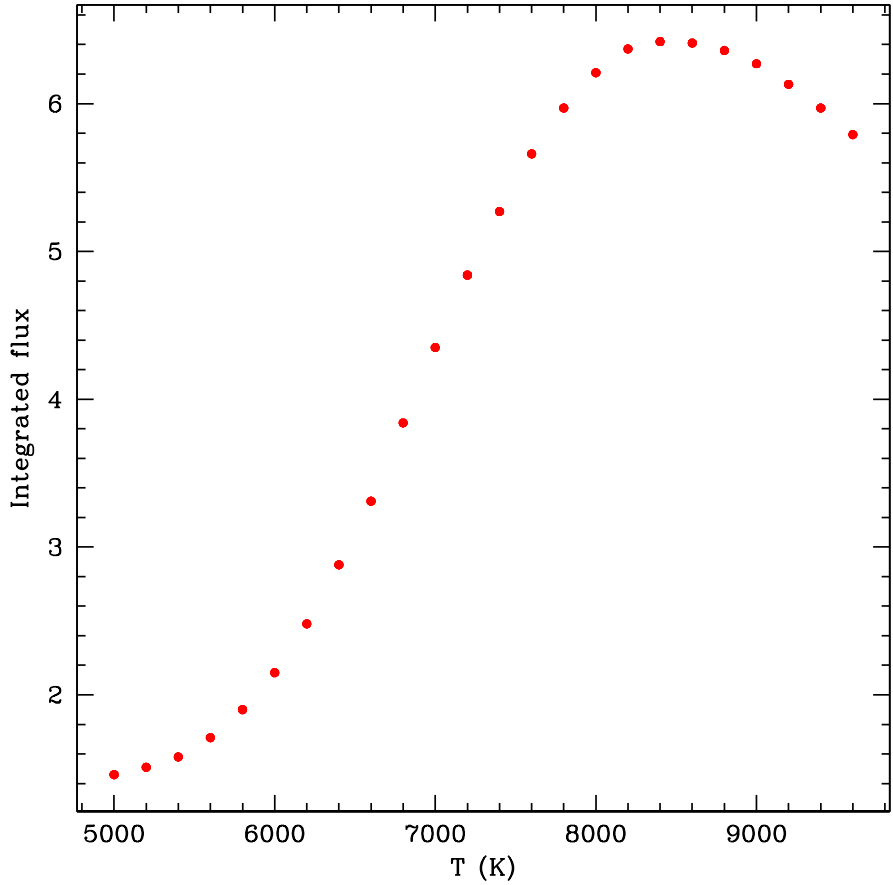

Figure 2.4: The integral of the Hα wings flux with respect to the continuum level for a sample of synthetic spectra with different T is plotted as a function of the temperature. For low temperatures ($\lesssim$ 5500 K), the Hα wings are not very sensitive to T variations. For high temperatures ($\gtrsim$ 8000 K) the trend with T is inverted and the Hα wings decrease for increasing T.

Mucciarelli et al. (2011a). Typical errors for RGB temperatures are of the order of 100 K.

Also gravities have been determined photometrically, with a precision of 0.2 dex. The adopted method to infer the BSS gravities may be affected by systematic uncertainties, since these stars could be underluminous for their masses (van den Berg et al., 2001; Sandquist et al., 2003; Mathieu & Geller, 2009). Unfortunately, we do not have direct spectroscopic indicators for gravities. However, gravity variations of 0.2 dex translate in abundances variations smaller than 0.1 dex. This procedure allows also to obtain the photometric masses of the targets: the mass distribution for the observed BSSs is peaked at $\sim 1 M_{\odot}$, with the most massive object being at $\sim 1.3 M_{\odot}$ (see Figure 2.5). For the micro-turbulent velocity we adopted $1.5 \pm 0.5\,\mathrm{km\,s^{-1}}$ for both the BSSs and SGBs. This quantity has been strictly measured for a few SGB stars only, since unfortunately the small number of metal lines does not allow to perform tha same calculation for the BSSs. However, the impact of such an assumption on the derived rotational velocities and chemical abundances is negligible. For the RGBs the microturbulent velocities have been derived spectroscopically, by erasing any trend between A(Fe) and the equivalent width (EW). Finally, the relation by Gray (1984) has been used to derive the macro-turbulence velocity for all the targets.

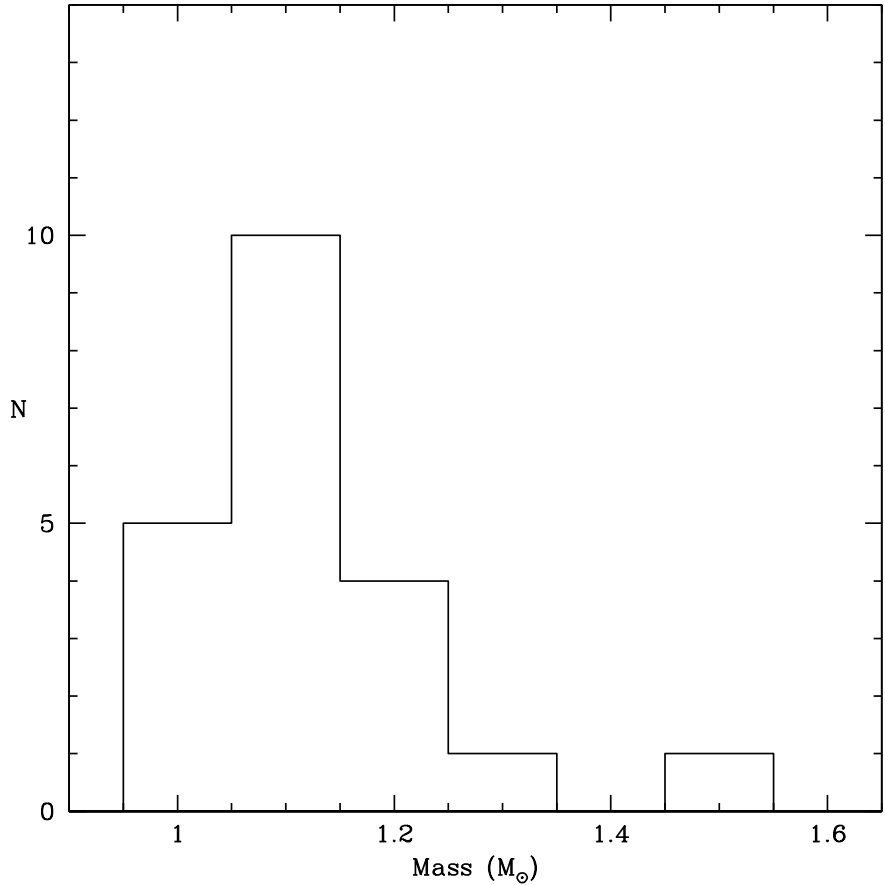

Figure 2.5: Photometrical masses for the BSS sample, derived by projecting the target position in the CMD on suitable tracks taken from the *BaSTI Library* (Pietrinferni et al. 2006). The peak of the distribution is at $\sim 1 M_\odot$ with the most massive object being at $\sim 1.3 M_\odot$.

### 2.2.3 Rotational velocities

We derived stellar rotational velocities by using the previously derived atmospheric parameters and the method described by Lucatello & Gratton (2003). In particular we computed the Doppler broadening due to the rotation of the object ($I_{\rm rot}$), which is linked to the rotational velocity $v\sin(i)$ by means of a coefficient $\alpha$: $I_{\rm rot} = v\sin(i)/\alpha$ (see equation 3 in Lucatello & Gratton 2003, where $I_{\rm rot}$ is indicated as "$r$"). In principle, the coefficient $\alpha$ should be calibrated by using standard stars with known rotational velocities observed with the same instrumental configuration. Although we did not perform such a calibration, the comparison with synthetic spectra computed for different rotational velocities indicates that $\alpha$ is close to unity. For this reason and for sake of clarity, we will call the derived rotational velocity indices $v\sin(i)$, by assuming that they are equal to the projected rotational velocities. Typical errors on $v\sin(i)$ are of the order of $2\,{\rm km\,s^{-1}}$. Since the adopted technique is efficient only for rotational velocities up to $\simeq 50-60\,{\rm km\,s^{-1}}$, the values of v sin($i$) for faster stars have been estimated by comparing the observed OI triplet region with rotationally broadened synthetic spectra.

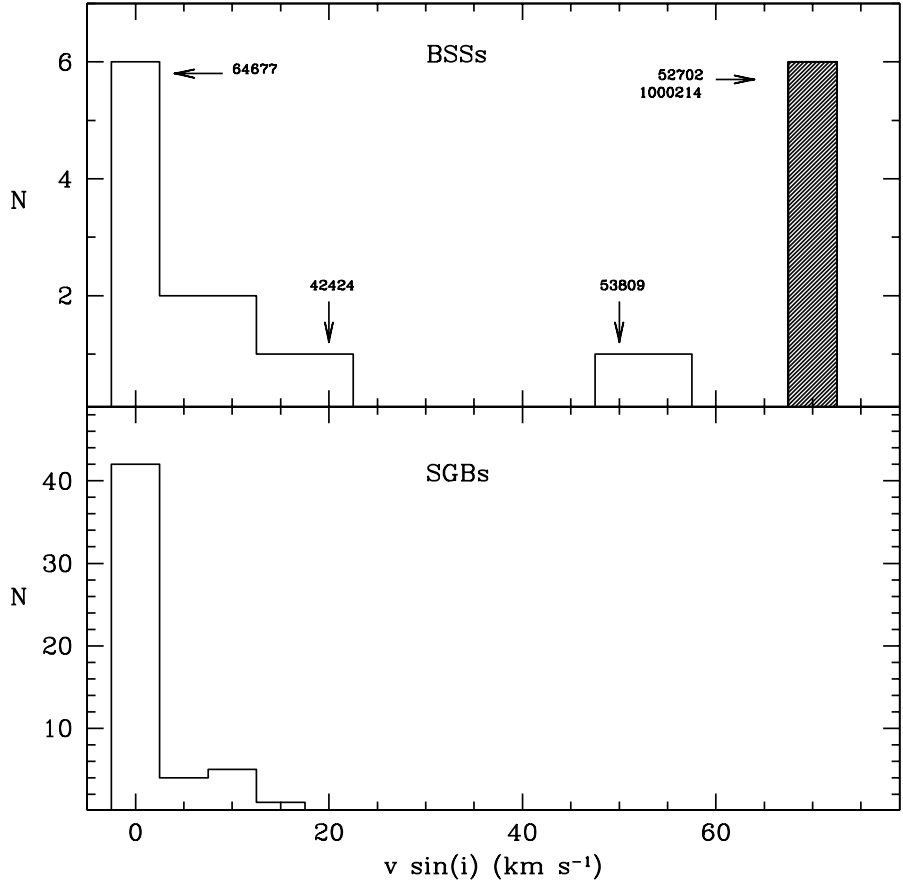

Figure 2.6: Rotational velocity distribution derived for the observed BSSs (upper panel) and SGBs (lower panel). Six BSSs with $v\sin(i) \geq 70$ km s$^{-1}$ have been all plotted within the same bin (shaded block), and five BSSs with anomalous radial velocity are labelled. All RGBs have $v\sin(i)=0\,{\rm km\,s^{-1}}$ (Mucciarelli et al. 2011a) and are not shown for sake of clarity.

Figure 2.6 shows the derived rotational velocity distributions. As can be seen, the SGB rotation distribution is peaked at v sin($i$)=0.0 km s$^{-1}$, with the highest value being $13.4 \pm 3.4$ km s$^{-1}$. The rotational velocity for all the RGBs is 0 km s$^{-1}$ (see Mucciarelli et al. 2011a). The distribution of rotational velocities for the BSS sample (see Table 2.1) is quite different, with eight stars (40% of the total) being fast rotators, i.e., rotating at more than 50 km s$^{-1}$ (while normal F-G type stars typically spin at less than $\sim 30$ km s$^{-1}$; Cortés et al. 2009). Interestingly, three (out of five) BSSs with anomalous RVs also are fast rotators.

The very high rotational velocity strongly changes the normal OI triplet profile. The top panel of

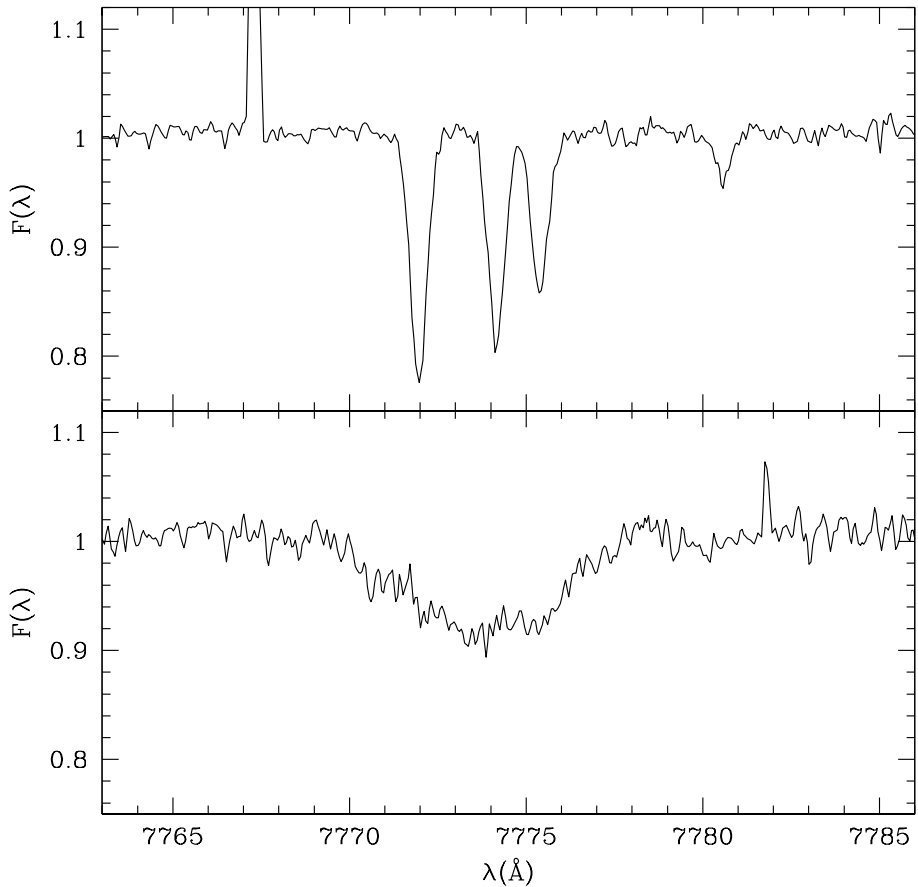

Figure 2.7: OI triplet region for a slow rotating BSS (top panel) and a fast rotating BSS (bottom panel). For high rotational velocities the three lines are totally blended together and the resulting profile can be fit with a rotational profile.

Figure 2.7 shows the spectral region around the OI triplet for a slow rotating BSS: the three lines have a Gaussian profile and their centroids are distinguishable and measureable. The same spectral region is shown in the bottom panel of Figure 2.7 for a fast rotating BSS: the line centroids are completely undistinguishable since they are totally blended together; the line profile is no longer Gaussian and the big dip can be well fit with a rotational profile.

### 2.2.4 Chemical abundances

As reference population necessary to identify possible anomalies in the BSS C/O surface abundances, we consider the SGB stars, since episodes of mixing and dredge-up may modify the primordial abundance patterns of the RGBs. Chemical abundances have been derived from the equivalent width measurements by using our own program GALA (Mucciarelli et al. 2013, ApJ submitted, see Appendix A) based on the WIDTH9 code (Kurucz, 1993; Sbordone et al., 2004). Abundance errors have been computed by taking into account the uncertainties on the atmospheric parameters and those on the equivalent width measurements. For each star they typically are of the order of $0.1 - 0.2$ dex.

The iron content for the SGBs and BSSs has been derived from the equivalent widths of about ten and 2–7 FeI lines, respectively. For the SGB stars the resulting average iron abundance is [Fe/H]$= -1.10 \pm 0.01$, with a dispersion $\sigma = 0.07$ about the mean, in good agreement with the values quoted in the literature, which range between $-1.20$ and $-1.07$ (Harris 1996, 2010 edition; Ivans et al. 1999; Marino et al. 2008; Carretta et al. 2009a). Because of the significant deformation of the spectral line profiles due to the high rotational velocity, no iron abundance has been derived for the eight fast rotators. Moreover, technical failures in the spectrograph fiber positioning prevented us to measure it for two additional objects (see Table 2.1). The iron abundance obtained for the remaining ten BSSs has a mean value of $-1.27$ and a dispersion $\sigma = 0.10$, consistent, within the errors, with the values derived for the SGBs.

Oxygen and carbon abundances have been computed from the measured equivalent widths of the OI triplet and the CI lines respectively, and the derived values have then been corrected for non-local thermodynamic equilibrium (NLTE) effects. For O abundances we derived NLTE corrections by interpolating the grid by Gratton et al. (1999); for C abundances we adopted the empirical relation obtained by interpolating the values listed by Tomkin et al. (1992). The resulting average abundances for the SGB sample are [C/Fe]$= -0.16 \pm 0.02$ ($\sigma = 0.17$) and [O/Fe]$= 0.29 \pm 0.02$ ($\sigma = 0.17$). No measurements have been possible for the very fast rotating BSSs and for a few other objects (see Table 2.1). Hence, we were able to measure both the C and O abundances only for 11 BSSs, out of 20 observed. Figure 2.8 shows the results obtained in the [C/Fe]−[O/Fe] plane. The values measured for the 11 BSSs are in agreement with those of the SGB stars, with no evidence of depletion either in carbon or in oxygen. We finally notice that also in the 3 cases for which only the oxygen or the carbon abundance has been measured (see Table 2.1), the values obtained are in agreement with those of the SGBs.

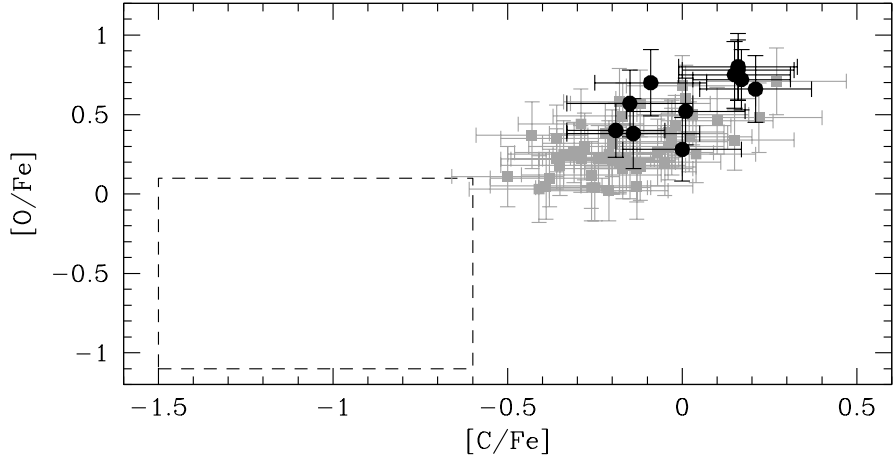

Figure 2.8: [O/Fe] ratio as a function of [C/Fe] for the measured BSSs (black dots) and SGB stars (grey dots). The dashed box marks the position of the CO-depleted BSSs observed in 47 Tuc (F06).

In all the BSS targets the Li line is not visible and we can only provide upper limits for the Li abundance. For the fast rotators the Li doublet cannot be distinguished from the noise. Unfortunately, we are not able to discriminate if this lack is due to a low Li abundance or (more likely) to the broadening effects of high rotation on the line profile. Moreover, the derived upper limits for the fast rotating BSSs are not meaningful (providing limits of A(Li) of 3-4 dex or higher). For this reason we reduce the discussion on the Li abundances to the BSSs with low ($<13$ km s$^{-1}$) values of v $\sin(i)$. Figure 2.9 shows the spectra of the sample in the region around the Li doublet. For sake of comparison the spectrum of the M4 SGB star 54711 (where the Li is well visible) is plotted: this star has an effective temperature of 6100 K and a surface Li abundance A(Li)= 2.44 dex (Mucciarelli et al., 2011a). $3\sigma$ detection upper limits have been estimated according to the formula of Battaglia et al. (2008, that slightly revises the classical formula by Cayrel 1988). This provides the uncertainty in the measurement of the equivalent width for a Gaussian line, $\sigma_{EW} = \sqrt{1.5 \cdot FWHM \cdot \delta x}/S/N$, where FWHM is the Full Width Half Maximum of the quoted spectral profile and $\delta x$ is the pixel-step of the used spectrograph (0.05 Å/pixel and 0.017 Å/pixel for GIRAFFE and UVES respectively). Note that the Li line is often fitted with a Gaussian profile at these resolutions, even if its profile is intrinsically more complex and not symmetric.

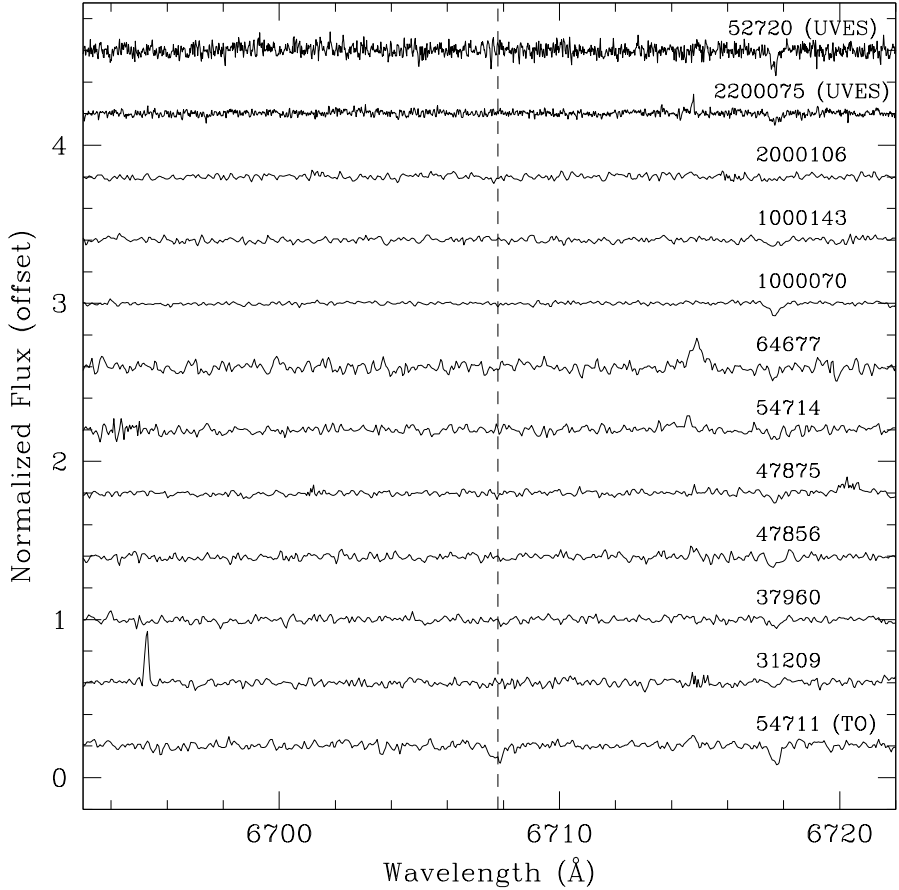

Figure 2.9: Spectra of the analysed BSSs in M4 in the spectral region around the Li doublet. The position of the Li doublet is marked with a vertical dashed line. The spectrum of the SGB star 54711 is shown for comparison. The spectra are shifted vertically for sake of clarity.

The abundance corresponding to the minimum EW detectable for a given S/N has been derived by interpolating, at that EW value, the curve of growth for the Li doublet. For each star, a curve of growth have been built by integrating the line profile for a grid of synthetic spectra calculated with the corresponding atmospheric parameters. Synthetic spectra have been computed with the code SYNTHE coupled with suitable ATLAS9 LTE plane-parallel model atmospheres. Atomic data for the Li doublet (including hyperfine structure and isotopic splitting) are from Yan & Drake (1995). This line is known to need corrections to take into account departures from LTE conditions. For sake of homogeneity with the analysis of dwarf/giant stars in M4 by Mucciarelli et al. (2011a), we adopted the NLTE corrections by Carlsson et al. (1994). The magnitude of the departure from LTE depends both on the atmospheric parameters and the line strength (therefore on the Li abundance). We interpolated the grid of corrections assuming for each star its own parameters, [M/H]=∼ −1 dex and the value of the upper limit as Li abundance. For these stars the magnitude of the NLTE corrections is very small, ranging from −0.01 up to −0.1 dex [1]. The behaviour of the derived A(Li) upper limits as a function of T is shown in Figure 2.10, where black points are the GIRAFFE targets and black squares the UVES targets. The reference position of the average A(Li) derived in the SGB stars of M4 is plotted as a horizontal grey line. The GIRAFFE targets exhibit ∼2 dex lower upper limits, with the exception of the hottest star of the sample (#31209) where the surface Li abundance is lower than 2.44 dex. For the GIRAFFE target with the highest spectral quality (#1000070 with S/N∼130) we can provide an upper limit of 1.61 dex. The two UVES targets (that have higher resolving power and smallest pixel-step) show lower values with respect to the GIRAFFE targets, providing Li abundances smaller than 1.54 and 1.69 dex, for #2000075 and #52720 respectively.

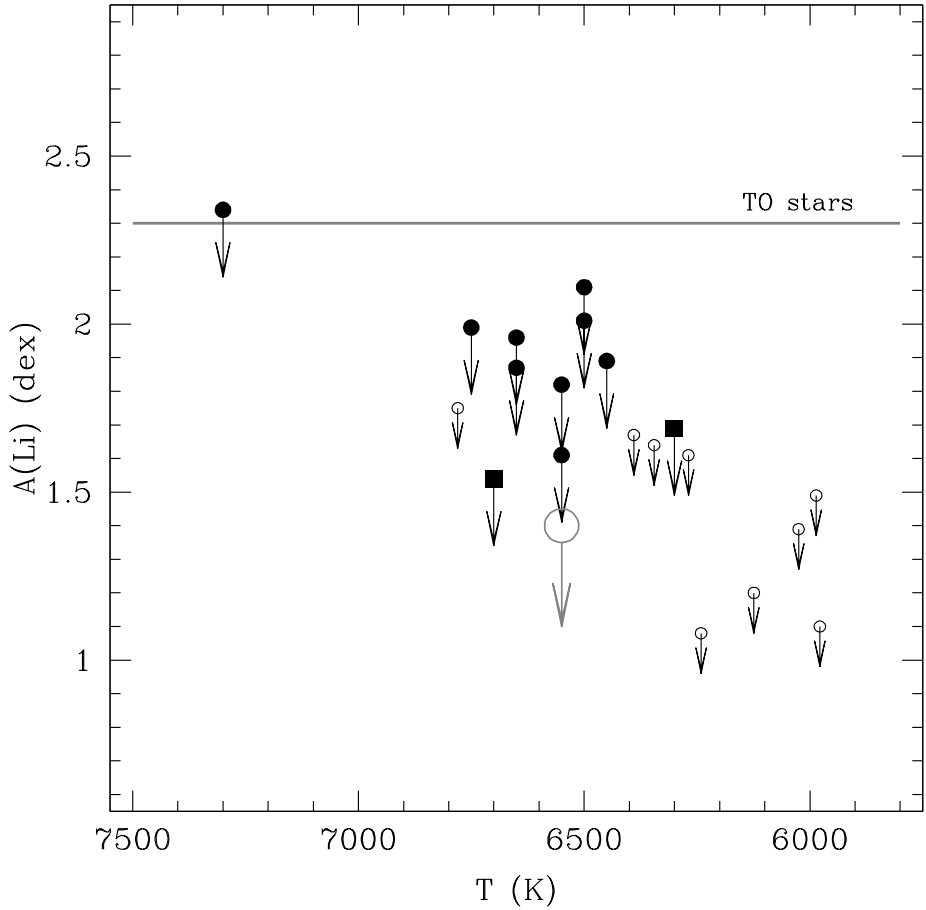

Figure 2.10: Upper limits of A(Li) for the analysed BSSs as a function of T: black points are the GIRAFFE targets and black squares the UVES targets. Horizontal grey line marks the average Li abundance derived from the MS stars in M4 by Mucciarelli et al. (2011a). Empty grey circle is the upper limit for the median spectrum calculated by using the GIRAFFE spectra of stars with T= 6450-6700 K. Small open circles are the ultra-Li-deficient Halo stars observed by Ryan et al. (2001).

---

[1]Note that the recent NLTE corrections presented by Lind et al. (2009) differ of a few hundredth of dex with respect to those by Carlsson et al. (1994): the use of values by Lind et al. (2009) do not affect our results.

As additional check to constrain the Li content in these stars, we combined together the GIRAFFE spectra of similar temperature (6450-6700 K, i.e. excluding the hottest stars) thus obtaining a median spectrum with enhanced S/N (∼200). Assuming that the amount of Li depletion with respect the original value of M4 is the same for all the BSSs (and assuming the median T of the individual stars), we derive an upper limit to the Li content of A(Li)<1.5 dex (see the large grey circle) very similar to the upper limits derived for the UVES spectra. We calculated also median spectra grouping the stars in two T bins, the first with T=6450-6550 K and the second with T=6600-6700 K, leading to S/N of ∼150 and ∼120, respectively. For these spectra we derived upper limits or 1.55 and 1.65 dex, respectively.

## 2.3 Discussion

Before discussing in details the main findings of the present work it is necessary to verify whether some of the investigated stars do not belong to the cluster. In particular, five BSSs display anomalous RVs, which may cast some doubts about their membership. However, BSSs #42424 and #64677 have measured proper motions well in agreement with those of the cluster members (Anderson et al. 2006). All the other objects have measured rotational velocities significantly larger than expected for normal stars of the same spectral type, thus making unlikely that they belong to the field. In order to further check for spurious contaminations, we have also used the Besançon Galactic model (Robin et al. 2003) to derive the RV and metallicity distributions of the Galactic field stars in the direction of M4, within the same magnitude and colour intervals covered by our BSS sample. The RV distribution turns out to be peaked at $RV = -14.6\,\mathrm{km\,s^{-1}}$ and has a dispersion $\sigma = 50.7$ km s$^{-1}$. As a consequence, the probability that the BSSs with anomalous radial velocity belong to the field is smaller than 1.7%. The metallicity distribution of field stars in the Besançon Galactic model is peaked at [Fe/H]$= -0.17 \pm 0.02$ ($\sigma = 0.45$). The iron abundance measured in two of the five BSSs with anomalous RV ([Fe/H]$= -1.35$ for #42424, and $-1.23$ for #64677) clearly is largely inconsistent with the field value and in agreement, within the errors, with that of M4 stars. For the other three (fast rotating) BSSs only upper limits of the iron abundance have been derived. However, such limits are significantly smaller than the peak value of the field distribution. Based on these considerations we therefore conclude that all the BSSs with anomalous RV are indeed members of M4.

We notice that if the discrepant RVs were caused by the orbital motion in binary systems, under realistic assumptions about the total mass, the orbital separations would range between a few and 10–20 AU which is reasonable for a GC. These BSSs do not show any evidence of photometric variability (Kaluzny et al. 1997), and variations of RV with full amplitudes exceeding $3 \, \mathrm{km \, s^{-1}}$ on a time interval of 72 hours are excluded by our observations for BSSs #42424 and #64677 (unfortunately no such information is available for the fast rotators). However, these results do not disprove that they are in binary systems. In fact, for instance, they are still consistent with what expected for roughly 90% of binaries characterized by an eccentricity-period distribution similar to that recently observed by Mathieu & Geller (2009) and populating the tail of the velocity distribution in M4.

The fact that none of the 14 BSSs for which we measured C and/or O abundances shows signatures of depletion is quite intriguing. Out of the 42 BSSs investigated in 47Tuc, F06 found that 6 (14%) are C-depleted, with 3 of them also displaying O-depletion. Accordingly, in M4 we could have expected 1-2 BSSs with depleted carbon abundance, and zero or 1 BSS with both C and O depletion. Hence, the resulting no-detection may just be an effect of low statistics and is still consistent with the expectations. Alternatively the lack of chemical anomalies in M4 BSSs might point to a different formation process: while at least 6 BSSs (the CO-depleted ones) in 47Tuc display surface abundance patterns consistent with the MT formation channel, all the (investigated) BSSs in M4 may derive from stellar collisions, for which no chemical anomalies are expected. Finally, it is also possible that, as suggested by F06, the CO-depletion is a transient phenomenon and (at least part of) the BSSs in M4 are indeed MT-BSSs which have already evolved back to normal chemical abundances.

The analysis of Li abundances for the slow rotating BSSs points out that these stars (with the exception of the hottest BSS) have a Li content well below (by a factor of 2-3 up to 6) that observed in the SGB stars. This is the first study focused on BSS Li abundances in a GC. Unfortunately, even if the theoretical predictions for A(Li) well distinguish between BSSs formed through different formation mechanisms (see Glebbeck et al. 2010), the detection of the Li doublet in hot stars (like BSSs) is an hard task to perform. In fact, the Li line strength is strictly linked to the stellar surface temperature and the EW deacreses rapidly for increasing T. Moreover, BSSs are predicted to have a Li content lower than that of the normal MS stars. In order to assess the capability to detect the Li line in BSSs (or alternatively to establish meaningful upper limits) we investigated the theoretical behaviour of the Li line as a function of T. A grid of synthetic spectra has been calculated assuming [M/H]$=-1$, log g=4.0 and $v_t$=1.5 km s$^{-1}$ [2] and with T ranging from 6200 to 8000 K (covering the entire T range of all our BSSs). The EWs have been computed by direct integration over the line profile without inclusion of instrumental brodening. Figure 2.11 shows the behaviour of the Li doublet EW as a function of T, assuming three different Li abundances, A(Li)= 1.3, 1.8 and 2.3 dex. The inner panel in the same figure shows the minimum measurable EW (calculated as a $3\sigma$ detection according to the formula by Battaglia et al. 2008) as a function of S/N for the HR15 GIRAFFE and the UVES Red Arm 580 setups. BSSs with A(Li)= 1.3 have Li lines with EW smaller than ~3 mÅ (depending on their T): such weak lines can be observed with spectra with S/R>70 for UVES and >150 for GIRAFFE. Note that for a close and metal-rich GC like M4, the bulk of the BSSs has typical temperatures ranging between ~6500 and ~6700 K, that lead to EWs of ~1.5-2 mÅ . In this case, S/N>100 is needed also for the UVES targets. Moreover, because the line weakens as the temperature increases, the detection of the Li line becomes harder and harder for hotter stars and the significance of the upper limits decreases. Therefore, high S/N (coupled with high spectral resolution) is needed to provide robust constrains to the Li abundance in such hot stars. For this reason, a reasonable detection of the Li doublet for a BSS with a surface A(Li) 1 dex lower than that of MS stars is virtually possible only for a handful of cold (T<6500 K) BSSs in close GCs. We therefore conclude that the Li abundance cannot be used to discriminate between the different BSS formation channels. The derived upper limits for the BSSs in M4 are compatible with both formation scenarios and we cannot assess if these objects are Li-poor (as predicted in collisional scenario) or Li-free (as expected in case of mass transfer).

The most intriguing result of this study is the discovery that a large fraction (40%) of the investigated BSSs in M4 are fast rotators, with rotational velocities ranging from ~ 50 km s$^{-1}$, up to more than 150 km s$^{-1}$. We emphasize that this is the largest population of fast rotating BSSs ever found in a cluster. Approximately 30% of the BSSs spinning faster (at $20-50$ km s$^{-1}$) than MS stars of the same colour has been recently found in the old open cluster NGC188 (Mathieu & Geller, 2009), while BSSs in younger open clusters are found to rotate slower than expected for their spectral type (e.g., Shetrone & Sandquist 2000; Schönberner et al. 2001). For GCs only scarce and sparse data have been collected to date. The most studied case is that of 47Tuc, where 3 (7%) BSSs out of the 45 measured objects (Shara et al. 1997; De Marco et al. 2005; F06) have rotational velocities larger than $50 \, \text{km s}^{-1}$, up to $\sim 155 \, \text{km s}^{-1}$. The object studied by Shara et al. (1997) is the second brightest BSS in 47Tuc, and all the others are located at the low-luminosity end of the BSS region in the CMD. In addition they span almost the entire range of surface temperatures and distances from the cluster centre. For comparison, apart from being more numerous, the fast rotating BSSs in M4 are also found at all luminosities, temperatures and radial distances (see Figure 2.12). There is a weak indication that the rapid rotating BSSs in M4 tend to be more centrally segregated than normal BSSs (see Figure 2.13), even if the small number of stars in our sample prevents a statistically robust result (following a Kolmogorov-Smirnov test, the probability that the two distributions are extracted from the same parent population is $\sim 44\%$).

---

[2]Note that the intensity of the Li line is virtually insensitive to log g and $v_t$ but sensitive to T.

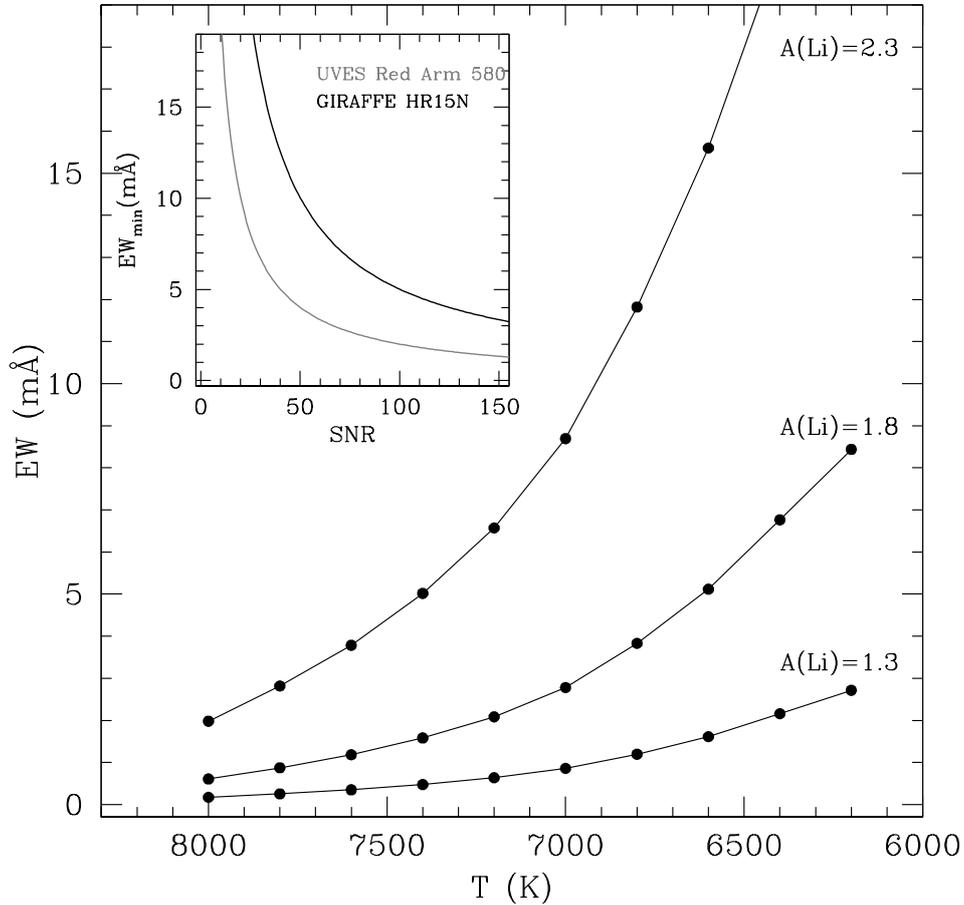

Figure 2.11: Expected behaviour of the Li line (6708 Å) EW as a function of T, assuming three different Li abundances: A(Li)=1.3, 1.8 and 2.3 dex. The inset panel shows the minimum EW measurable from the Li line as a function of the S/N, computed by using the formula by Battaglia et al. (2008) both for the HR15 and 580 Red Arm configurations.

The object with the largest rotational velocity in M4 (BSS #2000121, with v $\sin(i) \sim 150 - 200\,\mathrm{km\,s}^{-1}$) corresponds to star V53 of Kaluzny et al. (1997), which is classified as a W UMa contact binary. Interestingly, even in the F06 sample of 47Tuc the fastest spinning BSS (rotating at $\sim 100\,\mathrm{km\,s}^{-1}$) is a W UMa binary. These two stars also occupies a very similar position in the CMD, at the high-temperature and low-luminosity side of the BSS region. Unfortunately, from a theoretical point of view, rotational velocities cannot be easily interpreted in terms of BSS formation processes. In fact, fast rotation is expected for MT-BSSs because of angular momentum transfer, but some braking mechanisms may then intervene with efficiencies and characteristic time-scales that are still unknown. For the collisional formation scenarios the predictions about the BSS rotational velocities are controversial (Benz & Hills, 1987; Leonard & Livio, 1995; Sills et al., 2005). In particular, the latter models show that angular momentum losses through disk locking are able to decrease the BSS rotational velocity from values as high as $\sim 100\,\mathrm{km\,s^{-1}}$ to $20\,\mathrm{km\,s^{-1}}$. Hence, the fast rotating BSSs observed in M4 could be generated either through mass transfer activity or through stellar collisions, under the hypothesis that no significant braking has (still) occurred.

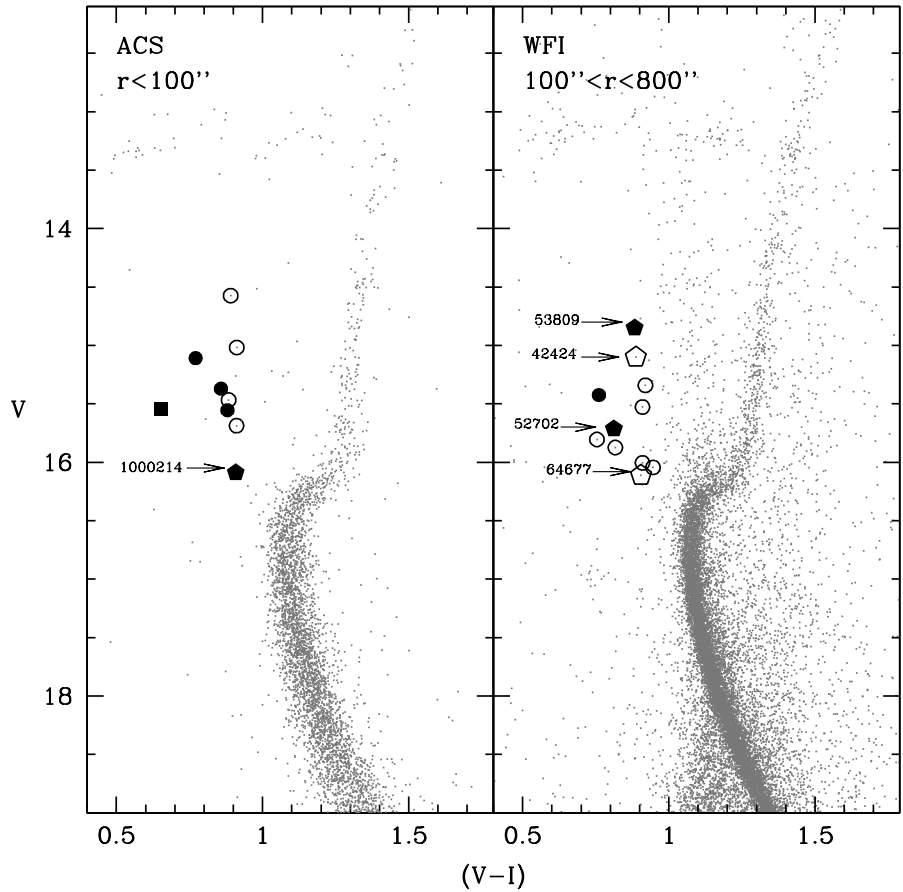

Figure 2.12: Colour-magnitude diagram of M4 for the ACS (left panel) and the WFI datasets (right panel). The observed BSSs are highlighted: empty circles mark normal BSSs, pentagons mark BSSs with anomalous radial velocity, the square indicates the W Uma star, black symbols mark fast rotators (v $\sin(i) > 50\,\mathrm{km\,s}^{-1}$).

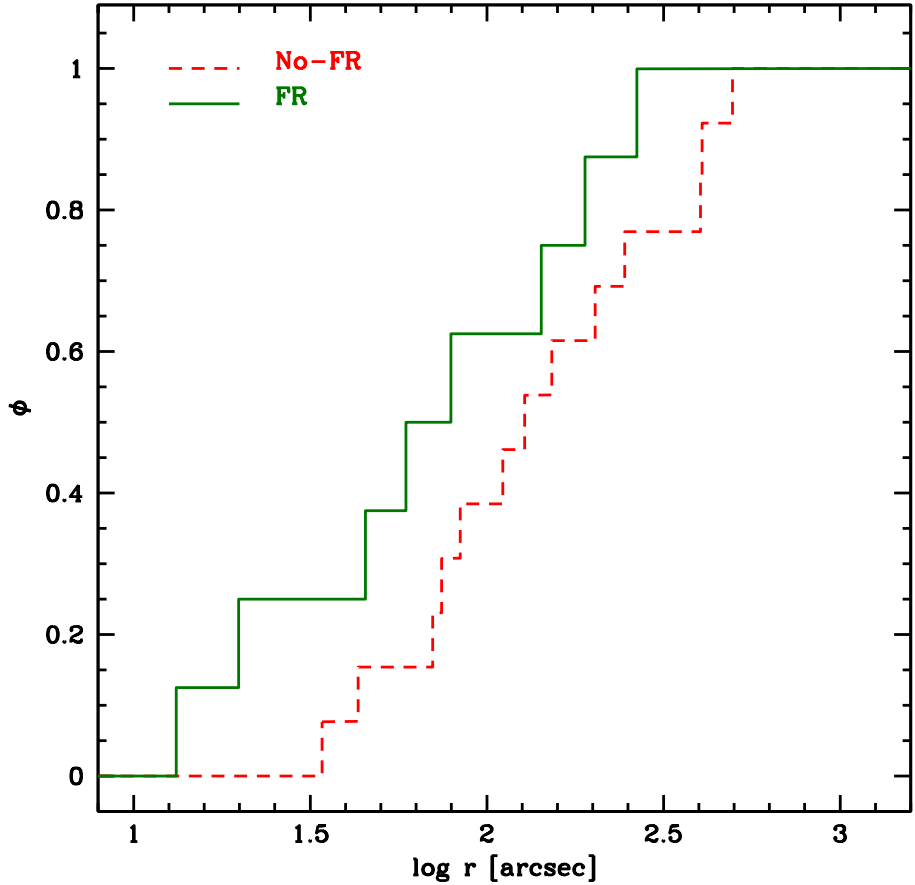

Figure 2.13: Cumulative radial distribution for slow rotating (red dashed line) and FR BSSs (green solid line). The latter ones are more centrally concentrated with respect to the other BSSs, even if at low statistical significance level.

Three out of the five BSSs with anomalous radial velocities are also fast rotators (namely #52702, #53809 and #1000214). Such a high rotation is difficult to account for by synchronization or mass transfer effects in binary systems, since the orbital separation would not be small enough. A fascinating alternative is that such anomalies in the radial and rotational velocities are due to three- and four-body interactions that occurred in the cluster core: these could have originated fast spinning BSSs and kicked them out to the external regions at high speed (apart from star #1000214, the other two are currently found well beyond the cluster core radius). Interestingly enough, the dynamical history of the cluster could support such a scenario. In fact, although its stellar density profile is well reproduced by a King model, Monte-Carlo simulations by Heggie & Giersz (2008) suggest that M4 could be a post-core collapse cluster, its core being sustained by binary burning activity. The fast rotating and high velocity BSSs could be the signature of such an activity.

| ID | T (K) | $\log(g)$ | RV (km s$^{-1}$) | $v\sin(i)$ (km s$^{-1}$) | [O/Fe] | [C/Fe] | [Fe/H] | A(Li) | Notes |
|---|---|---|---|---|---|---|---|---|---|
| 31209 | 7300 | 4.3 | 74.6±0.6 | 0.0±2.0 | 0.70 | −0.09 | −1.18 | <2.34 | — |
| 37960 | 6650 | 4.0 | 72.4±0.6 | 2.0±1.3 | 0.52 | 0.01 | −1.19 | <1.96 | — |
| 42424 | 6850 | 3.8 | 87.8±0.9 | 22.3±0.2 | 0.80 | 0.16 | −1.35 | — | RVa |
| 43765 | 6950 | 4.1 | 66.3±6.9 | 56.8±0.1 | 0.78 | 0.16 | — | — | FR |
| 44123 | 6900 | 4.2 | 71.2±4.6 | 10.3±0.4 | — | 0.10 | — | — | — |
| 47856 | 6450 | 4.1 | 66.1±0.6 | 0.0±2.5 | 0.57 | −0.15 | −1.20 | <1.89 | — |
| 47875 | 6550 | 3.9 | 71.3±0.5 | 0.0±7.8 | 0.73 | — | −1.37 | <1.82 | — |
| 52702 | 6900 | 4.2 | 95.1±7.8 | 70 - 120 | — | — | — | — | FR: RVa |
| 52720 | 6300 | 4.0 | 63.9±0.5 | — | — | — | — | <1.69 | — |
| 53809 | 6700 | 3.8 | 55.9±6.9 | 51.6±0.1 | 0.75 | 0.15 | — | — | FR: RVa |
| 54714 | 6500 | 4.1 | 74.5±0.6 | 0.0±6.4 | 0.28 | 0.00 | −1.20 | <2.01 | — |
| 64677 | 6500 | 4.2 | 58.6±0.6 | 0.0±2.0 | 0.38 | −0.14 | −1.23 | <2.11 | RVa |
| 1000070 | 6550 | 3.7 | 67.4±0.4 | 4.3±0.7 | 0.72 | 0.17 | −1.23 | <1.61 | — |
| 1000117 | 6700 | 4.0 | 72.2±4.6 | 70 - 100 | — | — | — | — | FR |
| 1000125 | 7100 | 4.0 | 69.5±6.9 | 80 - 130 | — | — | — | — | FR |
| 1000143 | 6650 | 4.0 | 68.4±0.3 | 3.2±0.9 | 0.40 | −0.19 | −1.29 | <1.87 | — |
| 1000214 | 6800 | 4.2 | 59.5±5.9 | 70 - 100 | — | — | — | — | FR: RVa |
| 2000075 | 6700 | 3.8 | 74.5±0.5 | 8.3±0.4 | 0.66 | 0.21 | −1.45 | <1.54 | FR |
| 2000085 | 7100 | 4.0 | 70.4±7.8 | 80 - 120 | — | — | — | — | FR |
| 2000106 | 6750 | 4.0 | 76.1±2.3 | 12.6±0.3 | — | 0.10 | — | <1.99 | — |
| 2000121 | 7350 | 4.3 | 76.4±8.2 | 150 - 200 | — | — | — | — | FR: WUMa |

Table 2.1: Identification numbers, temperatures, gravities, radial velocities, rotational indices, [O/Fe], [C/Fe] and [Fe/H] abundance ratios and A(Li) for the observed BSSs. "RVa" and "FR" flag the stars with anomalous radial velocities and high rotational speed ($v\sin(i) > 50\,\mathrm{km\,s^{-1}}$), respectively.

# Chapter 3

# BSSs in NGC 6397

– Based on the results of Lovisi et al. 2012 ApJ 754, 91

Star clusters have been long considered to host coeval and chemically homogeneous stars (Renzini & Buzzoni, 1986). The traditional paradigm of a striking chemical homogeneity in GC stars is still valid in terms of the iron content, with only two notable exceptions known to date: $\omega$ Centauri (Norris & Da Costa, 1995; Johnson & Pilachowski, 2010) and Terzan 5 (Ferraro et al., 2009b; Origlia et al., 2011), which are now thought to be the remnants of giant systems instead of genuine GCs [1]. Instead, large star-to-star abundance variations in the light elements (such as C, N, O, Na, Mg, Al; see Gratton et al. 2004 for a review) and strong anti-correlations (for example, between Na and O, or between Mg and Al) have been revealed in the last decade or so by high-resolution spectroscopic studies of MS and RGB stars in a large sample of GCs, both in the Milky Way (Carretta et al., 2010c) and in the Magellanic Clouds (Mucciarelli et al., 2009). The scenario generally invoked to explain these inhomogeneities is based on a second generation of stars formed during the first few 100 Myr of the cluster life, from intra-cluster medium polluted by the first stellar population. Intermediate-mass AGB stars and MS fast rotating massive stars have been proposed as the most likely polluters.

Large abundance anomalies have also been observed on the HB of some GCs: HB stars cooler than $\sim$11000 K show abundances consistent with those of giants (Glaspey et al., 1986, 1989; Lambert et al., 1992; Cohen & McCarthy, 1997; Behr et al., 1999, 2000a; Peterson et al., 2000) whereas for HB stars hotter than $\sim$11000 K, departures from the overall cluster abundances are found (Glaspey et al., 1989; Behr et al., 1999, 2000a; Peterson et al., 1995b, 2000; Moehler et al., 2000; Fabbian et al., 2005; Pace et al., 2006). In particular, a deficiency of He and an overabundance (up to solar and super-solar values) of elements heavier than Ne have been observed and they have been interpreted in terms of the gravitational settling of He and the radiative levitation of heavy elements (see Michaud et al. 1983), competing with mixing processes (like rotationally induced meridional circulation, mass loss and turbulent diffusion) able to moderate the final effects. The radiative levitation is a process affecting the stellar atmosphere and driven by the radiative acceleration. In a first approximation, the radiative acceleration for each chemical element may be calculated by using the fraction of the momentum flux that each element absorbs (see Richer et al. 1998)

$$g_R(A) = \frac{1}{4\pi r^2} \frac{L_r^{rad}}{c} \frac{\kappa_R}{X_A} \int_0^\infty \frac{\kappa_u(A)}{\kappa_u(total)} P(U) du \qquad (3.1)$$

---

[1]Significantly smaller dispersion (0.1-0.2 dex) in the iron content has been measured in M54 (Carretta et al., 2010a) M22 (Marino et al., 2009, 2011a) and NGC 1851 (Carretta et al., 2010b, 2011).

Here, $u$ is the dimensionless frequency variable

$$u \equiv \frac{h\nu}{kT}, \qquad (3.2)$$

$P(U)$ is the normalized black body flux, given by

$$P(U) \equiv \frac{15}{4\pi^4} \frac{u^4 e^u}{(e^u - 1)^2} \qquad (3.3)$$

and $L_{rad}/(4\pi r^2 c)$ is the total radiative momentum flux at radius $r$. When the radiative acceleration exceeds the gravity acceleration, the former pushes elements toward the stellar surface according to equation 3.1. The importance of radiative acceleration arises from the fact that it alters the surface chemical patterns in many stars. The extent of the predicted abundance variations varies with effective temperature, from none for HB stars cooler than about 5800 K (due to the very long time-scales for the process) to 2-4 dex in the hotter stars (T=20000 K), and also depends on the element considered. Observations of HB stars in GCs support the idea of radiative levitation being active above a certain temperature. Figure 3.1 (from Pace et al. 2006) shows the [Fe/H] ratio as a function of T for HB stars in NGC 2808: HB stars hotter than 12000 K have metallicities more than 1 dex higher than the cooler objects. This evidence supports the suggestion of Grundahl et al. (1999) that the onset of radiative levitation in stellar atmospheres may play a role in explaining the jump in magnitude toward brighter values for stars along the HB at effective temperatures of about 11500 K. This jump is seen in all CMDs of GCs that have Strömgren photometry of sufficient quality. The observed HB stars "return" to the theoretical ZAHB at temperatures between 15000 and 20000 K. The effective temperature of the jump is roughly the same for all GCs, irrespective of metallicity, central density, concentration, or mixing evidence. This photometric jump corresponds to that in the T-log(g) diagram of M15 discovered by Moehler et al. (1995). The mechanism suggested to be the most likely responsible for both features is the radiative levitation of heavy metals, that increases their atmospheric abundance. As a result the opacity due to the metallicity rises with respect to the hydrogen opacity. Hence there is an energy redistribution in which the Balmer jump is partially filled in, but the bolometric emission remains unchanged, and it is therefore seen as an increase in luminosity in the u Strömgren and U Johnson magnitudes. The occurrence of levitation was firstly argued by Michaud et al. (1983). They predicted that, if the outer envelope of hot HB stars is stable enough to allow He gravitational settling (as it was hypothesized by Greenstein 1967 to explain the observed helium underabundances), then there should be no mechanisms (like convection) able to prevent the metal enhancement caused by the radiation.

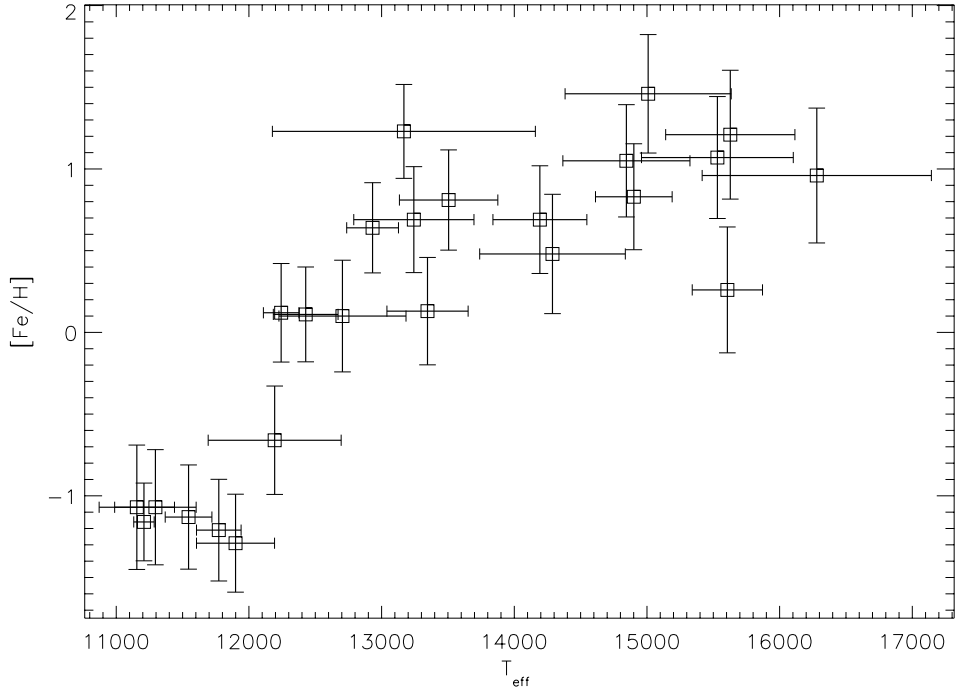

Figure 3.1: [Fe/H] ratio versus T from Pace et al. (2006). The coldest HB stars show abundances compatible with the cluster metallicity. The hottest ones shows Fe abundances up to more than 2 dex higher.

Analogous mechanisms are also invoked to explain the origin of the peculiar chemical patterns observed in Population I stars: MS stars of A and F spectral type in some open clusters (Varenne & Monier, 1999; Gebran et al., 2008, 2010; Gebran & Monier, 2008, OCs) and in the Ursa Major group (Monier, 2005) exhibit chemical abundance anomalies similar to those observed in GC HB stars. In particular, enhancement of iron peak elements has been observed, and large star-to-star scatter has been measured for many other elements, like C, O, Na, Mn, Ca, Sc, Sr and Ba. Finally, as already discussed, chemical anomalies can also be expected for BSSs, the MT-BSSs being C and O depleted.

Accurate high-resolution spectroscopic studies not only revealed the presence of all the chemical anomalies discussed above, but also showed that different stellar populations in GCs and OCs can behave differently also in terms of kinematical properties, in particular in terms of rotation. In fact, while TO and sub-giant branch stars in GCs always show low rotational velocities (few $\mathrm{km\,s^{-1}}$ only; Lucatello & Gratton 2003), a bimodal rotational velocity distribution has been observed for HB stars: the cooler HB stars (T<11000 K) have rotations between 10 and 40 $\mathrm{km\,s^{-1}}$ (much faster than old MS stars), while hotter HB stars show a markedly different behaviour and rotate at less than 10 $\mathrm{km\,s^{-1}}$ (Peterson et al. 1995b; Behr et al. 2000a; Behr et al. 2000b; Recio-Blanco et al. 2002, 2004). Even if the picture is still not completely clear, there are hints that rotational rates and gravitational settling are strictly related each other and possibly self-powered. In fact, the high rotation of cool HB objects could be explained by assuming that the progenitor red giant branch stars had rapidly rotating cores and differential rotation in their convective envelopes, and that angular momentum is redistributed from the former to the latter (Sills & Pinsonneault, 2000). Moreover, above a critical value of the equatorial rotational velocity, meridional circulation could prevent the gravitational settling of He and the radiative levitation of Fe-peak elements (which are already inefficient for these stars), thus contributing to explain the lack of chemical anomalies in HB stars with T<11000 K (see above). On the contrary, the decrease in rotation rates for HB stars toward higher temperatures is not predicted by the models. Vauclair (1999) suggests that it could be interpreted as a result of the gravitational settling, creating a mean molecular weight gradient that inhibits the angular momentum transport in the star. An alternative scenario to explain the decreasing rotation rates toward higher temperature for HB stars is provided by Vink & Cassisi (2002): according to their models, the radiative levitation of heavy elements triggers a stellar wind that can significantly remove angular momentum. Rotational velocities of A and F stars in OCs are typically lower than 100-150 $\mathrm{km\,s^{-1}}$, and, at odds with HB stars, no trend with stellar temperatures or chemical abundances is observed (Gebran & Monier, 2008; Gebran et al., 2008, 2010). Such a behaviour well agrees with the predictions of some models (Charbonneau & Michaud, 1991) where the time-scales of diffusion processes (including radiative levitation) are much shorter than those of rotational mixing. Concerning BSSs, a rapid rotation is expected in the MT scenario because of the angular momentum transfer (Sarna & De Greve, 1996). High rotational velocities are also expected for COL-BSSs (Benz & Hills, 1987), but some magnetic braking or disk locking mechanisms might intervene to significantly decrease the rotation (Leonard & Livio, 1995; Sills et al., 2005). Unfortunately, the efficiencies and time-scales of these mechanisms are still unknown, thus preventing a clear prediction of the expected rotational velocities. From an observational point of view, only one BSS rotating at $\simeq 80$ km s$^{-1}$ and no correlation between CO-depletion and rotational values have been found in 47 Tucanae (F06). Instead, the largest sample of fast rotating BSSs ever observed in a GC has been revealed in M4 by L10, $\sim 40\%$ of the surveyed BSSs have rotational velocities larger than 50 km s$^{-1}$.

## 3.1 Observations

The observations were performed with FLAMES at the ESO-VLT in the UVES+GIRAFFE combined mode during two different runs: 073.D-0058(A) and 081.D-0356(A) (hereafter P73 and P81, respectively). P73 was carried out during three nights between April and July 2004, whereas three nights in June 2008 have been devoted to P81. Spectra have been acquired for 33 BSSs and 42 HB stars, most of them being observed in both P73 and P81. In order to compare the results obtained for BSSs and HB stars with a sample of unevolved cluster stars, we also analysed archive GIRAFFE data obtained within program 075.D-0125(A) (hereafter P75), including spectra for 86 TO and SGB stars (hereafter both called TO stars), already discussed by Lind et al. (2008). The GIRAFFE gratings and exposure times used in the various observing runs are listed in Table 3.1. During P73 we also observed six BSSs with the UVES Red Arm 580nm, that provides a wavelength coverage from 4800 to 6800 Å with a spectral resolution of $R \simeq 40000$. The spectroscopic target selection has been performed from a photometric catalog obtained by combining ACS@HST data for the central region (within $r < 140''$ from the cluster centre) and WFI@ESO observations for the outer region (Contreras Ramos et al. 2013, in preparation). In order to avoid contamination from spurious light in the spectrograph fibers, only the most isolated stars have been selected: we conservatively excluded targets having stellar sources of comparable or brighter luminosity within 3″. Figure 3.2 shows the CMDs for our ACS and WFI datasets, with the position of all the analysed targets.

| period | grating | element | $N_{exp}$ | time (h) |
|--------|---------|---------|-----------|----------|
| P73 | HR5A | Fe, Mg | 6 | 4.5 |
| | HR18 | O | 3 | 2 |
| | UVES Red 580 | Fe, Mg | 9 | 6.5 |
| P81 | HR15N | H$\alpha$ | 2 | 1.5 |
| | HR18 | O | 3 | 2 |
| P75 | HR5B | Fe, Mg | 8 | 8 |

Table 3.1: Details about FLAMES observations for each observing run. The adopted gratings, the sampled elements, the number of exposures and the total integration time have been listed.

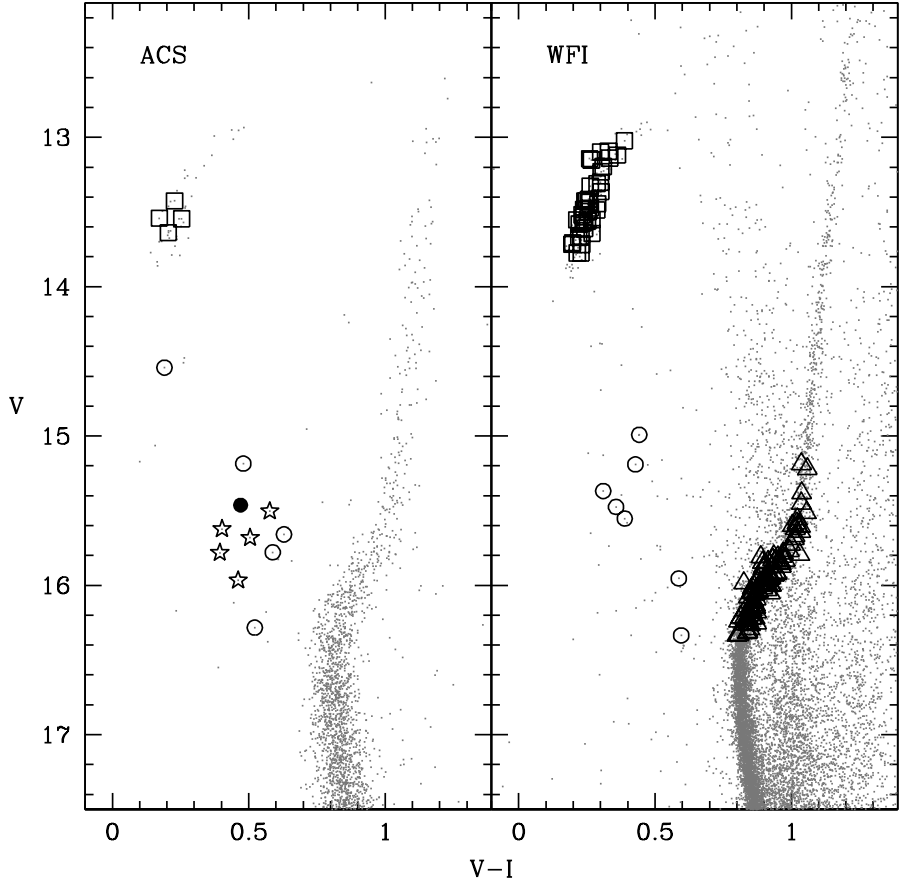

Figure 3.2: Spectroscopic targets in the CMDs of NGC 6397 for the ACS and WFI datasets. Open triangles and squares are TO and HB respectively, whereas large circles and stars are BSSs. In particular, star symbols mark the BSSs that show radial velocity variations and are also identified as SX Phe by Kaluzny & Thompson (2003). The filled black circle represents the fast rotating BSS #1100126 (that is also a SX Phe).

## 3.2 Data Reduction, radial velocities and cluster membership

All the spectra acquired during P73 and P81 have been reduced by using the last version of the GIRAFFE and UVES ESO pipelines [2], whereas for the P75 data pre-reduced spectra have been retrieved from the GIRAFFE archive maintained at the Paris Observatory [3]. The reduction procedure includes the bias subtraction, flat-field correction, computation of the dispersion solution by using a Th-Ar reference lamp and finally the extraction of the one-dimensional spectra. When it was possible, the accuracy of the wavelength calibration has been checked by measuring the position of some telluric emission lines (Osterbrock et al., 1996). The master sky spectrum, needed to sample the sky contribution, was computed by averaging several spectra of sky regions without bright stellar sources and was subtracted from each exposure. In order to measure the RVs of our targets, we have used the IRAF task *fxcor* and some tens of Fe I and Fe II lines, the Mg II line at $\lambda \approx 4481$ Å and the O I triplet at $\lambda \approx 7774$ Å . The H$\alpha$ Balmer line was used for stars that do not show significant lines in other gratings (because of the very high temperatures, high rotational velocities and/or low signal to noise ratio). As templates for *fxcor* we used synthetic spectra computed with the photometric parameters of each target (see details in Section 3.5). The RV distribution of the TO+HB stars is shown in Figure 3.3. The mean value is $\langle RV \rangle = 20.3 \pm 0.3$ ($\sigma = 3.8$) km s$^{-1}$, in good agreement with previous studies (see e.g. Harris 1996, 2010 edition; Milone et al. 2006; Lind et al. 2008; Hubrig et al. 2009; González Hernández et al. 2010). This has been adopted as the systemic velocity of NGC 6397 and used to assign the cluster membership to each star: only stars having radial velocity in agreement with this value within $3\sigma$ have been considered as members of NGC 6397. The inset of Figure 3.3 shows the radial velocity distribution for the entire sample of BSSs. As evident, BSSs span a wide range of values from $\sim -200$ km s$^{-1}$ up to $\sim +300$ km s$^{-1}$, with the bulk of population between 0 km s$^{-1}$ and 30 km s$^{-1}$. The application of the $\sigma$-rejection algorithm with respect to the cluster systemic velocity yields to exclude 15 objects, which probably are field stars. For all of them, the RV resulting from each exposure exceed the systemic velocity by more than $3\sigma$. The remaining 18 BSSs display the RV distribution shown (as shaded histogram) in the main panel of Figure 3.3, with a mean value $\langle RV \rangle = 20.1 \pm 1.5$ ($\sigma = 6.9$) km s$^{-1}$. Only two (namely #81828 and #1100063) out of six BSSs observed with UVES fibers result to be cluster members. Figure 3.4 shows the position with respect to the cluster centre for all member BSSs. We also found that BSSs #2200239, #1100127, #110126 and #1100170 correspond to stars V10, V11, V15 and V23, respectively, in Kaluzny & Thompson (2003) and are classified as variable SX Phe stars. For all the BSSs classified as cluster members, we checked the RV values (from each exposure) as a function of the Julian Day, to search for any relevant variability of RV. We found that only 4 BSSs (#1100126, #1100162, #1100208, #2200239) exhibit small RV variations. Furthermore, all RV variable stars are grouped in a restricted region of the CMD, close to the location of SX Phe stars. It is then probable that all these stars are pulsating variables. RV values together with other relevant information for the BSS, HB and TO samples are listed in Table 3.2, Table 3.3 and Table 3.4 respectively. For all the member stars we combined the rest-frame spectra of each exposure thus obtaining a medium spectrum with signal-to-noise ratio S/N $\simeq$ 50-100 for BSSs, S/N $\simeq$ 150-200 for HB stars and S/N $\simeq$ 50-70 for TO stars.

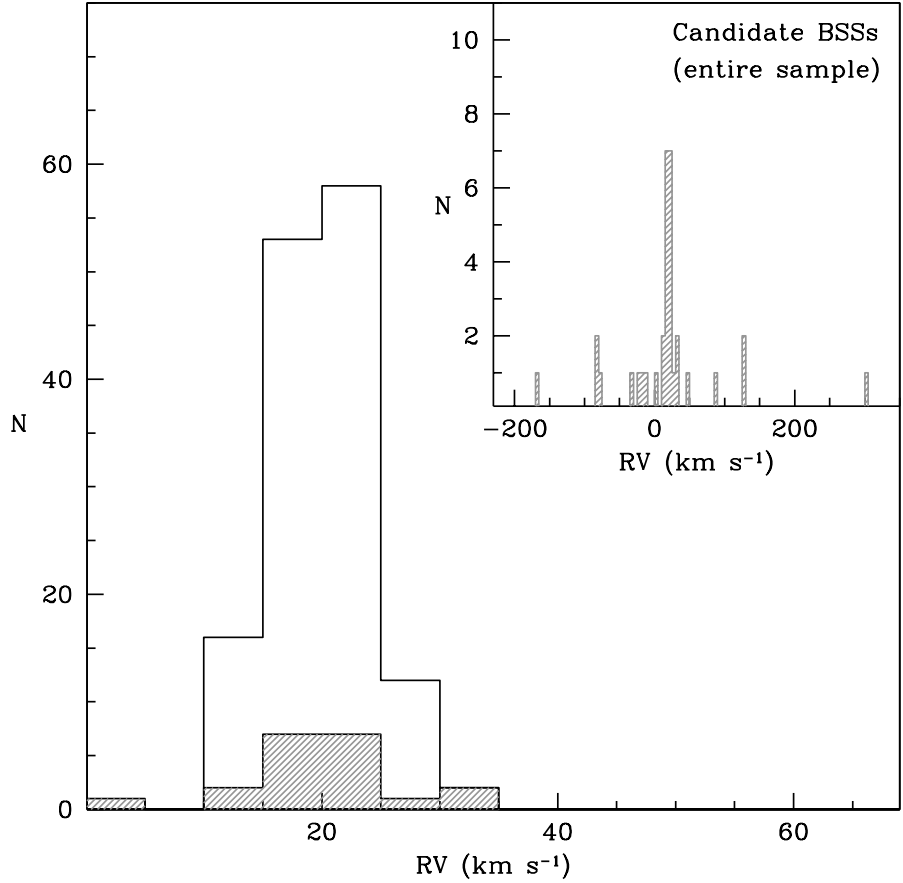

Figure 3.3: Radial velocity distribution for the TO+HB and the BSS samples. The inset shows the RV distribution for all the BSSs. The empty histogram in the main panel represents the RV distribution for the TO+HB stars, whereas the shade-histogram represents the BSS distribution of the BSS selected as cluster members.

_________________

[3]http://giraffe-archive.obspm.fr/

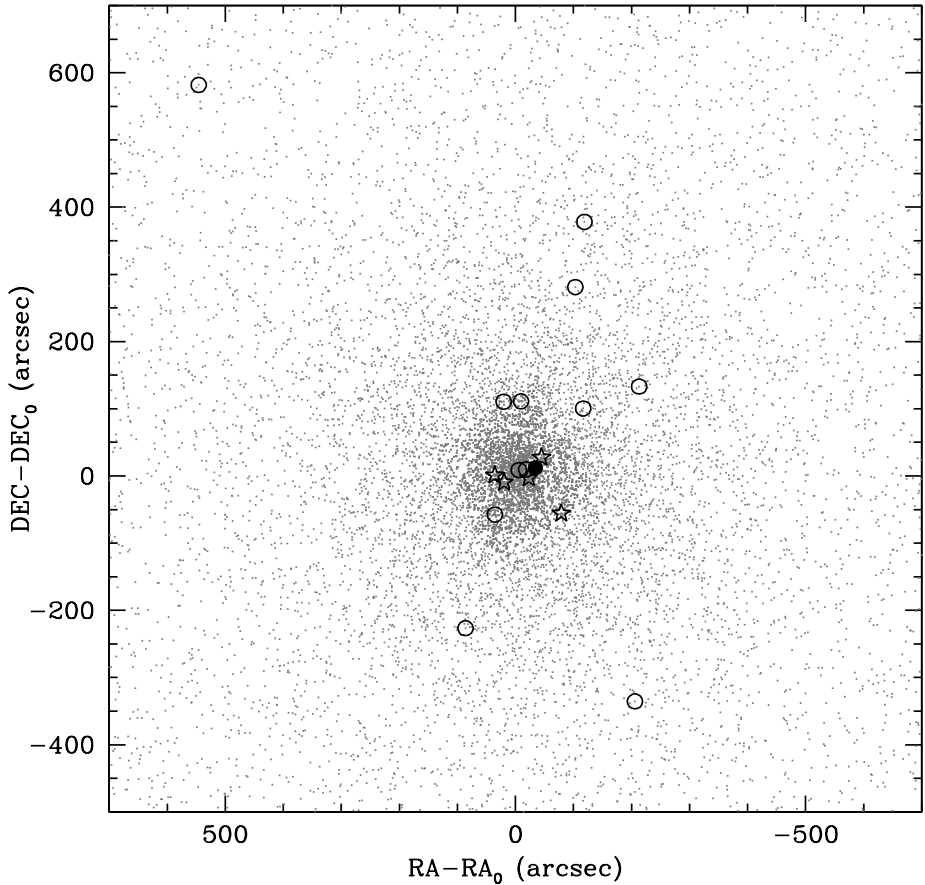

Figure 3.4: Position of member BSSs with respect to the cluster centre (symbols have the same meaning of Fig. 3.2).

## 3.3 Atmospheric parameters

The atmospheric parameters for our targets have been derived photometrically, according to their position in the CMD and the comparison with theoretical stellar models taken from the BaSTI database (Pietrinferni et al., 2006). The best-fit isochrone of the cluster sequences (with age=13.5 Gyr, metallicity Z=0.0003, $\alpha$-enhanced chemical mixture and assuming a distance modulus of 11.92 and E(B-V)=0.19, in agreement with Ferraro et al. 1999b) has been used to infer temperatures and gravities for TO stars (see Table 3.4). The ZAHB model taken from the same database has been used to infer parameters and masses of the HB stars (see Table 3.3). Finally, by projecting the star position on a set of isochrones with different ages, we derived T, $\log g$ and masses for the BSSs sample (see Table 3.2). In the case of M4, temperatures for BSSs have been derived by fitting the wings of the H$\alpha$ Balmer line. However this technique is more reliable in the range T$\sim$ 5500-8000 K, where the broadening of the wings is driven by the self-resonance broadening of the H atoms. For higher temperatures the H$\alpha$ wings mainly suffer from the Stark broadening and become less sensitive to temperature variations. According to the photometric determination, the majority of the BSSs in our NGC 6397 sample are hotter than 8000 K, their temperatures ranging from $\sim$ 7400 K up to $\sim$ 13000 K. Since this is the range where the sensitivity of the H$\alpha$ is reduced we did not apply that technique in the present study. Errors in temperatures and gravities were computed by assuming typical uncertainties in magnitude and colors for all the targets: we conservatively estimate $\delta$V $\lesssim$0.05 mag and $\delta$(V-I)$\lesssim$0.07 mag for TO and BSSs, and $\delta$V $\lesssim$0.02 mag and $\delta$(V-I)$\lesssim$0.028 mag for HB stars. Typical values for temperature errors are 50-150 K for BSSs, 100 K for TO and 50-80 K for HB. The uncertainty in the $\log g$ determination is substantially negligible (lower than 0.1) for all the targets. The microturbulent velocity is generally derived spectroscopically, by requiring that weak and strong lines of a given species (usually Fe) provide the same abundance within the uncertainties. In most of our targets, however, the number of available Fe lines is not large enough to allow a direct determination of this parameter. We therefore used the relation by For & Sneden (2010) to assign a microturbulent velocity value to our HB stars according to their temperature. For the BSS sample, we assumed 0 km s$^{-1}$ for all the targets. In fact, the analysis of the two BSSs observed with UVES (that have temperatures of 8299 K and 13183 K, covering almost the entire temperature range of the BSS sample) for which a large number of Fe lines with different strength are available, fully confirms that the adopted value is well appropriate for the BSSs. Finally, we assumed 1.5 km s$^{-1}$ as microturbulent velocity for TO stars (according to L10). A conservative error of 1 km s$^{-1}$ has been assumed for the microturbulent velocity of BSSs and HB, and 0.5 km s$^{-1}$ for TO stars.

## 3.4 Rotational velocities

Projected rotational velocities were measured from the analysis of the most prominent atomic lines (namely the Mg II line and the O I triplet in the GIRAFFE spectra and some Fe I line in the UVES ones) [4]. We performed a $\chi^2$ minimization between the observed spectrum and a grid of synthetic spectra, computed with different values of rotational velocities and by taking into account the instrumental profile, the microturbulent and the macroturbulent velocity and the Doppler broadening. The instrumental profile has been derived by measuring the FWHM of bright unsaturated lines in the reference Th-Ar calibration lamp (see i.e. Behr et al., 2000a).

---

[4]Only for two stars for which the only observed spectral feature is the H$\alpha$, we could not measure v sin($i$). In fact, the line profile is insensitive to rotational velocities smaller than 100 km s$^{-1}$.

The macroturbulent velocity was assumed to be 0 km s$^{-1}$ for BSSs and HB stars, whereas for TO stars the formula provided by Gray (1984) has been used. Finally, the Doppler and the microturbulent velocity broadenings have been computed by using the atmospheric parameters discussed in Section 3.3. In order to test the accuracy of our method we performed Monte Carlo simulations of 100 synthetic spectra with v sin($i$) = 0, 5, 10, 15 and 20 km s$^{-1}$, with the resolution and S/N typical of our observations. We recovered the original values within less than 2 km s$^{-1}$. Uncertainties on rotational velocities are 3-4 km s$^{-1}$ for TO, 1-2 km s$^{-1}$ for HB and 1 km s$^{-1}$ for the bulk of BSSs.

The values derived for 16 BSSs, HB and TO stars are listed in Table 3.2, 3.3 and 3.4, respectively. The distributions obtained for the different samples are compared in Figure 3.5. All the TO stars display very low rotational velocities, peaked at $\sim$ 7 km s$^{-1}$ and never exceeding $\sim$ 10 km s$^{-1}$ (in agreement with the values derived by Lucatello & Gratton 2003). In comparison, the HB sample exhibits a much broader distribution: only few HB stars show rotational velocity around $\sim 10 \, \mathrm{km \, s^{-1}}$ and v $\sin(i)$ can be as high as $\sim 42 \, \mathrm{km \, s^{-1}}$. The only previous analysis for HB stars in NGC 6397 is provided by Hubrig et al. (2009) that measured v $\sin(i) \sim 8 - 10 \, \mathrm{km \, s^{-1}}$ for three blue HB stars (T>11500 K).

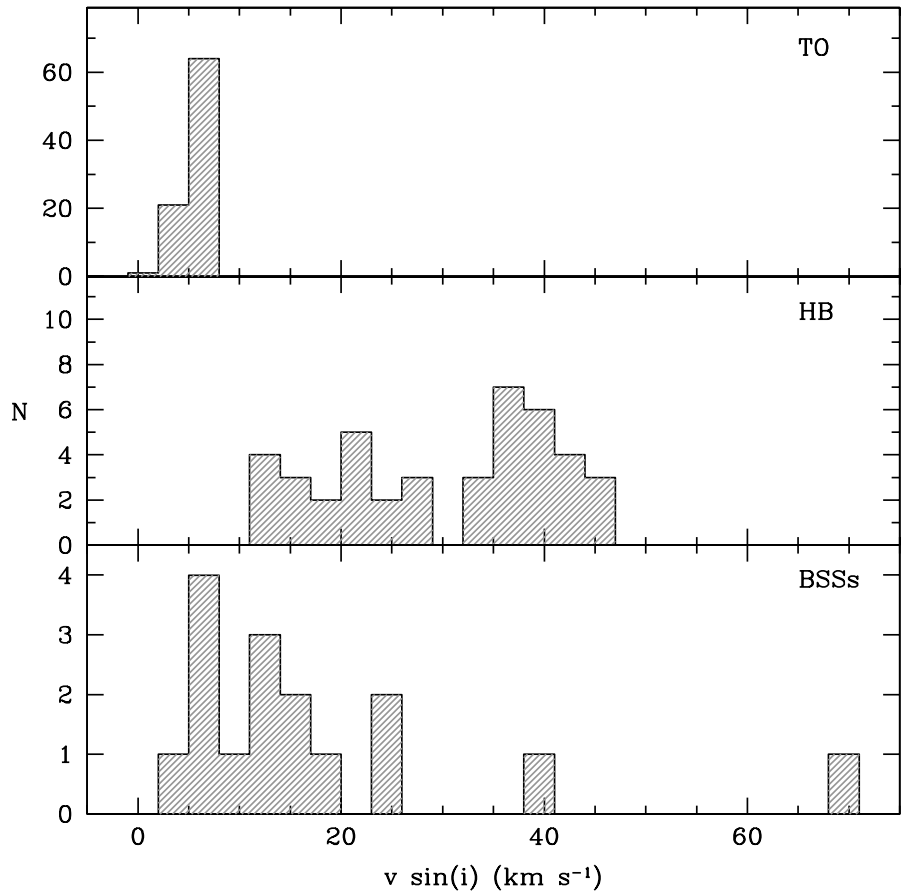

Figure 3.5: Rotational velocity distribution for TO, HB stars and BSSs.

The rotational velocities distribution for BSSs also results to be very broad, ranging from $\sim 0 \, \mathrm{km \, s^{-1}}$ up to $70 \, \mathrm{km \, s^{-1}}$, with an average value of $\langle$ v $\sin(i)\rangle$=18.2±1.0 $\mathrm{km \, s^{-1}}$ ($\sigma$=16.7) and a median value of 13 $\mathrm{km \, s^{-1}}$. In particular, BSS #1100126 (which is shown in Figure 3.2 as a filled circle) is the fastest rotating star in our sample, with v $\sin(i) = 70 \, \mathrm{km \, s^{-1}}$. Interestingly enough, among BSSs with T<8500 K RV variable stars (included SX Phe) show the largest rotational velocities, whereas the other ones have values compatible with the TO distribution. While no trend between rotational velocity and temperature is observed for TO and HB stars, a systematic increase is found for BSSs (Figure 3.6, top panel). The hottest BSS, however, show low rotation (less than 20 $\mathrm{km \, s^{-1}}$).

## 3.5   Chemical analysis

Iron, magnesium and oxygen abundances have been derived for almost all our targets (according to the available gratings): several tens of Fe I and Fe II lines, the Mg II line at 4481 Å and the O I triplet at $\sim 7774$ Å have been used. Chemical abundances have been derived by using both the measured EWs and spectral synthesis through the comparison with synthetic spectra. In order to calculate abundances for all the elements, our program GALA (Mucciarelli et al. 2013, ApJ submitted, see Appendix A) and the set of codes developed by R.L.Kurucz (Kurucz, 1993; Sbordone et al., 2004) have been used: model atmospheres have been computed by using ATLAS9, whereas WIDTH9 and SYNTHE have been used to obtain chemical abundances from the measured EWs and to compute synthetic spectra, respectively. In particular, ATLAS9 model atmospheres have been computed under the assumption of LTE plane-parallel geometry and by adopting the new opacity distribution functions by Castelli & Kurucz (2003), without the inclusion of the *approximate overshooting* (see Castelli et al. 1997). Atomic data for all the lines are from the most updated version of the Kurucz line list by F.Castelli [5]. Finally, the used reference solar abundances are from Grevesse & Sauval (1998) for iron and magnesium, and from Caffau et al. (2011) for the oxygen. The EWs of Fe I, Fe II and O I were measured with our own code that fits the absorption lines with a Gaussian profile (see L10 and Mucciarelli et al. 2011a); the abundance calculation was then performed with GALA. For the Mg II line (an unresolved blending of three close components belonging to the same multiplet) we derived abundances through a $\chi^2$ minimization between the observed spectra and a grid of synthetic spectra. Only for BSS #1100126 (the most fast rotating star of the sample) all the abundances were derived by comparison with synthetic spectra, because the observed line profile significantly deviates from the Gaussian approximation. An important point to recall (at least for the BSSs and HB stars analysed here) is the possible departure from the LTE assumption: in fact, the photospheric layers of hot stars are exposed to the strong UV radiation coming from the stellar interior and this effect is magnified in metal-poor stars because of the low opacity of their photospheres. The magnitude and the sign of the non-LTE corrections are still highly uncertain, due to the incompleteness of the used atom models (in particular for iron) and the uncertainties about the collision rate with the H I atoms. For the computation of the O abundances we included non-LTE corrections taken from the statistical equilibrium calculations of Takeda (1997). Instead, for the Mg and Fe abundances, no grid of non-LTE corrections are available in the literature for the range of parameters typical of our targets. Mashonkina et al. (2011) computed LTE and non-LTE abundances for some iron transitions in A and F spectral type stars, pointing out that non-LTE corrections are relevant for Fe I lines, whereas Fe II lines are negligibly affected by these effects (at the level of a few hundredths of dex). Thus, in the following we will use the abundances derived from Fe II lines as proxies of the iron content of BSSs and HB stars, in order to minimize the non-LTE effects. On the contrary, the iron content of TO stars is derived from Fe I lines, because these transitions are much more numerous with respect to the Fe II ones and because the non-LTE corrections for these lines should be negligible in dwarf stars (see e.g. Gratton et al. 1999). Concerning the magnesium, Przybilla et al. (2001) discussed the line formation for this feature, pointing out that large ($>0.2$-$0.3$ dex) corrections are expected at least for A type stars with solar metallicity. Since however no corrections are available for the Mg II line at 4481 Å , Mg abundances have not been corrected for non-LTE effects.

---

[5]http://wwwuser.oat.ts.astro.it/castelli/linelists.html

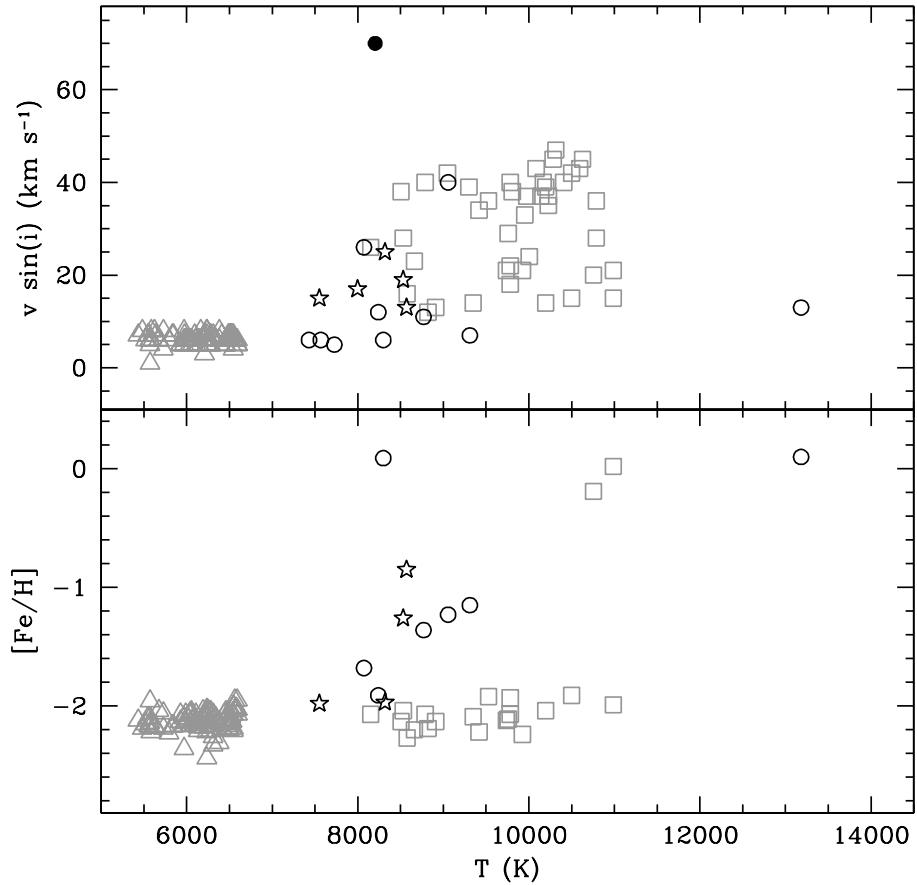

Figure 3.6: Rotational velocities (top panel) and [Fe/H] (bottom panel) as a function of the stellar temperatures for TO (empty grey triangles), HB (empty grey squares) and BSSs (same symbols used in the previous figures). Values and errors for all the targets are listed in tables 3.2, 3.3 and 3.4.

For BSS #1100063 (observed with UVES) we have been able to measure also the He I line at 5876 Å , a transition that is visible only in stars hotter than $\sim$9000 K. The He abundance has been derived through spectral synthesis. To generate the grid of synthetic spectra with different He abundances, we have computed for this star a number of suitable model atmospheres by using the ATLAS12 code (Castelli, 2005a) that allows to calculate model atmospheres with arbitrary chemical compositions through the use of the *opacity sampling* method, at variance with ATLAS9.

## 3.6 Chemical abundances

### 3.6.1 Iron

For the TO stars we find an average iron abundance [Fe/H]=$-2.12\pm0.01$ ($\sigma$=0.08) dex, in very good agreement with the values published in the literature (Harris, 1996, 2010 edition; Thévenin et al., 2001; Gratton et al., 2001; Lind et al., 2008). However, for BSSs and HB stars we obtain larger values and significantly higher dispersions. In fact, for the 11 BSSs for which we measured Fe II lines, we obtain [Fe/H] = $-1.20 \pm 0.22$ ($\sigma$ = 0.74) dex and for the entire HB sample we find [Fe/H]= −1.94 ± 0.14 (σ = 0.63) dex. The dispersions observed in the BSS and HB samples are completely incompatible with the uncertainties of the measures. In particular, as shown in Figure 3.6, the iron abundances of BSSs exhibit a systematic trend with the temperature: the coldest BSSs (with T∼7400-8200 K) have [Fe/H]=−1.88 ± 0.14 dex, that is marginally in agreement with the TO iron content, whereas for T>8200 K, [Fe/H] increases up to solar values. A milder but similar behaviour is observed also along the HB sequence: stars with T<10500 K have an average iron content of [Fe/H]=−2.16±0.02 dex (σ = 0.10), whereas two of the three stars with T >10500 K have an approximately solar iron abundance. No trend between [Fe/H] and rotational velocity is found for BSSs (see Figure 3.7).

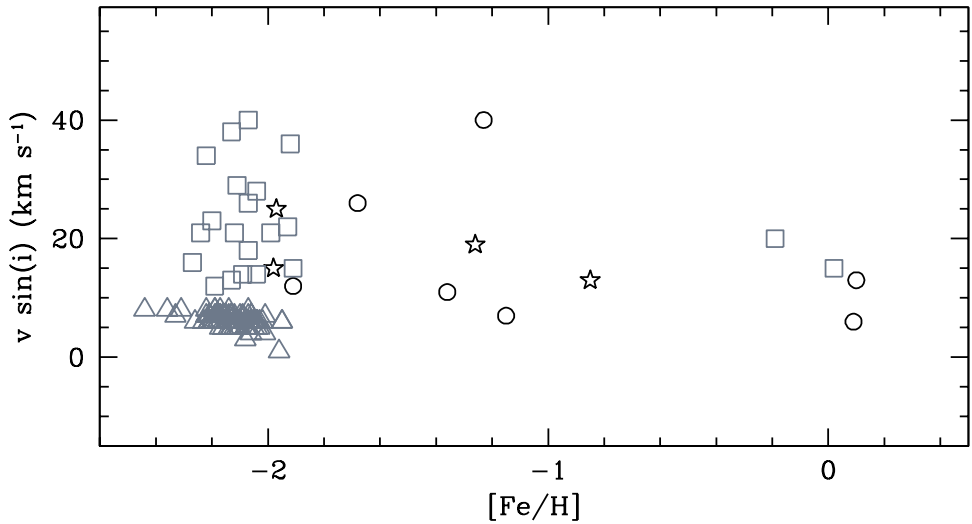

Figure 3.7: Rotational velocities as a function of the [Fe/H] for TO, HB stars and BSSs. Different symbols have the same meaning of Figure 3.2.

### 3.6.2 Magnesium

An increasing behaviour with T is also found for the BSS magnesium abundances, whereas no trend between [Mg/H] and T is detected for HB and TO stars (see Figure 3.8, top panel). In terms of [Mg/Fe] (lower panel in Figure 3.8) this evidence translate in a mild decrease of this parameter for increasing BSS and HB temperature, and a constancy of it for the TO population, with an average value of [Mg/Fe]∼0.2 dex. This value is ∼0.2 dex lower than the one provided by Carretta et al. (2009b). However, we remind that our Mg abundances do not include corrections for non-LTE, that could explain such a discrepancy.

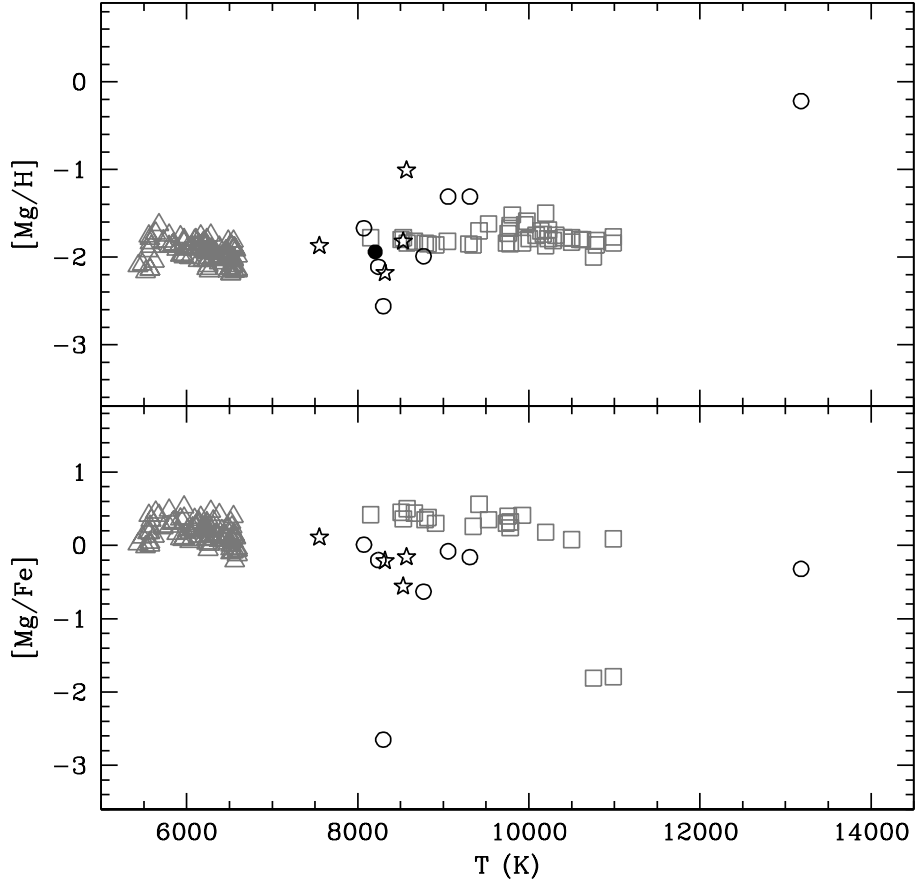

Figure 3.8: [Mg/H] and [Mg/Fe] as a function of stellar temperatures for TO, HB stars and BSSs (marked with the same symbols used in previous figures). Value and errors for BSS, HB and TO stars are listed in tables 3.2, 3.3 and 3.4.

### 3.6.3 Oxygen

HB stars show a large dispersion of [O/Fe] and a clear trend between O abundances and temperature (Figure 3.9): [O/Fe] decreases from enhanced values (0.75 dex) to sub-solar values (−0.47 dex) for increasing temperature. Almost all outliers in this temperature-oxygen anti-correlation are over luminous stars, likely evolved HB stars. Such a distribution highlights the intrinsic star-to-star scatter observed in all the GCs where light elements have been studied so far (see e.g. Carretta et al. 2009c) that is usually interpreted in the framework of self-enrichment processes occuring in the early stages of GC formation. We note that the range of the derived [O/Fe] ratios well matches the distribution recently discussed by Lind et al. (2011), who find

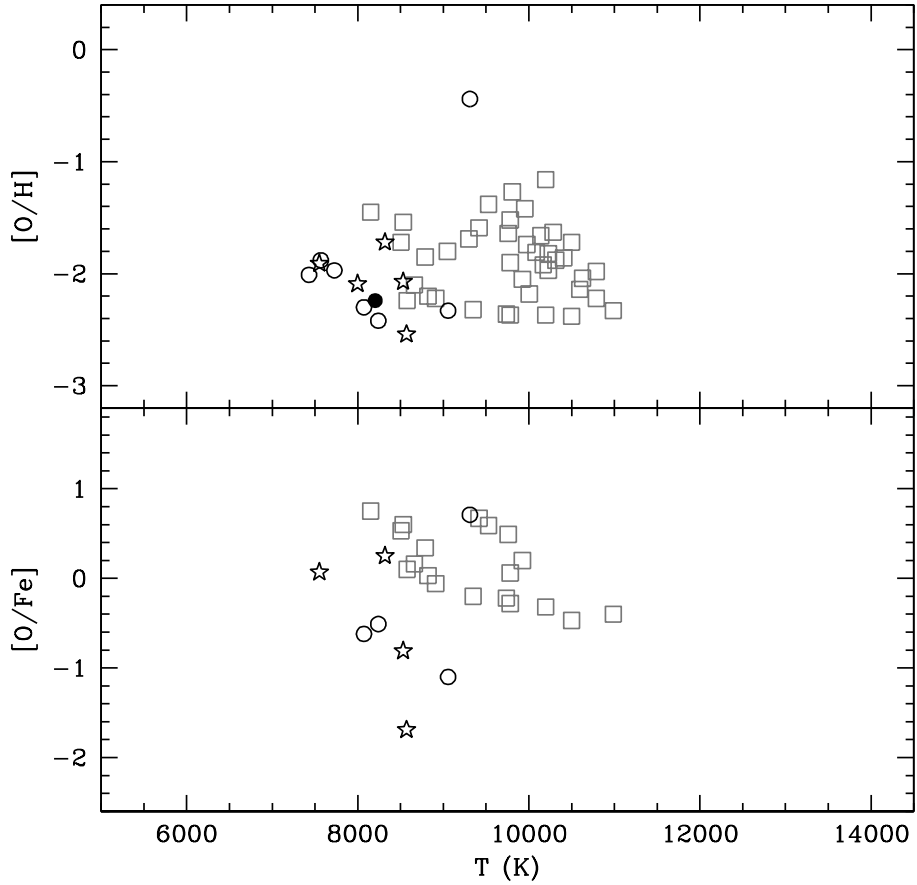

Figure 3.9: [O/H] and [O/Fe] as a function of the stellar temperatures for HB stars and BSSs (marked with the same symbols used in previous figures). Value and errors for BSS and HB stars are listed in tables 3.2 and 3.3.

[O/Fe]=+0.71 dex for the first generation stars and [O/Fe]=+0.56 dex for the second generation stars. The trend with effective temperature (corresponding to the star mass along the HB) is analogous to that observed in M4 (Marino et al., 2011b), NGC 2808 (Gratton et al., 2011) and NGC 1851 (Gratton et al., 2012), even if the scatter is higher. Note that low [O/Fe] abundance ratio observed in some of the hottest HB stars could be partially due to radiative levitation effects. This behaviour might be due to an anti-correlation between O and He abundances, taking into account the expected anti-correlation between the mass of evolving stars and the He content (see D'Antona & Caloi 2004). For BSSs we detect a mild decrease of [O/H] with T and one star (namely #79976) shows a very high oxygen abundance ([O/H]=−0.44 dex), higher than those measured for the other BSSs and the HB stars. Its spectrum is shown in Figure 3.10, top panel.

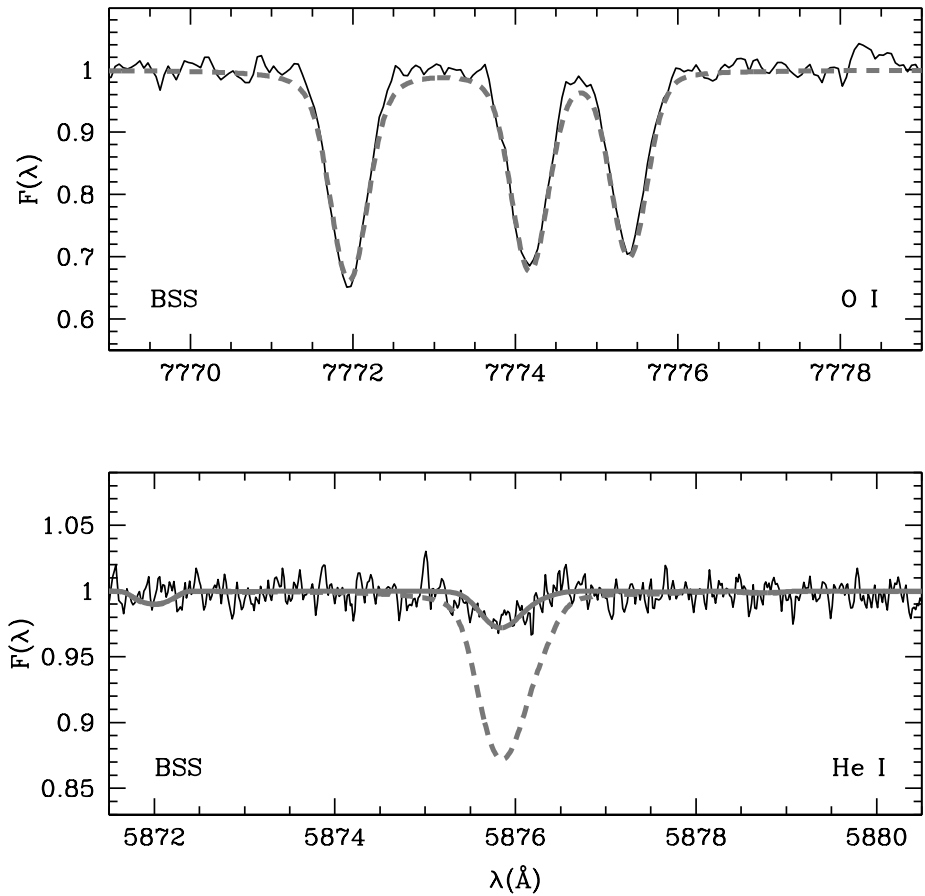

Figure 3.10: Top panel: comparison between the observed spectrum of the BSS #79976 (black line) and a synthetic spectrum with [O/H] = −0.44 dex (grey dashed line). Bottom panel: He I line at 5876 Å for BSS #1100063 compared with synthetic spectra with Y=0.25 (grey dashed line) and Y=0.001 (solid grey line).

### 3.6.4 Helium

BSS #1100063 exhibits a He I line significantly weaker than that predicted for a standard He abundance. In fact, the He I line (broadened by a moderate rotational velocity) is well reproduced with a helium mass fraction of Y=0.001, and it is totally inconsistent with Y=0.25 (Figure 3.10, bottom panel).

## 3.7 Discussion

The kinematical properties and chemical abundances here derived for the TO sample are in good agreement with previous determinations. They indicate that TO stars do not significantly rotate (v $\sin(i) \sim 7\,\mathrm{km\,s^{-1}}$) and are chemically homogeneous (at least in the iron and magnesium content), with very small dispersions around average values of [Fe/H]=−2.12 dex and [Mg/Fe]= 0.17 dex (see Section 3.6). On the other hand, the hot populations exhibit significantly different properties, in terms of both the rotational velocity and the chemical abundances.

### 3.7.1 Rotational velocities

The rotational velocities of HB stars in our sample range between $\sim 5$ and 45 $\mathrm{km\,s^{-1}}$, with a large spread for values lower than $\sim 25-30\,\mathrm{km\,s^{-1}}$ and a more peaked distribution at $\sim 35-40\,\mathrm{km\,s^{-1}}$. This is similar to that found by Peterson et al. (1995b) and Behr et al. (2000a). Unfortunately, there are no HB stars hotter than 11000 K in our sample so that we are not able to observe the bimodality of the rotational distribution already found in other GCs. A wide rotational distribution is also observed for the BSS population, with values of v $\sin(i)$ ranging from 0 to 70 $\mathrm{km\,s^{-1}}$. Although the statistics is low (only 16 stars in total), it is interesting to note that the two hottest stars in the sample (T > 9200 K) show rotational velocities lower than 20 $\mathrm{km\,s^{-1}}$, at odds with BSSs cooler than 9200 K which show a wide rotation distribution with a possible trend with temperature (the rotational velocity increasing with temperature). Additionally, no trend is observed between BSS rotational velocities and the [Fe/H] ratio (see Figure 3.7).

The distribution of v $\sin(i)$ obtained for the BSS population in NGC 6397 is qualitatively similar to that found in the other GCs studied so far, namely 47 Tuc and M4 (see Figure 3 in F06, and Figure 2 in L10, respectively; for 47 Tuc also see De Marco et al., 2005; Shara et al., 1997). In fact, in these GCs the rotation distribution is peaked at low values (consistent with the TO star velocities) and shows a long tail toward larger values. Quantitatively, however, significant differences can be recognized, especially in terms of the shape of the distribution, the highest measured values and the percentage of fast rotators. In fact, as shown in Figure 3.11, the distribution of v $\sin(i)$ between 0 and 20-30 km s$^{-1}$ is clearly peaked in 47 Tuc and M4, while in NGC 6397 it is more evenly large. In addition, while values as high as $100 - 150$ km s$^{-1}$ are found for a few BSSs in the two other clusters, the maximum measured value in NGC 6397 is v $\sin(i) = 70$ km s$^{-1}$. According to L10, by defining as "fast rotators" the BSSs spinning faster than $50$ km s$^{-1}$ , only 6-7% of such stars are found in NGC 6397 and 47 Tuc, while this fraction raises to 40% in M4. Clearly, the interpretation of such findings is not straightforward and more statistics is needed to explain these differences. The interpretation is unclear also in terms of the BSS formation channels, since conflicting predictions are still provided by the available theoretical models. In fact, high rotation rates are expected, at least in the early evolutionary stages, for BSSs formed by either direct collisions or mass transfer activity in binary systems. However, precise and solid theoretical predictions are still lacking, and possibly some (poorly constrained) braking mechanisms likely start to play a role in slowing down these stars during their subsequent evolution. Leonard & Livio (1995) predict that a convective zone able to slow down the star develops in the envelope of collisional BSSs, while the same phenomenon does not occurr in the models of Sills et al. (2005). The latter, instead, predict that a magnetically locked accretion disk forms around the star and is able to remove at least a fraction of the stellar angular momentum. Indeed, high rotational velocities could be just a transient phenomenon, characterizing only the early stages of (some) BSS life; then the rotation could slow down until those stars become indistinguishable from the BSSs generated with low rotational velocities. Such a scenario could partially explain the different fraction of fast rotating BSSs observed in 47 Tuc, M4 and NGC 6397.

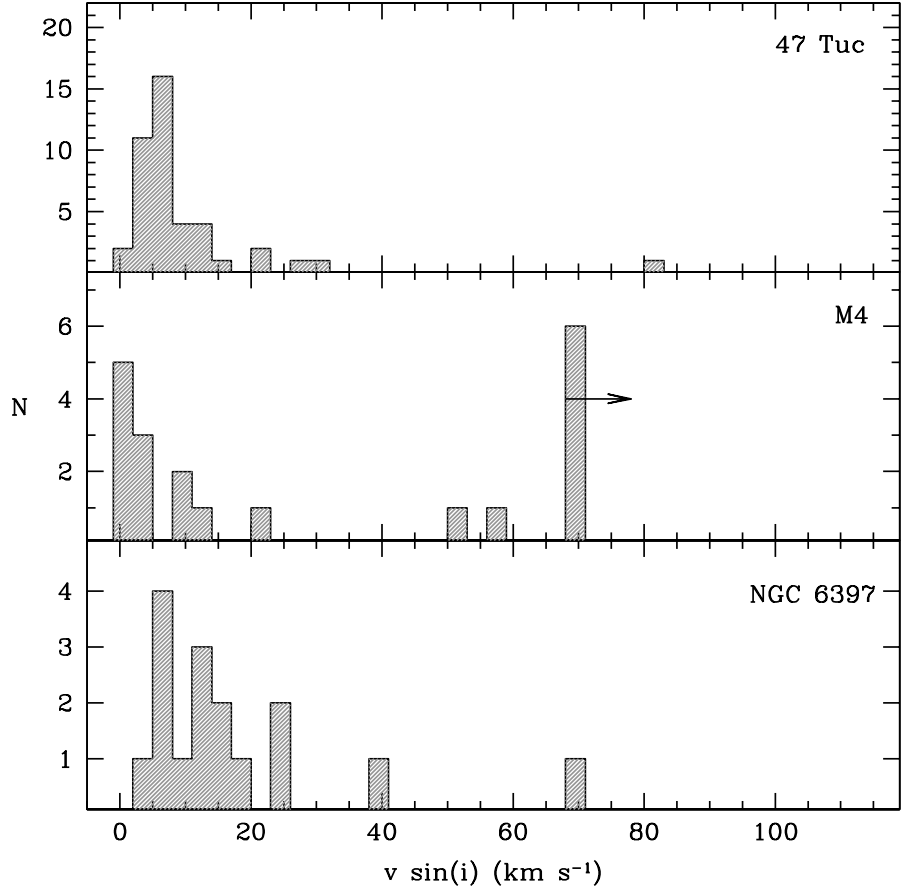

Figure 3.11: Rotational velocity distributions for the BSS populations in 47 Tuc, M4 and NGC 6397. The arrow in the central panel indicates the lower limit found for the fast rotators in M4.

### 3.7.2 Chemical abundances

As for the metallicity, with respect to the TO sample, the hot populations show a larger dispersion and a trend with the effective temperature (see Figures 3.6, 3.8, 3.9). Such features have been already observed in several GCs for HB stars hotter than $\sim 11000$ K (Behr et al. 1999, 2000a; Behr 2003; Moehler et al. 2000; Fabbian et al. 2005; Pace et al. 2006; Hubrig et al. 2009) and they are explained in terms of particle transport mechanisms (as radiative levitation and gravitational settling) occurring in the non-convective atmospheres of these stars (see, e.g., Richard et al., 2002; Michaud et al., 2008). Metal enhancements with respect to the initial composition are also observed in Population I main sequence stars hotter than $\sim 7000 - 8000$ K, where the surface convective zone starts to disappear (see Vick et al., 2010, and references therein). Although BSSs in NGC 6397 belong to a different stellar population (they are metal-poor, A-F type stars), a particle transport mechanism occurring in absence of convection could also explain the observational evidence presented here. Indeed, the observed behaviour of [Fe/H] and [Mg/H] as a function of T (with increasing abundances in the hotter stars) suggests that element transport mechanisms driven by radiative levitation occur at a threshold temperature of $\sim 8200$ K. Most of the BSSs cooler than this temperature shows Fe and Mg abundances similar to the pristine composition of the cluster (as derived from the TO population), while for hotter stars significant metal enhancements are detected. Such a scenario is strengthened by the significant He depletion observed in the hottest BSS of our sample. In fact, because of the gravitational settling, He is progessively diffused downward in the stellar interior, with a consequent reduction of its content on the surface. Since the convection is sustained by the He opacity, the convective zone starts to disappear and the elements with radiative acceleration larger than the gravitational one are diffused upward and enrich the photosphere with metals. Such a mechanism is particularly efficient in absence of stellar rotation, that otherwise would inhibit the metal levitation. Indeed, BSS #1100063 displays the highest temperature, a low rotational velocity and significant He depletion, all concurring to produce the observed remarkably high Fe abundance ([Fe/H]=0.10). Hence, while no theoretical models of levitation are currently available for stars similar to those analysed here (i.e. metal-poor, A-F type stars), these results and the fact that the onset of chemical anomalies in the BSS population occurs at temperatures similar to those predicted for solar abundance MS stars, suggest that BSSs behave like normal MS stars of the same spectral type. However, for the hot BSSs we also measured a significant enhancement of [Mg/H], that is not observed in HB and Population I MS stars, likely because of a balance between radiative and gravity accelerations.

While the observed chemical anomalies prevent us from drawing conclusions about the BSS formation mechanisms in this cluster, the present work provided us with the first information about the (still poorly understood) particle transport mechanisms in a range of metallicity and stellar mass not covered by other stellar systems. In this respect, similar studies of BSSs in metal-poor GCs, where the temperature of these stars are high enough for radiative levitation processes to occur, are highly desirable.

| ID(PSI) | RA (degrees) | DEC (degrees) | V | I | T (K) | log(g) | M (M☉) | RV (km s$^{-1}$) | v sin(i) (km s$^{-1}$) | [Fe/H] | [Mg/H] | [O/H] | Notes |
|---|---|---|---|---|---|---|---|---|---|---|---|---|---|
| 18705 | 264.9180785 | -53.5153682 | 15.55 | 15.16 | 8770 | 4.5 | 1.2 | 16.8 ± 0.7 | 11 ± 1 | -1.36 ± 0.17 | -1.99 ± 0.04 | — | — |
| 62594 | 265.2697513 | -53.7701740 | 16.33 | 15.74 | 7568 | 4.5 | 1.0 | 17.6 ± 0.8 | 6 ± 1 | — | — | -1.88 ± 0.07 | |
| 64782 | 265.1325939 | -53.7399209 | 14.99 | 14.55 | 8241 | 4.2 | 1.1 | 12.4 ± 0.4 | 12 ± 1 | -1.91 ± 0.09 | -2.11 ± 0.04 | -2.42 ± 0.09 | |
| 75241 | 265.2277155 | -53.6491197 | 15.95 | 15.37 | 7727 | 4.4 | 1.0 | 13.0 ± 0.7 | 5 ± 1 | — | — | -1.97 ± 0.14 | |
| 76278 | 265.2728458 | -53.6400582 | 15.47 | 15.12 | 9057 | 4.5 | 1.2 | 15.5 ± 0.6 | 40 ± 1 | -1.23 ± 0.16 | -1.31 ± 0.04 | -2.33 ± 0.09 | |
| 79976 | 265.2213165 | -53.5989412 | 15.37 | 15.06 | 9311 | 4.5 | 1.3 | 24.0 ± 0.2 | 7 ± 2 | -1.15 ± 0.07 | -1.31 ± 0.05 | -0.44 ± 0.06 | O-rich |
| 81828 | 265.2286114 | -53.5719809 | 15.19 | 14.76 | 8299 | 4.2 | 1.3 | 22.5 ± 0.4 | 6 ± 3 | 0.09 ± 0.06 | -2.56 ± 0.05 | — | |
| 100060 | 265.1757126 | -53.6745697 | 14.54 | 14.35 | 13183 | 4.7 | 2.0 | 17.5 ± 0.1 | 13 ± 6 | 0.10 ± 0.06 | -0.22 ± 0.18 | — | |
| 100126 | 265.1891195 | -53.6735768 | 15.46 | 14.99 | 8204 | 4.3 | 1.0 | 30.8 ± 3.4 | 70 ± 1 | — | -1.94 ± 0.04 | -2.24 ± 0.02 | FR, SX Phe |
| 100127 | 265.1837773 | -53.6778963 | 15.50 | 14.93 | 7551 | 4.1 | 0.9 | 23.8 ± 0.4 | 15 ± 1 | -1.98 ± 0.06 | -1.87 ± 0.05 | -1.91 ± 0.06 | SX Phe |
| 100157 | 265.1564245 | -53.6925979 | 15.66 | 15.03 | 7430 | 4.2 | 0.9 | 21.6 ± 0.4 | 6 ± 1 | — | — | -2.01 ± 0.13 | |
| 100162 | 265.2100723 | -53.6924561 | 15.62 | 15.22 | 8531 | 4.5 | 1.1 | 22.2 ± 0.5 | 19 ± 1 | -1.26 ± 0.12 | -1.82 ± 0.05 | -2.07 ± 0.13 | |
| 100170 | 265.1638009 | -53.6796128 | 15.68 | 15.18 | 7998 | 4.4 | 1.0 | 23.1 ± 0.5 | 17 ± 1 | — | — | -2.09 ± 0.13 | SX Phe |
| 100191 | 265.1816976 | -53.6743268 | 15.78 | 15.19 | 7638 | 4.3 | 0.9 | 15.9 ± 1.8 | — | — | — | — | |
| 100208 | 265.1940391 | -53.6693609 | 15.78 | 15.39 | 8570 | 4.6 | 1.2 | 26.6 ± 0.3 | 13 ± 1 | -0.85 ± 0.11 | -1.01 ± 0.05 | -2.54 ± 0.10 | |
| 2200110 | 265.1775032 | -53.6461613 | 15.18 | 14.70 | 8072 | 4.2 | 1.0 | 19.7 ± 0.3 | 26 ± 1 | -1.68 ± 0.09 | -1.67 ± 0.05 | -2.30 ± 0.12 | |
| 2200239 | 265.1563541 | -53.6767039 | 15.97 | 15.50 | 8318 | 4.6 | 1.2 | 10.8 ± 0.8 | 25 ± 4 | -1.97 ± 0.13 | -2.18 ± 0.03 | -1.72 ± 0.08 | SX Phe |
| 2200356 | 265.1635196 | -53.6463086 | 16.28 | 15.76 | 7816 | 4.6 | 1.1 | 23.4 ± 4.0 | — | — | — | — | |

Table 3.2: Coordinates, magnitudes, atmospheric parameters, masses, radial and rotational velocities, Fe, Mg and O abundances of the BSS sample.

| ID(P81) | RA (degrees) | DEC (degrees) | V | I | T (K) | log(g) | M (M☉) | RV (km s⁻¹) | v sin(i) (km s⁻¹) | [Fe/H] | [Mg/H] | [O/H] |
|---|---|---|---|---|---|---|---|---|---|---|---|---|
| 474 | 265.3579826 | -53.5451451 | 13.48 | 13.24 | 9978 | 3.6 | 0.6 | 19.2 ± 1.0 | 38 ± 1 | – | -1.59 ± 0.04 | -1.74 ± 0.10 |
| 1655 | 265.392097 | -53.5002154 | 13.14 | 12.80 | 8663 | 3.3 | 0.7 | 14.1 ± 0.4 | 21 ± 1 | -2.26 ± 0.07 | -1.82 ± 0.09 | -2.10 ± 0.17 |
| 9731 | 265.2902124 | -53.5461587 | 13.36 | 13.06 | 9529 | 3.5 | 0.7 | 19.9 ± 0.3 | 38 ± 1 | -1.97 ± 0.05 | -1.62 ± 0.05 | -1.38 ± 0.10 |
| 10485 | 265.1828792 | -53.5167585 | 13.72 | 13.49 | 10789 | 3.7 | 0.6 | 25.6 ± 0.9 | 32 ± 1 | – | -1.81 ± 0.03 | -1.98 ± 0.06 |
| 11063 | 265.2707910 | -53.4935456 | 13.49 | 13.20 | 9952 | 3.6 | 0.6 | 18.2 ± 0.4 | 33 ± 1 | – | -1.63 ± 0.04 | -1.42 ± 0.10 |
| 11363 | 265.1442501 | -53.4802352 | 13.73 | 13.51 | 10789 | 3.7 | 0.6 | 18.5 ± 0.5 | 23 ± 1 | – | -1.86 ± 0.03 | -2.22 ± 0.05 |
| 12158 | 265.1336942 | -53.4423799 | 13.72 | 13.52 | 10757 | 3.7 | 0.6 | 15.5 ± 0.5 | 9 ± 1 | -0.19 ± 0.14 | -2.00 ± 0.03 | – |
| 18540 | 265.0135736 | -53.5217368 | 13.23 | 12.93 | 9050 | 3.4 | 0.7 | 19.5 ± 0.5 | 40 ± 1 | – | -1.82 ± 0.04 | -1.80 ± 0.10 |
| 51874 | 264.9632133 | -53.6890207 | 13.77 | 13.56 | 10989 | 3.8 | 0.6 | 14.6 ± 0.2 | 12 ± 2 | 0.02 ± 0.16 | -1.77 ± 0.04 | – |
| 53805 | 264.9638040 | -53.6514855 | 13.30 | 13.00 | 9299 | 3.4 | 0.7 | 22.6 ± 0.9 | 36 ± 1 | – | -1.85 ± 0.03 | -1.69 ± 0.10 |
| 54043 | 265.0640413 | -53.6464338 | 13.64 | 13.37 | 10501 | 3.7 | 0.6 | 28.5 ± 0.9 | 38 ± 2 | – | -1.78 ± 0.03 | -1.72 ± 0.08 |
| 56322 | 264.9978322 | -53.5980665 | 13.44 | 13.15 | 9805 | 3.5 | 0.6 | 21.7 ± 0.4 | 41 ± 1 | – | -1.52 ± 0.05 | -1.27 ± 0.10 |
| 57042 | 265.0286135 | -53.5790632 | 13.55 | 13.28 | 10198 | 3.6 | 0.6 | 20.3 ± 0.9 | 42 ± 1 | – | -1.50 ± 0.05 | -1.16 ± 0.10 |
| 60940 | 265.2943802 | -53.7943407 | 13.14 | 12.88 | 8787 | 3.3 | 0.7 | 24.0 ± 0.9 | 38 ± 1 | -2.19 ± 0.07 | -1.84 ± 0.01 | -1.85 ± 0.11 |
| 61653 | 265.1776640 | -53.7844834 | 13.55 | 13.34 | 10227 | 3.6 | 0.6 | 28.4 ± 0.3 | 31 ± 1 | – | -1.79 ± 0.04 | -1.97 ± 0.07 |
| 63769 | 265.1643436 | -53.7537368 | 13.10 | 12.80 | 8578 | 3.3 | 0.7 | 20.6 ± 0.1 | 14 ± 1 | -2.34 ± 0.07 | -1.84 ± 0.04 | -2.24 ± 0.14 |
| 64937 | 265.2327151 | -53.7377808 | 13.43 | 13.17 | 9782 | 3.5 | 0.6 | 15.7 ± 0.4 | 41 ± 1 | – | -1.67 ± 0.05 | -1.52 ± 0.10 |
| 65798 | 265.1531778 | -53.7272727 | 13.50 | 13.24 | 10004 | 3.6 | 0.6 | 22.8 ± 0.4 | 20 ± 1 | – | -1.79 ± 0.04 | -2.18 ± 0.14 |
| 70707 | 265.2944846 | -53.6828826 | 13.33 | 13.07 | 9418 | 3.5 | 0.7 | 22.0 ± 0.3 | 34 ± 1 | -2.26 ± 0.05 | -1.70 ± 0.04 | -1.59 ± 0.10 |
| 72734 | 265.2751355 | -53.6692360 | 13.43 | 13.19 | 9782 | 3.5 | 0.6 | 13.8 ± 0.2 | 13 ± 1 | -2.09 ± 0.05 | -1.85 ± 0.04 | -2.37 ± 0.10 |
| 73084 | 265.2395639 | -53.6667402 | 13.12 | 12.76 | 8533 | 3.3 | 0.7 | 19.9 ± 0.2 | 27 ± 1 | -2.14 ± 0.07 | -1.78 ± 0.05 | -1.54 ± 0.12 |
| 73278 | 265.2576032 | -53.6653473 | 13.66 | 13.43 | 10596 | 3.7 | 0.6 | 30.0 ± 1.1 | 38 ± 1 | – | -1.80 ± 0.03 | -2.14 ± 0.05 |
| 74139 | 265.3052140 | -53.6585939 | 13.58 | 13.36 | 10316 | 3.6 | 0.6 | 22.8 ± 0.9 | 42 ± 1 | – | -1.75 ± 0.04 | -1.88 ± 0.08 |
| 74430 | 265.2862434 | -53.6562185 | 13.02 | 12.64 | 8151 | 3.2 | 0.7 | 19.8 ± 0.2 | 26 ± 1 | -2.20 ± 0.08 | -1.78 ± 0.05 | -1.45 ± 0.13 |
| 74883 | 265.2546642 | -53.6522206 | 13.47 | 13.21 | 9926 | 3.6 | 0.6 | 20.0 ± 0.2 | 16 ± 1 | -2.25 ± 0.06 | -1.84 ± 0.04 | -2.05 ± 0.13 |
| 76169 | 265.3280914 | -53.6409329 | 13.67 | 13.45 | 10628 | 3.7 | 0.6 | 27.0 ± 1.1 | 40 ± 2 | – | -1.80 ± 0.04 | -2.04 ± 0.06 |
| 76330 | 265.1989537 | -53.6396150 | 13.53 | 13.29 | 10141 | 3.6 | 0.6 | 30.6 ± 0.4 | 35 ± 1 | – | -1.70 ± 0.03 | -1.66 ± 0.10 |
| 76738 | 265.1755658 | -53.6357717 | 13.61 | 13.37 | 10407 | 3.7 | 0.6 | 31.6 ± 1.0 | 36 ± 1 | – | -1.78 ± 0.03 | -1.86 ± 0.08 |
| 77082 | 265.2399386 | -53.6322805 | 13.57 | 13.33 | 10286 | 3.6 | 0.6 | 14.0 ± 0.6 | 40 ± 1 | – | -1.81 ± 0.03 | -1.63 ± 0.10 |
| 77542 | 265.1136890 | -53.6274217 | 13.31 | 13.02 | 9349 | 3.4 | 0.6 | 13.2 ± 0.1 | 10 ± 2 | -2.12 ± 0.04 | -1.86 ± 0.04 | -2.32 ± 0.13 |
| 77637 | 265.2727433 | -53.6263464 | 13.09 | 12.76 | 8506 | 3.3 | 0.7 | 19.6 ± 0.7 | 37 ± 1 | -2.25 ± 0.12 | -1.80 ± 0.05 | -1.72 ± 0.11 |
| 80792 | 265.1226697 | -53.5878451 | 13.41 | 13.16 | 9737 | 3.5 | 0.7 | 14.7 ± 0.3 | 17 ± 1 | -2.14 ± 0.05 | -1.84 ± 0.04 | -2.36 ± 0.06 |
| 81455 | 265.1657796 | -53.5774433 | 13.77 | 13.55 | 10989 | 3.8 | 0.6 | 13.3 ± 0.2 | 15 ± 1 | -1.93 ± 0.08 | -1.84 ± 0.03 | -2.33 ± 0.08 |
| 82001 | 265.1649280 | -53.5693729 | 13.20 | 12.89 | 8911 | 3.4 | 0.7 | 18.2 ± 0.2 | 9 ± 1 | -2.16 ± 0.07 | -1.86 ± 0.04 | -2.22 ± 0.13 |
| 82561 | 265.2503688 | -53.5602131 | 13.56 | 13.31 | 10227 | 3.6 | 0.6 | 23.9 ± 1.0 | 35 ± 1 | – | -1.69 ± 0.04 | -1.82 ± 0.08 |
| 89014 | 265.4087478 | -53.7085078 | 13.15 | 12.87 | 8820 | 3.3 | 0.7 | 25.2 ± 0.1 | 9 ± 2 | -2.23 ± 0.05 | -1.85 ± 0.04 | -2.20 ± 0.17 |
| 90302 | 265.4928341 | -53.6737610 | 13.43 | 13.18 | 9759 | 3.5 | 0.6 | 22.5 ± 0.4 | 28 ± 1 | -2.13 ± 0.11 | -1.73 ± 0.04 | -1.64 ± 0.09 |
| 91315 | 265.4124151 | -53.6457029 | 13.52 | 13.29 | 10085 | 3.6 | 0.6 | 22.4 ± 0.8 | 41 ± 1 | – | -1.75 ± 0.04 | -1.81 ± 0.09 |
| 100024 | 265.1680268 | -53.6816479 | 13.54 | 13.37 | 10198 | 3.6 | 0.6 | 23.7 ± 0.2 | 7 ± 2 | -2.05 ± 0.04 | -1.87 ± 0.03 | -2.37 ± 0.09 |
| 100029 | 265.1822376 | -53.6849407 | 13.64 | 13.43 | 10501 | 3.7 | 0.6 | 19.5 ± 0.2 | 8 ± 1 | -1.91 ± 0.05 | -1.83 ± 0.04 | -2.38 ± 0.07 |
| 2200024 | 265.1991228 | -53.6643097 | 13.43 | 13.20 | 9782 | 3.5 | 0.6 | 21.9 ± 0.3 | 23 ± 1 | -1.96 ± 0.05 | -1.64 ± 0.04 | -1.90 ± 0.16 |
| 2200028 | 265.1423624 | -53.6557501 | 13.54 | – | 10169 | 3.6 | – | 18.9 ± 0.4 | 38 ± 1 | – | -1.74 ± 0.03 | -1.92 ± 0.08 |

Table 3.3: Coordinates, magnitudes, atmospheric parameters, masses, radial and rotational velocities, Fe, Mg and O abundances of the HB sample.

| ID(P81) | RA (degrees) | DEC (degrees) | V | I | T (K) | log(g) | M (M$_\odot$) | RV (km s$^{-1}$) | v sin(i) (km s$^{-1}$) | [Fe/H] | [Mg/H] |
|---|---|---|---|---|---|---|---|---|---|---|---|
| 9407 | 265.1440384 | -53.5557399 | 15.56 | 14.54 | 5546 | 3.5 | 0.7 | 17.2 ± 0.10 | 7 ± 4 | −2.10 ± 0.20 | −1.93 ± 0.10 |
| 17909 | 264.9203310 | -53.5488131 | 16.26 | 15.43 | 6546 | 4.1 | 0.7 | 21.9 ± 0.10 | 6 ± 3 | −2.21 ± 0.22 | −1.82 ± 0.04 |
| 41395 | 264.8446941 | -53.6477354 | 15.79 | 14.76 | 5728 | 3.6 | 0.7 | 18.4 ± 0.12 | 4 ± 3 | −2.06 ± 0.20 | −1.82 ± 0.10 |
| 41662 | 264.8421933 | -53.6398126 | 15.83 | 14.85 | 5848 | 3.7 | 0.7 | 19.2 ± 0.12 | 7 ± 4 | −2.17 ± 0.20 | −1.85 ± 0.10 |
| 43151 | 264.8440820 | -53.5979440 | 16.05 | 15.13 | 6281 | 3.9 | 0.7 | 20.3 ± 0.12 | 7 ± 3 | −2.21 ± 0.21 | −1.75 ± 0.06 |
| 46551 | 265.0716357 | -53.8067644 | 16.22 | 15.35 | 6501 | 4.0 | 0.7 | 18.7 ± 0.19 | 7 ± 3 | −2.15 ± 0.22 | −2.00 ± 0.04 |
| 48483 | 265.0570850 | -53.7609905 | 15.77 | 14.78 | 5728 | 3.6 | 0.7 | 19.2 ± 0.11 | 8 ± 4 | −2.19 ± 0.20 | – |
| 48487 | 265.0266095 | -53.7599815 | 16.00 | 15.09 | 6237 | 3.9 | 0.7 | 24.8 ± 0.11 | 7 ± 3 | −2.22 ± 0.21 | −1.93 ± 0.06 |
| 48562 | 265.0198484 | -53.7584877 | 15.96 | 15.03 | 6138 | 3.8 | 0.7 | 19.9 ± 0.11 | 6 ± 3 | −2.18 ± 0.21 | −1.86 ± 0.06 |
| 48646 | 265.0818209 | -53.7564627 | 15.95 | 15.05 | 6138 | 3.8 | 0.7 | 23.5 ± 0.10 | 6 ± 3 | −2.21 ± 0.21 | −1.91 ± 0.06 |

Table 3.4: Coordinates, magnitudes, atmospheric parameters, masses, radial and rotational velocities, Fe and Mg abundances of the TO sample. A complete version of the table is available in electronic form.

# Chapter 4

# BSSs in M30

It has been known since mid-1970s that some GCs have extremely compact cores. The observed central surface brightness profiles of these clusters are characterized by power-law cusps rather than resembling single-mass King models with resolved cores (Djorgovski & King, 1984; Lugger et al., 1987). The theory of GC dynamical evolution predicts the devolopement of such power-law structure during core collapse (Cohn, 1980). Djorgovski & King (1984) have thus classified clusters with a central surface brightness cusp as "post-core collapse" (PCC) and Djorgovski & King (1986) identified 21 GCs as being PCC, in a large-scale survey of core structure of 123 GCs. Modification of the stellar populations are very likely to occur in the central regions of these objects mainly through stellar interactions (Benz & Hills, 1987) and mass segregation (Elson et al., 1987). Color and population gradients are also observed in some GCs, in the sense of having bluer color inward (Djorgovski et al., 1991; Djorgovski, 1993), and radial changes in bright stellar populations are observed in those clusters exhibiting a visible color gradient: the brightest RGB stars seem to be depleted near the cluster centre when compared to the underlying light and to the blue HB stars (Stetson, 1994; Lauzeral et al., 1992; Piotto et al., 1988).

M30 (NGC 7099) is one of the 21 Galactic GCs that show evidence of having undergone core collapse (Djorgovski & King, 1986; Lugger et al., 1995). The collapsed state of the core has been confirmed by HST imaging (Yanny et al., 1994; Sosin, 1997; Guhathakurta et al., 1998). Sosin (1997) has placed an upper limit of 1.9" (0.08 pc) on the core radius of M30 from high resolution HST Faint Object Camera imaging. The extraordinarily high central density of M30, which may exceed $\sim 10^6$ $M_\odot$ pc$^{-3}$, makes its core and the surrounding power-law cusp one of the highest density environments in the Galaxy. A high rate of stellar interactions is expected in the core and cuspy regions, resulting in BSS formation via stellar mergers, binary formation via tidal capture and/or 3-body interactions, and binary interactions such as exchange encounters (Hut et al., 1992a). M30 shows one of the strongest evidence of stellar interactions ever seen in any GC: it has a very high BSS frequency, a very high bluer-inward color gradient and a large deficit of bright giants in the inner region (Guhathakurta et al., 1998; Howell et al., 2000). Moreover, recently two distinct and well defined BSS sequences have been found by Ferraro et al. (2009a, see Section 1.5). All these reasons make M30 a very interesting target for our research.

## 4.1 Observations

The observations were performed at the ESO-VLT by using the multi-object high-resolution spectrograph FLAMES in the UVES+GIRAFFE combined mode during 3 nights in August 2011.

The sample includes 12 BSSs and 52 RGBs. The spectroscopic target selection has been performed on a photometric catalog obtained by combining WFPC2@HST data for the central region and MegaCam (at the Canada-France-Hawaii telescope) observations for the outer region. We conservatively excluded targets having stellar sources of comparable or brighter luminosity within 3 arcsec. The FLAMES fibres allocation has been made in the attempt to maximize the number of the observed BSSs both in the blue and the red sequence. Unfortunately this was a very hard task: in fact, the majority of the BSSs are concentrated in the inner 30″ and some of them have close bright companions; moreover, the physical size of the FLAMES magnetic buttons, prevents us to simultaneously observe BSSs that are too close to each other. For all these reasons, we finally were able to take spectra only for 12 BSSs, 4 in the blue sequence and 8 in the red one. Figure 4.1 shows the CMD of M30: the analysed BSSs are marked with circles whereas RGBs with triangles.

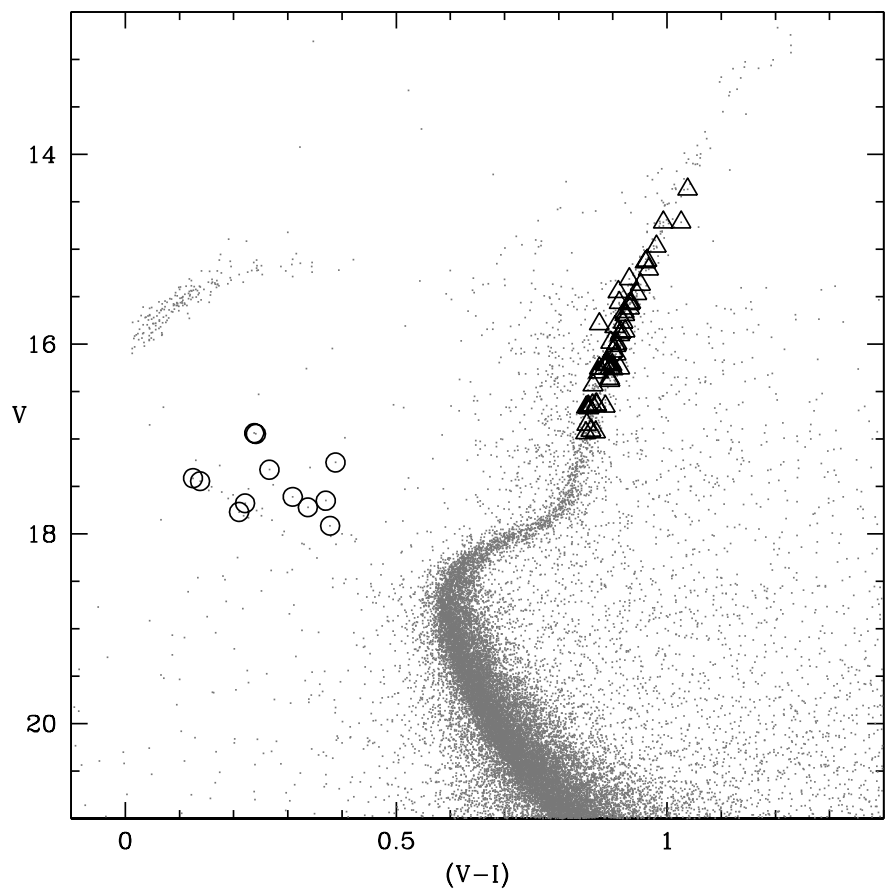

Figure 4.1: CMD of M30. The observed BSSs are marked with empty circles whereas RGBs are triangles.

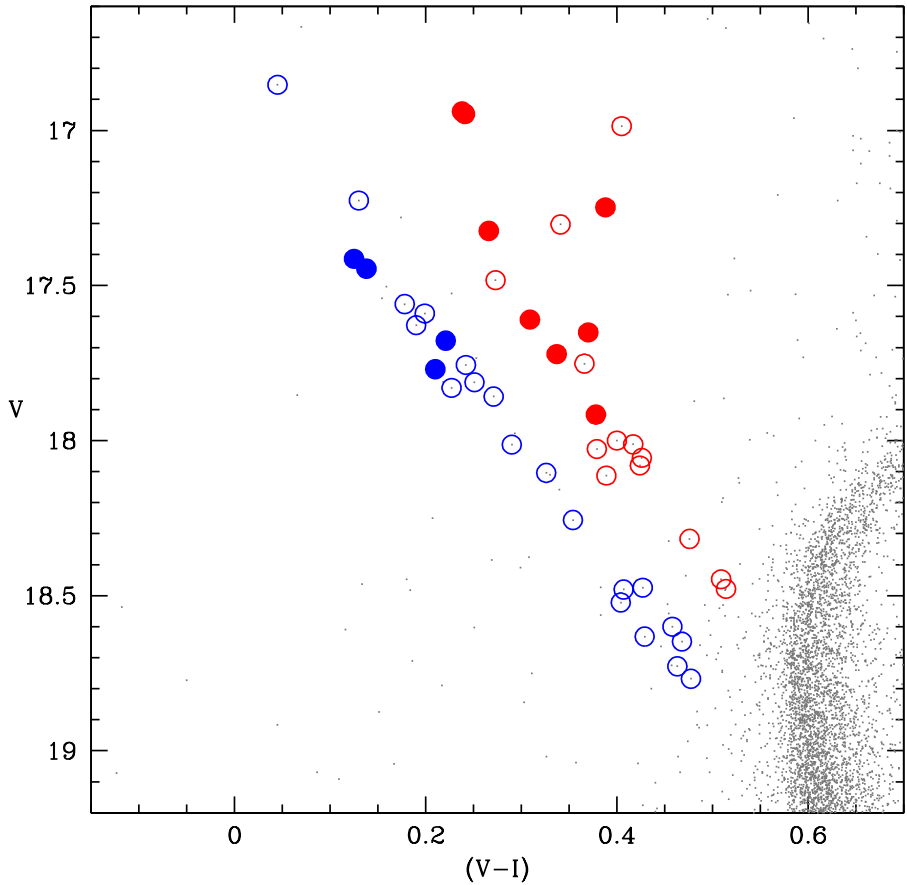

Figure 4.2: CMD of M30 zoomed in the BSS region. All the candidate BSSs in the two sequences are marked with empty (blue and red) circles, whereas filled circles mark the BSSs analysed in this work.

Figure 4.2 shows a zoom in the BSS region: empty circles label all the candidate BSSs in the GC M30, whereas filled circles mark the position in the two sequences of the observed BSSs. Finally, in Figure 4.3 the BSS positions with respect to the cluster centre are shown. The usual setups have been used for the observations: HR5A to sample some metallic lines and HR18 for the OI triplet at $\lambda \simeq 7774$ Å . Exposure times amount to 4.5 hours for the HR5A setup, and 3 hours each for the HR18. Spectra pre-reduction has been done by using the standard ESO pipeline that includes the bias subtraction, the flat-field correction, the wavelength calibration and the one-dimensional spectra extraction. The accuracy of the wavelength calibration has been checked by measuring the wavelength position of a number of emission telluric lines (Osterbrock et al. 1996). Then, we subtracted the mean sky spectrum from each stellar spectrum. By combining the exposures, we obtained GIRAFFE median spectra with signal-to-noise ratios S/N$\simeq 30 - 70$ for the selected

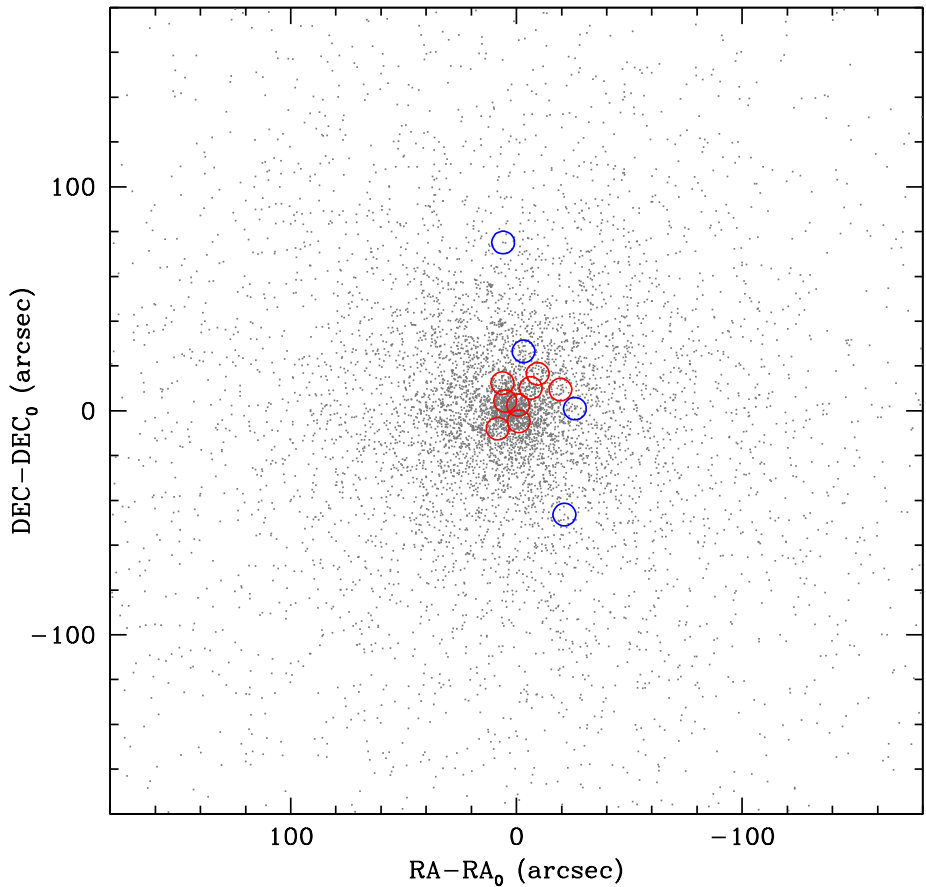

Figure 4.3: Position, shown as empty circles, of the observed BSSs.

BSSs, and S/N$\gtrsim$ 80 for the RGBs. Finally we normalized the combined spectra.

## 4.2 Radial velocities

In order to assess the cluster membership for each target, we measured the RVs with the IRAF task *fxcor*. As template for the cross-correlation, synthetic spectra with atmospheric parameters similar to the analysed targets have been used. Only for the BSS #12005407 we are not able to derive the RV because the spectrum of this star appears featureless. Nevertheless, we verified that the spectral counts are similar to those of the other BSSs, proving that star #12005407 has been correctly observed even if no line in the spectrum can be seen. The derived RV distribution is shown in Figure 4.4: the empty histogram shows the distribution for the RGB stars, whereas the grey shaded one is for the BSSs. The mean radial velocity of the RGB stars is $\langle RV \rangle = -189.2 \pm 0.4$ ($\sigma = 3.5$) km s$^{-1}$, in agreement with previous results by Harris (1996, 2010 edition) and Carretta et al. (2009a,b). We consider this value as the cluster systemic velocity and we used it to infer the cluster membership for all the observed targets through a $\sigma$-rejection algorithm: stars having RVs within $3\sigma$ with respect to the mean RV value have been considered as members of M30. The BSS RV distribution is in good agreement with that of the RGB stars, with a mean value $\langle RV \rangle = -188.9 \pm 0.7$ ($\sigma = 2.3$) km s$^{-1}$. All the observed BSSs (but the #12005407 for which we can say nothing more) turn out to be cluster members. The blue BSSs have <RV>=$-189.7 \pm 0.7$ ($\sigma$=1.3), whereas for the red BSSs <RV>=$-188.6 \pm 0.9$ ($\sigma$=2.5). Concerning star #12005407, it has been classified as W Uma variable by Pietrukowicz & Kaluzny (2004) and we attribute the complete lack of lines in all the observed setups to the very high rotational velocity of this kind of stars. We finally decided to consider also this star as cluster member according to its position (the BSS is located at $\simeq 75''$ from the cluster centre), even if we are not able to derive its RV and chemical abundances.

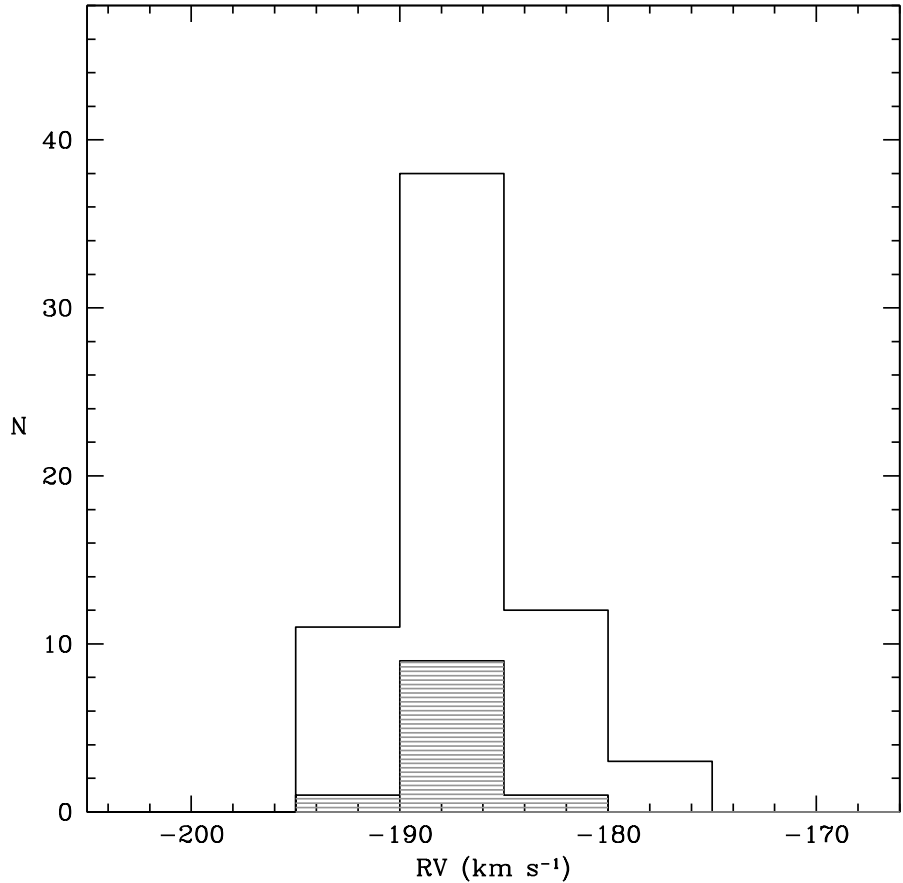

Figure 4.4: Radial velocity distributions for the RGB stars (empty histogram) and BSSs (grey shaded histogram). The two distributions are similar and their mean values agree quite well with previous results in literature.

## 4.3 Atmospheric parameters

Temperatures and gravities for the RGB stars and BSSs have been derived photometrically from the position of the targets in the CMD. Theoretical isochrones from the PEL database (Cariulo et al., 2004) have been used for this purpose. For the RGB stars, an isochrone of 12 Gyr, Z=0.0002 and $\alpha$-enhanced chemical mixture has been superimposed on the CMD of M30 (see Figure 4.5), assuming a distance modulus of 14.71 and E(B-V)=0.03 (Ferraro et al., 1999b). For the BSSs, different isochrones with different ages have been used to sample the BSS region (both the red and the blue sequence) in the best way. Then, given its position in the CMD, each observed target has been orthogonally projected on the closest theoretical isochrone, and T and log g have been derived for each star. The atmospheric parameters for the BSS and RGB samples are shown in Tables 4.1 and 4.2. Conservative errors in temperatures and gravities have been assumed based on the typical photometric uncertainties of the targets: 100 K and 0.2 dex for the BSSs, and 50 K and 0.1 dex for the RGB stars. Concerning the microturbulent velocity, the values for the RGB stars have been derived from the relation presented by Kirby et al. (2009), obtained by using a sample of RGB stars in the Sculptor dwarf spheroidal galaxy. This relation expresses the microturbulent

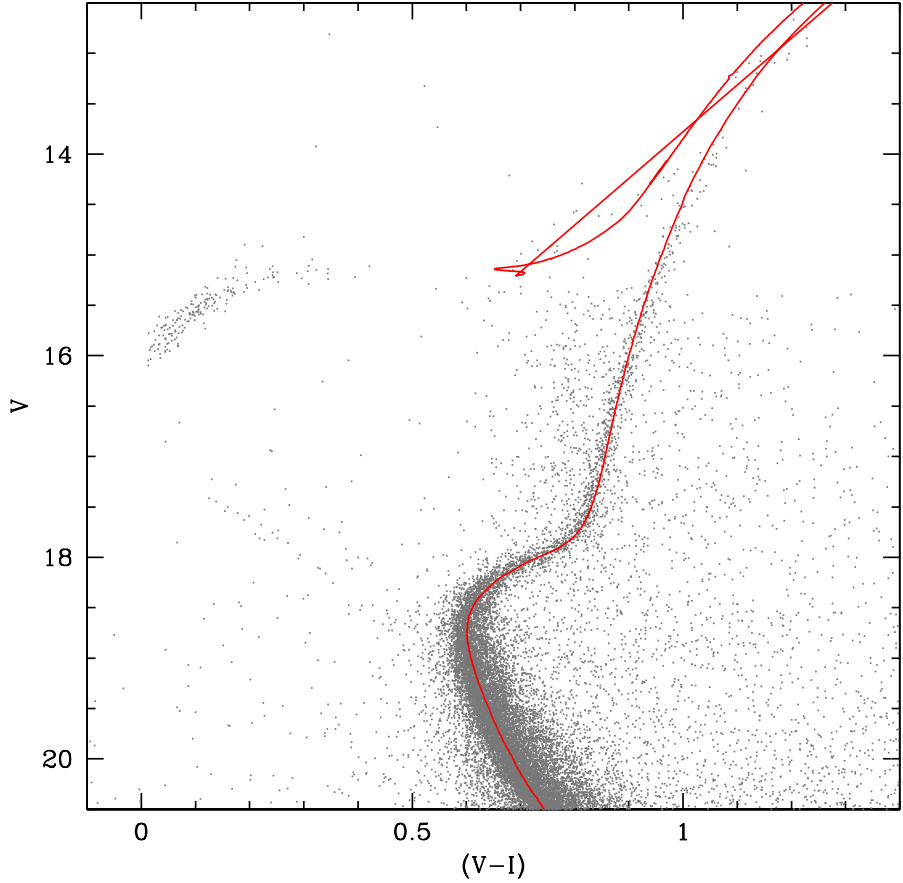

Figure 4.5: CMD of M30: the best-fit isochrone (calculated with 12 Gyr, Z=0.0002 and $\alpha$-enhanced chemical mixture) has been superimposed.

velocity as a function of the surface gravity:

$$v_t = (2.13 \pm 0.05) - (0.23 \pm 0.03)log(g) \qquad (4.1)$$

As shown in Table 4.2 the derived microturbulent velocities for all the RGB stars are between 1 and 2 $\mathrm{km\,s^{-1}}$. Unfortunately, the small amount of metallic lines in the BSS spectra (due to the low metallicity of the cluster and to the high temperatures of these stars) prevents us to derive the microturbulent velocity values spectroscopically. We therefore assumed 2 $\mathrm{km\,s^{-1}}$ for the BSSs. However the assumption of a different value of microturbulent velocity has a negligible impact on the derivation of the rotational velocities and the chemical abundances and it does not change our results. We finally adopted a conservative error of 0.5 $\mathrm{km\,s^{-1}}$ both for the RGB stars and BSSs. No macroturbulent velocity has been considered either for the RGB stars nor for the BSSs.

## 4.4 Rotational velocities

Projected rotational velocities were measured through the fitting of the most prominent metallic lines. For each star, a grid of synthetic spectra with different rotational broadenings has been computed (see Section 4.5 for more details) by taking into account the instrumental profile, the microturbulent velocity and the Doppler broadening. The instrumental profile has been derived by measuring the FWHM of bright unsaturated lines in the reference Th-Ar calibration lamp, according to the procedure described in Behr et al. (2000a). Finally, the Doppler and the microturbulent velocity broadenings have been computed by using the atmospheric parameters derived for each star (as described in Section 4.3). We then performed a $\chi^2$ minimization between the observed spectrum and the computed synthetic grid. For the BSS #12005407, it was possible to obtain only an upper limit of v sin($i$). Rotational velocity values for the RGB stars and BSSs are listed in Tables 4.1 and 4.2. The v sin($i$) distributions for the RGB stars and BSSs are shown in Figure 4.6. All the RGB stars display low v sin($i$) values ($< 10$ km s$^{-1}$) and the distribution is peaked at 0 km s$^{-1}$. To our knowledge, no previous determination of v sin($i$) for M30 stars exists in the literature. Nevertheless, the RGB distribution agrees very well with results found for both field and cluster RGB stars (Cortés et al., 2009; Carney et al., 2008). In comparison, the v sin($i$) distribution of BSSs is larger, ranging from 0 km s$^{-1}$ (#11000416 and #12000384) up to 25 km s$^{-1}$ (#11000978). The upper limit derived for BSS #12005407 is marked with an arrow in Figure 4.6 and it corresponds to a rotational velocity higher than 90 km s$^{-1}$. Interestingly enough, this star was identified as a W Uma by Pietrukowicz & Kaluzny (2004). In Figure 4.7 (upper panel) v sin($i$) as a function of T is shown: while no trend between rotational velocity and temperature is observed for the RGB stars, a slight increase is found for BSSs. Apparently the hottest BSSs, both in the red and the blue sequence, rotate faster than the coldest one. Moreover, there are indications that the BSSs in the blue sequence rotate faster than the ones in the red sequence.

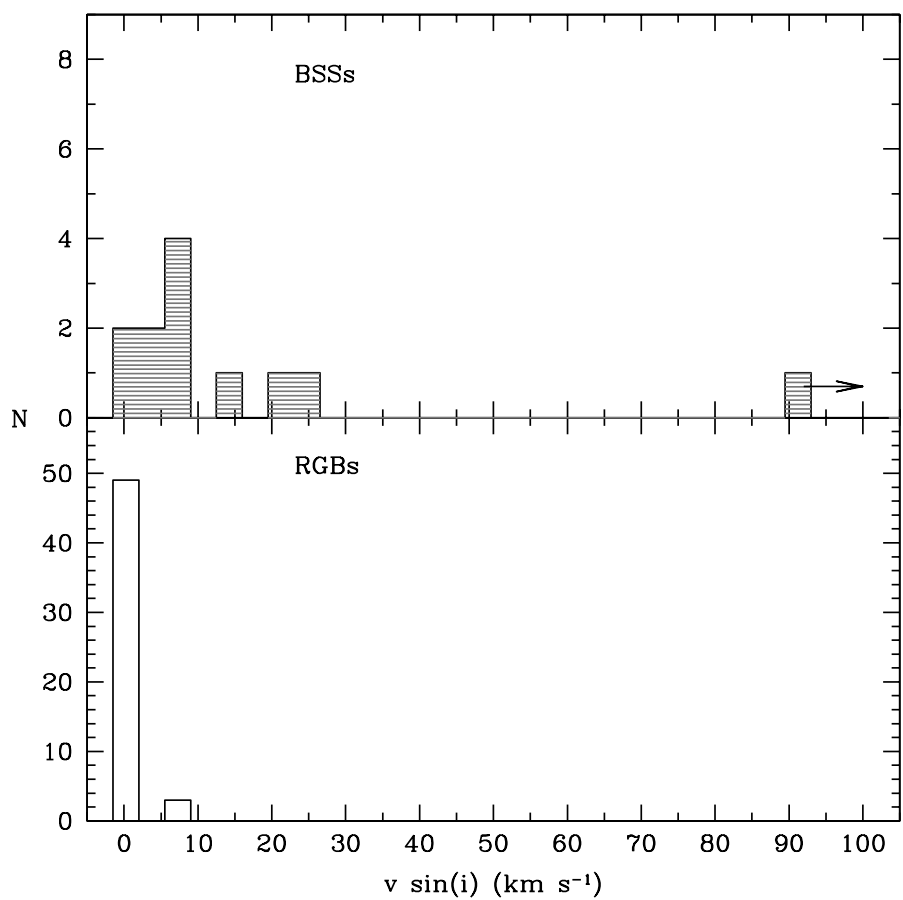

Figure 4.6: Rotational velocity distributions for the BSSs (upper panel) and the RGB stars (lower panel).

## 4.5  Chemical abundances

We derived iron, magnesium and titanium abundances for almost all the targets. Just in the case of a few BSSs only upper limits for these abundances could be estimated. In all spectra the OI triplet at $\lambda \simeq 7774$ Å is so weak that only upper limits could be derived. All the abundances have been obtained either by measuring the EW of each line (as in the case of Fe and Ti), or through the spectral synthesis (as in the case of Mg). Chemical abundances (from EWs) and synthetic spectra (for the spectral synthesis and the rotational velocity determination, see Sect. 4.4) have been computed by using the R. L. Kurucz's codes ATLAS9, WIDTH9 and SYNTHE. ATLAS9 model atmospheres have been computed under the assumption of LTE plane-parallel geometry and by adopting the new opacity distribution functions by Castelli & Kurucz (2003), without the inclusion of the approximate overshooting (Castelli et al., 1997). Atomic data for all the lines are from the most updated version of the Kurucz line list by F.Castelli [1]. The used reference solar abundances are from Grevesse & Sauval (1998) for Fe, Mg and Ti and from Caffau et al. (2011) for the O. The EWs of Fe and Ti lines were measured with our own code that fits the absorption lines with a Gaussian profile (see L10 and Mucciarelli et al. 2011a). Then chemical abundances have been computed with our program GALA (Mucciarelli et al. 2013, ApJ submitted, see Appendix A) based on WIDTH9. The Mg line at $\sim 4481$ Å is an unresolved blending of multiple components. In this case, the approximation of the line profile with a Gaussian profile (the assumption under which we compute chemical abundances from EWs) is no longer valid and we derived the abundances from the spectral synthesis. We obtain Mg abundances by performing a $\chi^2$ minimization between the observed spectra and a grid of synthetic spectra with different rotational broadenings and chemical abundances. For low $v \sin(i)$ values, the rotational velocity and the chemical abundance are (at least partially) degenerate: a similar line profile can be reproduced with a change in either the chemical abundance or the rotational velocity. In order to obtain the more reliable values for both quantities, we used our own code that computes a grid of synthetic spectra with different rotational broadenings and chemical abundances (by using ATLAS9 and SYNTHE) and iteratively performs a $\chi^2$ minimization until the best line fit is obtained. Finally, $3\sigma$ detection upper limits for the O have been estimated according to the formula by Cayrel (1988). An important issue to take into account is the problem of the possible deviation from the LTE assumption, particularly for the (relatively) hot BSSs. For the O limits we included non-LTE corrections taken from the statistical equilibrium calculations of Takeda (1997). For Fe, Mg and Ti abundances, no grid of non-LTE corrections are available in the literature for the range of parameters typical of our targets. Based on the fact that non-LTE corrections for ionized elements in hot (A and F type) stars are negligible with respect to those for neutral elements, whereas the opposite is true for cold stars, we obtained Fe and Mg abundances for the BSSs by using the FeII and MgII lines, whereas FeI and MgI lines have been used for the RGB stars. We did the same for the Ti abundances in BSSs whereas, unfortunately, this was not possible for RGB stars since only lines from ionized elements transitions are present both in the RGB stars and BSSs spectra. Chemical abundances for all the elements are listed in Tables 4.2 and 4.1 for the RGB stars the BSSs, respectively.

---

[1]http://wwwuser.oat.ts.astro.it/castelli/linelists.html

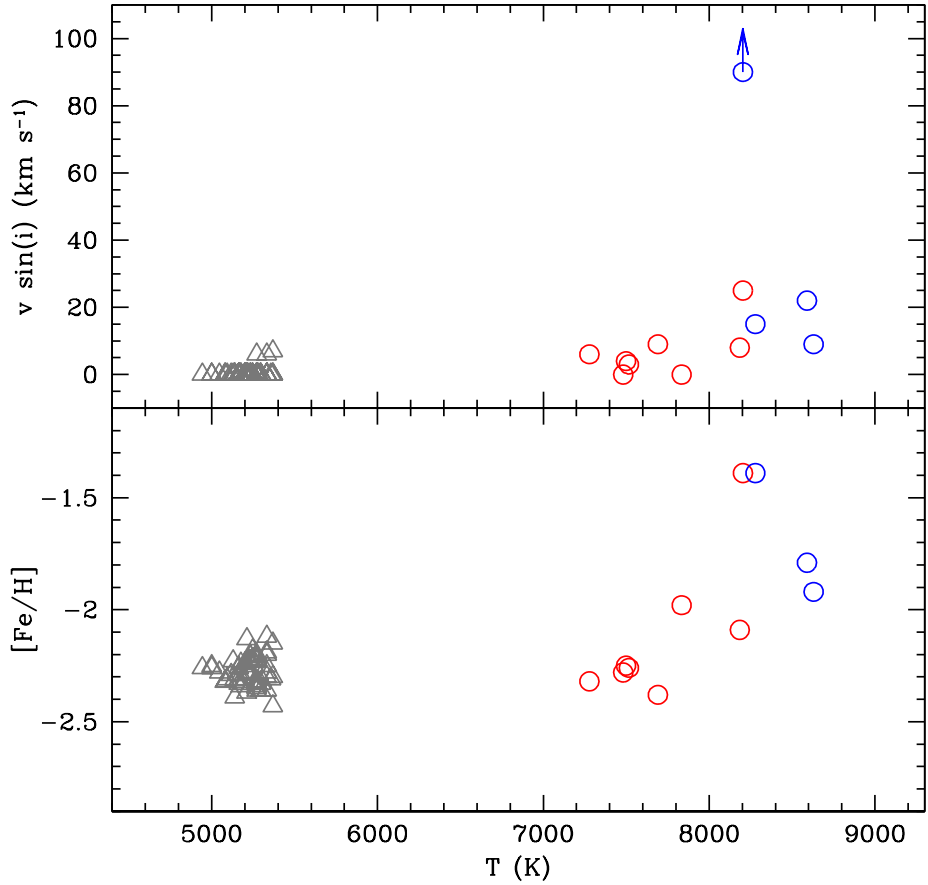

Figure 4.7: Upper panel: rotational velocities for RGBs (grey triangles) and BSSs in the two sequences (blue and red dots) as a function of T. Lower panel: [Fe/H] ratio for RGBs and BSSs.

### 4.5.1 Iron

For the RGB stars we find an average iron abundance of [Fe/H]=$-2.28\pm0.01$ ($\sigma$=0.07) dex in very good agreement with results by Carretta et al. (2009a,b,c) and Harris (1996, 2010 edition). On the contrary, the BSSs show a larger distribution with an average iron abundance of [Fe/H]=−2.00±0.11 and a dispersion $\sigma$=0.35 that is not compatible with the uncertainties. Figure 4.7 (lower panel) shows the [Fe/H] ratio as a function of T: at variance with the RGB stars, BSSs show a trend of [Fe/H] ratio as a function of T with the metallicity increasing at higher temperatures. The 5 coldest BSSs (T $\lesssim$ 7800 K) have [Fe/H]=−2.30±0.02 ($\sigma$=0.05) in very good agreement with the RGB stars, whereas the hottest ones (T $\gtrsim$ 7800 K) have [Fe/H]=−1.76±0.12 ($\sigma$=0.30), sensibly higher than the mean cluster metallicity. Moreover, BSSs in the red sequence have [Fe/H]=−2.12±0.11 ($\sigma$=0.32), whereas the BSSs in the blue sequence have [Fe/H]=−1.70±0.16 ($\sigma$=0.28). No trend between v $\sin(i)$ and [Fe/H] exists for the RGB stars, whereas a clear increase of v $\sin(i)$ for higher metallicity is observed for the BSSs (see Figure 4.8).

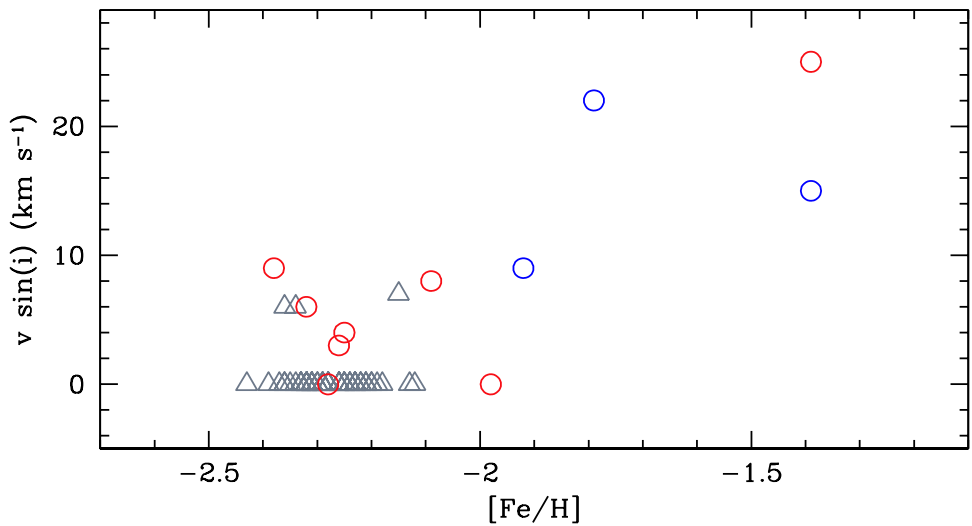

Figure 4.8: Rotational velocity for RGBs and BSSs in two sequences as a function of the [Fe/H] ratio.

### 4.5.2 Magnesium

RGB stars have [Mg/Fe]=0.4±0.02 ($\sigma$=0.16) in good agreement with results by Carretta et al. (2009b). For the BSSs we were able to measure Mg abundances for 5 BSSs, whereas only upper limits were estimated for the remaining 5 stars. The mean [Mg/Fe] ratio is 0.17±0.12 ($\sigma$=0.27), sensibly lower than that of RGB stars and with an higher dispersion. The Mg dispersion for the BSSs is intrinsic and only partially due to the dispersion in the iron content: in fact, BSSs show [Mg/H]=−1.85±0.17 with a dispersion $\sigma$=0.38, whereas for the RGBs [Mg/H]=−1.88±0.02 with a dispersion $\sigma$=0.14. For the BSSs in the red sequence the mean magnesium abundance corresponds to [Mg/H]=−2.06±0.12 ($\sigma$=0.21), whereas for the BSSs in the blue sequence it is [Mg/H]=−1.53±0.28 ($\sigma$=0.39). Figure 4.9 shows [Mg/H] and [Mg/Fe] ratios as a function of T for the RGBs and BSSs. A possible trend with T could exists (Mg abundances could be higher for colder BSSs), but the statistics is too low and the upper limits are not very strong constraints.

### 4.5.3 Oxygen

Concerning O abundances, only upper limits could be determined for BSSs. The comparison with the results found by Carretta et al. (2009c) for RGB stars. The comparison is shown in Figure 4.10. With the only exception of BSS #11000416, the upper limits for almost all the BSSs are incompatible with those of the RGB stars. Figure 4.11 shows the spectrum of BSS #11002171 zoomed in the region of the OI triplet. For sake of comparison, the figure also shows the same region for synthetic spectra with solar and alpha-enhanced mixture (red and green, respectively).

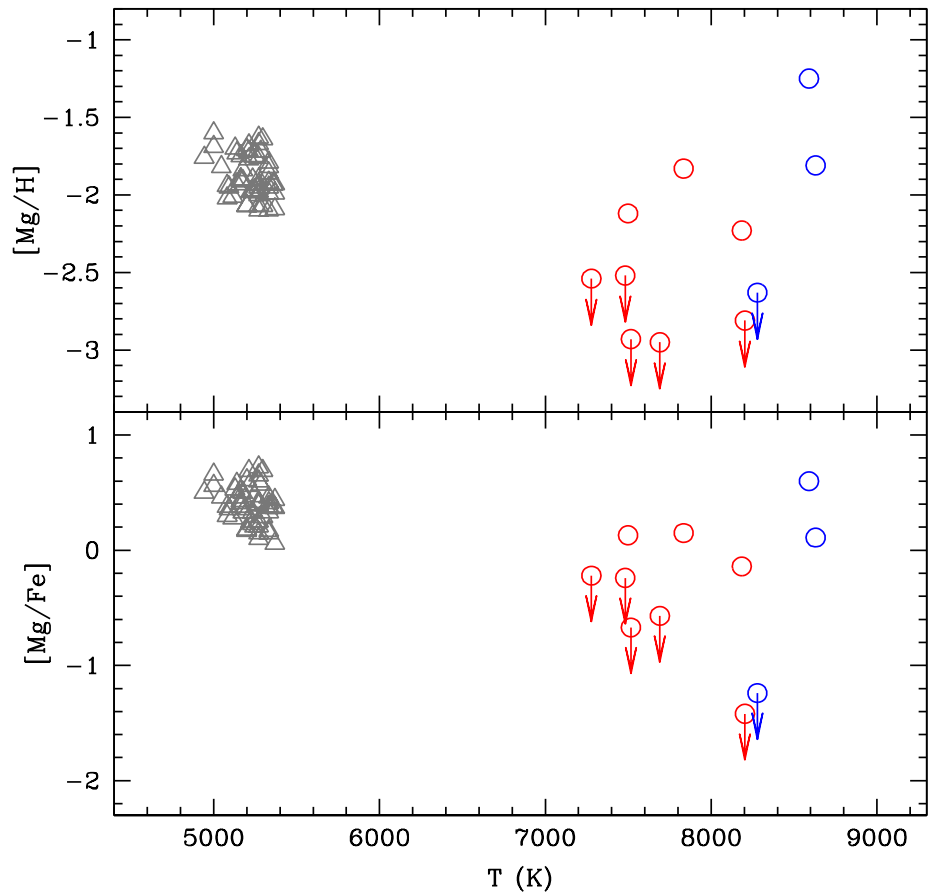

Figure 4.9: Mg abundances for RGB stars (black dots) and BSSs in the two sequences (red and blue dots)

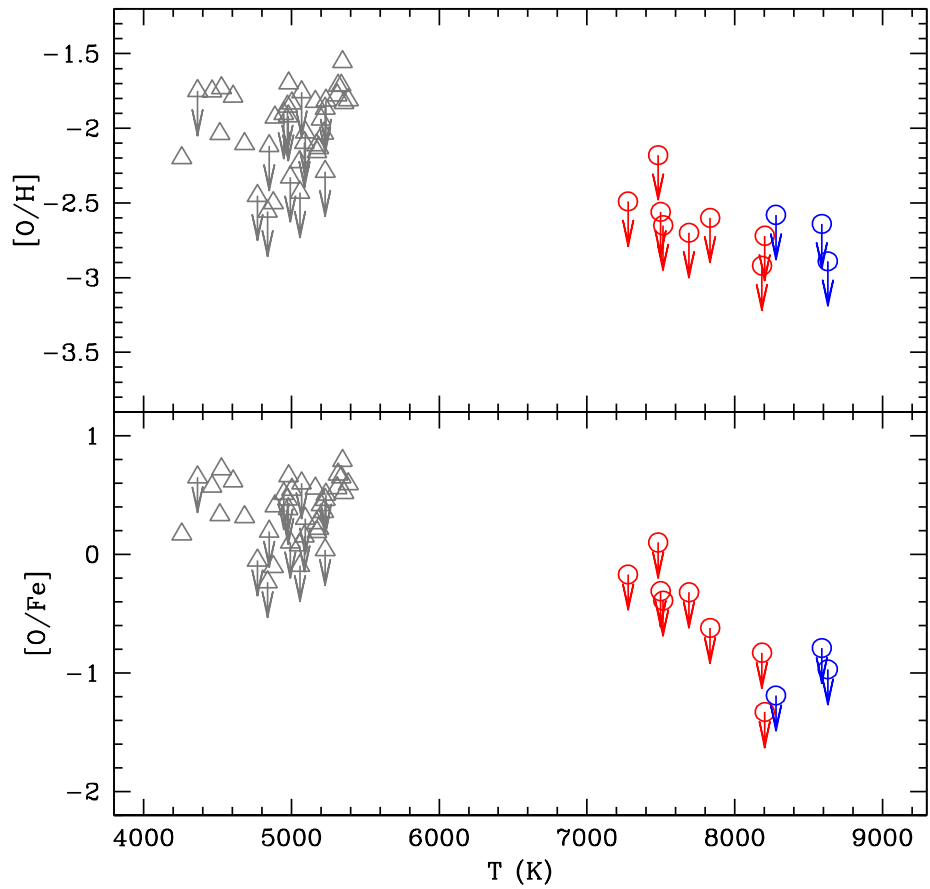

Figure 4.10: O abundances for BSSs in the two sequences compared with RGB stars (grey dots) from Carretta et al. (2009c) for RGB stars.

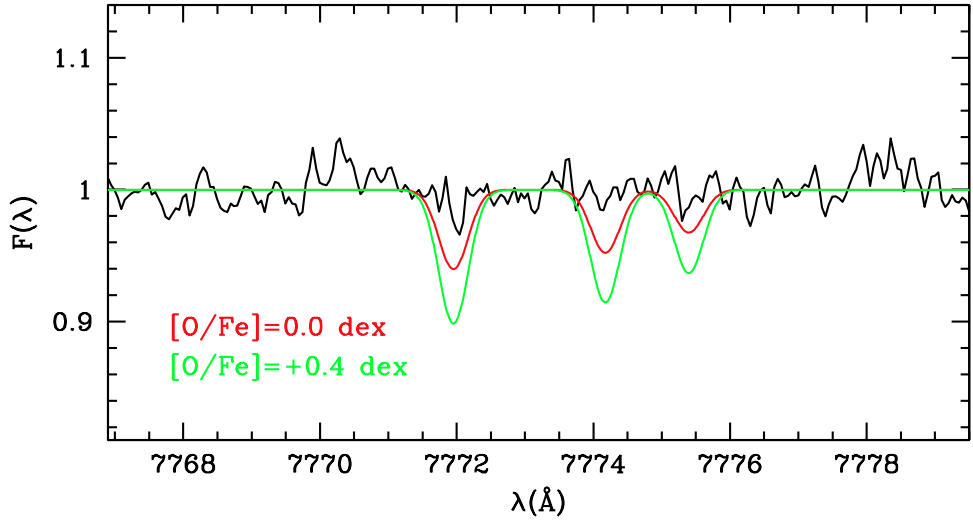

Figure 4.11: BSS #11002171 spectrum zoomed in the OI triplet region. The red and green line are synthetic spectra with solar and alpha-enhanced abundances, respectively.

### 4.5.4   Titanium

Ti abundances have been derived both for RGB stars and BSSs by measuring the EWs of a tens of TiII lines. Results are shown in Figure 4.12. The Ti mean abundance for the RGB is <[Ti/H]>=−2.07±0.01 ($\sigma$=0.08). To our knowledge, no determination of Ti abundances for M30 stars exists in the literature. Nevertheless, Ti is commonly known to be a good tracer of the $\alpha$-enhancement. For the RGB stars the mean Ti abundance is <[Ti/Fe]>=0.21±0.02 ($\sigma$=0.11) and it is compatible with an $\alpha$-enhanced scenario. For the BSSs we derive <[Ti/H]>=−1.70±0.11 ($\sigma$=0.36) that is ∼0.3 dex higher and more spread than the RGB value. Moreover, a strong trend with T exists, even if the two hottest (blue) BSSs have not the higher Ti abundance.

## 4.6   Discussion

Radial velocities, rotational velocities and chemical abundances have been derived for a sample of BSSs in the metal poor GC M30 and they have been compared with the results found for a sample of RGB stars in the same cluster. The analysed BSSs have been chosen in order to sample both the red and the blue sequence (see Ferraro et al. 2009a), with the aim of investigating possible differences in their kinematical and chemical properties. Our results confirm that RGB stars do not rotate fast, as already shown by other authors (Cortés et al., 2009; Carney et al., 2008). They are also chemically homogeneus and have an iron content perfectly in agreement with the literature ([Fe/H]=−2.28±0.01, $\sigma$=0.07). On the contrary, different results have been found for the BSSs: they show very large chemical abundance distributions for many elements, and higher rotational velocities with respect to the RGB stars.

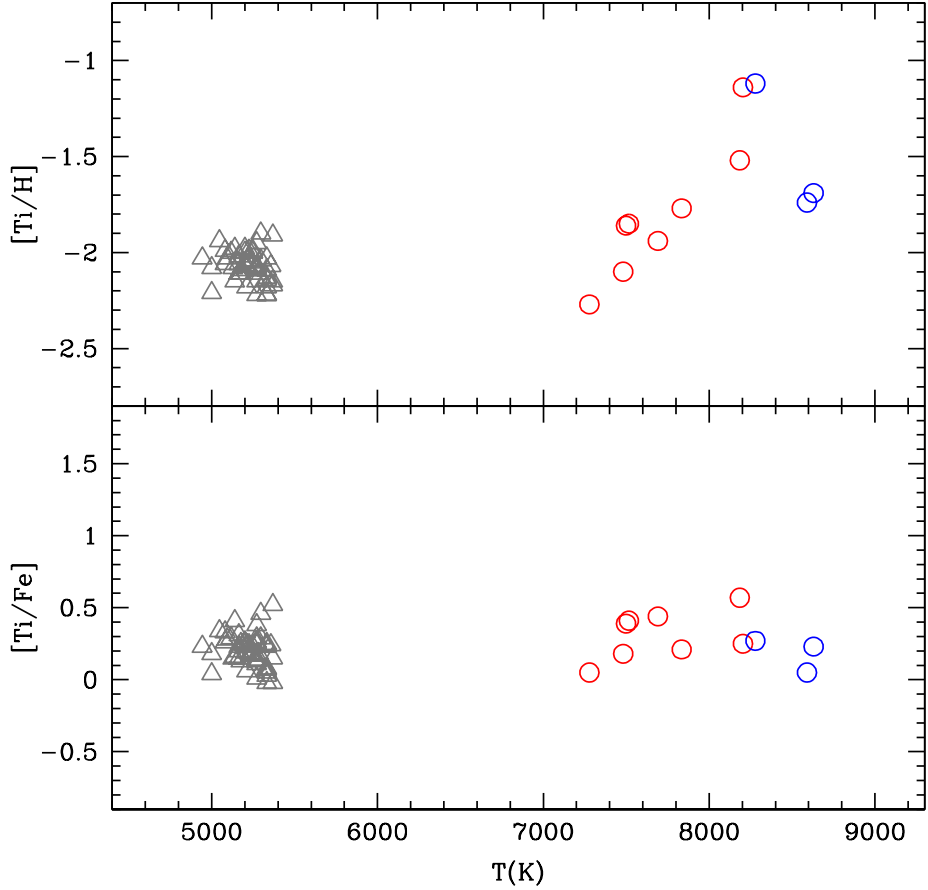

Figure 4.12: Ti abundances for RGB stars (triangles) and BSSs (circles).

### 4.6.1  Radial velocities

We used the RV values to infer the cluster membership for each star. The mean value measured for the RGB stars well agrees with other results in the literature and it has been considered as the systemic velocity of the cluster. It has been used, with a $\sigma$-rejection algorithim, to infer the cluster membership for the BSS sample. At odds with the results by L10 for the BSS sample in M4, no outliers have been found M30 and all the observed BSSs turn out to be members of the cluster. Finally, no difference between the RV distribution of the BSSs in the blue and in the red sequences has been found.

### 4.6.2 Rotational velocities

The rotational velocity distribution for BSSs is wider than the RGB one, with values ranging between 0 and 25 km s$^{-1}$. Moreover a fast rotating BSS (v sin($i$) >50 km s$^{-1}$) with v sin($i$) > 90 km s$^{-1}$, has been identified. Comparing our results with the ones obtained in other GCs, the rotational velocity distribution of the BSSs in M30 is very similar to what found in 47 Tuc by F06, and in NGC 6397 by Lovisi et al. (2012). In fact, most of the BSSs rotate slowly (even if faster than the reference population) and only one fast rotating star is detected. Interestingly enough, the fast rotator in M30 is also a W Uma variable. Also F06 and L10 found that the fastest rotating BSSs in 47 Tuc and M4 (namely #700912 and #2000121, respectively) are W Uma variables, suggesting a high rotation for this kind of objects. In Figure 4.13 we compare the v sin($i$) distribution for the blue and the red BSSs: even if the statistics is too low (especially for the blue BSSs), there are hints that stars in the blue sequence rotate faster than the ones in the red sequence. To further investigate this possibility, additional observations of the BSSs in the blue sequence are needed.

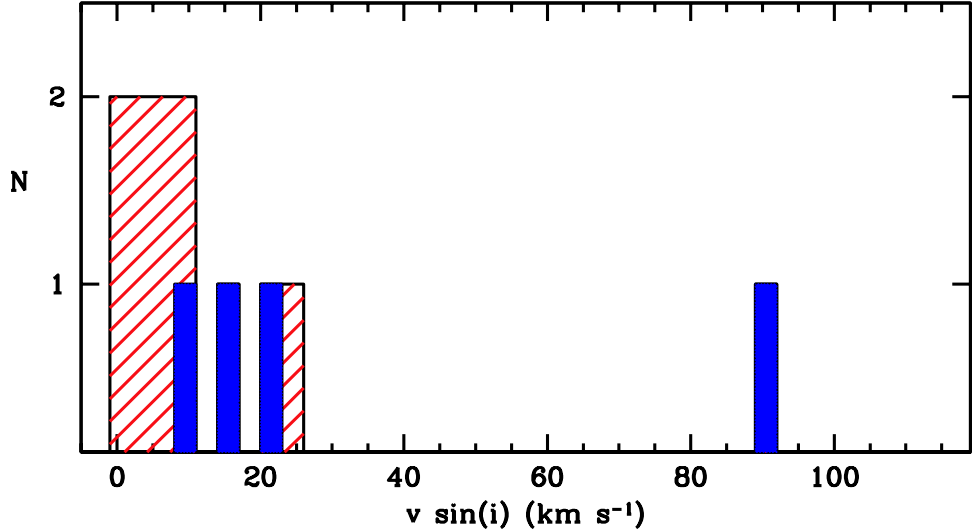

Figure 4.13: Rotational velocity distribution for the BSSs in the blue sequence (blue histogram) and the red one (red sequence).

### 4.6.3 Chemical abundances

We measured Fe, Mg, O and Ti abundances and upper limits for all the BSSs in our sample, with the only exception of the fast rotating BSS #12005407. From the Fe abundance analysis, it is clear that some of the BSSs in the sample (the hottest ones) suffer from radiative levitation. In particular, all the BSSs in the blue sequence and the hottest BSSs in the red one are affected by this kind of mechanism. As already shown in Lovisi et al. (2012) for NGC 6397, the radiative levitation alters the surface chemical abundances of the stars with shallow or no convective envelopes. This prevents us to observe the real chemical composition of these stars and thus to derive any information on the BSS formation mechanisms. Consequently, this effect has to be taken into account at the moment of the observation planning, in order to avoid the observation of stars with altered chemical patterns with a consequent waiste of observing time. The results obtained for M30 confirm that radiative levitation alters the chemical abundances of the BSSs hotter than $\simeq$ 7800-8000 K. Moreover, this phenomenon afflicts not only Fe, Mg and O abundances (as already found by Lovisi et al. 2012) but also Ti abundances.

The most important and fascinating result of our study concerns O abundances. We are able to obtain only upper limits for the BSSs and we compare them with the results obtained by Carretta et al. (2009c) for RGBs. Radiative levitation and gravitational settling affect the O abundances, so that we can discuss only the five BSSs where the iron abundance well matches the metallicity of the cluster (indicating that the surface chemical abundances of these stars are not modified by the occurrence of the radiative levitation). In Figure 4.14 the position of these five BSSs is marked with large red triangles. Clearly all these stars belong to the red sequence, since the observed BSSs along the blue sequence are too hot and their surface chemical abundances are already altered by the radiative levitation. Figure 4.10 shows that the O abundances derived by Carretta et al. (2009c) for a sample of RGB stars are not compatible with the upper limits derived for four out the five coldest BSSs of our sample. The O distribution derived from the giants ranges from ∼0.0 dex up to ∼0.5 dex and sub-solar upper limits are available only for three stars. The O abundances for these BSSs seem to be lower than that of the RGB stars, suggesting that there is a depletion on their surface that cannot be attributed to a self-enrichment process. Interestingly, the O depletion is observed in all the red BSSs of our sample that are not affected by radiative levitation. According to Ferraro et al. (2009a), the red sequence should include only BSSs produced through MT and for which tO depletion is indeed expected.

It is worth to note that this is the second evidence of O depletion in the BSS atmospheres. This supports the results found in 47 Tuc by F06, where two different subpopulations of BSSs were identified: one population (likely formed by MT-BSSs) with C and O depletion, and the other one (possibly formed by COL-BSSs) with C and O abundances compatible with those of TO stars. In the case of M30, due to the very low metallicity of the cluster ([Fe/H]=$-2.28\pm0.01$, $\sigma$=0.07), we are not able to derive C abundances. Moreover, the size of our sample is lower than that discussed by F06: in fact, the radiative levitation prevent us to derive the real O abundances for most of the BSSs in our sample. However, this finding is an important clue that seems to confirm the hypothesis that red BSSs are linked to the MT formation channel.

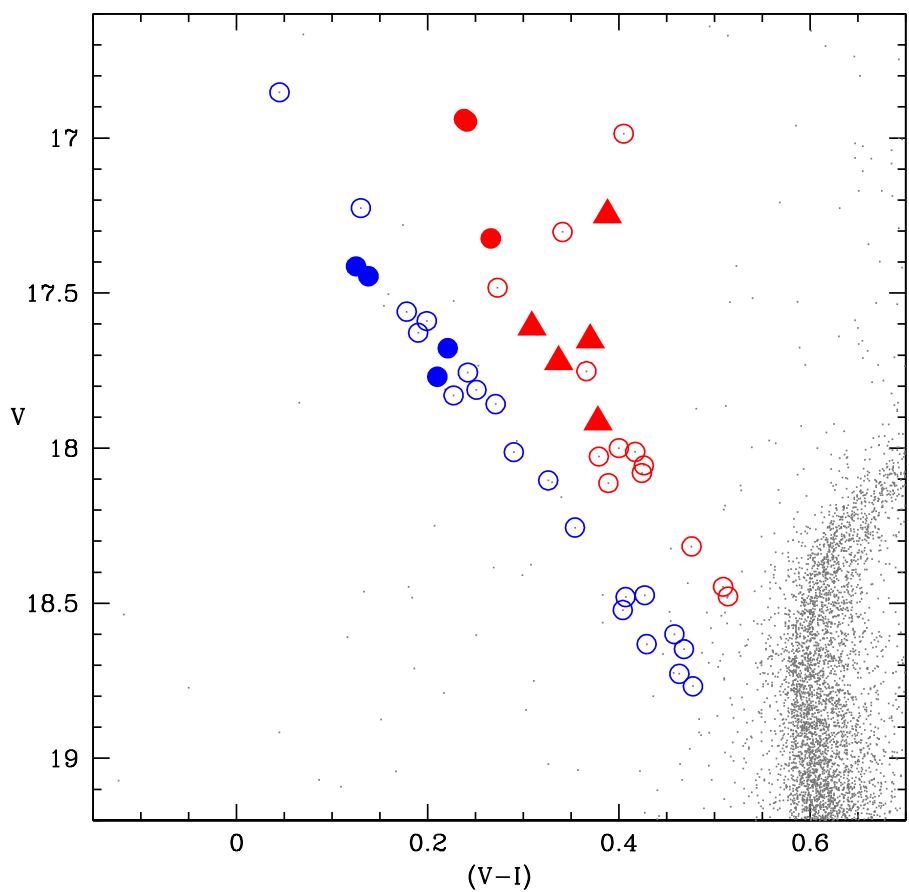

Figure 4.14: CMD of M30 zoomed in the BSS region. Filled simbols are the observed targets. Stars not affected by radiative levitation are marked with triangles.

| ID | RA (degrees) | DEC (degrees) | V | I | T (K) | $\log(g)$ | M ($M_\odot$) | RV (km s$^{-1}$) | v sin($i$) (km s$^{-1}$) | [Fe/H] | [Mg/H] | [O/H] | [Ti/H] | Notes |
|---|---|---|---|---|---|---|---|---|---|---|---|---|---|---|
| 11000416 | 325.09480740 | -23.17706600 | 17.9160 | 17.5380 | 7482 | 4.2 | 1.0 | -190.4±1.7 | 0 | -2.28 | <2.52 | <2.18 | -2.10±0.10 | |
| 11000978 | 325.09102120 | -23.17645980 | 16.9470 | 16.7060 | 8204 | 4.0 | 1.1 | -187.7±3.3 | 25 | -1.39±0.10 | <2.81 | <2.72 | -1.14±0.25 | |
| 11001895 | 325.09313210 | -23.17917410 | 16.9390 | 16.7010 | 8185 | 4.0 | 1.1 | -188.8±0.8 | 8 | -2.09±0.08 | -2.23±0.07 | <2.92 | -1.52±0.09 | |
| 11002171 | 325.09138800 | -23.17869200 | 17.7210 | 17.3840 | 7516 | 4.1 | 1.0 | -189.1±1.0 | 3 | -2.26±0.10 | <2.93 | <2.65 | -1.85±0.10 | |
| 11003263 | 325.09321560 | -23.18119670 | 17.6100 | 17.3010 | 7691 | 4.1 | 1.0 | -193.3±1.2 | 9 | -2.38±0.18 | <2.95 | <2.70 | -1.94±0.25 | |
| 11004551 | 325.09039300 | -23.18210920 | 17.6510 | 17.2810 | 7499 | 4.1 | 1.0 | -184.6±0.8 | 4 | -2.25±0.12 | -2.12±0.07 | <2.56 | -1.54±0.20 | |
| 12000384 | 325.09878460 | -23.17726520 | 17.3240 | 17.0580 | 7834 | 4.1 | 1.0 | -187.3±1.1 | 0 | -1.98±0.18 | -1.83±0.06 | <2.60 | -1.77±0.11 | |
| 12001595 | 325.09574750 | -23.17530400 | 17.2480 | 16.8600 | 7278 | 3.9 | 1.0 | -187.4±1.6 | 6 | -2.32±0.08 | <2.54 | <2.49 | -2.27±0.09 | |
| 12002629 | 325.09385260 | -23.17252470 | 17.4140 | 17.2890 | 8630 | 4.3 | 1.1 | -188.3±0.7 | 9 | -1.92±0.09 | -1.81±0.06 | <2.89 | -1.69±0.08 | |
| 12005407 | 325.09111420 | -23.15903090 | 17.6780 | 17.4570 | 8204 | 4.3 | 1.1 | - | >90 | - | - | - | - | W Uma |
| 14000585 | 325.10071720 | -23.17963720 | 17.7700 | 17.5600 | 8279 | 4.4 | 1.1 | -190.0±1.4 | 15 | -1.39±0.12 | <2.63 | <2.58 | -1.12±0.16 | |
| 14002580 | 325.09932290 | -23.19277030 | 17.4460 | 17.3080 | 8590 | 4.3 | 1.2 | -190.8±0.4 | 22 | -1.79±0.15 | -1.25±0.14 | <2.64 | -1.74±0.20 | |

Table 4.1: Coordinates, magnitudes, atmospheric parameters, masses, radial and rotational velocities, Fe, Mg, O and Ti abundances of the BSS sample.

| ID | RA (degrees) | DEC (degrees) | V | I | T (K) | log(g) | $v_t$ (km s$^{-1}$) | RV (km s$^{-1}$) | v sin(i) (km s$^{-1}$) | [Fe/H] | [Mg/H] | [Ti/H] |
|---|---|---|---|---|---|---|---|---|---|---|---|---|
| 10201736 | 325.09865910 | -23.16353260 | 15.3122 | 14.3817 | 5117 | 2.3 | 1.6 | -181.3±0.3 | 0 | -2.29 | -2.01 | -2.00 |
| 10202312 | 325.09906680 | -23.15721880 | 15.4704 | 14.5258 | 5140 | 2.4 | 1.6 | -187.8±0.4 | 0 | -2.39 | -1.93 | -1.98 |
| 10300934 | 325.11683240 | -23.18025420 | 15.6906 | 14.7685 | 5176 | 2.5 | 1.6 | -182.6±0.4 | 0 | -2.34 | -1.84 | -2.09 |
| 10301439 | 325.11496840 | -23.17490070 | 14.9721 | 13.9918 | 5047 | 2.2 | 1.6 | -192.7±0.4 | 0 | -2.28 | -1.82 | -1.94 |
| 10400043 | 325.09078220 | -23.19370770 | 15.7876 | 14.9125 | 5200 | 2.6 | 1.5 | -183.4±0.4 | 0 | -2.32 | -1.71 | -2.07 |
| 12001703 | 325.10283330 | -23.16267250 | 16.6671 | 15.8160 | 5333 | 3.0 | 1.4 | -183.4±0.4 | 6 | -2.36 | -1.99 | -2.13 |
| 13002556 | 325.11976810 | -23.17711570 | 16.4342 | 15.5707 | 5297 | 2.9 | 1.5 | -189.6±0.5 | 0 | -2.36 | -2.07 | -1.90 |
| 13003187 | 325.11866010 | -23.17273810 | 16.9133 | 16.0535 | 5370 | 3.0 | 1.4 | -188.4±0.7 | 0 | -2.30 | -1.93 | -2.15 |
| 13004219 | 325.12641650 | -23.16881780 | 16.2132 | 15.3220 | 5272 | 2.8 | 1.5 | -188.9±0.4 | 0 | -2.22 | -2.07 | -2.08 |
| 14000326 | 325.09158620 | -23.19347730 | 16.9374 | 16.0874 | 5370 | 3.1 | 1.4 | -196.6±0.4 | 0 | -2.43 | -1.99 | -1.91 |
| 14004027 | 325.10665670 | -23.19099970 | 16.2081 | 15.3187 | 5272 | 2.8 | 1.5 | -194.6±0.4 | 0 | -2.20 | -2.10 | -2.09 |
| 14004178 | 325.10073240 | -23.20218400 | 16.6457 | 15.7755 | 5333 | 3.0 | 1.4 | -187.1±0.4 | 0 | -2.30 | -1.92 | -2.22 |
| 20100040 | 325.07682510 | -23.20379530 | 15.1210 | 14.1580 | 5082 | 2.2 | 1.6 | -186.7±0.3 | 0 | -2.32 | -1.94 | -1.99 |
| 20100069 | 325.06742010 | -23.17563310 | 15.4480 | 14.5380 | 5140 | 2.4 | 1.6 | -186.8±0.4 | 0 | -2.31 | -1.73 | -2.15 |
| 20100145 | 325.07189260 | -23.20879970 | 16.2510 | 15.3560 | 5272 | 2.8 | 1.5 | -194.7±0.4 | 0 | -2.21 | -2.00 | -2.08 |
| 20100146 | 325.08291560 | -23.20829610 | 16.2620 | 15.3620 | 5272 | 2.8 | 1.5 | -186.8±0.4 | 0 | -2.36 | -1.76 | -2.07 |
| 30000021 | 325.03630520 | -23.16163890 | 15.5760 | 14.6450 | 5164 | 2.5 | 1.6 | -183.4±0.4 | 0 | -2.29 | -1.91 | -2.06 |
| 30000024 | 325.04319280 | -23.19972420 | 15.6610 | 14.7400 | 5176 | 2.5 | 1.6 | -189.2±0.4 | 0 | -2.24 | -1.91 | -2.11 |
| 30000032 | 325.16977920 | -23.27236360 | 15.8200 | 14.9160 | 5200 | 2.6 | 1.5 | -188.0±0.3 | 0 | -2.25 | -2.07 | -2.00 |
| 30000045 | 325.09900890 | -23.06656590 | 16.2200 | 15.3210 | 5272 | 2.8 | 1.5 | -189.5±0.3 | 0 | -2.32 | -1.90 | -2.15 |
| 30000046 | 325.05886780 | -23.25877090 | 16.2540 | 15.3790 | 5272 | 2.8 | 1.5 | -194.1±0.3 | 0 | -2.33 | -1.93 | -1.95 |
| 30000047 | 325.05401260 | -23.11081450 | 16.2550 | 15.3700 | 5272 | 2.8 | 1.5 | -188.0±0.4 | 6 | -2.34 | -1.95 | -2.08 |
| 30000053 | 325.10810800 | -23.06418910 | 16.3850 | 15.4900 | 5297 | 2.8 | 1.5 | -191.5±0.4 | 0 | -2.33 | -1.64 | -2.04 |
| 30000061 | 325.02346330 | -23.14591650 | 15.6380 | 15.7690 | 5333 | 2.9 | 1.5 | -187.2±0.4 | 0 | -2.20 | -1.79 | -2.22 |

Table 4.2: Coordinates, magnitudes, atmospheric parameters, radial and rotational velocities, Fe, Mg and Ti abundances of the RGB sample.

| ID | RA (degrees) | DEC (degrees) | V | I | T (K) | log(g) | $v_t$ (km s$^{-1}$) | RV (km s$^{-1}$) | $v \sin(i)$ (km s$^{-1}$) | [Fe/H] | [Mg/H] | [Ti/H] |
|---|---|---|---|---|---|---|---|---|---|---|---|---|
| 30000063 | 325.12908900 | -23.28113370 | 16.6580 | 15.8030 | 5333 | 3.0 | 1.4 | -188.5±0.4 | 0 | -2.25 | -2.10 | -2.18 |
| 30000064 | 325.09997320 | -23.25570890 | 16.6730 | 15.8170 | 5333 | 3.0 | 1.4 | -192.2±0.5 | 0 | -2.28 | -1.82 | -2.03 |
| 30007099 | 325.14366090 | -23.08405120 | 14.3670 | 13.3290 | 4943 | 1.9 | 1.7 | -192.0±0.4 | 0 | -2.26 | -1.76 | -2.03 |
| 30007103 | 325.00351240 | -23.25430390 | 14.7180 | 13.6920 | 5000 | 2.0 | 1.7 | -190.4±0.3 | 0 | -2.26 | -1.60 | -2.08 |
| 30007162 | 325.20083670 | -23.16984280 | 16.0160 | 15.1100 | 5236 | 2.7 | 1.5 | -190.4±0.4 | 0 | -2.23 | -1.90 | -2.05 |
| 30007164 | 325.16884440 | -23.20773110 | 16.0750 | 15.1750 | 5248 | 2.7 | 1.5 | -191.6±0.3 | 0 | -2.18 | -1.97 | -2.00 |
| 30007165 | 325.15376100 | -23.11702380 | 16.1070 | 15.2010 | 5248 | 2.7 | 1.5 | -190.9±0.4 | 0 | -2.22 | -1.75 | -1.99 |
| 30007180 | 325.06520100 | -23.26229700 | 16.2980 | 15.4240 | 5272 | 2.8 | 1.5 | -190.6±0.4 | 0 | -2.35 | -1.63 | -2.09 |
| 30007184 | 325.12630150 | -23.11433800 | 16.3640 | 15.4700 | 5284 | 2.8 | 1.5 | -191.0±0.3 | 0 | -2.29 | -1.72 | -2.11 |
| 30007188 | 325.13516080 | -23.19361180 | 16.6450 | 15.7830 | 5333 | 3.0 | 1.4 | -195.0±0.5 | 0 | -2.19 | -1.86 | -2.14 |
| 30007192 | 324.99767420 | -23.25180400 | 16.6590 | 15.7730 | 5333 | 3.0 | 1.4 | -189.3±0.4 | 0 | -2.12 | -1.95 | -2.09 |
| 30007210 | 325.00291600 | -23.21504060 | 16.9300 | 16.0620 | 5370 | 3.1 | 1.4 | -186.1±0.4 | 7 | -2.15 | -2.09 | -2.17 |
| 30011118 | 325.11614470 | -23.20792620 | 15.7730 | 14.8540 | 5200 | 2.6 | 1.5 | -188.2±0.4 | 0 | -2.24 | -2.07 | -1.98 |
| 30011153 | 325.08519540 | -23.13326840 | 16.8460 | 15.9940 | 5358 | 3.0 | 1.4 | -188.4±0.4 | 0 | -2.31 | -1.93 | -2.07 |
| 30012890 | 325.00798070 | -23.19626430 | 14.7160 | 13.7230 | 5000 | 2.0 | 1.7 | -192.7±0.4 | 0 | -2.25 | -1.69 | -2.21 |
| 30012898 | 325.23730570 | -23.26909710 | 15.5640 | 14.6310 | 5164 | 2.5 | 1.6 | -191.9±0.4 | 0 | -2.31 | -1.90 | -2.11 |
| 30012904 | 325.15144620 | -23.24496780 | 15.8780 | 14.9630 | 5212 | 2.6 | 1.5 | -188.7±0.4 | 0 | -2.13 | -1.77 | -2.07 |
| 30012905 | 325.03089770 | -23.21436580 | 15.9040 | 14.9910 | 5212 | 2.6 | 1.5 | -189.6±0.4 | 0 | -2.37 | -1.68 | -2.18 |
| 30013293 | 325.22173160 | -23.20713000 | 15.3730 | 14.4220 | 5129 | 2.4 | 1.6 | -192.1±0.4 | 0 | -2.23 | -1.70 | -2.08 |
| 30013317 | 325.15449850 | -23.22281560 | 15.9820 | 15.0860 | 5224 | 2.6 | 1.5 | -186.0±0.4 | 0 | -2.23 | -1.98 | -1.99 |
| 30014808 | 325.20234900 | -23.07233400 | 15.6230 | 14.6920 | 5164 | 2.5 | 1.6 | -189.4±0.4 | 0 | -2.26 | -1.75 | -2.11 |
| 30014827 | 325.08162500 | -23.15216580 | 16.2140 | 15.3280 | 5272 | 2.8 | 1.5 | -179.3±0.4 | 0 | -2.23 | -1.98 | -2.22 |
| 30015977 | 325.15129360 | -23.19554590 | 15.2140 | 14.2480 | 5093 | 2.3 | 1.5 | -187.5±0.4 | 0 | -2.31 | -1.95 | -2.03 |
| 30017833 | 325.06178010 | -23.18780190 | 15.5650 | 14.6530 | 5164 | 2.5 | 1.6 | -189.6±0.3 | 0 | -2.33 | -1.90 | -2.02 |
| 30017843 | 325.10227870 | -23.21021300 | 15.9950 | 15.0870 | 5236 | 2.7 | 1.5 | -188.4±0.4 | 0 | -2.21 | -1.98 | -2.04 |
| 30021349 | 325.09084140 | -23.14720680 | 15.8650 | 14.9420 | 5212 | 2.6 | 1.5 | -192.1±0.4 | 0 | -2.28 | -1.72 | -2.04 |
| 30030119 | 324.99771490 | -23.26930720 | 16.2540 | 15.3410 | 5272 | 2.8 | 1.5 | -187.7±0.3 | 0 | -2.32 | -1.67 | -2.08 |
| 30037611 | 325.03441030 | -23.27397540 | 15.1360 | 14.1760 | 5082 | 2.2 | 1.6 | -188.2±0.4 | 0 | -2.32 | -2.02 | -2.06 |

Table 4.2: Continued.

# Chapter 5

# BSSs in $\omega$ Centauri

The GC $\omega$ Centauri is one of the most studied objects in the Milky Way since the 1960s. Among GC-like stellar systems that populate the Galactic Halo, it is beyond any doubt the most surprising, harboring a complex stellar content. All the evidence collected so far - kinematics, spatial distribution and chemical composition - make $\omega$ Centauri a unique object if classed as usual among the GCs of the Milky Way. It is more massive (M $\sim$ 2.5 $10^6$ M$_\odot$; van de Ven et al. 2006) and luminous than any other GC. It differs dynamically from ordinary GCs: it is one of the most flattened clusters (Meylan, 1987), being also elongated (Geyer et al., 1983; Pancino et al., 2003), partially supported by rotation (Meylan, 1987; Merritt et al., 1997) and not completely relaxed dynamically (Anderson, 1997; Ferraro et al., 2006a). The most astonishing peculiarity of $\omega$ Centauri is its chemical inhomogeneity, not only in the light elements but also in the iron-peak elements: a large metallicity scatter ($\sim$1 dex) has been measured both spectroscopically (Norris et al., 1996; Suntzeff & Kraft, 1996) and photometrically (Lee et al., 1999; Pancino et al., 2000; Hilker & Richtler, 2000; Frinchaboy et al., 2002; Sollima et al., 2005). Being the only Halo stellar system with such a large chemical inhomogeneity, it has been suspected not to be a "genuine" GC, but the remnant of a dwarf galaxy that merged with the Milky Way in the past.

The particular dynamical status of $\omega$ Centauri has been confirmed also by the radial distribution of its BSS population. Typically, in almost all the properly studied GCs, this appears bimodal, with a peak in the central regions, a minimum at intermediate radii and a rising branch toward the outer regions. On the contrary, $\omega$ Centauri is one of the three GCs (together with NGC 2419 and Palomar 14, see Dalessandro et al. 2008a and Beccari et al. 2011) where the radial distribution of BSSs has been found to be completely flat and similar to that of RGB and HB stars (Ferraro et al., 2006a). Thus, the BSSs are not centrally segregated, suggesting that $\omega$ Centauri is not relaxed yet and the observed BSSs are the progeny of primordial binaries, whose radial distribution is not yet significantly altered by stellar collisions and by the dynamical evolution of the cluster (Ferraro et al., 2012).

## 5.1 Observations

The observations have been performed in two runs in June 2008 and August 2012 at the ESO-VLT with the high-resolution spectrograph FLAMES in the UVES+GIRAFFE combined mode. The sample includes 80 BSSs. The target selection has been performed on a photometric catalog obtained by combining ACS@HST data for the central region and WFI@ESO data for the outer region. Stars with other sources of comparable or brighter luminosity within $3''$ have been discarded. Figures 5.1, 5.2 and 5.3 show the CMDs for our ACS and WFI datasets and the position of all the analyzed targets with respect to the cluster centre ($\alpha$(J2000) = $13^{\rm h}\,26^{\rm m}\,47.24^{\rm s}$, $\delta$(J2000) = $-47^{\rm o}\,28'\,46.50''$, according to Harris 1996, 2010 edition): all the observed BSSs are marked with circles.

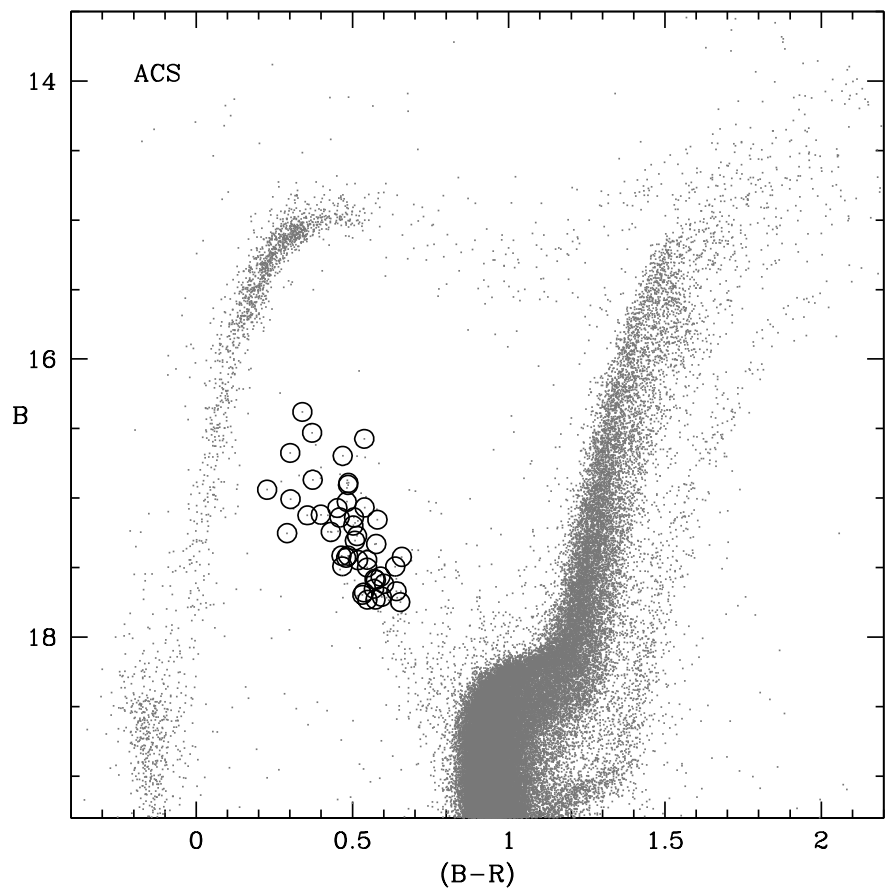

Figure 5.1: CMD of $\omega$ Centauri for the ACS dataset: all the observed BSSs are marked with circles.

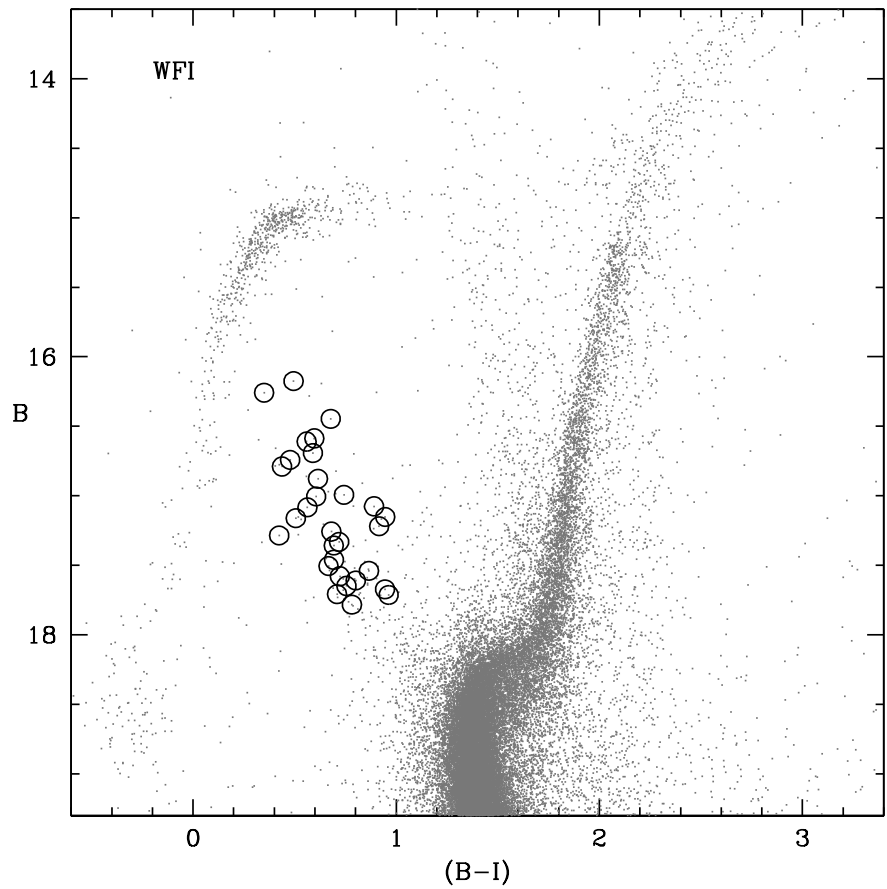

Figure 5.2: CMD of ω Centauri for the WFI dataset: all the observed BSSs are marked with circles.

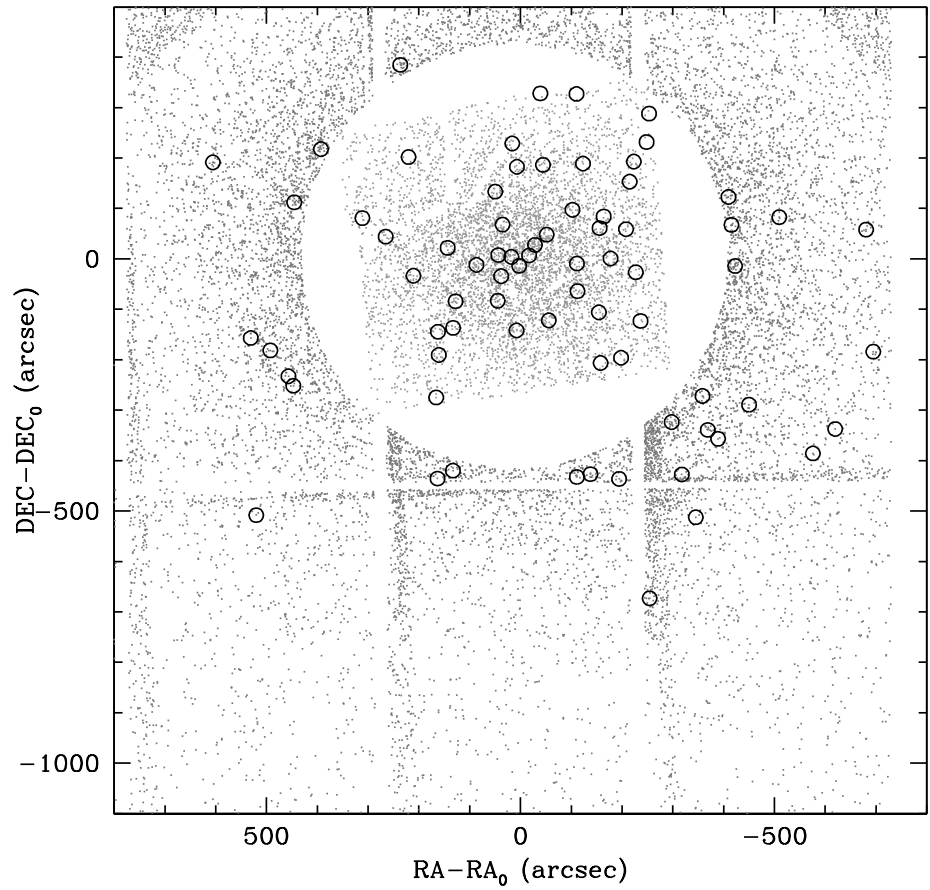

Figure 5.3: Position of all the observed BSSs (marked with circles) with respect to the cluster centre.

Four different setups have been used for the analysis: HR5A (with spectral resolution R=17400), HR14A (R=17700), HR18 (R=18300) and HR22 (R=11700) suitable to sample some metallic lines, the H$\alpha$ line, the OI triplet at $\lambda \simeq 7774$ Å, and the CI line at $\lambda \simeq 9094.8$ Å, respectively. Exposure times (obtained with multiple exposures of 45 minutes each) amount to about 2.5 hours for the HR14 and HR5A setups, almost 4 hours for the HR18 and 4.5 hours for the HR22. Spectra have been pre-reduced by using the standard ESO pipelines, including bias subtraction, flat-fielding, wavelength calibration and spectrum extraction. The accuracy of the wavelength calibration has been checked by measuring the wavelength position of a number of emission telluric lines (Osterbrock et al. 1996). For each exposure we obtained a mean sky spectrum that has been subtracted from each stellar spectrum to properly remove the sky contribution.

## 5.2 Radial velocities

Radial velocities have been computed by using the IRAF task *fxcor*: all the observed spectra have been cross-correlated with synthetic spectra computed with atmospheric parameters suitable for the analyzed targets. RVs have been inferred for the different setups and all values are in agreement with each other. However we finally decided to adopt for almost all the targets the RV values obtained from the HR5A setup becuase of the highest number of lines available in this grating. For some stars that appear featureless in the HR5A setup, the RV has been computed through the cross-correlation with the H$\alpha$ line in the HR14 setup. Due to the detection of cosmic rays on the core of the H$\alpha$ line, we are not able to infer the RV value for BSS #330829.

The RV distribution of the entire sample ranges from -20 up to $\sim$314 km s$^{-1}$, but almost all the stars have RV>180 km s$^{-1}$. In the case of $\omega$ Centauri the identification of interlopers is easy, because the RV distribution of the GC is peaked at $\sim$230 km s$^{-1}$ (see e.g. Meylan et al. 1995; Harris 1996, 2010 edition; Sollima et al. 2009; Scarpa & Falomo 2010), while that of the field stars along the line of sight is peaked at $\sim -8$ km s$^{-1}$ (see e.g. Figure 3 in Sollima et al. 2009), in agreement with the predictions of the model by Robin et al. (2003). In this way, we recognize in our sample 2 stars that certainly are outliers in the observed RV distribution. The RV distribution of our sample (excluding the 2 field stars) is shown in Figure 5.4. From the remaining sample (assuming that all the stars in Figure 5.4 are cluster members), we derived a mean RV value of 237.2±2.2 km s⁻¹ with a dispersion $\sigma$=19.3 km s⁻¹. The value of the mean RV nicely agrees with the mean values provided by other authors: for instance, Pancino et al. (2007) estimate an average value of 233.4 km s⁻¹, while Sollima et al. (2009) find 233.2 km s⁻¹. Sollima et al. (2009) adopted as membership criterion the RV range between 190 and 290 km s⁻¹ (in fact, as they noted, a small fraction of field Galactic stars can have RV higher than ∼290 km s⁻¹). According to this criterion, 2 BSSs (#195451 and #307749) should be discarded from our sample. Moreover, by applying a 3$\sigma$-rejection algorithm, BSS #195451 should be considered as a field star. When we exclude these 2 stars, the derived RV distribution has a mean value of 236.8±1.9 km s⁻¹ and a dispersion $\sigma$=16.3 km s⁻¹.

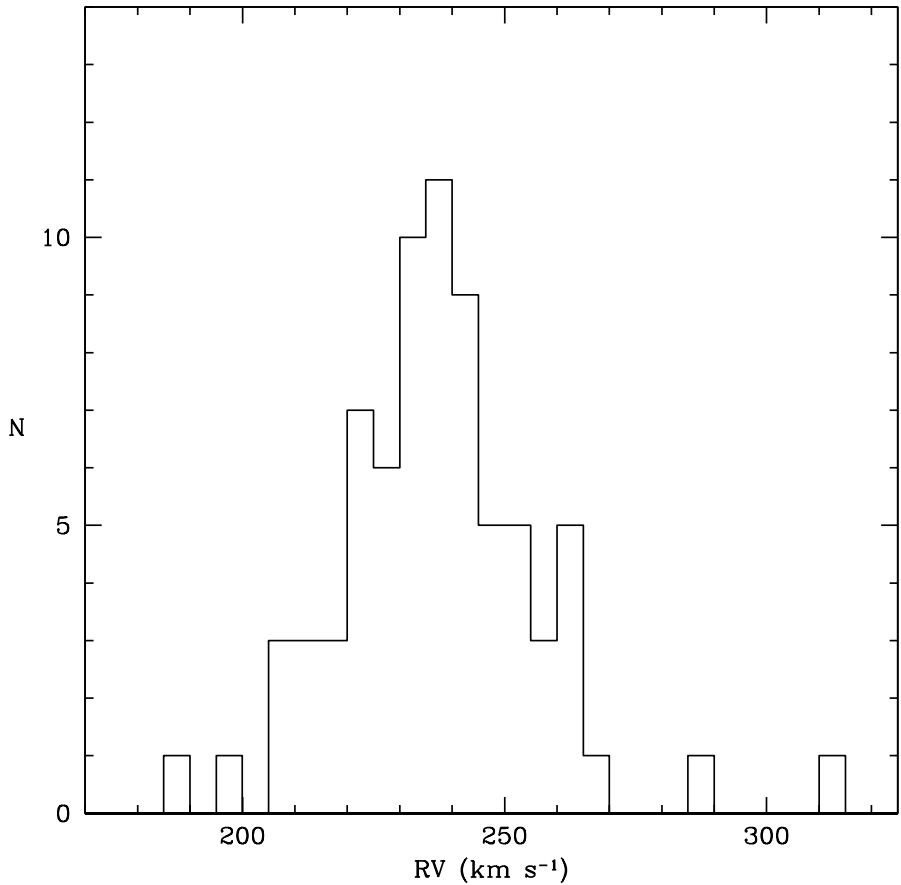

Figure 5.4: RV distribution of the observed BSSs.

## 5.3 Atmospheric parameters

Temperature and gravities for our targets have been derived photometrically, according to the position of the BSSs in the CMD. Isochrones of different ages, extracted from the Padova database (Marigo et al., 2008) with Z=0.0006, E(B-V)=0.11 and distance modulus (m-M)$_0$=13.7 (according to Ferraro et al. 1999b) have been superimposed to the CMD in order to fit the BSS region. Then, the position of all the observed BSSs has been orthogonally projected on the closest isochrone and temperatures and gravities have been inferred for each target. Results are shown in Tables 5.2 and 5.4. The microturbulent velocity has been assumed equal to 0 km s$^{-1}$ for all the BSSs (according to the fact that these stars are hot enough to have shallow or no convective envelope). Finally, no macroturbulent velocity has been considered.

## 5.4 Rotational velocities

Projected rotational velocities for all the BSSs have been measured by using the MgII line at ∼ 4481 Å because this feature is strong enough to be well detectable in each spectrum, regardless of the temperature, the metallicity and the S/N ratio. For each star, a grid of synthetic spectra with different rotational broadenings has been computed by taking into account the instrumental profile (derived as in Behr et al. 2000a), the microturbulent velocity, the Doppler broadening (computed by using the atmospheric parameters derived for each star) and the metallicity of the star (derived as described in Section 5.5). Then a $\chi^2$ minimization between the observed spectrum and the computed synthetic grid has been performed, thus obtaining the best rotational velocity. Six BSSs show spectra completely featureless so that we were not able to infer the rotational velocity. For these stars, we derived only upper limits by a visual comparison with synthetic spectra computed with different v sin($i$). For these 6 BSSs we estimate that v sin($i$) should be larger than 100 km s$^{-1}$. The rotational velocity distribution for all the BSSs is shown in Figure 5.5. Most of the BSSs rotate less than 50 km s$^{-1}$, but the distribution is very large ranging between 0 up to 120 km s$^{-1}$ (for the BSS #15822). The mean v sin($i$) value is 34.3 ± 3.5 km s$^{-1}$ with a very high dispersion $\sigma$ = 29.4 km s$^{-1}$. Figure 5.6 shows the v sin($i$) distribution as a function of T: the coldest BSSs (T$\lesssim$ 9000 K) show a large distribution covering the entire extension of values in our sample; on the contrary, the distribution shows an abrupt cut off for the hottest BSSs (T$\gtrsim$ 9000 K) that have v sin($i$) lower than 50 km s$^{-1}$.

We have matched the list of our targets with the catalog of variable stars identified by Kaluzny et al. (2004). The cross-identification is listed in Tables 5.2 and 5.4. We find that 13 BSSs are in common, 8 are recognized as SX Phe and 5 as W Uma.

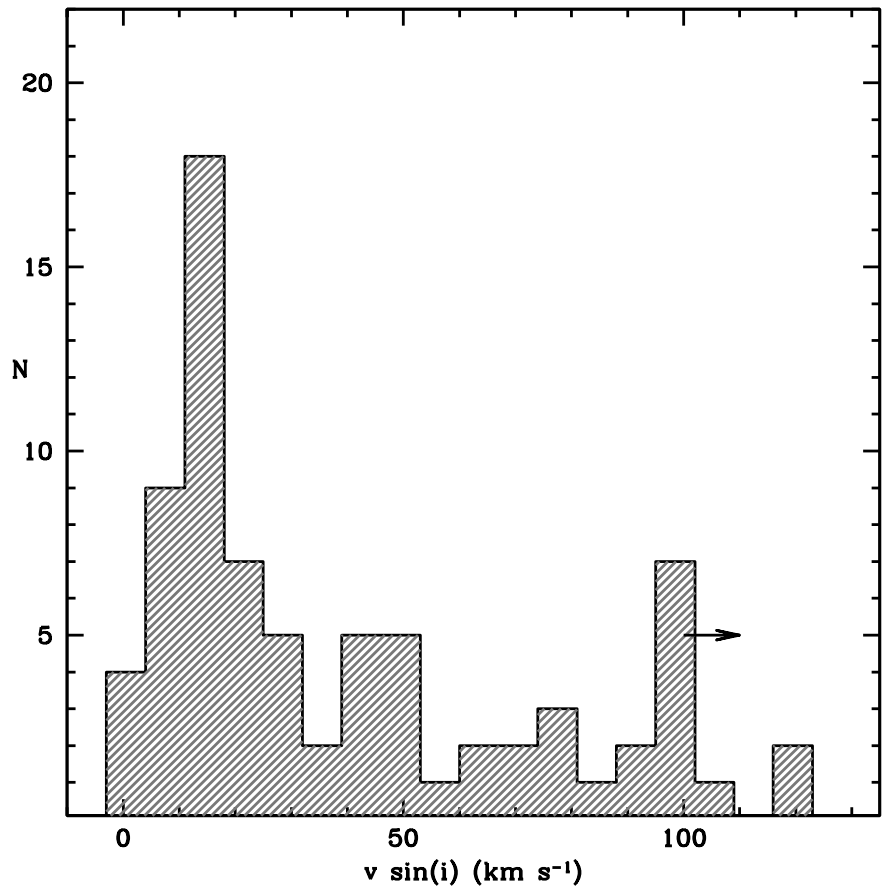

Figure 5.5: Rotational velocity distribution of the observed BSSs. Upper limits are marked with an arrow.

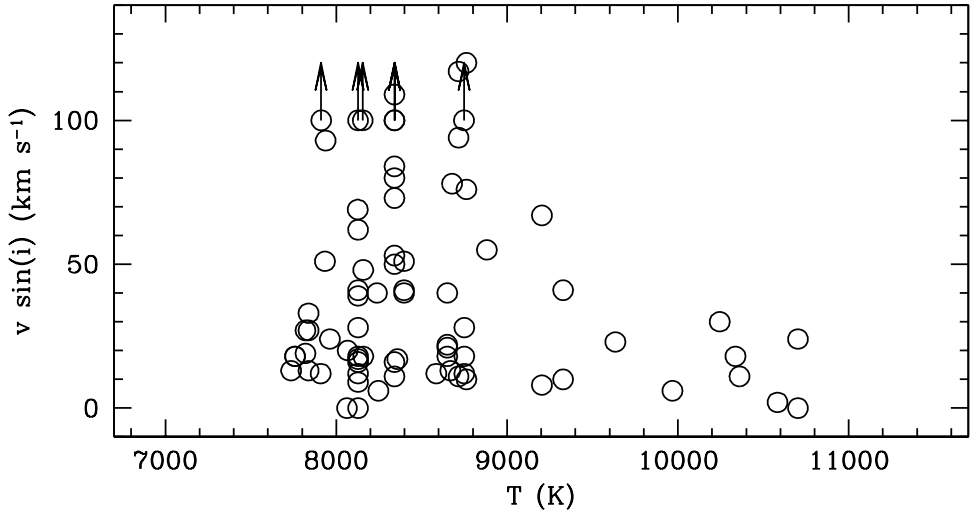

Figure 5.6: Rotational velocities as a function of T. Upper limits obtained for some star are marked with arrows.

Finally, we checked the radial distribution of the fast rotating BSSs, with respect to the normal ones. Results are shown in Figure 5.7. The fast rotating BSSs appear to be less concentrated toward the centre with respect to the other BSSs in the sample. Following a Kolmogorov-Smirnov test, the probability that the two distributions are extracted from the same parent population is $\sim$ 7%.

## 5.5 Chemical analysis: preliminary results

The iron content has been derived for a sub-sample of 42 BSSs with projected rotational velocity lower than $\sim 30$ km s$^{-1}$. For higher values of v $\sin(i)$, the iron lines are not detectable in the noise envelope (the same holds for other lines with the exception of the Mg II lines at 4481 Å that remains visible also at high v $\sin(i)$ values). We measured both neutral and single ionized Fe lines. FeI abundances turn out to be higher than those obtained from FeII lines because of the relevant NLTE effects occuring in such hot stars (see Chapter 3). For three targets (namely #127397, #178458 and #200941) we provide only $3\sigma$ detection upper limits, because of the combination of low metallicity ([Fe/H]<-1.8 dex) and low S/N. The derived [Fe/H] distribution ranges from [Fe/H]$\sim$-2 dex up to -0.5 dex, with only three stars with solar or oversolar metallicity: #112043 ([Fe/H]$\simeq$-0.12 dex), #509168 ([Fe/H]$\simeq$+0.5 dex) and #258947 ([Fe/H]$\simeq$+1.04 dex).

Figure 5.8 shows the behaviour of the entire distribution as a function of the temperature.

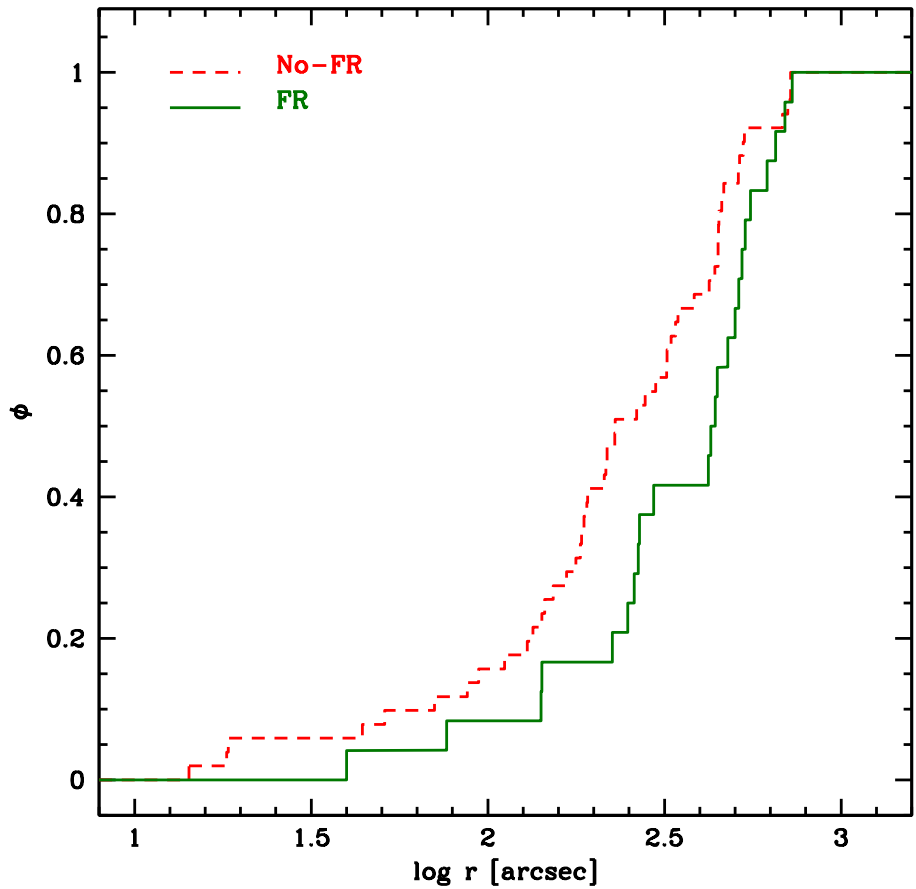

Figure 5.7: Cumulative radial distribution for the fast rotating BSSs ($v \sin(i) > 50 \ \mathrm{km\,s^{-1}}$, green solid line) and the normal BSSs (red dashed line).

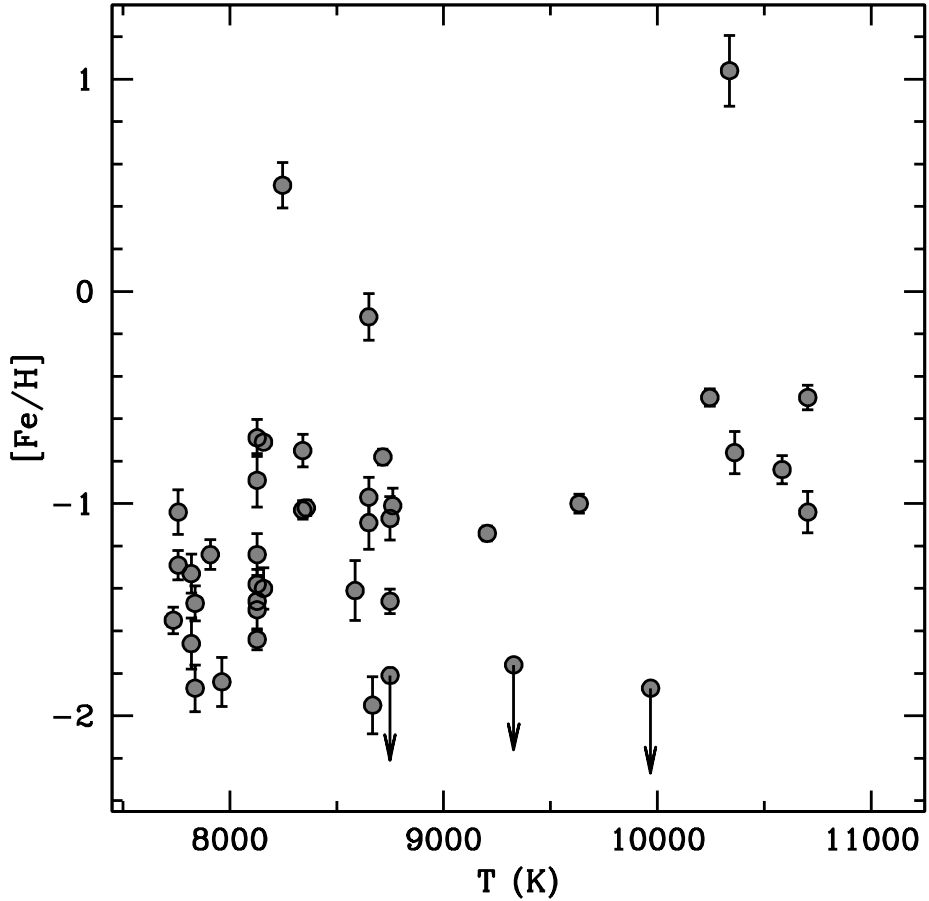

Figure 5.8: Iron content as a function of the temperature for the total BSS sample with v sin(*i*) <30 km s$^{-1}$.

A clear trend of the metallicity as a function of temperature is observed. Because of the large metallicity distribution of ω Centauri, the occurence of radiative levitation is more difficult to disentangle than on the case of single-metallicity GCs (as NGC 6397 and M30). The BSS metallicity distribution is shown in the upper panel of Figure 5.9, while the other two panels show the histograms of the metallicity calculated including only BSSs colder (central panel) and hotter (bottom panel) than 8200 K (we assume this temperature as the boundary for the occurrence of the radiative levitation). The metallicity distribution of the BSSs colder than 8200 K (despite the small statistics) resembles the metallicity distribution derived from the giant stars of ω Centauri (Johnson & Pilachowski, 2010; Marino et al., 2012). In particular, the distribution of the coldest BSSs is peaked at [Fe/H]∼-1.40 dex, slightly higher than the main peak of the distribution (this small offset can be easily ascribed to the small statistics of this sample, but also to the inclusion in the sample of BSSs affected by levitation). On the other hand, the iron distribution for the hottest BSSs is very broad with a median value at [Fe/H]≃-1.01 dex, including also the three stars with [Fe/H]>-0.2 dex. The evidence that the metallicity distribution is shifted toward higher value for higher temperature is consistent with the occurence of the radiative levitation in the photosphere of these stars. However, the behaviour results to be noisy and less definite with respect to the one observed in NGC 6397 and M30, because of the complex stellar content of $\omega$ Centauri, in terms of metallicity and age.

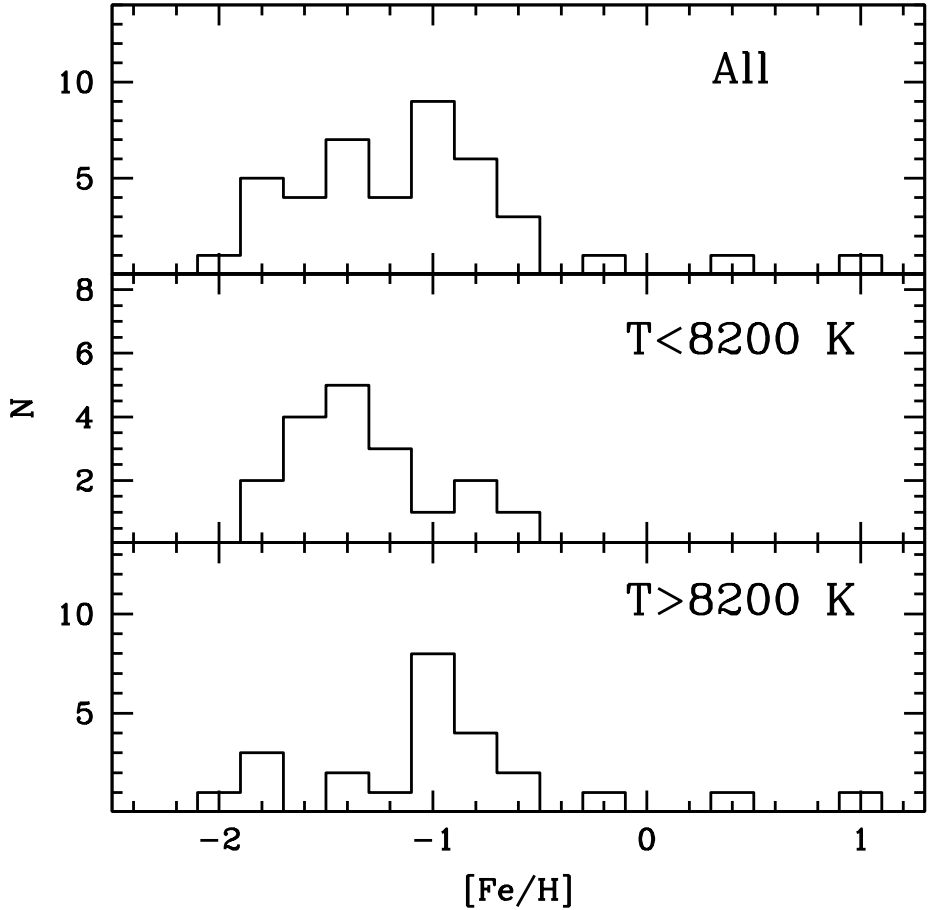

Figure 5.9: Iron distribution for all the BSSs with v sin($i$) <30 km s$^{-1}$ (top panel), BSSs with T < 8200 K (central panel) and BSSs with T > 8200 K (bottom panel).

We derived C and O abundances only for 9 BSSs colder than ∼8000 K and with low rotational velocity, because for the other stars the radiative levitation (as well as the gravitational settling processes) modifies the surface abundances and prevents us to identify real chemical anomalies due to MT mechanisms.

The C abundances have been inferred by measuring the EW of the CI line at 9094.8 Å. This feature results to be the only transition belonging to the C multiplet available in this spectral region not contaminated by telluric features (it is also strong enough to be measured in the majority of the stars). The derived abundances have been corrected to take into account the effects of NLTE by applying the equation provided by F06. The O abundances have been derived from the O triplet at ∼7774 Å. Corrections for the departure from NLTE have been applied by interpolating the grid of corrections calculated by Takeda (1997).

The derived [C/Fe] and [O/Fe] abundance ratios for the 9 BSSs are listed in Table 5.1 and shown in Figure 5.10 in comparison with the C and O abundances derived by Marino et al. (2012) for a sample of giant stars of the cluster. Even if possible systematic offsets could exist between the two datasets due to the use of different C and O diagnostics, we found that the abundances of the BSSs well match the mean locus described by the giants, with the exception of the BSS #70908 that shows a lower C abundance ([C/Fe]∼-1.5 dex). This C abundance is ∼0.5 dex lower than the most C-poor stars of ω Centauri. Thus, this star seems to be very similar to the C-depleted BSSs observed in 47 Tuc and its depletion can be interpreted in terms of MT processes.

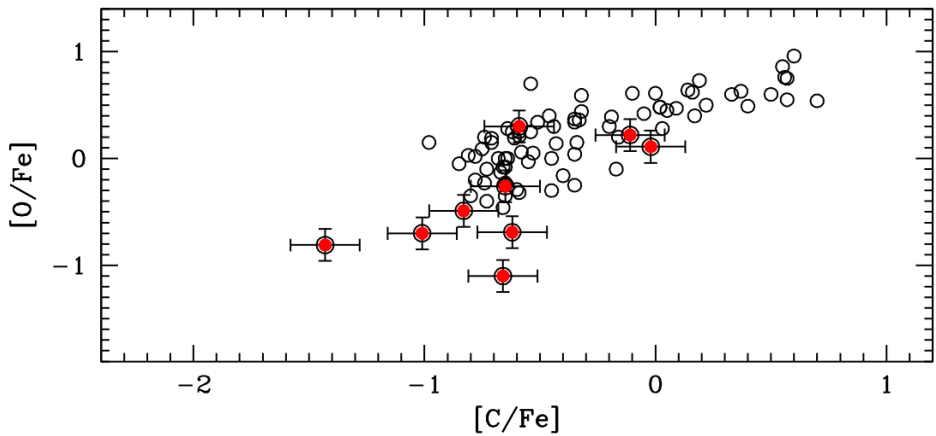

Figure 5.10: [C/Fe] and [O/Fe] ratios for the BSSs (red dots) compared with those of the RGB stars (empty circles) extracted from Marino et al. (2012).

| ID | [C/Fe] | [O/Fe] |
|------|--------|--------|
| 45771 | -0.66 | -1.10 |
| 70908 | -1.01 | -0.70 |
| 85967 | -0.83 | -0.49 |
| 100581 | -0.11 | 0.22 |
| 138528 | -0.62 | -0.69 |
| 178689 | -0.65 | -0.26 |
| 308593 | -0.02 | 0.11 |
| 332149 | -0.59 | 0.30 |
| 112188 | -1.43 | -0.81 |

Table 5.1: [C/Fe] and [O/Fe] ratios for the 9 coldest BSSs that are not affected by the radiative levitation.

## 5.6 Discussion

We measured radial and rotational velocities and Fe abundances for all the observed BSSs. For the coldest ones we obtained also preliminary results on the C and O abundances. We compare our findings with results in the literature and with our previous results obtained for other BSS samples in different GCs.

### 5.6.1 Radial velocities

Before discussing in details the main results of the present work it is necessary to verify whether some of the investigated stars do not belong to the cluster. In particular, we were not able to measure RV for BSS #330829 (its spectra appear featureless and there are cosmic rays on the core of the Hα line). This star is positioned at ∼ 636″ from the cluster centre, in a region where the contamination by a field population is relatively high. Moreover, this star is featureless in all the observed setups and we are not able to derive any metallicity. However, by assuming the mean metallicity of ω Centauri ([Fe/H]=-1.53, Harris 1996, 2010 edition), the lack of metallic lines can be reproduced only by assuming a rotational velocity larger than 100 $km\,s^{-1}$. Such a high value of v sin(i) is very unlikely for field stars (that commonly have v sin(i) lower than 50 $km\,s^{-1}$, see Cortés et al. 2009 and Carvalho et al. 2009). Thus we consider BSS #330829 as a cluster member. The radial velocity of BSS #195451 is discrepant with respect to the mean cluster value. Nevertheless, this star is located close to the cluster centre (r ≃ 280″) and we measured a metallicity [Fe/H]= -0.8 dex, compatible with the secondary peak in the distribution found by Sollima et al. (2009). For these reasons we decided to consider this BSS as a cluster member.

Also BSS #307749 has a discrepant RV. This star is at r ≃ 510 ″ and we are not able to derive any metallicity value. Moreover its value of v $\sin(i)$ (50 km s$^{-1}$) is compatible with both a cluster BSS and a field star. Hence we do not have any strong indication of its membership but we decided to include this star in the sample of cluster members.

This choice is also dictated by the result of L10, who found 5 BSSs with anomalous RVs in M4 and proposed that they were due to the stellar interactions suffered in the central regions of the cluster. Indeed, this could be also the case of BSS #307749 (and even BSS #195451) in ω Centauri. In conclusion, the mean RV value obtained from all the selected members is 237.2±2.2 km s$^{-1}$ with a very large dispersion σ=19.3 km s$^{-1}$. Both the mean value and the dispersion are in agreement with results in the literature.

### 5.6.2 Rotational velocities

The main results concerning the BSS rotational velocities in ω Centauri can be summarysed as follows:

- the v $\sin(i)$ distribution is very large and span a range between 0 up to 120 km s$^{-1}$.

- the fast rotating BSSs are less centrally segregated with respect to the BSSs with low v $\sin(i)$ values.

- the rotational velocity decreases suddenly for BSS temperatures exceeding ∼9000 K.

This is the first time that the rotational velocities have been obtained for such a large sample of BSSs in a GC. The large v $\sin(i)$ distribution is very similar to that found by L10 for the BSSs of M4, even if in that case the size of the sample was smaller. The majority of the targets (∼ 70% of the entire sample) shows rotational velocities lower than 50 km s$^{-1}$ and a distribution with a median value equal to 16.5 km s$^{-1}$. Nevertheless, some fast rotating BSSs (with rotational velocities higher than 50 km s$^{-1}$) are present. Moreover, we are able to derive only upper limits for some BSSs that rotate faster than 100 km s$^{-1}$.

This is the second time that a high percentage (∼ 30%) of fast rotating BSSs has been found (the first case being ∼ 40% in the GC M4, see L10). According to Ferraro et al. (2006a), the BSS population in ω Centauri shows a radial distribution completely flat (instead of centrally peaked or bimodal, as it is observed in the majority GCs). This evidence has been interpreted in terms of the dynamical status of ω Centauri, indicating that it is not relaxed yet (Ferraro et al. (2012, not even in the central regions)). The BSS population has been suggested to be no-collisional, but mainly formed from primordial binaries, whose radial distribution is not yet significantly altered by stellar collisions and by the dynamical evolution of the cluster. As a consequence, most of (if not all) the BSSs should be formed through MT. Hence they are expected to have high rotational velocities because the angular momentum is transferred together with the mass (Sarna & De Greve, 1996). Unfortunately, however, accurate simulations are lacking. Concerning the COL-BSSs, models by Leonard & Livio (1995) and Sills et al. (2005) suggest that some braking mechanisms can act, reducing the initial $v \sin(i)$. These kinds of mechanisms are expected to be at work also for the MT-BSSs. Our work shows for the first time, the rotational velocity distribution for a completely no-collisional sample of BSSs. As already noticed, the distribution is very large and covers also very low values of $v \sin(i)$ (down to 0 $\mathrm{km\,s^{-1}}$). This suggests that, if present, the braking mechanisms are very efficient in reducing the initially high rotation of the BSSs in $\omega$ Centauri.

We also checked the radial distribution of the fast rotating BSSs ($v \sin(i) > 50$ $\mathrm{km\,s^{-1}}$) with respect to the other stars in the sample, finding that they are less centrally segregated. This result is at odds with what found in the case of M4 (L10), where the fast rotators are closer to the cluster centre. In addition, a Kolmogorov-Smirnov test indicates that there is a very small probability ($\sim$ 7%) that the two samples have been extracted from the same population, while the probability was significantly higher ($\sim 44\%$) for M4.

Figure 5.6 shows the rotational velocities as a function of the stellar temperatures. The distribution shows a clear cut-off for the stars hotter than 9000 K. These stars are possibly affected by radiative levitation effects (see Figure 5.8). Hints of the same behaviour have already been observed by Lovisi et al. (2012) for a very small sample of BSSs in NGC 6397: also in that case, the two hottest BSSs (suffering from radiative levitation) have rotational velocities lower than 20 $\mathrm{km\,s^{-1}}$. A bimodal rotational velocity distribution is also observed in HB stars for which many authors (Peterson et al. 1995b; Behr et al. 2000a; Behr et al. 2000b; Recio-Blanco et al. 2002, 2004) observe a decrease in the rotational velocities for the hottest stars, that are also affected by radiative levitation. The larger sample (with respect to the previous investigations) studied in this work, suggest that BSSs behave like HB stars for what concern the $v \sin(i)$ distribution and further suggests that a link between the radiative levitation and the rotational velocity could exist.Finally, all the W Uma stars in the BSS sample of $\omega$ Centauri are also fast rotators.

### 5.6.3 Chemical abundances

Iron abundances have been measured for all the BSSs having v $\sin(i)$ < 30 km s$^{-1}$. In fact, for higher rotational velocities, the Gaussian line profile is severely altered, so that the EW measurement is not reliable anymore. The analysis of the iron content as a function of the temperature reveals an increase of the metallicity for higher temperature. As already discussed in the case of NGC 6397 and M30 (see Chapter 3 and 4), this is commonly interpreted as due to the occurrence of the radiative levitation mechanism that alters the surface chemical abundances and, in particular, it increases the surface metallicity. Even if our previous studies (see NGC 6397 and M30) show that all the BSSs hotter than ∼8000 K are affected by the radiative levitation, it is difficult to clearly recognize the effects of this transport mechanism, in the BSS sample of ω Centauri. In fact, ω Centauri shows a large metallicity distribution covering a wide range of values (from [Fe/H]≃-2 dex up to [Fe/H]≃-0.5 dex) that prevents us to clearly distinguish the BSSs with a metallicity enhanced by radiative levitation from the BSSs that formed from an iron-enriched population. According to the results obtained for the BSS samples in NGC 6397 and M30, we fix the boundary temperature for the occurrence of the radiative levitation at 8200 K. The differences found in the metallicity distribution for BSSs colder and hotter than ∼ 8200 K suggest that radiative levitation does indeed occur for stars hotter than this temperature threshold.

Finally, preliminary results have been obtained for the O and C abundances. We compared the derived BSS C and O abundances with those obtained by Marino et al. (2012) for a sample of RGB stars: most of the BSSs show [C/Fe] and [O/Fe] well in agreement with the values of giant stars. Nevertheless, one BSS (#70908) shows a lower C abundance ([C/Fe]∼-1.5 dex). Even if this evidence could be interpreted in terms of the MT-process, more investigations are necessary in order to exclude the possibility that such a kind of depletion is due to other transport mechanisms, for instance the gravitational settling.

Table 5.2: Coordinates, magnitudes, atmospheric parameters, radial and rotational velocities and Fe abundances of the ACS BSS sample.

| ID | RA (degrees) | DEC (degrees) | B | R | T (K) | log(g) | RV (km s$^{-1}$) | v sin(i) (km s$^{-1}$) | [Fe/H] | Notes |
|---|---|---|---|---|---|---|---|---|---|---|
| 6365 | 201.6079938 | -47.4235140 | 17.243 | 16.812 | 8762 | 4.3 | 222.5±0.3 | 10 | -1.01 ±0.08 | |
| 15741 | 201.6117617 | -47.4889112 | 16.530 | 16.159 | 9204 | 4.1 | 219.8±0.3 | 8 | -1.14 ±0.04 | |
| 15822 | 201.5894328 | -47.4673384 | 17.119 | 16.720 | 8762 | 4.3 | 248.5±0.3 | 120 | – | |
| 15913 | 201.5705488 | -47.4571373 | 17.564 | 16.974 | 8128 | 4.3 | 220.3±1.4 | 28 | -1.38 ±0.07 | |
| 25168 | 201.6458578 | -47.5029603 | 16.890 | 16.403 | 8158 | 4.1 | 231.6±1.1 | 48 | -0.71 ±0.00 | SX Phe |
| 25256 | 201.6394925 | -47.4735261 | 17.139 | 16.634 | 8341 | 4.2 | 243.8±0.6 | 16 | -1.03 ±0.04 | |
| 45771 | 201.6301436 | -47.5559563 | 16.573 | 16.035 | 7909 | 3.9 | 230.7±0.3 | 12 | -1.24 ±0.07 | |
| 45842 | 201.6315878 | -47.5197093 | 17.124 | 16.768 | 9634 | 4.5 | 233.6±0.5 | 23 | -1.00 ±0.04 | |
| 45881 | 201.6323122 | -47.5325120 | 17.330 | 16.754 | 8341 | 4.2 | 209.3±1.7 | 53 | – | SX Phe |
| 45952 | 201.6438653 | -47.5175900 | 17.652 | 17.083 | 8128 | 4.3 | 228.9±1.6 | 39 | – | SX Phe |
| 55728 | 201.6952662 | -47.5189999 | 16.380 | 16.040 | 0361 | 4.3 | 215.6±0.4 | 11 | -0.76 ±0.10 | |
| 55838 | 201.6800314 | -47.5026468 | 16.939 | 16.712 | 0703 | 4.6 | 236.5±2.6 | 24 | -1.04 ±0.10 | |
| 70908 | 201.7214003 | -47.5134752 | 17.492 | 16.855 | 7838 | 4.1 | 245.7±0.9 | 13 | -1.47 ±0.08 | |
| 85555 | 201.7101843 | -47.4718895 | 16.696 | 16.227 | 8882 | 4.2 | 207.9±1.4 | 55 | – | |
| 85967 | 201.7054542 | -47.4777023 | 17.423 | 16.765 | 7736 | 4.1 | 242.3±2.7 | 13 | -1.55 ±0.06 | |
| 86024 | 201.7196512 | -47.4662258 | 17.499 | 16.952 | 8128 | 4.3 | 223.4±1.2 | 17 | -1.50 ±0.10 | |
| 86148 | 201.6973886 | -47.4834818 | 17.678 | 17.141 | 8397 | 4.4 | 237.3±3.5 | 40 | – | |
| 11709 | 201.6628008 | -47.4828820 | 16.675 | 16.374 | 10245 | 4.4 | 251.8±0.6 | 30 | -0.50 ±0.04 | |
| 11781 | 201.6805045 | -47.4774985 | 16.905 | 16.419 | 8158 | 4.1 | 206.4±0.6 | 18 | -1.40 ±0.10 | |
| 11909 | 201.6839648 | -47.4607277 | 17.155 | 16.575 | 8341 | 4.2 | 235.2±1.0 | >100 | – | W Uma |
| 12043 | 201.6828741 | -47.4891104 | 17.414 | 16.949 | 8650 | 4.4 | 221.5±0.6 | 18 | -0.12 ±0.11 | |
| 12188 | 201.6910951 | -47.4784625 | 17.616 | 17.015 | 7838 | 4.1 | 240.4±1.6 | 27 | -1.87 ±0.11 | |
| 138386 | 201.6918172 | -47.4160236 | 17.139 | 16.680 | 8750 | 4.2 | 237.6±0.8 | 28 | -1.46 ±0.06 | SX Phe |
| 138442 | 201.7144872 | -47.3884409 | 17.429 | 16.949 | 8650 | 4.4 | 244.4±0.6 | 21 | -0.97 ±0.09 | |

| ID | RA (degrees) | DEC (degrees) | B | R | T (K) | $\log(g)$ | RV (km s$^{-1}$) | v sin(i) (km s$^{-1}$) | [Fe/H] | Notes |
|---|---|---|---|---|---|---|---|---|---|---|
| 138528 | 201.7166414 | -47.4278205 | 17.709 | 17.115 | 7759 | 4.2 | 239.5±0.7 | 18 | -1.04±0.11 | |
| 138533 | 201.6955695 | -47.4288758 | 17.731 | 17.183 | 8358 | 4.4 | 225.4±0.9 | 17 | -1.02±0.04 | |
| 157543 | 201.8003110 | -47.4152200 | 17.582 | 17.011 | 8128 | 4.3 | 240.1±0.9 | 9 | -1.64±0.05 | |
| 157564 | 201.7900914 | -47.4259965 | 17.695 | 17.163 | 8397 | 4.4 | 288.7±0.2 | 51 | – | |
| 157567 | 201.8023800 | -47.3995839 | 17.730 | 17.156 | 8063 | 4.3 | 251.9±3.4 | 0 | – | SX Phe |
| 161635 | 201.7490050 | -47.4271204 | 16.868 | 16.495 | 8716 | 4.2 | 254.7±2.0 | 117 | – | |
| 161724 | 201.7437092 | -47.3887751 | 17.446 | 16.928 | 8128 | 4.3 | 261.7±0.4 | 18 | -0.69±0.09 | |
| 168924 | 201.7838484 | -47.4632348 | 17.026 | 16.544 | 8716 | 4.2 | 261.0±0.4 | 11 | -0.78±0.04 | |
| 169012 | 201.7917715 | -47.4869316 | 17.414 | 16.928 | 8650 | 4.4 | 260.0±1.0 | 22 | -1.09±0.12 | |
| 169022 | 201.7867477 | -47.4370575 | 17.444 | 16.897 | 8128 | 4.3 | 230.5±0.7 | 12 | -1.46±0.06 | |
| 178368 | 201.7441258 | -47.4822019 | 17.007 | 16.705 | 10703 | 4.6 | 234.6±1.3 | 0 | -0.50±0.06 | |
| 178393 | 201.7405677 | -47.4525833 | 17.067 | 16.528 | 8341 | 4.2 | 233.1±0.4 | >100 | – | FEATURELESS |
| 178458 | 201.7713350 | -47.4793453 | 17.252 | 16.961 | 9968 | 4.6 | 228.9±2.4 | 6 | -1.87±0.00 | |
| 178463 | 201.7657691 | -47.4562230 | 17.272 | 16.757 | 8341 | 4.2 | 238.3±0.4 | 11 | -0.75±0.08 | |
| 178689 | 201.7624184 | -47.4626381 | 17.748 | 17.095 | 7759 | 4.2 | 233.4±0.6 | 18 | -1.29±0.07 | SX Phe |
| 195364 | 201.7957226 | -47.5138077 | 17.197 | 16.694 | 8341 | 4.2 | 257.5±0.1 | 73 | – | |
| 195451 | 201.7799168 | -47.5340856 | 17.669 | 17.028 | 7838 | 4.1 | 314.4±0.5 | 33 | -0.80±0.00 | |
| 203136 | 201.7445921 | -47.4973876 | 17.071 | 16.619 | 8750 | 4.2 | 257.6±0.7 | 12 | -1.07±0.10 | SX Phe |
| 203255 | 201.7619501 | -47.5090240 | 17.490 | 17.022 | 8586 | 4.4 | 240.5±3.1 | 12 | -1.41±0.14 | |
| 203281 | 201.7633330 | -47.5369467 | 17.588 | 17.013 | 8128 | 4.3 | 232.4±2.0 | >100 | – | |
| 516544 | 201.6780197 | -47.4425657 | 17.306 | 16.798 | 8341 | 4.2 | 261.6±2.8 | 109 | – | |

Table 5.3: Continued.

| ID | RA (degrees) | DEC (degrees) | B | I | T (K) | log(g) | RV (km s⁻¹) | v sin(i) (km s⁻¹) | [Fe/H] | Notes |
|---|---|---|---|---|---|---|---|---|---|---|
| 100581 | 201.8293171 | -47.5983161 | 17.783 | 17.001 | 7963 | 4.4 | 237.5±0.8 | 24 | -1.84 ± 0.12 | SX Phe |
| 102431 | 201.9356024 | -47.5867305 | 17.005 | 16.399 | 8762 | 4.3 | 246.4±0.1 | 76 | — | |
| 103804 | 201.8586653 | -47.5787055 | 16.789 | 16.351 | 10583 | 4.6 | 243.8±0.3 | 2 | -0.84 ± 0.07 | |
| 104641 | 201.8503340 | -47.5738003 | 17.505 | 16.838 | 8678 | 4.5 | 240.2±0.1 | 78 | — | |
| 104716 | 201.9535903 | -47.5732270 | 17.648 | 16.893 | 8067 | 4.4 | 248.9±3.4 | 20 | — | |
| 105432 | 201.8209694 | -47.5694759 | 17.257 | 16.576 | 8128 | 4.3 | 241.7±1.1 | 0 | -0.89 ± 0.13 | |
| 107232 | 201.8836798 | -47.5599300 | 17.460 | 16.766 | 8239 | 4.4 | 224.1±0.1 | >100 | — | FEATURELESS |
| 108182 | 201.8459107 | -47.5550924 | 17.076 | 16.185 | 8128 | 4.3 | 254.4±1.0 | 16 | -1.24 ± 0.10 | |
| 113023 | 201.9842670 | -47.5307156 | 17.358 | 16.665 | 8397 | 4.4 | 231.9±0.9 | 41 | — | |
| 123157 | 201.8721599 | -47.4835152 | 17.083 | 16.520 | 8650 | 4.4 | 239.5±1.1 | 40 | — | |
| 127397 | 201.9780442 | -47.4634932 | 17.285 | 16.861 | 9328 | 4.6 | 245.3±0.6 | 10 | -1.76 ± 0.00 | |
| 128030 | 201.8690663 | -47.4608768 | 16.692 | 16.102 | 8748 | 4.2 | 253.6±0.8 | >100 | — | W Uma |
| 128922 | 201.9077314 | -47.4566234 | 17.219 | 16.303 | 8128 | 4.3 | 238.4±1.7 | 41 | -1.69 ± 0.06 | |
| 131236 | 201.8660028 | -47.4455165 | 16.446 | 15.768 | 7938 | 4.0 | 228.2±0.1 | 93 | — | W Uma |
| 131407 | 201.9612096 | -47.4444442 | 16.971 | 16.305 | 8341 | 4.2 | 225.4±2.2 | 84 | — | |
| 200361 | 201.7784712 | -47.6007958 | 17.609 | 16.808 | 7912 | 4.3 | 265.5±1.7 | >100 | — | W Uma FEATURELESS |
| 200467 | 201.6311363 | -47.6006105 | 17.161 | 16.655 | 9328 | 4.6 | 226.9±2.5 | 41 | — | |
| 200606 | 201.7441538 | -47.5997050 | 16.610 | 16.051 | 8716 | 4.2 | 256.5±1.4 | 94 | — | |
| 200941 | 201.7551824 | -47.5981764 | 16.742 | 16.264 | 8750 | 4.3 | 210.8±1.2 | 18 | — | |
| 201398 | 201.6437122 | -47.5962351 | 17.152 | 16.206 | 8128 | 4.3 | 236.5±2.7 | 62 | -1.81 ± 0.00 | |
| 258947 | 201.6013857 | -47.3727477 | 16.258 | 15.908 | 10337 | 4.4 | 235.9±0.9 | 18 | 1.04 ± 0.17 | |
| 307749 | 201.5143975 | -47.5495173 | 16.993 | 16.250 | 8341 | 4.2 | 187.2±1.3 | 50 | — | |
| 308593 | 201.5104586 | -47.5442354 | 17.672 | 16.727 | 7820 | 4.3 | 222.7±0.6 | 19 | -1.33 ± 0.09 | |
| 311051 | 201.4958250 | -47.5300342 | 17.540 | 16.674 | 7934 | 4.3 | 213.7±2.1 | 51 | — | |
| 312293 | 201.4800185 | -47.5231779 | 16.175 | 15.680 | 9204 | 4.1 | 217.8±0.1 | 67 | — | |
| 326661 | 201.5154661 | -47.5484552 | 16.587 | 15.990 | 8668 | 4.2 | 214.8±1.9 | 13 | -1.95 ± 0.13 | |
| 330829 | 201.4497342 | -47.4264523 | 17.579 | 16.856 | 8155 | 4.4 | — | >100 | — | FEATURELESS |
| 332149 | 201.5371929 | -47.4191479 | 17.713 | 16.750 | 7820 | 4.3 | 196.8±0.7 | 27 | -1.66 ± 0.12 | |
| 509168 | 201.8033053 | -47.1665462 | 17.707 | 16.998 | 8247 | 4.5 | 224.8±0.7 | 6 | 0.50 ± 0.11 | |
| 514020 | 201.8407061 | -47.6219334 | 16.877 | 16.262 | 8341 | 4.2 | 263.6±0.2 | 80 | — | W Uma |
| 711338 | 201.4840039 | -47.6207373 | 17.331 | 16.612 | 8126 | 4.3 | 233.3±0.1 | 69 | — | |

Table 5.4: Coordinates, magnitudes, atmospheric parameters, radial and rotational velocities and abundances of the WFI BSSs sample.

# Chapter 6

# Conclusions

The understanding of the formation mechanisms of BSSs in GCs remains an open issue in modern stellar astrophysics, linking different aspects of stellar evolution and dynamics. Theoretical models identify the MT processes and stellar collisions as the main BSS formation channels. However some controversial predictions are provided by different models concerning several properties of the BSSs, for instance their rotational velocities (and the possible role played by braking mechanisms) and the occurence of surface chemical anomalies (and their timescales). From the observational point of view, a complete characterization of the spectroscopic properties (as radial and rotational velocities and chemical abundances) of BSSs is still lacking and only few works are devoted to this kind of investigation. This Thesis discusses the spectroscopic properties of BSSs in 4 GCs obtained from an extensive use of FLAMES@VLT high resolution spectra. The main results are summarysed as follows:

- The discovery of a large population of fast-rotating BSSs (with $v \sin(i) > 50$ km s$^{-1}$) in M4 and $\omega$ Centauri, corresponding to 40% and 30% of the entire sample, respectively.

- The discovery of elemental transport mechanisms driven by radiative levitation in the surface of BSSs hotter than $\sim$8000 K (in NGC 6397, M30 and $\omega$ Centauri).

- Some hints about the CO-depletion (probably linked to MT processes) in the red BSS sequence of M30 and in one BSS of $\omega$ Centauri.

## 6.1   Rotational velocities

Concerning rotational velocities, systematic studies were lacking before we started this Thesis work: results for only 7 BSSs were published by De Marco et al. (2005) and Shara et al. (1997), who used low resolution spectra. The first analysis with high resolution spectra for a large BSS sample, has been performed by F06 in 47 Tuc. Most of the stars in the sample do not rotate faster than 30 $km\,s^{-1}$ and only one BSS has $v\sin(i) \simeq 80$ $km\,s^{-1}$. By considering all the entire sample of BSSs observed in 47 Tuc, $\sim$7% of fast rotating ($v\sin(i) > 50$ $km\,s^{-1}$) BSSs has been found. Similar results have been obtained also for NGC 6397 and M30, where the percentage of fast rotating BSSs is 6% and 8%, respectively. On the contrary, very different results have been found for M4 and $\omega$ Centauri. The computed percentage of fast rotating BSSs is very high, corresponding to 40% and 30%, for M4 and $\omega$ Centauri respectively. These are the highest fractions of fast rotating BSSs ever observed in any GCs. No significative evidence of difference in the radial distribution of fast and slow rotators has been found in M4: a Kolmogorv-Smirnov test gives a probability of 49% that the two populations are extracted from the same parent distribution. Instead, the fast rotating BSSs in $\omega$ Centauri have been found to be less concentrated than "normal" BSSs with a level of statistical significance (the probability that the two populations are extracted from the same distribution is 7%). Unfortunately, the currently available theoretical models do not provide any explanation for these findings.

By considering the BSSs in all the analysed samples, we computed also the cumulative distribution of the rotational velocities. This is shown in Figure 6.1 and turns out to be peaked at $\sim 6$ $km\,s^{-1}$, showing that the vast majority ($\sim$78%) of the observed BSSs rotate less than 50 $km\,s^{-1}$. The distribution for low values ($v\sin(i) < 50$ $km\,s^{-1}$) can be fit with a Gaussian, but a very long tail towards high values ($> 100$ $km\,s^{-1}$) is present. Nevertheless, it is worth to keep in mind that the sizes of the various samples are very different, so that the distribution can be essentially dominated by the samples observed in 47 Tuc and (particularly) $\omega$ Centauri. We investigated also the presence of a possible link between the rotational velocity and the BSS formation mechanisms. No difference is found between the CO-depleted BSSs and the normal ones in 47 Tuc and both the samples have the same distribution, peaked towards very low values of $v\sin(i)$. Concerning M30, the size of the sample is very small, counting only 12 BSSs among which only 4 belong to the blue sequence. Nevertheless, some preliminary hints exist that the rotational velocity distribution for the BSSs in the blue sequence is shifted toward higher values. In order to further investigate this possibility, additional observations of the BSSs in the blue sequence are necessary. Finally, a very broad rotational velocity distribution has been found in the case of $\omega$ Centauri, where all the BSSs are expected to be originated through MT. Consequently, no conclusive statements can be done concerning the link between rotational velocities and formation mechanisms and more studies have to be performed in this direction.

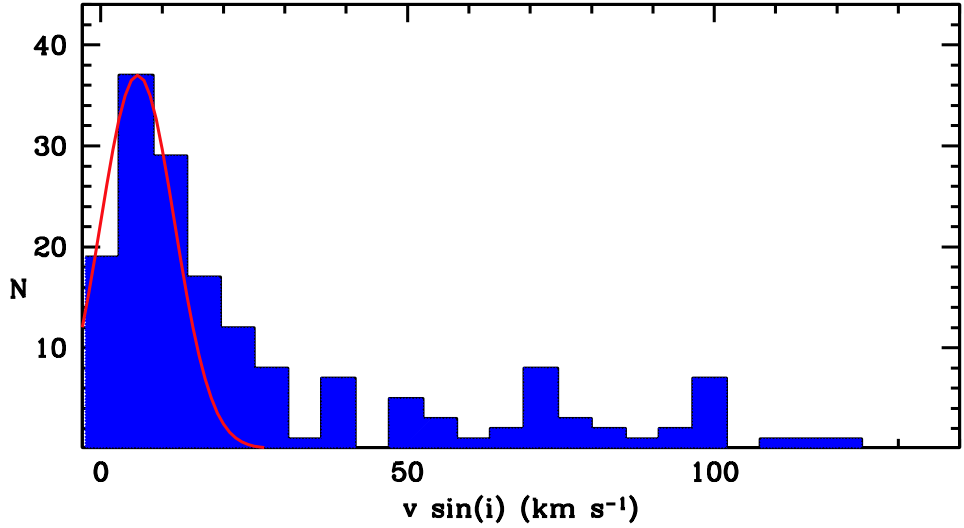

Figure 6.1: Cumulative distribution of the rotational velocities of the analysed BSSs for the studied GCs (47 Tuc, M4, NGC6397, M30 and $\omega$ Centauri). The red curve is the Gaussian function that best fits the main peak of the distribution.

Finally, for the largest studied sample (in $\omega$ Centauri) it is possible to observe that the rotational velocity distribution shows a cut off for T $\gtrsim$ 9000 K. The same peculiar behaviour has been observed also for HB stars in several GCs and it has been suspected to be produced by gravitational settling and radiative levitation, even if the theoretical models are not able to reproduce a similar decrease in the v $\sin(i)$ value. We suggest that the same process acting for the HB stars is the responsible of the abrupt decrease in the rotational velocity of BSSs in $\omega$ Centauri. Since the hottest BSSs in this sample are affected by the radiative levitation, we suggest that it could also be considered as a braking mechanism, able to reduce the inizitially high rotational velocity of BSSs.

## 6.2 Chemical anomalies and radiative levitation

One of the main findings of this work is the discovery that many BSSs are affected by radiative levitation. When the radiative acceleration exceeds the gravity acceleration, the former pushes elements toward the stellar surface, altering the chemical abundances of many elements. The effects of this transport mechanism are usually observed in HB stars in GCs and in MS stars of A and F spectral type in some OC. In this work we present the first evidence of the occurrence of this phenomenon also in the BSS atmospheres.

The effects of radiative levitation on BSSs have been observed for the first time in NGC 6397,

where an increase of the stellar metallicity has been found for BSSs hotter than $\sim$ 8000 K. The same effect has been observed also in the blue BSS sequence of the M30. Moreover, there is evidence that also the hottest BSSs in $\omega$ Centauri are experiencing radiative levitation. Nevertheless, $\omega$ Centauri is well known to have a large dispersion in the Fe abundance so that, at odds with NGC 6397 and M30, it is difficult to distinguish stars affected by the radiative levitation, from stars that are genuinely more metallic. No models including radiative levitation exist in the literature for BSSs, although this effect is very relevant in order to predict reliable chemical patterns for their atmospheres. Moreover, the occurrence of this phenomenon has to be taken into account also at the moment of the observations planning: in fact, if the scientific goal is to infer some information on the formation mechanisms from the surface chemical abundances, hot BSSs affected by the radiative levitation should be discarded, because their chemical patterns are mainly altered by this diffusion process.

## 6.3 C and O abundances

The first detection of C and O depletion has been found by F06 in 47 Tuc. Two sub-populations of BSSs have been discovered in a sample of 42 BSSs, one population showing C and O abundances in agreement with the cluster MS stars and 6 BSSs showing evidence of C and O depletion. That finding has been interpreted as the first spectroscopic evidence of the occurrence of the MT process. Nevertheless, the origin of this peculiar chemical pattern is not completely clear and it has been suggested that it could be explained by the large star-to-star abundance variations in the light elements (such as C, N, O, Na, Mg, Al) observed in all GCs. In fact, it is now well known that GCs are not simple stellar populations, but they host at least two different stellar populations, where the second one has been formed during the first few 100 Myr of the cluster life, from intra-cluster medium polluted by the first generation of stars. As a consequence of this process, a large scatter in the light element abundances is expected (and it is also observed in many GCs). In spite of this, the very low C and O abundances found by F06 for 6 BSSs in 47 Tuc is completely incompatible with a second generation scenario: in fact, the observed depletion covers $\sim$ 1 dex both in [C/Fe] and [O/Fe] and the abundances of both elements are lower than the minimum abundances observed in MS stars (see Briley et al. 2004; Carretta et al. 2010c for comparison). A depletion similar to that observed for some BSSs in 47 Tuc has never been observed in "normal" cluster stars (such as TO or RGB stars) and the only possible explanation (at the moment) is linked to the BSS formation mechanisms .

In our study we analyzed chemical abundances for BSSs having rotational velocities lower than $\sim$ 30 km s$^{-1}$ (in fact, for BSSs with higher rotational velocities, the line profile is severely altered by the high rotation and it prevents reliable abundance measurements). No evidence of CO-depletion has been found in the GC M4: all the studied BSSs show C and O abundances in good agreement with the cluster TO stars. No conclusive explanation has been found yet for this finding. According to the percentage of CO-depleted BSSs in 47 Tuc, in M4 we could have expected 1-2 BSSs with depleted C abundance, and zero or 1 BSS with both C and O depletion. Hence, the resulting no-detection may just be an effect of low statistics and it is still consistent with the expectations. In alternative, the lack of CO-depleted BSSs in M4 could be the evidence of a complete COL-BSS sample. Finally, it is also possible that, as suggested by F06, the CO-depletion is a transient phenomenon and (at least part of) the BSSs in M4 are indeed MT-BSSs that have already evolved back to normal chemical abundances. Concernig M30, we can discuss only the 5 coldest BSSs (because of radiative levitation affects the hottest ones): unfortunately they all belong to the red sequence so that no comparison can be drawn with the BSSs in the blue one. Only upper limits have been obtained and 4 out of 5 BSSs show upper limits not compatible with the O abundance distribution of the RGB stars, suggesting a depletion on the BSS surface. The depleted stars all belong to the red sequence that, according to Ferraro et al. (2009a), should include only BSSs produced through MT and for which the O depletion is expected. It is worth to note that, in spite of all the observational difficulties that have to be faced in the BSSs study (relative low stellar brightness, very high stellar crowding, frequent contamination by field stars and finally radiative levitation), this is the second evidence of O-depletion in the BSS surface. This supports the results found in 47 Tuc by F06 (according to which MT-BSSs are CO-depleted) and the results by Ferraro et al. (2009a), according to which the red BSS sequence in M30 is formed by MT-BSSs. Evidence of C-depletion has been also found for one BSS in $\omega$ Centauri. This stellar system is thought to be not a "genuine" GC but the remnant of a dwarf galaxy accreted by the Milky Way. As a consequence, a very large intrisic spread both in light and iron-peak elements is present. By considering only the 9 coldest BSSs that do not suffer from radiative levitation, we discovered that one of them is not compatible with the large [C/Fe] distribution observed for the RGB stars. As in the case of the O abundances for the M30 sample, this could be the second evidence of a C-depleted BSS. Nevertheless, more investigations are necessary.

# Appendix A

# GALA: an automatic tool for the abundance analysis of stellar spectra

– Extracted from Mucciarelli A., Pancino E., Lovisi L., Ferraro, F. R. 2013, ApJ submitted

## A.1   Introduction

The last decade has seen a significant improvement in the study of the chemical composition of the stellar populations (in our Galaxy and its satellites), thanks to the 8-10 meters class telescopes, coupled with the design of several multi-object mid/high-resolution spectrographs, e.g. FLAMES (Pasquini et al., 2002) mounted at the Very Large Telescope. These instruments have allowed to enlarge the statistical significance of the acquired stellar spectra, but also they have required a relevant effort to manage such large databases.

The next advent of new surveys aimed to collect huge samples of mid-high resolution stellar spectra, as for instance the European Space Agency *Gaia* mission, the Gaia-ESO Spectroscopical Survey committed by the European Southern Observatory, the APOGEE Survey at the Apache Point Observatory, and the RAVE Survey at the Anglo-Australian Observatory, will make available in real time to the astronomical community an enormous volume of data. This perspective, coupled with the huge amount of high quality spectra available in the main on-line archives (and not yet totally analysed) clarifies the urgency to develop automatic tools able to rapidly and reliably manage such samples of spectra.

In the last years several codes aimed at the automatic measurements of the chemical abundances have been already developed. They are mainly based on the comparison between the observed spectrum and grids of synthetic spectra, for instance ABBO (Bonifacio & Caffau, 2003), MATISSE (Recio-Blanco, Bijaoui & de Laverny, 2006), SME (Valenti & Piskunov, 1996),

SPADES (Posbic et al., 2012), MyGIsFOS (Sbordone et al., 2010). In particular, in these codes the main effort has been devoted to robustly determine the atmospheric parameters (and hence the elemental abundances) for low (<50) signal-to-noise (SNR) spectra and generally to develop an algorithm able to accurately treat different kind of stars (in terms of metallicity and stellar parameters).

In this paper we present and discuss a new code (named GALA) specifically designed for automatically determining the atmospheric parameters by using the observed equivalent widths (EWs) of metallic lines in stellar spectra, at variance with the majority of the available automated codes. GALA is a tool developed within Cosmic-Lab, a 5-years project funded by the European Research Council and it is freely available at the website of the project *http://www.cosmic-lab.eu/Cosmic-Lab/Products.html.*

The paper is structured as follows: Section 2 discusses the outline of the classical method to derive the main parameters and the interplay occurring among them; Section 3 describes the algorithm; Section 4 describes the identification and the rejection of the outliers and Section 5 discusses other aspects of the code. Section 6 provides a complete description of the uncertainties calculation. Finally, Sections 7, 8 and 9 discuss a number of tests performed to check the stability and the performances of GALA.

## A.2  The method

The main advantage of infering the stellar atmospheric parameters from the EWs is its reproducibility: any researcher can directly compare its own results about a given star with other analysis based on the same approach. This allows to distinguish between discrepancies due to the method (i.e. the measured EWs) and those due to the physical assumptions of the analysis (model atmospheres, atomic data...). On the other side, one of the most critical aspects of this method is the particular accuracy needed in the definition of the linelist, by excluding blended lines. In fact, the codes developed to calculate the abundance from the measured EW compare the latter with the theoretical strength of the line, changing the abundance until the observed and theoretical EW match within a convergence range. The theoretical line profile is usually calculated including the continuum opacity sources but neglecting the contribution of the neighbouring lines (see Castelli, 2005b, for details), hence the spectral lines to be analysed with this technique are to be checked accurately against blending (a practice not always performed). Otherwise, the use of synthetic spectra allows to use also blended features (and in principle to exploit the information derived from all the pixels), but it is more expensive in terms of computing time, because large grids of spectra must to be computed, at different parameters and with different chemical compositions, and each change in the atomic data leads to a recomputation of the synthetic spectra. This approach does not allow a direct comparison with other analysis.

### A.2.1 The classical spectroscopic method

The main parameters that define the model atmosphere, namely the effective temperature ($T_{\text{eff}}$), the surface gravity (log g), the microturbulent velocity ($v_{\text{t}}$) and the overall metallicity ([M/H][1]) are constrained through four *levers*:

1. *Temperature:* the best value of $T_{\text{eff}}$ is derived by imposing the so-called *excitation equilibrium*, requiring that there is no correlation between the abundance and the excitation potential $\chi$ of the neutral iron lines. The number of electrons populating each energy level is basically a function of $T_{\text{eff}}$, according to the Boltzmann equation. If we assume a wrong $T_{\text{eff}}$ in the analysis of a given stellar spectrum, we need different abundances for matching the observed profile of transitions with different $\chi$. For instance, the use of a value of $T_{\text{eff}}$ too large will lead to under-populate the lower energy levels, thus the predicted line profile for low-$\chi$ transitions will be too shallow and a higher abundance will be needed to match the line profile. On the other hand, a wrong (too low) $T_{\text{eff}}$ will lead to a deeper line profile for the low $\chi$ transitions.

   For this reason, a wrong, too large value of $T_{\text{eff}}$ will introduce an anticorrelation between abundances and $\chi$, and in the same way, a positive correlation is expected in the case of the adoption of a value of $T_{\text{eff}}$ too small;

2. *Surface gravity:* the best value of log g is derived with the so-called *ionization equilibrium* method, requiring that for a given species, the same abundance (within the uncertainties) has been obtained from lines of two ionization states (typically, neutral and singly ionized lines). Being the gravity a direct measure of the pressure of the photosphere, variations of log g lead to variations of the ionized lines (very sensitive to the electronic pressure), while the neutral lines are basically insensitive to this parameter. This method assumes implicitly that the energy levels of a given species are populated according to the Boltzmann and Saha equations (thus, under Local Thermodynamical Equilibrium (LTE) conditions).

---

[1]We adopted the classical *bracket* notation where [X/H]= $A(X)_{star}$-$A(X)_{\odot}$, where A(X)= $\log N_A/N_H + 12$.

Possible departures from this assumption (especially critical for metal-poor and/or low-gravity stars) could alter the derived gravity when it is derived from the ionization balance, because non-LTE effects affect mainly the neutral lines (even if the precise magnitude of the departures from LTE for the iron lines is still matter of debate). As sanity check, following the suggestion by Edvardsson (1988), surface gravities determined from the ionization equilibria have to be *checked - when possible - with gravities determined from the wings of pressure broadened metal lines*;

3. *Microturbulent velocity:* $v_t$ is computed by requiring that there is no correlation between the iron abundance and the line strength (see Mucciarelli, 2011b, for a discussion about different approaches). The microturbulent velocity affects mainly the moderate/strong lines located along the flat regime of the curve of growth, while the lines along the linear part of the curve of growth are mainly sensitive to the abundance instead of the velocity fields. The necessity to introduce the microturbulent velocity as an additional broadening (added in quadrature to the Doppler broadening) arises from the fact that the non-thermal motions (basically due to the onset of the convection in the photosphere) are generally not well described by the 1-dimensional, static model atmospheres. Citing Kurucz (2005), *microturbulent velocity is a parameter that is generally not considered physically except in the sun*, because in the sun the velocity fields can be derived as a function of the optical depth through the analysis of the intensity spectrum (as performed by Fontenla, Avrett & Loeser, 1993). For the other stars $v_t$ represents only a corrective factor that minimizes the line-to-line scatter for a given species and it compensates (at least partially) to the incomplete description of the convection as implemented in the 1-dimensional model atmospheres;

4. *Metallicity:* [M/H] is chosen according to the average iron content of the star, assuming [Fe/H] as a proxy of the overall metallicity. Generally, [Fe/H] is adopted as a good proxy of the metallicity because of its large number of available lines, but it does not indicate necessarily the overall metallicity of the studied star. In fact, iron is generally not the most abundant element in the stars, while elements as C, N and O would be the best tracers of the stellar metallicity (but they are difficult to measure).

Because of its statistical nature, the spectroscopic optimization of all the parameters simultaneously can be performed only if we have a sufficient number of Fe lines, distributed in a large range of EW and $\chi$ and in two levels of ionization. Alternatively, $T_{\rm eff}$ and log g can be inferred from the photometry (for instance with the isochrone-fitting technique or employing empirical or theoretical $T_{\rm eff}$–color relations) or by fitting the wings of damped lines (as the hydrogen Balmer lines or the Mg $b$ triplet) sensitive to the parameters, and only $v_{\rm t}$ needs to be tuned spectroscopically (following the approach described above). Note that some authors consider the method to derive the parameters based on these *levers* only as sanity checks performed *a posteriori* on the photometric parameters, while other authors rely on these constraints to infer the best parameters.

## A.2.2 The interplay among the parameters

In light of the method described above, it is worth to bear in mind that the atmospheric parameters are correlated with each other. In fact, the strongest lines are typically those with low $\chi$: Fig. A.1 plots all the transitions in the range $\lambda$= 4000-8000 $\mathring{A}$ and with $\chi$ <10 eV in the Kurucz/Castelli database[2] in the plane $\chi$ vs $\log(gf) - \theta\chi$ [3]. As mentioned in Section A.2.1, $T_{\rm eff}$ and $\chi$ are strictly linked, as well as there is a link between $v_{\rm t}$ and the line strength. Hence, the statistical correlation between $\chi$ and the line strength leads to a correlation between $T_{\rm eff}$ and $v_{\rm t}$. Thus, a variation of $T_{\rm eff}$ implies a variation of $v_{\rm t}$. Also, variations of $T_{\rm eff}$ and $v_{\rm t}$ will change differently the abundances derived from different levels of ionization (hence, the gravity).

Let us consider an ATLAS9 model atmosphere computed with $T_{\rm eff}$=4500 K, log g=1.5, $v_{\rm t}$=2 km/s and [M/H]=−1.0 dex (Castelli, 2004) and a set of neutral and singly ionized iron lines (predicted to be unblended through the inspection of a synthetic spectrum calculated with the same parameters and convoluted at a spectral resolution of 45000). The EWs of these transitions are computed by integrating the theoretical line profile through the WID subroutine implemented in the WIDTH9 code (Castelli, 2005b). This means that each of these EWs will provide exactly [Fe/H]=−1.0 dex when they are analysed by using the model atmosphere described above.

The analysis of these lines (adopting always the same set of EWs) is repeated investigating a regular grid of the atmospheric parameters, namely $T_{\rm eff}$= 3600–5400 K, log g= 0.5–2.5, $v_{t}$= 1.0–3.0 km/s, steps of $\delta T_{\rm eff}$= 200 K, $\delta logg$= 0.2, $\delta v_{t}$= 0.5 km/s and assuming for all the models [M/H]= −1 dex. Fig. A.2, A.3 and A.4 display the quite complex interplay occurring among the atmospheric parameters.

---

[2]http://wwwuser.oat.ts.astro.it/castelli/linelists.html
[3]The term $\log(gf) - \theta\chi$ is used as theoretical proxy of the line strength, where log(gf) is the oscillator strength and $\theta$= 5040/$T_{\rm eff}$ $eV^{-1}$.

- Fig. A.2 shows the behaviour of the slope $S_\chi$ of the A(Fe)–$\chi$ relation for the above sample of lines as a function of $T_{\text{eff}}$, keeping gravity fixed, but varying the microturbulent velocity. The thick grey line connects points calculated with the original $v_{\text{t}}$ of the model. The global trend is basically linear, at least if we consider a range of $\pm 1000$ K around the original $T_{\text{eff}}$. The inset panel shows the behaviour of $T_{\text{eff}}$ for which $S_\chi$ is zero (thus, the best $T_{\text{eff}}$) as a function of $v_{\text{t}}$ and considering different gravities. The derived best temperature increases increasing $v_{\text{t}}$ (at fixed log g); the gravity has only a second-order effect and it does not change the general behaviour of the best $T_{\text{eff}}$ as function of $v_{\text{t}}$.

- Fig. A.3 summaryzes the behaviour of the slope $S_{EWR}$ of the A(Fe)–EWR relation as a function of $v_{\text{t}}$, keeping gravity fixed at the original value but varying $T_{\text{eff}}$. For a given temperature, the slope decreases increasing $v_{\text{t}}$, with a behaviour that becomes less steep at $v_{\text{t}}$ larger than the original value. The effect of the $T_{\text{eff}}$ is appreciable for $T_{\text{eff}}$ larger than the original value, while for lower $T_{\text{eff}}$ all the curve are very similar to each other. The inset shows the change of the best value of $v_{\text{t}}$ (for which the slope of the A(Fe)–EWR relation is zero) as a function of $T_{\text{eff}}$ and for different log g. The observed trend is quite complex: basically, we note that the best value of $v_{\text{t}}$ is very sensitive to $T_{\text{eff}}$, when the latter is overestimated with respect to the true temperature, but with a negligible dependence from gravity, while the behaviour is the opposite when $T_{\text{eff}}$ is under-estimated, with a degeneracy between $T_{\text{eff}}$ and the best $v_t$ but a consistent dependence from gravity.

- Fig. A.4 shows the behaviour of the difference between A(Fe I) and A(Fe II) as a function of log g, assuming $v_{\text{t}}$= 2 km/s and for different $T_{\text{eff}}$. The general behaviour is linear and the iron difference increases sensibly increasing the temperature. The best value of gravity (see the inset in Fig. A.4) is highly sensitive to changes in $T_{\text{eff}}$, with $\delta \log g / \delta T_{\text{eff}} \simeq 1$ dex/300 K in the investigated case, but for a fixed $T_{\text{eff}}$ turns out to be marginally sensitive to $v_{\text{t}}$ (this is due to the fact that the Fe II lines are basically distributed in strength in a similar way to the Fe I lines).

It is important to bear in mind that the these considerations are appropriate for the investigated case of a late-type star but the dependencies among the parameters can be different for different regimes of atmospheric parameters and/or metallicity. However, the example presented above demonstrates that an analytic approach to derive the best model atmosphere is discouraged because it needs to know the precise topography of the parameters space and requires the inspection of a large number of model atmospheres.

## A.3 GALA

GALA is a program written in standard Fortran 77, that uses the WIDTH9 code developed by R. L. Kurucz in its Linux version (Sbordone et al., 2004) to derive the chemical abundance of single, unblended absorption lines starting from their measured EWs. We used our own version of WIDTH9, modified in order to have a more flexible format for the input/output files with respect to the standard version of the code available in the website of F. Castelli, while the input physics and the method to derive the abundances are unchanged.

GALA is specifically designed to

1. choose the best model atmosphere by using the observed EWs of metallic lines;

2. manage the input/output files of the WIDTH9 code;

3. provide statistical and graphical tools to evaluate the quality of the final solution and the uncertainty of the derived parameters.

GALA is designed to handle both ATLAS9 (Castelli & Kurucz, 2004) and MARCS (Gustafsson et al., 2008) model atmospheres, that are the most popular employed dataset of models. The current version has been compiled with the Fortran Intel Compiler (versions 11 and 12) and tested on the Leopard, Snow Leopard and Lion Mac OSX systems, and on the Ubuntu, Fedora and Mandriva Linux platforms.

### A.3.1 Optimization parameters

GALA has been designed to perform the classical chemical analysis based on the EWs in an automated way. The user can choose to perform a full spectroscopic optimization of the parameters or to optimize only some parameters, keeping the other parameters fixed to the specified input values.

The algorithm optmizes one parameter at a time, checking continuously if the new value of a given parameter changes the validity of the previously optmized ones.

For each atmospheric parameter $X$ (corresponding to $T_{eff}$, log g, $v_t$ and [M/H]) we adopt a specific *optimization parameter $C(X)$*, defined in a way that it turns out to be zero when the best value of the $X$ parameter has been found. Hence, GALA varies $X$ until a positive/negative pair of the $C(X)$ is found, thus bracketing the zero value corresponding to the best value. Thus, the condition $C(\tilde{X})= 0$ identifies X=$\tilde{X}$ as best value of the given parameter. Finally, the best solution converges to a set of parameters which verifies simultaneously the constraints described in Section A.2.1.

According to the literature, the adopted *C(X)* have been defined to parametrize the conditions listed in Section A.2.1:

1. the angular coefficient of the A(Fe)–$\chi$ relation ($S_\chi$) to constrain $T_{eff}$; the lower panel of Fig. A.5 shows the change of this slope for a set of theoretical EWs computed with the correct $T_{eff}$ (grey points) and with temperatures varied by $\pm 500$ K (empty points). The variation of $T_{eff}$ produces a change of the slope (but also a change in the y-intercept);

2. the angular coefficient of the A(Fe)–EWR relation ($S_{EWR}$) to constrain $v_t$ (where EWR indicates the reduced EW, defined as EWR=lg($EW/\lambda$)). The upper panel of Fig. A.5 shows the same set of theoretical EWs analysed with different $v_t$;

3. the difference of the mean abundances obtained from Fe I and Fe II lines to constrain the gravity;

4. the average Fe abundance to constrain the metallicity of the model.

If the errors in the EW measurement ($\sigma_{EW}$) are provided as an input, the slopes are computed taking into account the abundance uncertainties of the individual lines; the uncertainty on the iron abundance of a given line is estimated from the difference of the iron abundance computed for the input EW and for EW+$\sigma_{EW}$ [4]. In the A(Fe)–$\chi$ plane the uncertainties of $\chi$ are reasonably assumed negligible, because the uncertainties of $\chi$ are typically less than 0.01 eV, while in the A(Fe)–EWR plane the least square fit takes into account the uncertainties in both the axis (following the prescriptions by Press et al., 1992).

The flexibility of GALA permits to simultaneously deal with stars of different spectral types. This is made by considering that a large number of Fe I lines is generally available for F-G-K spectral types stars, whereas they are less numerous (or lacking) in O-B-A stars, for which a large number of Fe II lines is typically available. Also, in some spectral regions there is a large number of lines for other iron-peak elements (mainly Ni, Cr and Ti). For this reason, GALA is designed to optimize the parameters also using other lines instead of Fe I lines, by appropriately configuring the code. In the following, we will refer ever to the optimization made by using Fe I for $T_{\mathrm{eff}}$ and $v_{\mathrm{t}}$, but our considerations are valid also for other elements with a sufficient number of lines.

---

[4] The uncertainties in A(Fe) are assumed symmetric with respect to the variations of the EW ($\pm\sigma_{EW}$); we checked that this assumption is correct at a level of $\sim 0.01$ dex.

## A.3.2    The main structure

GALA is structured in three main Working-Blocks:

(1) the *Guess Working-Block* is aimed at finding the guess atmospheric parameters in a fast way (this is especially useful in cases of large uncertainties or in lacking of first-guess value for the parameters);

(2) the *Analysis Working-Block* which finds the best model atmosphere through a local minimization and starting from the guess parameters provided by the user or obtained through the previous block;

(3) the *Refinement Working-Block* that refines the solution, starting from the atmospheric parameters obtained in the previous block.

GALA can be flexibly configured to use different combinations of the three main Working-Blocks. We defer the reader to Section A.8.4 for a discussion of the effects of the Working-Blocks. In the following we describe the algorithm of each block and the cases in which they are recommended.

**Guess Working-Block**

If the atmospheric parameters are poorly known, this Working-Block verifies them quickly by exploring the parameters space in a coarse grid. Thus, it saves a large amount of time if the initial parameters are far away from the correct solution.

1. As a first step, the abundances for each line are derived with the input parameters and some lines are labelled as outliers and excluded from the analysis (the criteria of the rejection are described in Sect. A.4). The surviving lines will be used in this Working-Block and no other line rejection will be performed until convergence.

2. The metallicity of the model is eventually readjusted according to the average iron abundance.

3. $S_\chi$ is computed with the input $T_{\rm eff}$ and with a $T_{\rm eff}$ varied by $+500$K (if $S_\chi$ is positive) or $-500$ K (if $S_\chi$ is negative). This procedure is repeated until a pair of positive/negative values of $S_\chi$ is found, thus to bracket the $T_{\rm eff}$ value for which $S_\chi = 0$. The behaviour of $S_\chi$ as a function of $T_{\rm eff}$ is described with a linear relation, finding the value of $T_{\rm eff}$ for which $S_\chi = 0$. The description of this relation with a linear fit is legitimate inasmuch as the employed $T_{\rm eff}$ range is relatively small (in this case 500 K), because for larger range the behaviour of $S_\chi$ as a function of $T_{\rm eff}$ could become non linear (mainly due to the interplay with the other parameters).

4. The new value of $T_{\rm eff}$ is adopted to find a new value of $v_{\rm t}$, following the same approach used for $T_{\rm eff}$ and searching for a positive/negative pair of $S_{EWR}$ over a range of 0.5 km/s.

5. Finally, a new value of log g is found, starting from the $T_{\rm eff}$ and $v_{\rm t}$ derived above, by searching for a positive/negative pair of $\Delta(Fe)$, over a range of 0.5 dex in gravity.

The entire procedure (from the optimization of [M/H] to that of log g) is repeated for a number of iterations choosen by the user and finally a new set of input parameters is found. Generally three/four iterations are sufficient to find a good solution. The final solution is accurate enough to identify the neighbourhood of the real solution in the parameters space but it could be unreliable since it needs to be checked for the covariances among the parameters (which is the task of the *Analysis Working-Block*.

**Analysis Working-Block**

This Working-Block performs a complete optimization starting from the input parameters provided by the users or from those obtained with the *Guess Working-Block*. This block is developed to find a robust solution under the assumption that the input values are reasonably close to the real solution. When good priors are available (for instance, in the case of stellar cluster stars) this block is sufficient to find the solution, without the use of the guess block. Otherwise, when the guess model is uncertain (for instance, in the case of field stars for whose reddening, distance and evaluative mass could be highly uncertain, or, generally, in the case of inaccurate photometry), the analysis block is recommended to be used after the guess block. Fig. A.6 shows as a flow chart the main steps of this Working-Block. The following iterative procedure is performed:

1. The procedure starts by computing the abundances using the guess parameters and performing a new line rejection (independent from that of the previous block). At variance with the previous Working-Block, now the parameters are varied by little steps, configured by the user.

2. The model metallicity is refined to match the average iron abundance.

3. A new model with different $T_{\rm eff}$ is computed, according to the sign of $S_\chi$ of the previous model (i.e., a negative slope indicates a overestimated $T_{\rm eff}$ and vice versa). New models, varying only $T_{\rm eff}$, are computed until a pair of negative/positive $S_\chi$ is found. Thus, these two values of $T_{\rm eff}$ identify the range of $T_{\rm eff}$ where the slope is zero. $T_{\rm eff}$ corresponding to the minimum $|S_\chi|$ is adopted.

4. The same procedure is performed for $v_{\rm t}$. If the final value of $v_{\rm t}$ is different from that used in the previous loop, GALA goes back to item (2), checking if the new value of $v_{\rm t}$ needs a change in [M/H] and $T_{\rm eff}$. Otherwise, the procedure moves on to the next loop.

5. The surface gravity is varied until a positive/negative pairs of $\Delta(Fe)$ is found. If the output log g differs from the input value, GALA returns to (2) with the last obtained model atmosphere and the entire procedure is repeated.

6. When a model that sastisfies simultaneously the four constraints is found, the procedure ends and the next star is analysed.

We stress that the method employed in this Working-Block is very robust but it has the disadvantage of being slow if the guess parameters are far from the local solution.

**Refinement Working-Block**

This block allows to repeat the previous Working-Block using the solution obtained in the previous block as a starting point. A new rejection of the outliers is performed and the same approach of the *Analysis Working-Block* is used. This block can be useful to refine locally the solution when the first block is swichted off.

The main advantage of the *Refinement Working-Block* is that the new line-rejection is performed by using accurate atmospheric parameters (because obtained from the *Analysis Working-Block*). As will be discuss in Section A.4, the line-rejection performed on abundance distributions obtained with wrong parameters can be risky, losing some useful lines.

## A.4   Weeding out the outliers

The detection and the rejection of lines with discrepant abundances are crucial aspects of the procedure and require some additional discussion. Before ruling out a line from the line list, we need to understand the origin of the detected discrepancy. Basically, the main reasons for a discrepant abundance are:

(1) inaccurate atomic data (i.e. oscillator strengths) that can under- or over-estimate the abundance;

(2) unrecognized blends with other lines (providing systematically overestimated abundances);

(3) inaccurate EW measurement.

The first two cases can be partially avoided with an effort during the definition of the adopted linelist, including only transitions with accurate log gf and checking each transition against blending, according to the atmospheric parameters and the spectral resolution.

GALA rejects the lines according to the following criteria:

1. lines weaker or stronger than the input EWR thresholds are rejected, in order to exclude either weak and/or strong lines. In fact, weak lines can be heavily affected by the noise, whereas strong lines can be too sensitive to $v_t$ and/or they can have damping wings for which the fit with a Gaussian profile could be inappropriate, providing a systematic under-estimate of the EW;

2. lines whose uncertainty on the EW measurement is larger than an input threshold chosen by the user and expressed as a percentage. Note that not all the codes developed to measure EWs provide an estimate of the EW error, despite the importance of this quantity. For instance, among the publicly available codes aimed to measure EWs, DAOSPEC (Stetson & Pancino, 2008) and EWDET (Ramirez et al., 2001) provide accurate uncertainty evaluations for each line, while SPECTRE (Fitzpatrick & Sneden, 1987) and ARES (Sousa et al., 2007) do not include EW error calculations. For this reason GALA works even if $\sigma_{EW}$ are not provided, although this affects the final solution accuracy, because all the transitions will be weighed equally, despite their different measurement quality;

3. lines are rejected according to their distance from the best-fit lines computed in the A(Fe)–$\chi$ and A(Fe)–EWR planes through a $\sigma$-rejection algorithm. A $\sigma$-rejection from the best-fit lines in the planes used for the optimization is more robust with respect to a simple $\sigma$-rejection based on the abundance distribution. In the latter case, there is the risk to lose some lines important for the analysis, thus biassing the results. Fig A.7 explains this aspect: we consider a synthetic spectrum of a giant star ($T_{\mathrm{eff}}$= 4500 K) and we measure the EWs after the injection of Poissonian noise in the spectrum in order to reproduce a reasonable good SNR ($\sim$30). Fig. A.7 shows the distribution of the Fe I lines in the A(Fe)–$\chi$ plane when the chemical analysis is performed by using a wrong model atmosphere with $T_{\mathrm{eff}}$= 5200 K (thus leading to an anticorrelation between A(Fe I) and $\chi$). In the upper panel the outliers were rejected according to the median value of the abundance distribution, shown as gray solid line, while the two dashed lines mark $\pm 3\sigma$ level and black points are the surviving lines. In the lower panel the rejection is performed according to the distance from the best-fit line (shown as solid line while the two dashed lines mark $\pm 3\sigma$ level). It is evident that in the first case the majority of the discarded lines are those with low $\chi$ (thus, the most sensitive to the $T_{\mathrm{eff}}$ changes), with the risk of introducing a bias in the $T_{\mathrm{eff}}$ determination. On the other hand, the method of rejection shown in the lower panel of Fig. A.7 preserves the low-$\chi$ lines, guaranteeing the correctness of the final solution.

An important point is that the outliers rejection in GALA is not performed independently in each iteration of the code, but only at the beginning of each Working-Block. This is especially important, because it allows to use always the same sample of lines during the optimization process, avoiding the risk to introduce spurious trends in the behaviour of the given *optimization parameter* as a function of the corresponding atmospheric parameters. In fact, the values of C(X) derived from two different sets of lines of the same spectrum, but for which an independent rejection of the outliers has been performed, cannot be directly compared to each other to derive X. In particular, this effect is magnified in cases of small number of lines, where the impact of the lines rejection can be critical.

## A.5 More details

### A.5.1 A comment about the gravity

The most difficult parameter to be constrained with the classical spectroscopic method is the gravity. This because of the relatively small number of available Fe II lines, which can vary in the visual range from a handful of transitions up to $\sim$20, depending on the spectral region and/or the metallicity (for instance, some high-resolution spectra with a small wavelength coverage, as the GIRAFFE@VLT or the Hydra@BlancoTelescope spectra, can be totally lacking in Fe II lines).

GALA is equipped with different options to optimize log g:

(1) the normal optimization by using the difference between the average abundances from neutral

and singly ionized iron lines (as described above);

(2) the gravity is computed from the Stefan-Boltzmann equation, by providing in input the term $\epsilon = \log(4\text{GM}\pi\sigma/\text{L})$, where G is the gravitational constant, $\sigma$ the Boltzmann constant and M and L are the mass and the luminosity of the star. Thus, during the optimization process, the gravity is re-computed (as log g= $\epsilon$-4log $T_{\text{eff}}$) in each iteraction according to the new value of $T_{\text{eff}}$;

(3) the gravity is computed by assuming a quadratic relation log g= A+B·$T_{\text{eff}}$+C·$T_{\text{eff}}^2$ and providing in input the coefficients A, B, C. This option is useful when the investigated stars belong for instance to the same stellar cluster and log g and $T_{\text{eff}}$ can be parametrized by a simple relation (i.e. as that described by a theoretical isochrone for a given evolutionary stage).

The user can choose the way to treat log g (fixed or optimized following one of the methods described above); if the optimization of log g from the iron lines is requested but no Fe II lines are available, GALA will try to use the second option (lines of other elements in different stages of ionization), or eventually will fix log g to the input value.

### A.5.2   Model atmospheres

The algorithm used in GALA is basically independent from the code adopted to derive the abundances and from the model atmospheres. GALA is designed to manage both the two most used and publicly available models atmospheres, namely ATLAS9 and MARCS:

- **ATLAS9** The suite of Kurucz codes represents the only suite of open-source and free programs to face the different aspects of the chemical analysis (model atmospheres, abundance calculations, spectral synthesis), allowing any user to compute new models and upgrade parts of the codes. GALA includes a dynamic call to the ATLAS9 code [5]. Any time GALA needs to investigate a given set of atmospheric parameters, ATLAS9 is called, a new model atmosphere is computed and finally it is stored in a directory. The latter is checked by GALA whenever a model atmosphere is requested, and ATLAS9 called only if the model is lacking. The convergence of the new model atmosphere is checked for each atmospheric layer. Following the prescriptions by Castelli (1988), we require errors less than 1% and 10% for the flux and the flux derivative, respectively. Additional information about the calculation for each model atmosphere is saved.

---

[5]The ATLAS9 source code is available at the website http://wwwuser.oat.ts.astro.it/castelli/grids.html

- **MARCS** At variance with ATLAS9, for the MARCS models the code to compute new model atmospheres is not released to the community. However, the Uppsala group provides a large grid of the MARCS models on their website[6]. When GALA works with these grids (including both plane-parallel and spherical symmetry), new models are computed by interpolating into the Uppasala grid by using the code developed by T. Masseron (Masseron, 2006) [7]. This code has been modified in order to put the interpolated MARCS models in ATLAS9 format to use with WIDTH9.

Note that the automatization of the chemical analysis based on EWs needs a wide grid of model atmospheres (both to interpolate and compute new models) linked to the code in order to freely explore the parameter space. Thus, GALA is linked to the ATLAS9 grid both with solar-scaled and $\alpha$-enhanced chemical composition and to the MARCS grid with standard composition. Also, the use of other models or model grids can be easily implemented in the code. Sometimes, peculiar analysis or tests need to use specific models, for instance for the Sun (see the set of solar model atmospheres available in the website of F. Castelli) or with arbitrary chemical compositions, as those computed with the ATLAS12 code (Castelli, 2005a). When a specific, single model is called, all the optimization options are automatically swichted off.

### A.5.3 Exit options

GALA is equipped with a number of exit flags in order to avoid infinite loops or unforeseen cases stopping the analysis of the entire input list of stars. We summarize here the main exit options:

(1) the user can set among the input parameters the maximum number of allowed iterations for each star. When the code reaches this value it stops the analysis, moving to the next star. Generally this parameter depends on the adopted grid steps and if the input atmospheric parameters are close or not to the real parameters. When the *Guess Working-Block* is used, typically the analysis block converges in 3-5 iterations;

(2) if the dispersion around the mean of the abundances of the lines used for the optimization (after the line rejection) exceeds a threshold value GALA skips the star. In fact, very large dispersions can possibly suggests some problems in the EW measurements;

(3) if the number of the lines used for the optimization (after the line rejection) is smaller than a threshold, the optimization is not performed and the atmospheric parameters are fixed to the input guess values;

---

[6]http://marcs.astro.uu.se/
[7]The original code is available at http://marcs.astro.uu.se/software.php.

(4) the procedure is stopped if the requested atmospheric parameter is outside the adopted grid of model atmospheres;

(5) GALA skips the analysis of the star if the call to the model atmosphere fails (problems in the ATLAS9 models computation or in the MARCS models interpolation).

## A.6 Uncertainties

The code is equipped with different recipes to compute the uncertainties on each derived abundance. Several sources of error can affect the determination of chemical abundances, mainly the uncertainties due to the EW measurements and to the adopted log$gf$ (that are random errors from line to line) and those arising from the choice of the atmospheric parameters (that are random errors from star to star but systematic from line to line in a given star). These uncertainties are quantified by GALA while other sources of errors (as the choice of the abundance calculation code or the adoption of the grid of model atmospheres) are neglected because considered *external* errors.

### A.6.1 Abundances statistical errors

The statistical uncertainty on the abundance of each element is computed by considering only the surviving lines after the rejection process (see Sect. A.4). When the uncertainty on the EW is provided for each individual line (thus allowing to compute the abundance error for each transition), the mean abundance is computed by weighing the abundance of each line on its error, otherwise simple average and dispersion are computed. For those elements for which only one line is available, the error in abundance is obtained by varying the EW of $1\sigma_{EW}$ (if the EW uncertainties are provided). Otherwise, the adopted value is zero. As customary, the final statistical error on the abundance ratios is defined as $\sigma/\sqrt{N_{lines}}$.

### A.6.2 Uncertainties on the atmospheric parameters

GALA estimates the internal error for each stellar parameter that has been derived from the spectroscopic analysis. The uncertainties of $T_{\mathrm{eff}}$, $v_{\mathrm{t}}$ and log g are estimated propagating the errors of the corresponding optimization parameter:

$$\sigma_{\mathrm{X_i}} = \frac{\sigma_{C(\mathrm{X_i})}}{\left(\frac{\delta C(X_i)}{\delta X_i}\right)}$$

where $X_i$ are $T_{eff}$, log g and $v_t$, $C(X_i)$ indicates the optimization parameters defined in Section A.3.1, and $\sigma_{C(X_i)}$ are the corresponding uncertainties.

The terms $\frac{\delta C(X_i)}{\delta X_i}$ (which parametrize how the slopes and the iron difference vary with the appropriate parameters) are calculated numerically, by varying $X_i$ locally around the final best value, assuming the step used in the optimization process and recomputing the corresponding $C(X_i)$.

The terms $C(X_i)$ are computed by applying a Jackknife bootstrapping technique (see Lupton, 1993). The quoted quantities are recomputed by leaving out from the sample each time one different spectral line (thus, given a sample of N lines, each C(X) is computed N times by considering a sub-sample of N-1 lines). The uncertainty on the parameter X is $\sigma_{Jack} = \sqrt{N-1}\sigma_{sub}$, where $\sigma_{sub}$ is the standard deviation of the C(X) distribution derived from N sub-samples. The $\sigma_{Jack}$ takes into account the uncertainty arising from the sample size and the line distribution, and this resampling method is especially useful to estimate the bias arising from the lines statistics. Note that the computation of the slopes is performed taking into account the effect of the EW and abundance uncertainties of each individual line. Thus, the uncertainty in the atmospheric parameter $X_i$ becomes

$$\sigma_{X_i} = \frac{\sigma_{C(X_i)}^{Jack}}{\left(\frac{\delta C(X_i)}{\delta X_i}\right)}$$

It is worth to notice that these uncertainties represent the internal error in the derived parameters and are strongly dependent on the number of used lines and on the distribution of the lines (weak and strong transitions for the estimate of $v_t$ and low and high $\chi$ lines for $T_{eff}$). Other factors that can affect the determination of the parameters (for instance, the threshold adopted in the EWs and in $\sigma_{EW}$) are not included in the error budgets and they can be considered as external errors. Finally, the error due to the adopted grid size could be considered as a systematic uncertainty (being the same for all the analysed stars) and eventually added in quadrature to the internal error estimated by GALA.

### A.6.3  Abundances uncertainties due to the atmospheric parameters

The evaluation of the uncertainties arising from the atmospheric parameters is a more complex task. Generally these errors are referred to as "systematic" uncertainties but this nomenclature is rather imprecise. In fact, the variation of a given parameter changes the abundance derived from the lines of the same element in a similar way (for instance, an increase of $T_{eff}$ increases the abundance of all the iron lines). However this error will be different from star to star, due to the different number of lines, strength and $\chi$ distributions, EWs quality and so on. Thus, the uncertainties from the atmospheric parameters should be considered as random errors, when we compare different stars (but they are systematic uncertainties from line to line).

Several recipes are proposed in the literature. The most common method is to re-compute the abundances changing each time one parameter only, and keeping fixed the other ones to their best estimates. Then, the corresponding variations in the abundances are added in quadrature. This approach is the most conservative, because it neglects the covariance terms arising from the interplay among the parameters (see Sect. A.2.2), providing only an upper limit for the total error budget.

GALA follows the approach described by Cayrel et al. (2004) to naturally take into account the covariance terms. When the optimization process is ended, the analysis is repeated by altering the final $T_{\text{eff}}$ by $+\sigma_{T_{\text{eff}}}$ and $-\sigma_{T_{\text{eff}}}$ (these uncertainties are calculated as described in Section 6.2), and re-optimizing the other parameters. The net variation of each chemical abundance with respect to the original value is assumed as final uncertainty due to the atmospheric parameters and including naturally the covariance terms. Additionally, under request, also the abundance variations following the classical approach to vary one only parameter each time (keeping the other parameters fixed) are calculated, leaving the user free to use this information as preferred.

### A.6.4 Quality parameter for the final solution

GALA provides also a check parameter, useful to judge the quality of the global solution and to identify rapidly stars with unsatisfactory solutions. For each model used during the optimization process a merit function $F_{merit}$ is defined as:

$$F_{merit} = \sqrt{\left(\frac{S_\chi}{\sigma_{Jack}^{S_\chi}}\right)^2 + \left(\frac{S_{EWR}}{\sigma_{Jack}^{EWR}}\right)^2 + \left(\frac{\Delta Fe}{\sigma_{Jack}^{\Delta Fe}}\right)^2},$$

taking into account the values of the optimization parameters and the corresponding uncertainties. In an ideal case, $F_{merit}$ is zero if the three optimization parameters are exactly zero. Generally, all the solutions with $F_{merit} \simeq 1$ are valid and equally acceptable, while values of $F_{merit} >> 1$ are suspect and point out that at least one of the parameters is not well constrained within the quoted uncertainty. Note that $F_{merit}$ provides only an indication if the solution is acceptable or not, but it does not specify which parameter is not well defined.

Summaryzing, when the full optimization process is completed, GALA will provide for each analysed element the (weighted) mean abundance, the dispersion and the number of used lines (which provide the statistical uncertainty), the net variation in abundance due to the new optimization with $T_{eff} + \sigma_{T_{eff}}$ and that with $T_{eff} - \sigma_{T_{eff}}$ (which provide the uncertainty owing to the choice of stellar parameters). Also, for each atmospheric parameters the quoted internal uncertainties are computed. Finally, the quality parameter $F_{merit}$ is provided to evaluate the goodness of the solution as whole.

## A.7 Dependence on SNR

We performed a number of experiments to test the stability and reliability of the derived atmospheric parameters with GALA at different noise conditions. We performed two kind of experiments, described in the following: the first based on a grid of synthetic spectra of abundances and atmospheric parameters known a priori, in order to estimate the reliability of the code as a function of the parameters and the signal-to-noise; the second group of tests is based on real spectra already analysed in literature. In the following, the EWs were measured by means of DAOSPEC (Stetson & Pancino, 2008) adopting a Gaussian profile for the line fitting.

### A.7.1 Synthetic spectra at different noise conditions

We analysed with GALA a grid of synthetic spectra, computed to mimic the UVES@VLT high resolution spectra with the 580 Red Arm setup. The grid of synthetic spectra includes SNR of 20, 30, 50, 100 per pixel for two different sets of atmospheric parameters: $T_{eff}$= 4500 K, logg=1.5, $v_t$= 2 km/s, [M/H]= −1.0 dex to simulate a giant star, and $T_{eff}$= 6000 K, logg=4.5, $v_t$= 1 km/s, [M/H]= −1.0 dex to simulate a dwarf star. The spectra were computed with the following procedure:

(1) for a given model atmosphere, two synthetic spectra were calculated with the SYNTHE code over the wavelength range covered by the two CCDs of the 580 UVES Red Arm grating and then convoluted with a Gaussian profile in order to mimic the formal UVES instrumental broadening;

(2) the spectra were rebinned to a constant pixel-size ($\delta\lambda$= 0.0147 and 0.0174 pixel/$\mathring{A}$ for the lower and upper chips respectively);

(3) the synthetic spectra (normalized to unity) were multiplied with the efficiency curve computed by the FLAMES-UVES ESO Time Calculator in order to model the shape of the templates as realistically as possible;

(4) Poissonian noise was injected in the spectra to simulate different noise conditions. Basically, the SNR varies along the spectrum, as a function of the efficiency (and thus of the wavelength). The noise was added in any spectrum according to the curve of SNR as a function of $\lambda$ provided by the FLAMES-UVES ESO Time Calculator. For each SNR a sample of 200 synthetic spectra were generated.

Fig. A.8 summarizes the average values obtained for each MonteCarlo sample for each atmospheric parameter as a function of SNR; the errorbars indicate the dispersion around the mean. Results of the simulations of the giant star model atmosphere are shown in the upper panels of each window, while the lower panels summaryze the results for the dwarf star simulations. Basically, the original parameters of the synthetic spectra (marked in Fig. A.8 as dashed horizontal lines) are recovered with small dispersions and without any significant bias. The major departure from the original values is observed in the microturbulent velocities (both dwarf and giant) at SNR= 20, because of the loss of weak lines.

### A.7.2 Real echelle spectra at different noise conditions

The previous test provides an indication about the stability of the method against the SNR, starting from spectra whose parameters are known *a priori*. However, the employed noise model is a simplification because it does not take into account some effects that can also heavily affect the measurement of the EWs, as the correlation of the noise among adjacent pixels, flat-fielding residuals, failures in the echelle orders merging, presence of spectral impurities. Also, the atomic data of the analysed lines are the same used in the computation of the synthetic spectra, thus excluding from the final line-to-line dispersion the random error due to the uncertainty on the atomic data.

In order to provide an additional test about the performance of GALA in conditions of different noise, we performed a simple experiment on the spectra acquired with UVES@FLAMES of the giants star NGC 1786-1501 in the Large Magellanic Cloud globular cluster NGC 1786 (see Mucciarelli et al., 2009, 2010). This is a sample of 8 spectra with the same exposure time ($\sim$45 min) obtained under the same seeing conditions was secured, with a typical SNR per pixel of 20 for each exposure. We used rhis dataset in order to obtain 8 spectra with different SNR ranging from $\sim$20 to $\sim$60 depending on the number of exposures averaged: the spectrum with the lowest SNR is just one exposure, the spectrum with the largest SNR the average of all the 8 acquired exposures. The derived parameters for each spectrum are showed in Fig. A.9 as a function of SNR. The errorbars are derived by applying the Jackknife bootstrap technique for $T_{\rm eff}$, log g and $v_{\rm t}$, while for [Fe/H] I used the dispersion by the mean as estimate of the error.

We note that, also for spectra with low SNR, the parameters are well constrained, with small uncertainties: this is not a numerical artifact of the code but it is due to the large number of transitions available in the UVES spectra, coupled with an accurate rejection of the outliers and the use of the uncertainties for each individual line in the slopes computations. Also, we note that major departures from the final parameters are again in the determination of $v_t$ at SNR= 20: the derived low value of $v_t$ is due to the fact that at low SNR several weak lines (useful to constrain the microturbulent velocity) are not well measured (then discarded by GALA) or not identified in the noise envelope by DAOSPEC. Note that the derived trend of $v_t$ as a function of SNR shows the same behaviour found and discussed by Mucciarelli (2011b).

## A.8   An efficient approach: Arcturus, Sun, HD 84937 and $\mu$ Leonis

In this section we describe a convenient and robust method for performing abundance analysis with GALA, applied to the case of 4 stars (namely, the Sun, Arcturus, HD 84937 and $\mu$ Leonis) of different metallicity and evolutionary stage and whose parameters are well established among the closest F-G-K stars. We retrieved high-resolution ($\sim$45000) spectra from the ESO[8] (for Sun, Arcturus and HD 84937) and ELODIE[9] (for $\mu$ Leonis) archives.

### A.8.1   Selection of the lines

For each star we defined a suitable linelist of Fe I and Fe II transitions, starting from the most updated version of the Kurucz/Castelli lines dataset [10]. We apply an iterative procedure to define the linelist. Assuming that the parameters of the targets are not known a priori, we performed a first analysis by using a preliminary linelist including only laboratory transitions with $\chi$ <6 eV and log (gf)>−5 dex. Such a linelist is not checked against the spectral blendings arising from the adopted spectral resolution and the atmospheric parameters and it is used only to perform a preliminary analysis. With the derived new parameters we define a new linelist for each star. The lines are selected by the inspection of synthetic spectra computed with the new parameters and convoluted with a Gaussian profile in order to reproduce the observed spectral resolution. At this step, only iron transitions predicted to be unblended are taken into account and used for the new analysis.

---

[8]http://archive.eso.org/eso/eso_archive_main.html
[9]http://atlas.obs-hp.fr/elodie/
[10]http://wwwuser.oat.ts.astro.it/castelli/linelists.html

### A.8.2 EWs measurements

EWs are measured by using the code DAOSPEC which adopts a saturated Gaussian function to fit the line profile and an unique value for the full width half maximum (FWHM) for all the lines. We start from the FWHM derived from the nominal spectral resolution of the spectra, leaving DAOSPEC free to re-adjust the value of FWHM according to the global residual of the fitting procedure. The measurement of EWs is repeated by using the optmized FWHM value as a new input value, until convergence is reached at a level of 0.1 pixel. The formal error of the fit provided by DAOSPEC is used as $1\sigma$ uncertainty on the EW measurement.

### A.8.3 Analysis with GALA

The programme stars are analysed by employing all the three Working-Blocks, requiring a spectroscopic optimization of $T_{\mathrm{eff}}$, log g, $v_{\mathrm{t}}$ and [M/H] and starting in all cases from the same set of guess parameters (namely $T_{\mathrm{eff}}$= 5000, logg= 2.5, [M/H]= −1.0 dex and $v_{\mathrm{t}}$= 1.5 km/s). The optimization is performed by exploring the parameters space in small steps of $\delta T_{\mathrm{eff}}$= 50 K, $\delta$log g= 0.1 and $\delta v_{\mathrm{t}}$= 0.1 km/s; the metallicity is investigated by adopting the step of the ATLAS9 grids ($\delta$[M/H]= 0.5 dex).

In the first run, we analysed the programme stars assuming the same configuration for the input parameters of GALA, in particular we included only lines with $\sigma_{EW}$ <10% and with EWR>−5.8 (corresponding to ∼10 mÅ at 6000 Å). After a first run of GALA, we refined the maximum allowed EWR, that depends mainly by the onset of the saturation along the curve of growth (and thus it is different for stars with different atmospheric parameters). We adopted as maximum allowed value, EWR= −4.65 for Arcturus and μLeonis, and −4.95 for the Sun and HD 84937. These values are chosen on the basis of the visual inspection of the curve of growth, in order to exclude too strong lines, for which the Gaussian approximation can fail. After a first run of GALA, the linelist is refined by using the new parameters obtained by GALA as described above and the procedure repeated. Table 1 summaryzes the derived atmospheric parameters (with the corresponding Jackknife uncertainties) and the [Fe/H] I and [Fe/H] II abundance ratios, together with the two errorbars due to the atmospheric parameters and the internal error computed as $\sigma/\sqrt{(N_{lines})}$.

We compare our results with those available in literature. The results obtained for the Sun agree very well with those listed by Stix (2004) and derived from mass, radius and luminosity.

Ramirez & Allende Prieto (2011) provide an accurate determination of the atmospheric parameters and the chemical composition of Arcturus; in particular $T_{\rm eff}$ and log g are derived in an independent way with respect to our approach, finding $T_{\rm eff}$= 4286±30 K (by fitting the observed spectral energy distribution), log g=1.66±0.05 (through the trigonometric parallax), while $v_{\rm t}$ turns out to be 1.74 km/s (by using the same approach used in GALA). The final iron abundance is [Fe/H]= –0.52±0.04 dex. In both cases, the agreement with the literature values is good. Finally, Fig. A.10 shows as example the graphical output of GALA for Arcturus.

For HD 84937 and $\mu$Leonis several determinations are available in literature and we decide to use as reference the average of the values listed in the classical compilation by Cayrel de Strobel, Soubiran & Ralite (2001), providing $T_{\rm eff}$= 6251±94 K, log g= 3.97±0.18 and [Fe/H]= –2.14±0.17 dex for HD 84937 and $T_{\rm eff}$= 4504±121 K, log g= 2.33±0.27 and [Fe/H]= +0.28±0.13 dex for $\mu$Leonis.

### A.8.4   Stability against the initial parameters

A relevant feature of an automatic procedure to infer parameters and abundances is its stability against the input atmospheric parameters. In order to assess the effect of different first guess parameters, we analyse the spectrum of Arcturus by investigating a regular grid of input parameters with $T_{\rm eff}$ ranging from 3800 to 4800 K (in steps of 100 K) and log g from 1.0 to 2.2 (in steps of 0.1). Fig. A.11 shows the grid of the input parameters in the $T_{\rm eff}$–log g plane (empty points) with the position of the derived parameters (black points); the upper panel summarizes the results when GALA is used without the *Refinement Working-Block*, while the lower panel shows the results obtained by employing also the refinement option. The recovered parameters cover a small range: in the first run the dispersion of the mean is of 46 K for $T_{\rm eff}$ and 0.09 for log g, while these values drop to 25 K and 0.05 respectively, when the *Refinement Working-Block* is enabled.

## A.9   A test on globular clusters

Globular clusters are ideal templates where to check the capability of our procedure to derive reliable atmospherical parameters, because of the homogeneity (in terms of metallicity, age and distance) of their stellar content. Thus, the derived parameters for stars in a given globular cluster can be easy compared with theoretical isochrones in the $T_{\rm eff}$–log g plane.

We apply the same procedure described in Sect. A.8 to analyse a set of high-resolution spectra for stars in the globular cluster NGC 6752, ranging from the Turn-Off up to the bright portion of the Red Giant Branch. The spectra have been retrieved by the ESO archive[11] and reduced with the standard ESO pipeline[12]. They are from different observing programmes and with different SNR, including giant stars crossing the Red Giant Branch Bump region observed with UVES@VLT (slit mode) within the ESO Large Program 65.L-0165 with very high (>200) SNR, the stars in the bright portion of the Red Giant Branch observed with UVES-FLAMES@VLT (fiber mode) within the Galactic globular clusters survey presented by Carretta et al. (2009b) and the dwarf/subgiant stars observed with UVES@VLT (slit mode) within the ESO Large Program 165.L-0263.

Fig. A.12 shows the position of the final parameters derived with GALA in the $T_{\text{eff}}$−log g plane. Also, two theoretical isochrones with an age of 12 Gyr and a metallicity of Z= 0.0006 (assuming an $\alpha$-enhanced chemical mixture) are shown as reference (grey curve is from BaSTI database by Pietrinferni et al. (2006) and black curve from Padua dabase by Girardi et al., 2000)

The parameters derived with GALA well reproduce the behaviour predicted by the theoretical models for a old simple stellar population with the same metallicity of the cluster, confirming the physical reliability of the final solution. Also, we note that the errorbars, both in $T_{\text{eff}}$ and log g, change according to the quality of the spectra, ranging from ∼30 K and ∼0.15 for the giants with the highest SNR up to ∼200 K and ∼0.5 for the dwarf stars with the lower SNR.

## A.10   Summary

In this paper we have presented a new, automatic tool to perform accurate analysis of stellar spectra. GALA is designed to perform automatically the search for the best atmospheric parameters for moderate and high resolution stellar absorption spectra, by using the EWs of metallic lines. The main advantages of the code are the capability:

1. to optimize all the parameters or only part of them. The code is versatile in order to perform different kind of analysis (full or partial spectroscopic analysis, experiments about the guess parameters...) and adopting different recipes to derive log g;

2. to perform a careful rejection of the outliers according to the line strength, the EW quality and the line distribution in the A(Fe)–$\chi$ and A(Fe)–EWR planes;

3. to estimate for each individual star the internal errors for *(a)* the optimized parameters by adopting the Jackknife bootstrapping technique, and *(b)* the derived uncertainties due to the

---

[11]http://archive.eso.org/cms/eso-data.html
[12]http://www.eso.org/sci//software/pipelines/

choice of atmospheric parameters, following both the prescriptions by Cayrel et al. (2004)

and the classical method of altering one parameter at a time.

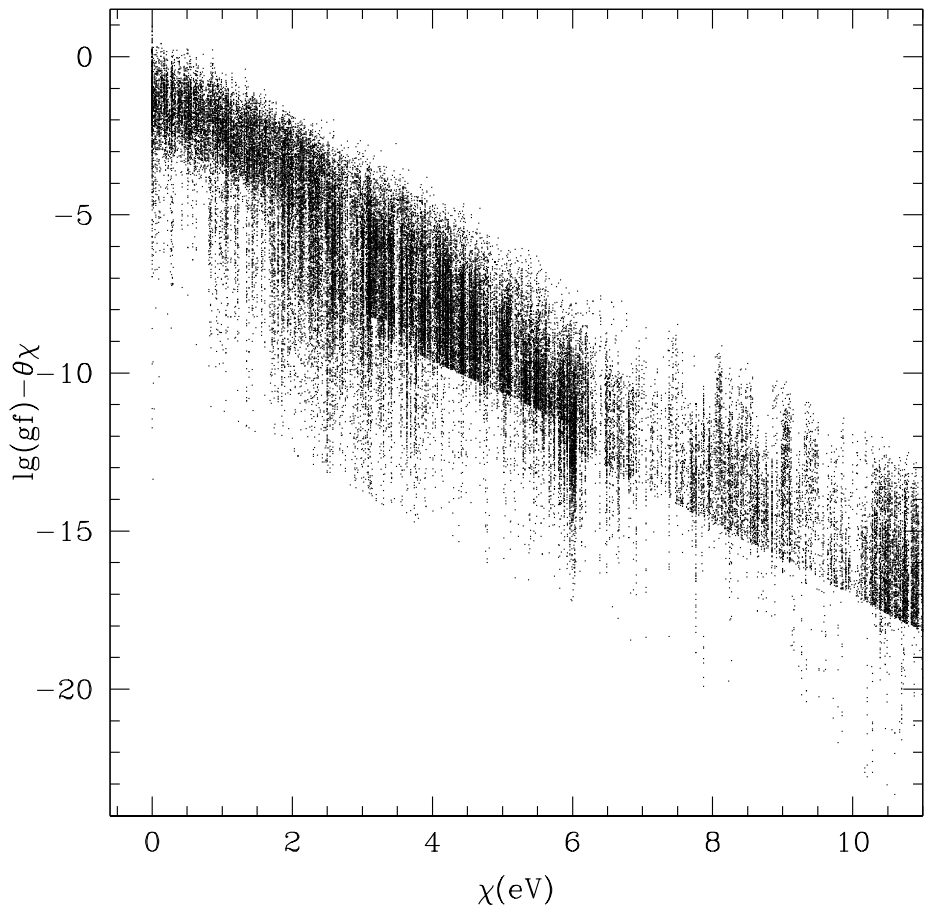

Figure A.1: Behavior of the line strength (computed assuming $T_{\rm eff}$=4500 K) as a function of the excitation potential $\chi$ for all the transitions available in the Kurucz/Castelli linelist in the range of wavelength $\lambda$=4000-8000 $\mathring{A}$ and with $\chi$ <10 eV.

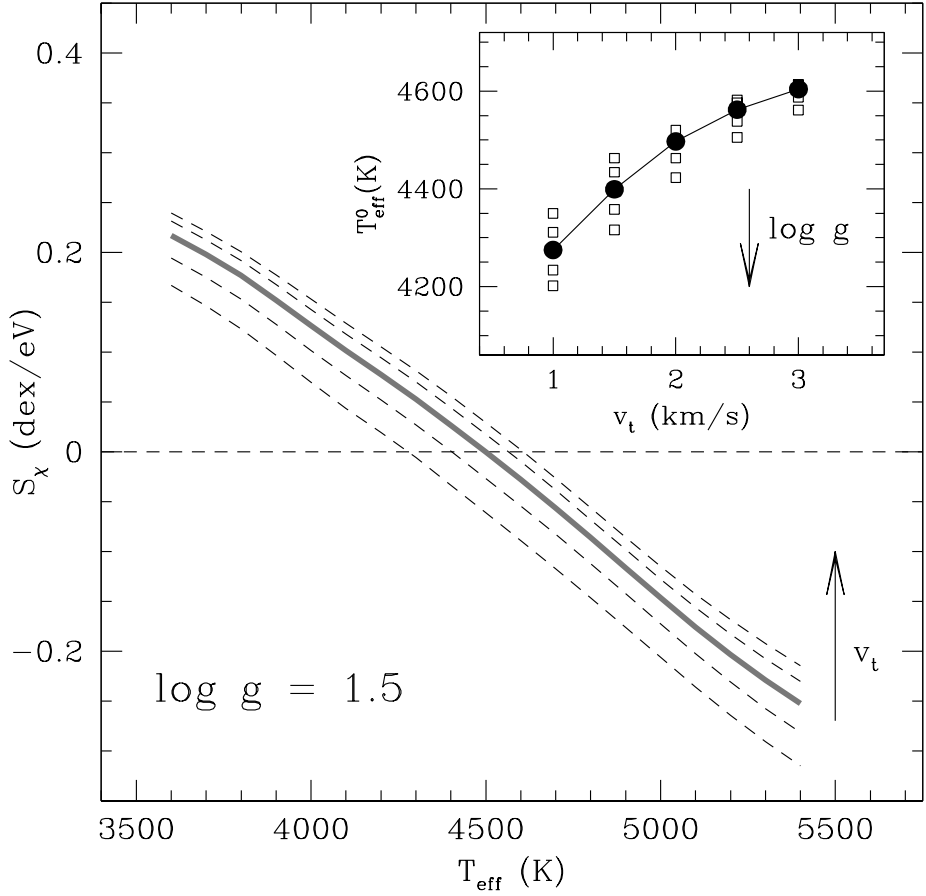

Figure A.2: Main panel: behaviour of the slope $S_\chi$ of the A(Fe)–$\chi$ relation as a function of $T_{\text{eff}}$ assuming log g= 1.5 and for different values of $v_t$ (dashed curves). The thick grey curve represents the behaviour computed for the original microturbulent velocity ($v_t$= 2.0 km/s). Horizontal dashed line is the zero value (according to the excitation equilibrium). The inset panel shows the behaviour of the best value of $T_{\text{eff}}$ (for which $S_\chi$= 0) as a function of the used $v_t$ (empty squares) and for different gravities; black points are referred to the original gravity (log g=1.5).

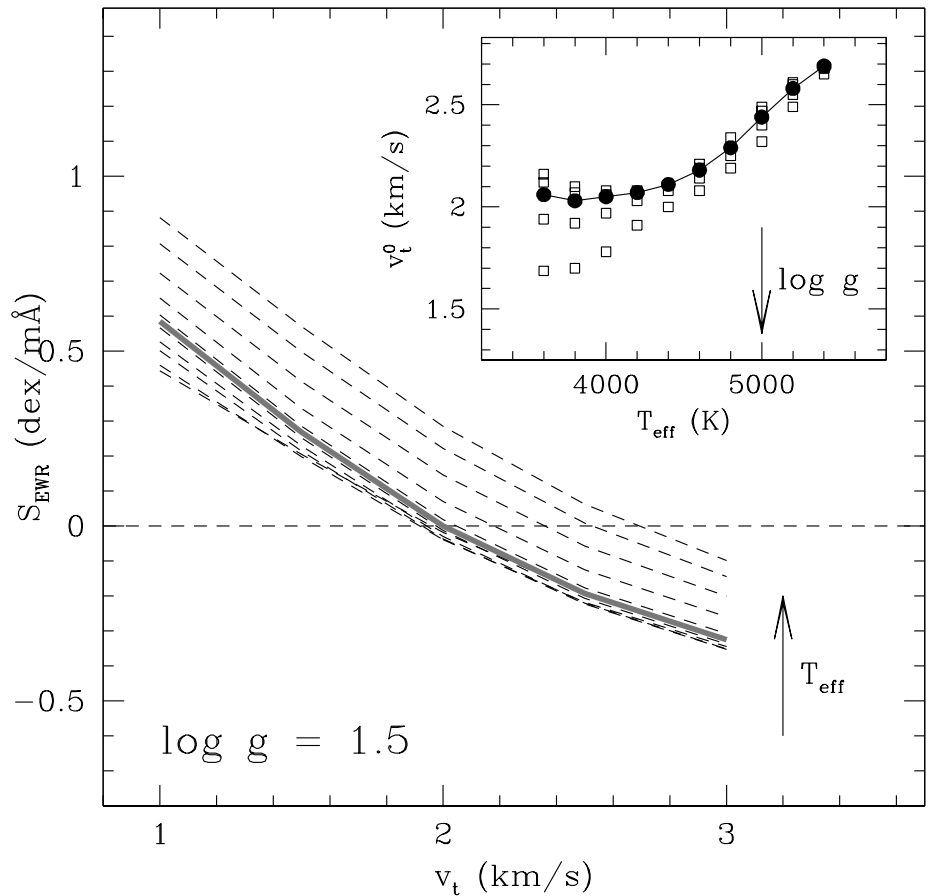

Figure A.3: Main panel: behaviour of the slope $S_{EWR}$ of the A(Fe)–EWR relation as a function of $v_t$ assuming log g= 1.5 and for different $T_{\rm eff}$ (dashed curves). The thick grey curve represents the behaviour computed for the original temperature ($T_{\rm eff}$= 4500 K). Horizontal dashed line is the zero value. The inset panel shows the behaviour of the best value of the microturbulent velocity $v_t$ (for which $S_{EWR}$= 0) as a function of the used $T_{\rm eff}$ (empty squares) and for different values of gravities; black points are reffered to the original gravity (log g=1.5).

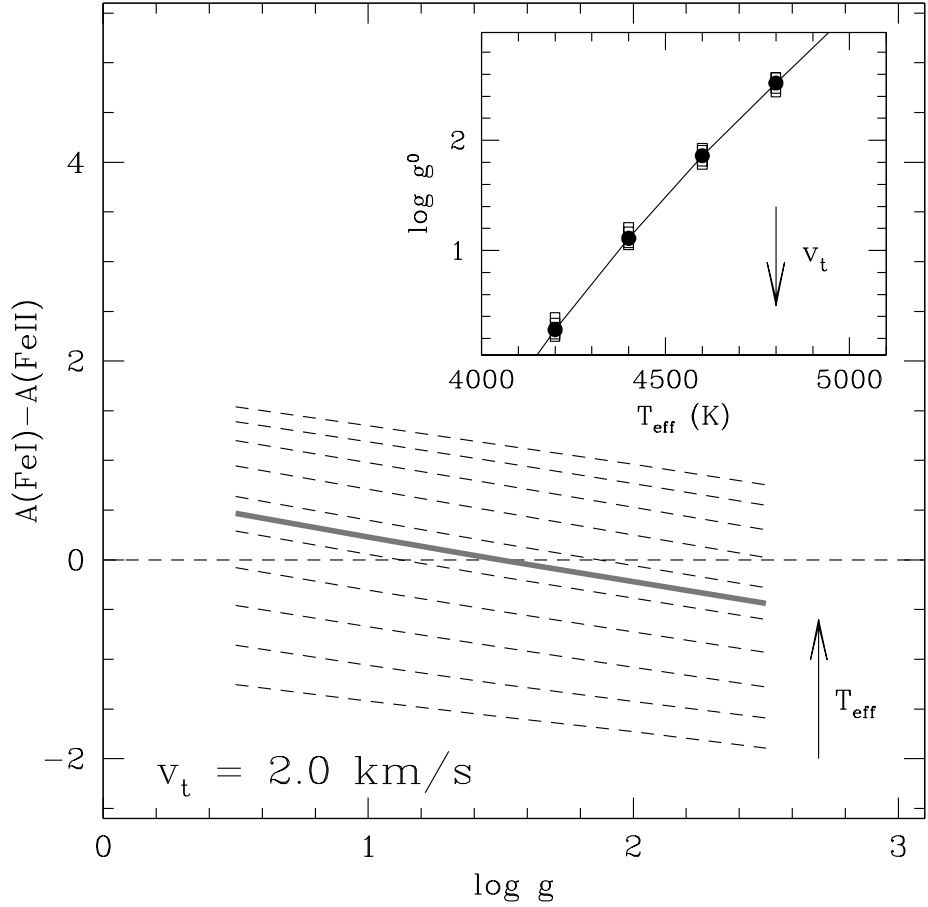

Figure A.4: Main panel: behaviour of the difference between A(Fe I) and A(Fe II) as a function of log g assuming $v_t$= 2.0 km/s and for different $T_{eff}$ (dashed curves). The thick grey curve represents the behaviour computed for the original temperature ($T_{eff}$= 4500 K). Horizontal dashed line is the zero value (according to the ionization equilibrium). The inset panel shows the behaviour of the best log g (for which A(Fe I)= (Fe II)) as a function of the used $T_{eff}$ (empty squares) and for different gravities; black points are referred to the original microturbulent velocity ($v_t$= 2.0 km/s).

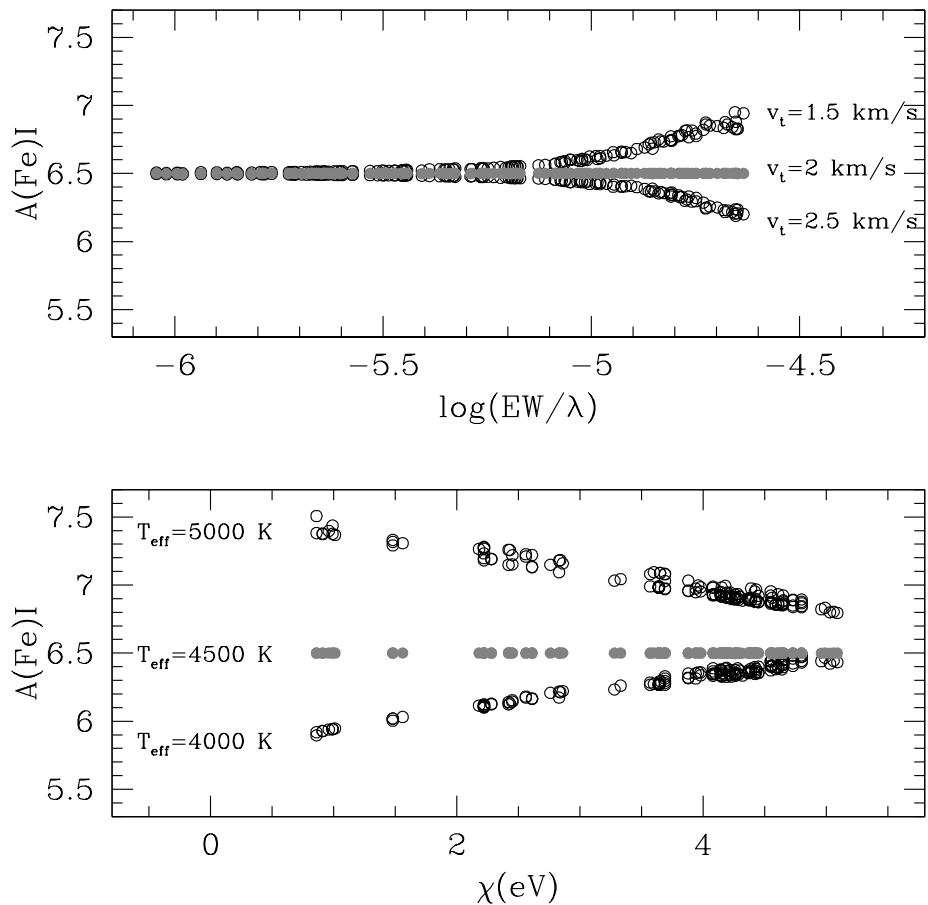

Figure A.5: Upper panel: behaviour of A(Fe I) as a function of the reduced equivalent widths for a set of theoretical EWs obtained from a model atmosphere computed with $T_{\rm eff}$= 4500 K, log g= 2.0 and $v_{\rm t}$= 2 km/s. The derived abundances are obtained by adopting the correct value of $v_{\rm t}$ (grey points) and two *wrong* values of $v_{\rm t}$ (open circles). Lower panel: behaviour of A(Fe I) as a function of the excitational potential for the same set of theoretical EWs. Grey points are the results obtained by analyzing the lines with the correct value of $T_{\rm eff}$, while the empty circles are obtained by over/under-estimate $T_{\rm eff}$ by $\pm$500 K.

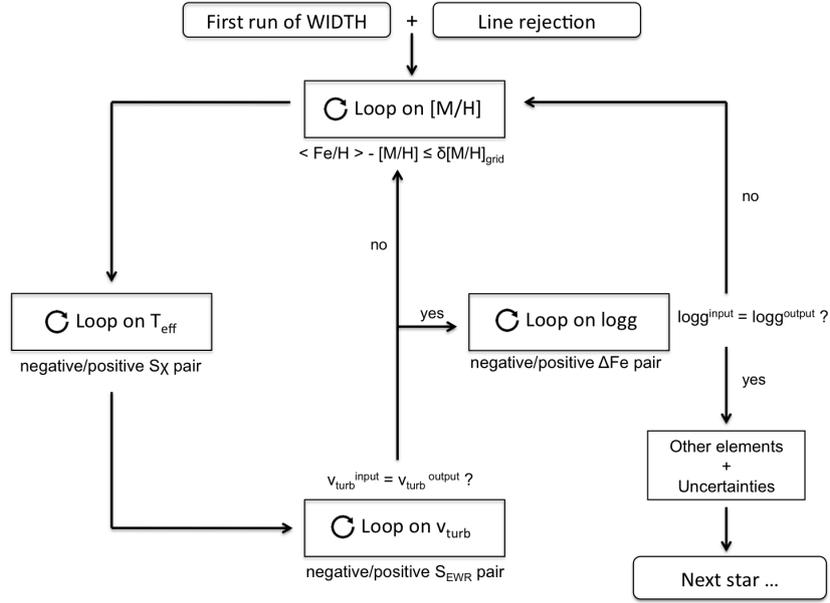

Figure A.6: Flow diagram for the *Analysis Working-Block* of GALA (see Section 3.2 for details).

Table A.1: Iron abundance (from neutral and singly ionized lines) and atmospheric parameters derived with GALA for Arcturus, $\mu$Leonis, the Sun and HD 84937. For the abundances, the first two errorbars are the uncertainties arising from the atmospheric parameters following the prescriptions by Cayrel et al. (2004), while the last one is the internal error calculated as $\sigma/\sqrt{(N_{lines})}$. The uncertainties in the derived atmospheric parameters are computed with a Jackknife bootstrap technique.

| Star | [Fe I/H] | [Fe II/H] | $T_{\text{eff}}$ | log g | $v_{\text{t}}$ |
|------|----------|-----------|------------------|-------|----------------|
|      | (dex)    | (dex)     | (K)              |       | (km/s)         |
| Arcturus | $-0.51^{+0.05}_{-0.04} \pm 0.007$ | $-0.52^{+0.03}_{-0.05} \pm 0.015$ | 4300±60 | 1.60±0.06 | 1.50±0.06 |
| $\mu$Leonis | $+0.37^{+0.04}_{-0.06} \pm 0.011$ | $+0.38^{+0.04}_{-0.08} \pm 0.033$ | 4500±81 | 2.40±0.26 | 1.40±0.07 |
| Sun | $-0.01^{+0.02}_{-0.03} \pm 0.009$ | $+0.01^{+0.03}_{-0.07} \pm 0.003$ | 5800±64 | 4.50±0.18 | 1.20±0.13 |
| HD 84937 | $-2.28^{+0.06}_{-0.04} \pm 0.007$ | $-2.27^{+0.05}_{-0.04} \pm 0.020$ | 6150±56 | 3.20±0.13 | 0.70±0.24 |

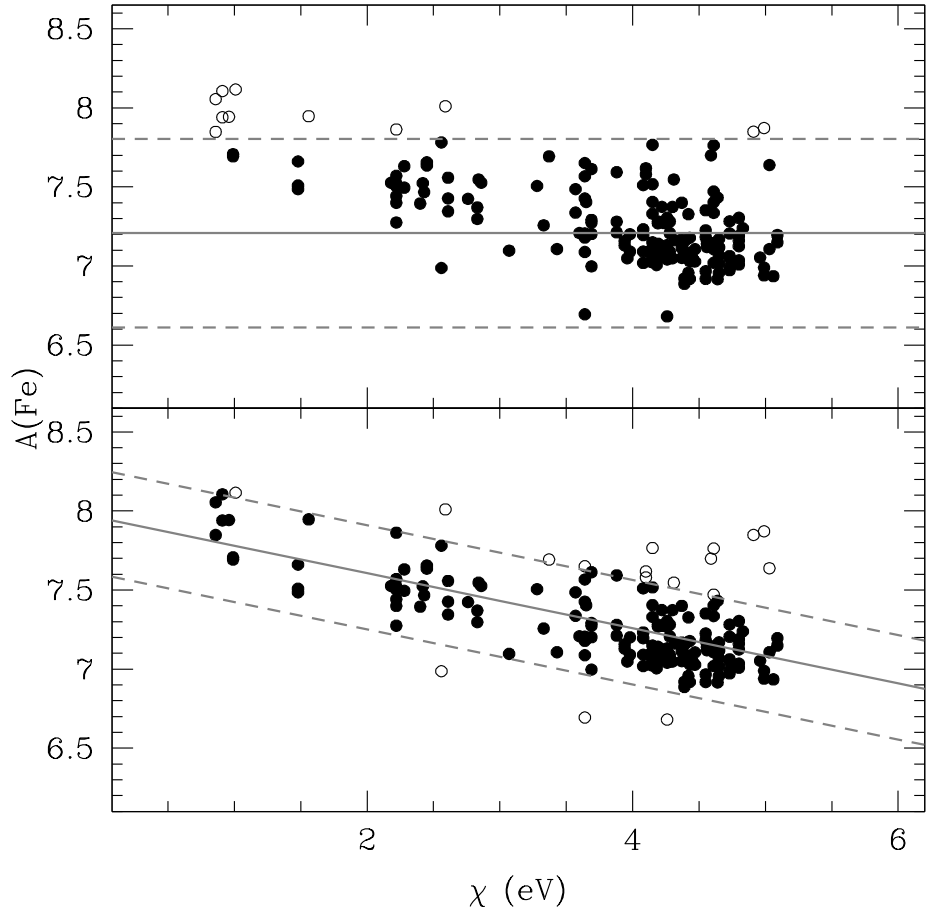

Figure A.7: Behaviour of the Fe I abundances as a function of the excitation potential for a synthetic spectrum computed assuming $T_{\mathrm{eff}}$= 4500 K but analysed with a model atmosphere with $T_{\mathrm{eff}}$= 5200 K. Black circles are the lines survived after the line-rejection procedure and the empty points the rejected lines. Upper panel shows the results by adopting a rejection based on the abundance distribution; the solid line indicates the median abundance and the dashed lines mark $\pm 3\sigma$ level. The lower panel shows the results by using the procedure employed by GALA: solid line is the best-fit line and the dashed lines mark $\pm 3\sigma$ from the best-fit line.

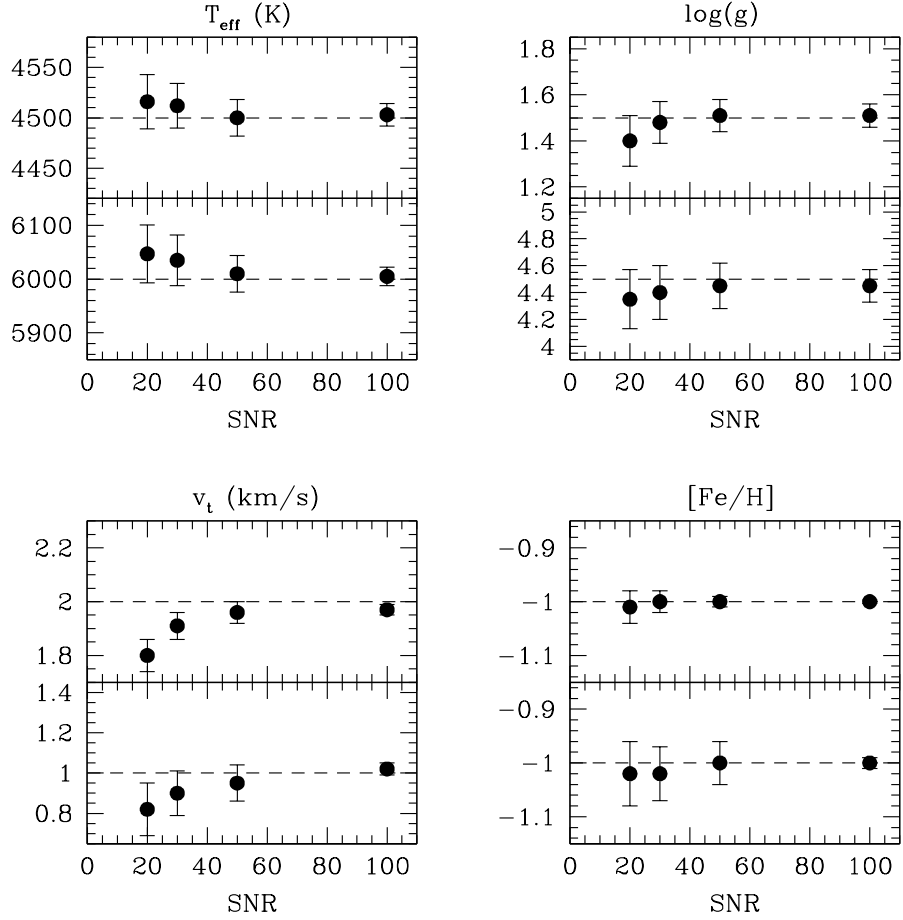

Figure A.8: Average values for the recovered atmospheric parameters as a function of SNR for the MonteCarlo samples described in Sect. A.7.1: upper panels of each window show the results for the giant star model, while lower panels show the results for the dwarf stars. Errorbars are the dispersion by the mean. Dashed lines are the original value for each parameter.

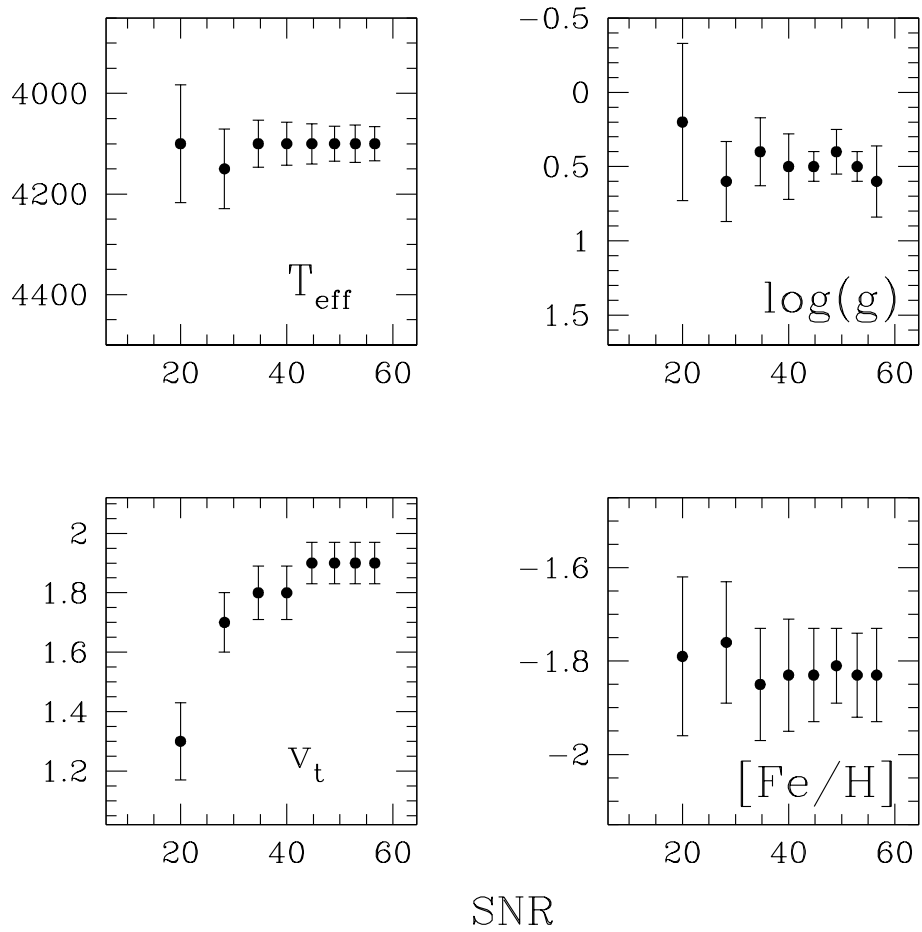

Figure A.9: Behaviour of the derived atmospheric parameters for the giant star NGC 1786-1501 as a function of SNR obtained by using different UVES co-added spectra. Errorbars are derived from the Jackknife bootstrap technique for $T_{\mathrm{eff}}$, log g and $v_{\mathrm{t}}$, and as dispersion by the mean for [Fe I/H].

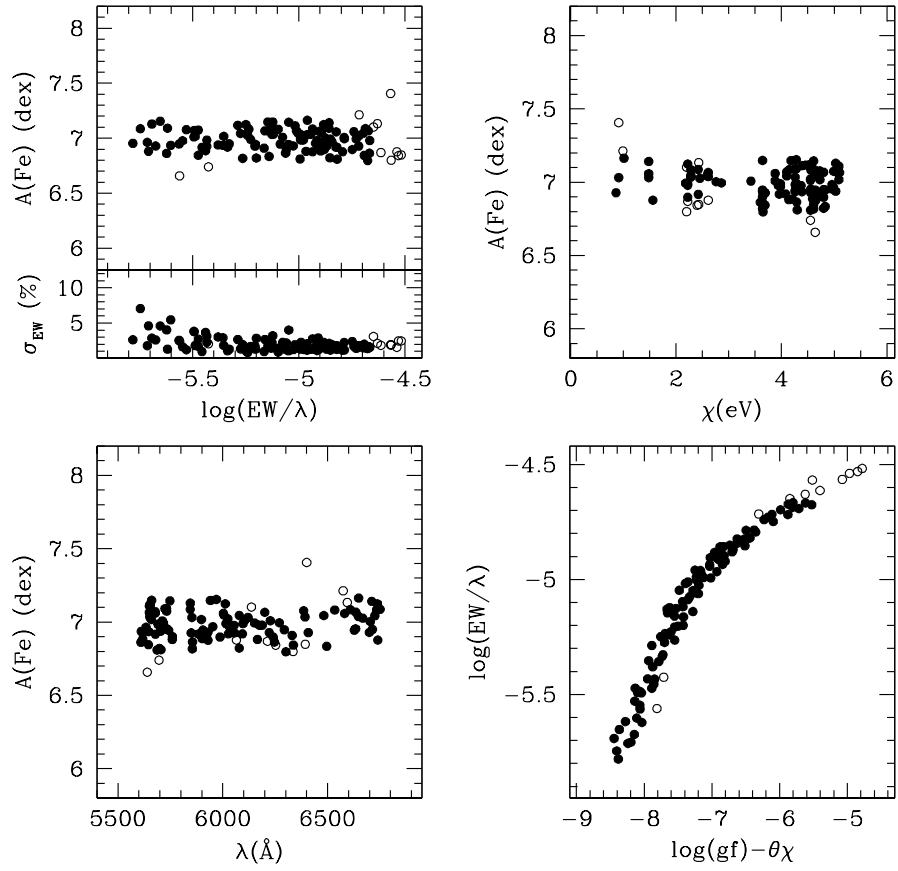

Figure A.10: Example of the graphical output of GALA for Arcturus (see Section A.8): black circles are the Fe I lines used in the analysis and the empty circles those rejected.

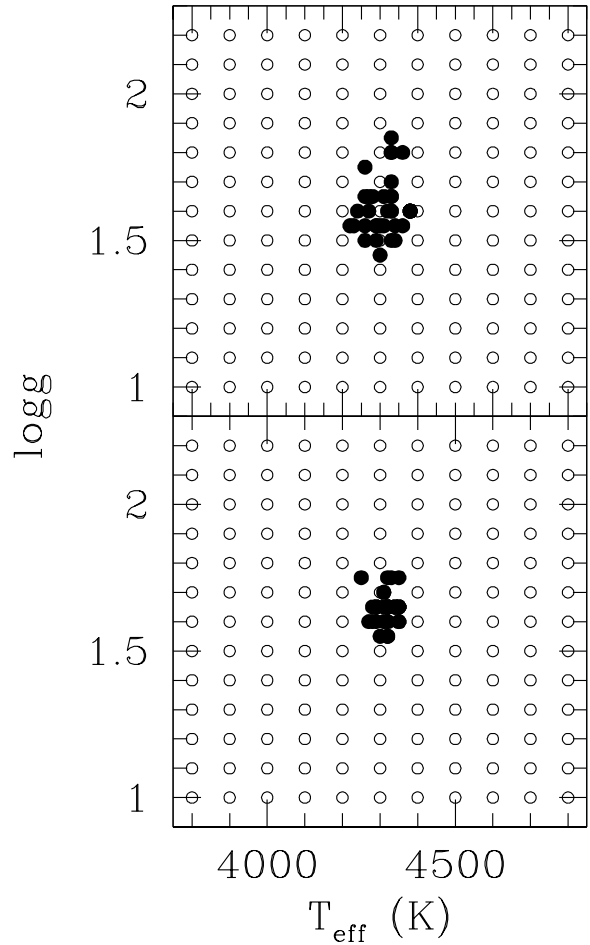

Figure A.11: Position of the final parameters for Arcturus (black points) in the $T_{eff}$–logg plane, in comparison with the input parameters (empty circles), obtained by using GALA without (upper panel) and with (lower panel) the *Refinement Working-Block* (upper and lower panel, respectively).

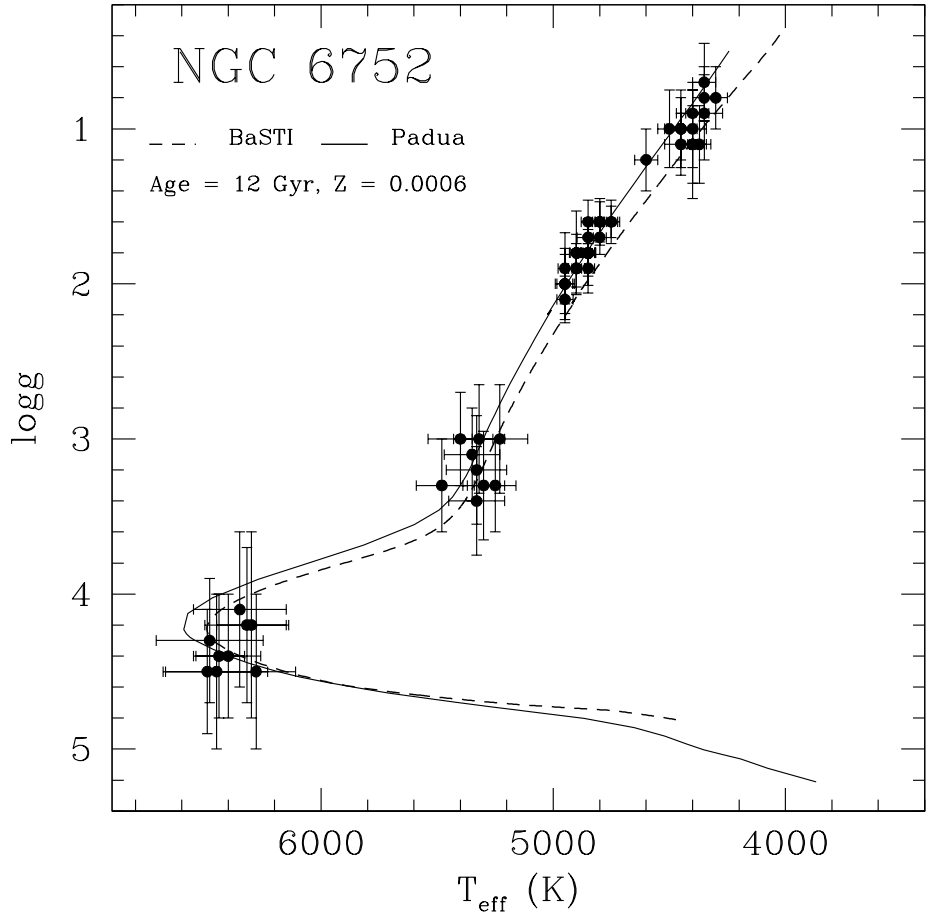

Figure A.12: Position in the $T_{\text{eff}}$–log g plane of the stars in the globular cluster NGC 6752 analysed with GALA . We plotted as references two isochrones computed with an age of 12 Gyr and a metallicity of Z= 0.0006 (assuming an $\alpha$-enhancement chemical mixture), from the BaSTI (Pietrinferni et al., 2006, dotted curve) and Padua (Girardi et al., 2000, solid curve) database.

# Appendix B

# List of publications

1. **Lovisi L.**, Mucciarelli, A., Ferraro F.R., Lucatello S., Lanzoni B., Dalessandro E., Beccari G., Rood R.T., Sills A., Fusi Pecci F., Gratton R., Piotto G., 2010 Apj 719, L121: Fast rotating Blue Stragglers in the Globular Cluster M4

2. Mucciarelli A., Salaris M., **Lovisi L.**, Ferraro F.R., Lanzoni B., Lucatello S., Gratton R., MNRAS 412, 81: Lithium abundance in the globular cluster M4: from the turn-off to the red giant branch bump

3. **Lovisi L.**, Mucciarelli A., Lanzoni B., Ferraro F.R., Gratton R., Dalessandro E., Contreras Ramos R., 2012 ApJ 754, 91: Chemical and kinematical properties of Blue Straggler stars and Horizontal Branch stars in NGC 6397

4. Mucciarelli A., Pancino E., **Lovisi L.**, Ferraro, F. R. 2013, ApJ submitted: GALA: an automatic tool for the abundance analysis of stellar spectra

5. **Lovisi L.**, Mucciarelli A., Lanzoni B., Ferraro F.R., Dalessandro E., 2013 arXiv:1301.3295: Blue straggler stars in Globular Clusters: chemical and kinematical properties

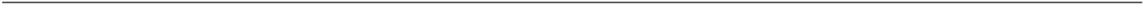